\documentclass[a4paper,11pt]{article}
\pdfoutput=1  
\usepackage{jcappub} 
\usepackage{verbatim}
\usepackage[T1]{fontenc}  
\usepackage{graphicx}  
\usepackage{array,multirow}
\usepackage[usenames]{xcolor} 
\usepackage{dcolumn}
\usepackage{bm}
\usepackage{hyperref}
\usepackage{pbox}
\usepackage{color} 
\usepackage[normalem]{ulem} 
\usepackage[section]{placeins} 

  \DeclareMathOperator{\mev}{MeV}    \DeclareMathOperator{\fm}{fm}          
 
\def\lag{\mathcal{L}}
\def\lagint{\lag_\text{int}}
\def\op{\mathcal{O}}

\def\chibar{\bar{\chi}}
\def\Nbar{\bar{N}}
\def\W{\widetilde{W}}

\def\qsq{\vec{q}^{\,2}}

\def\vqsq{\vec{q}^{\,2}}
\def\mag{\tilde{\mu}}

     \newcommand{\cL}{{\cal L}}

\newcommand{\pL}{\left(} \newcommand{\pR}{\right)} \newcommand{\bL}{\left[} \newcommand{\bR}{\right]} \newcommand{\cbL}{\left\{}   
\newcommand{\beq}{\begin{equation}} \newcommand{\eeq}{\end{equation}}
\newcommand{\bea}{\begin{eqnarray}} \newcommand{\eea}{\end{eqnarray}}
\newcommand{\alg}[1]{\begin{align} \begin{split} #1 \end{split}  \end{align}}

  \newcommand{\Eqm}[2]{Eqs.~(\ref{#1}) through (\ref{#2})}

\newcommand{\Tab}[1]{Tab.~\ref{#1}}

\newcommand{\eq}[1]{{Eq.~(\ref{#1})}}

\def\@preprint{\@empty}
\newcommand\preprint[1]{\gdef\@preprint{\hfill #1}}
\preprint{YITP-SB-15-16\\
\vspace{-0.65cm}
\begin{flushright}
\end{flushright}
}

\title{Identifying the Theory of Dark Matter with Direct Detection}
\author[1]{Vera Gluscevic,\note{Corresponding author.}}
\affiliation[1]{School of Natural Sciences, Institute for Advanced Study,\\Einstein Drive, Princeton NJ 08540, USA}
\author[2]{Moira I. Gresham,}
\affiliation[2]{Whitman College, Walla Walla, WA 99362}
\author[3]{Samuel D. McDermott,}
\affiliation[3]{C. N. Yang Institute for Theoretical Physics, Stony Brook, NY 11794}
\author[4, 5]{Annika H. G. Peter,}
\affiliation[4]{CCAPP and Department of Physics, The Ohio State University,\\191 W. Woodruff Ave., Columbus, OH 43210, USA}
\affiliation[5]{Department of Astronomy, The Ohio State University, \\140 W. 18th Ave., Columbus, OH 43210, USA}
\author[6]{Kathryn M. Zurek}
\affiliation[6]{Theoretical Physics Group, Lawrence Berkeley National Laboratory, Berkeley, CA 94720}
\affiliation{Berkeley Center for Theoretical Physics, University of California, Berkeley, CA 94720}

\emailAdd{verag@ias.edu}
\emailAdd{gresham@whitman.edu}
\emailAdd{samuel.mcdermott@stonybrook.edu}
\emailAdd{apeter@physics.osu.edu}
\emailAdd{kzurek@berkeley.edu}

\abstract{Identifying the true theory of dark matter depends crucially on accurately characterizing interactions of dark matter (DM) with other species.  In the context of DM direct detection, we present a study of the prospects for correctly identifying the low--energy effective DM--nucleus scattering operators connected to UV--complete models of DM--quark interactions.  We take a census of plausible UV--complete interaction models with different low--energy leading--order DM--nuclear responses.  For each model (corresponding to different spin--, momentum--, and velocity--dependent responses), we create a large number of realizations of recoil--energy spectra, and use Bayesian methods to investigate the probability that experiments will be able to select the correct scattering model within a broad set of competing scattering hypotheses.
We conclude that agnostic analysis of a strong signal (such as Generation--2 would see if cross sections are just below the current limits) seen on xenon and germanium
experiments is likely to correctly identify momentum dependence of the dominant response, ruling out models
with either ``heavy'' or ``light'' mediators, and enabling downselection of allowed models. However, a unique determination of
the correct UV completion will critically depend on
the availability of measurements from a wider variety of nuclear
targets, including iodine or fluorine. We investigate how model--selection prospects depend on the energy window available for the analysis. In addition, we discuss accuracy of the DM particle 
mass determination under a wide variety of scattering models, and investigate impact of the specific types of particle--physics uncertainties on prospects for model selection.}

\begin{document}
\@preprint
\maketitle
\flushbottom

\section{Introduction}
\label{sec:introduction}

Identifying the particle nature of dark matter (DM) is one of the most important open problems in modern physics. 
There is a world--wide effort to build increasingly large experiments to search for scattering of the weakly interacting massive particle (WIMP) off nucleons, with the next--generation experiments cutting into theoretically favored WIMP parameter space. Direct detection experiments of Generation 2 (G2) feature increased exposures and sensitivities as compared to current experiments, and their aim is to establish a first confirmed detection of DM particles in the coming decade \cite{Cushman:2013zza}. Once DM is detected, the immediate goal will be to infer details about DM properties, cross--compare those results between searches, and confront various theories of DM--nucleon scattering with data.

Direct detection data analysis has so far focused on the simplest scattering scenarios, in which the cross section is momentum-- (i.e.~energy--) and velocity--independent. In these scenarios, DM effectively couples to target nuclei through a coherent spin--independent (SI) interaction (so that the scattering rate scales as the size of the nucleus squared), or through a spin--dependent (SD) interaction (where the scattering rate scales as the square of the average spin of the unpaired nucleons in the target). However, a number of studies have pointed out that direct detection can access a richer phenomenology, manifesting as a nontrivial momentum or velocity dependence of the scattering cross section \cite{Bagnasco:1993st,Pospelov:2000bq,Sigurdson:2004zp,Masso:2009mu,Chang:2009yt,Feldstein:2009tr,Chang:2010yk,An:2010kc,Fitzpatrick:2010br,DelNobile:2013cta,Gresham:2014vja}, and potentially triggering new types of nuclear responses \cite{Fitzpatrick:2012ib,Gresham:2014vja,Catena:2014epa}.

Though the standard SI or SD interactions often naturally dominate scattering rates, it is possible for other kinds of low-energy interactions to contribute at comparable levels, or even to dominate if the standard responses are suppressed or forbidden. For example, if the DM has a magnetic dipole moment, couplings dependent on the spin and orbital angular momentum contribute the observed event rate. These new types of interactions give rise to different nuclear responses than those appearing in the standard SI or SD cases \cite{Fitzpatrick:2012ib,Gresham:2014vja}, and have been invoked in the literature to try to reconcile apparently inconsistent results reported by some experiments \cite{Masso:2009mu,Chang:2009yt,Feldstein:2009tr,Chang:2010yk,An:2010kc,Fitzpatrick:2010br,DelNobile:2013cta,Gresham:2013mua}. 

With the complexity of direct detection phenomenology in mind, an effective field theory (EFT) approach has recently been proposed to categorize the entirety of available theories accessible through direct detection measurements \cite{Fitzpatrick:2012ix,Anand:2013yka}. These studies give a complete basis of non-relativistic operators allowed by the symmetries of the scattering process that may describe the scattering of
DM through mediators of spin 1 or less.  They also provide the correct mapping of these operators onto the appropriate nuclear responses for a wide variety of targets, introducing some novel responses that have previously been neglected in the literature and data analyses.

This study addresses future extraction of the particle nature of DM from direct detection experiments in the context of the rich array of possible scattering phenomenologies.  Several studies have previously explored different aspects of the direct detection inverse problem: Ref.~\cite{McDermott:2011hx} focused on a subset of UV completions and limited to standard nuclear responses; Refs.~\cite{Peter:2013aha,Gluscevic:2014vga} investigated an incomplete set of EFT operators that provoke only the standard nuclear response; Ref.~\cite{Catena:2014epa} looked at a broad set of operators, turning on one operator at a time; Ref.~\cite{Schneck:2015eqa} looked at a complete EFT but isolated a single nuclear response at a time; and Ref.~\cite{Dent:2015ly} explored ``simplified models''. 
In this study, we focus only on elastic scattering of a fermion, and for the choice of UV completions to consider, follow Ref.~\cite{Gresham:2014vja}. We then investigate phenomenology from a wide variety of realistic models, including models that are are either {\it theoretically well motivated (have natural UV completions), plausible (have viable UV completions), or otherwise phenomenologically distinct}.    

Specifically, we investigate how different DM scattering theories can be distinguished with direct detection data in the presence of significant Poisson noise, as this is the regime in which the first signals would arise. The first study to analyze distinguishability of scattering phenomenologies in a statistical way was presented in Ref.~\cite{Gluscevic:2014vga}; we adopt the methods presented there and apply them to a wide range of scattering scenarios. Concretely, we choose a set of competing scattering ``hypotheses'', compute the expected signals for each hypothesis on a variety of targets, and analyze such simulations to establish whether the underlying model can be correctly identified in light of future data. 
We thereby address two main questions: (a) How likely is direct detection to correctly identify the underlying theory of DM--nucleon interactions? (b) What experimental strategies maximize chances for success of such a pursuit? 

\pagebreak
To answer these and related questions, three prerequisites are necessary: 
\begin{enumerate}
\item A broad class of available theoretical models (hypotheses) and corresponding phenomenologies for DM--nucleon scattering, discussed in \S\ref{sec:census}.
\item A statistical representation of possible experimental outcomes, {\em i.e.}~an ensemble of simulations that include Poisson noise, described in \S\ref{sec:simulations}.
\item An analysis framework for evaluating how well a given hypothesis ``fits'' a single data realization, as compared to other theories in consideration, detailed in \S\ref{sec:analysis}.
\end{enumerate}
After introducing each of these prerequisites, we present the results of our analysis. The key results of this study are in Figures \ref{fig:model_selection_gexe_50gev_select}, \ref{fig:class_selection_gexe_50gev_select}, and \ref{fig:model_selection_2_50gev_select}. They show the following: G2 xenon and germanium targets (with several ton-year, and several hundred kilogram-year exposures, respectively) are sufficient to correctly identify the momentum dependence of the scattering cross section for a wide range of DM masses, if the signal is close to the current upper limit, regardless of the underlying scattering model. Thus, if a strong signal is seen with G2 experiments, we will be able to successfully discriminate between widely different phenomenologies (for example, the standard SI coupling of DM will be distinguishable from a dipole coupling, etc.). However, xenon and germanium on their own cannot generally distinguish amongst models that give rise to similar momentum and velocity dependence, even if novel nuclear responses contribute to the observed scattering rate. For this purpose, new targets with different spin and mass structure, such as iodine and fluorine, will be necessary. For example, $\sim200$ kilogram-years of exposure on an iodine target helps break degeneracies between a wide variety of theories that would otherwise be indistinguishable with xenon and germanium alone. We quantify these statements in detail in \S \ref{sec:results}. 

Since this is the first study to explore model selection for a wide variety of well-motivated scattering models in a statistical way, we fix both the astrophysical and the nuclear--response models for DM--nuclear interactions in \S \ref{sec:analysis} and beyond.  While our results make a case for a future comprehensive analysis including astrophysical and nuclear uncertainties, in \S \ref{sec:results}, Appendix \ref{app:nuclear}, and Appendix \ref{app:astrophysical}, we demonstrate arguments for optimism for model--selection results to be qualitatively robust with respect to these uncertainties. 

The rest of this paper is organized as follows. In \S\ref{sec:scattering}, we present the basic definitions relevant for describing direct detection scattering. In \S \ref{sec:census}, we assemble a representative list of DM-nucleon scattering operators. In \S \ref{sec:simulations} we describe our simulations of the recoil energy spectra under each of these scenarios. In \S \ref{sec:analysis} we describe the analysis of simulated data. In \S\ref{sec:results} we present and discuss our results. We summarize and conclude in \S\ref{sec:conclusions}. Appendix \ref{app:model_selection} includes a more complete set of results of model selection, a subset of which is presented in \S\ref{sec:results}. Appendix \ref{app:nuclear} and Appendix \ref{app:astrophysical} include a qualitative investigation of the impact of astrophysical and nuclear uncertainties on the results of model selection.

\section{DM--nucleus scattering}
\label{sec:scattering}

The nuclear recoil energy spectrum is the number count of nuclear recoil events observed per recoil energy $E_R$, per unit time, per unit target mass,
\begin{equation}
\frac{dR}{dE_R}(E_R) =  \frac{\rho_\chi}{m_T m_\chi} \int\limits_{v_{\mathrm{min}}}^{v_{\mathrm{esc, lab}}}  v f(\mathrm{\mbox{\bf{v}}}) \frac{d\sigma_T}{dE_R} (E_R,v) d^3v .
\label{eq:dRdEr_general}
\end{equation}
This quantity is the observable output of most direct detection experiments. It is a function of the experimental parameters, astrophysics inputs, particle properties of DM, and nuclear properties  of target material. In the above Equation, $\rho_\chi$ is the local DM density; $m_\chi$ and $m_T$ are the DM particle and target--nucleus mass, respectively\footnote{Note that, unlike some of the related literature, throughout this paper we use $T$ to denote the nuclear target and we reserve $N$ to denote a nucleon ($n$ or $p$).}; $v_\mathrm{min} = \sqrt{m_T E_R/2\mu_T^2}$ is the minimum velocity a DM particle of mass $m_\chi$ needs to produce a nuclear recoil of energy $E_R$, where $\mu_T=\frac{m_Tm_\chi}{m_T+m_\chi}$ is the DM--nucleus reduced mass; $v_{\mathrm{esc, lab}}$ is the escape velocity from the Galactic halo in the lab frame; $v$ is the relative velocity of the DM in the lab frame; and $f(\mathrm{\mbox{\bf{v}}})$ is the local DM velocity distribution. For the purposes of this study, we set the astrophysical parameters to the following values \cite{Bovy:2013raa}: $\rho_\chi=0.3$ GeV/cm$^3$; $v_{\mathrm{esc}} = 544$ km/sec in the Galactic frame; we assume that $f(\mathrm{\mbox{\bf{v}}})$ is a Maxwellian distribution, with an rms speed of $155$ km/sec, and a mean speed equal to the Sun's rotational velocity around the Galactic center, $v_\textrm{lag}=220$ km/sec. The choice of scattering model enters the calculation of the recoil rate through the differential scattering cross section ${d\sigma_T}/{dE_R}$, as discussed in \S\ref{sec:census}. The ultimate goal of direct detection analysis is to reconstruct both the normalization and functional form of ${d\sigma_T}/{dE_R}$ from nuclear recoil data.

The total rate $R$ of events (per target mass, per time) is an integral of the differential rate within the nuclear-recoil energy window of a given experiment\footnote{For simplicity, in this work we assume unit efficiency of detection, at all energies within the analysis window, for each experiment considered here, and rescale the exposures to take this assumption into account when choosing experimental parameters to represent the capabilities of G2 experiments.}. The total expected number of events for exposure $T_\text{obs}$ (typically in kilogram-years) is $N\equiv RT_\text{obs}$.
\section{Scattering operators and responses}
\label{sec:census}

In the following, we restrict our attention to spin--1/2 DM. This allows us to sample DM-nucleus interactions that depend on spin, but introduces (minimal) model--building limitations. Furthermore, we focus only on $t$--channel elastic scattering, and assume that scattering changes neither the mass nor the flavor of the incoming DM particle $\chi$, or nucleon $N$.

The SI and SD cross sections are standard choices for interpreting experimental data because, given similar coupling strengths for most interactions, these responses dominate at low energy. These cross sections arise if any of the contact operators 
\begin{align}
\chibar \chi  \Nbar N &,\label{SI 1}\\
\chibar \gamma^\mu \chi  \Nbar \gamma_\mu N&,~\text{~~~or} \label{SI 2} \\
\chibar \gamma^\mu \gamma_5 \chibar \Nbar \gamma_\mu \gamma_5 N&, \label{SD}
\end{align}
are generated by the theory. The interactions of \Eqm{SI 1}{SD} make simple predictions for experiments: cross sections produced by Eqs.~(\ref{SI 1}) and (\ref{SI 2}) scale coherently with nucleon number, and so nuclei of larger atomic mass are always more effective in direct detection searches; the scattering cross section produced by \eq{SD}
is most effectively probed by a nucleus with large spin. Despite these generic expectations, a key point is that these leading interactions may be suppressed or only co--leading. For example, the operator in \eq{SI 2} does not arise if the DM is a Majorana particle.  Exploring nonstandard nuclear responses can thus be crucial for understanding the full range of possible particle physics interactions and extracting the correct one.

As a first step away from the conventional case, we consider models with scattering mediated by the photon of electromagnetism, or by kinetic mixing of a massive gauge field with the photon \cite{Sigurdson:2004zp,Banks:2010eh,Chang:2010aa,Barger:2010gv,Fitzpatrick:2010br,Fortin:2011hv}. These models are well motivated and provide a low--mass degree of freedom that can couple the DM to the SM. Interestingly, (dark) photon--mediated models give rise to momentum-- and velocity--dependent detection rates. Likewise, scattering mediated by a pseudoscalar particle is characterized by pronounced momentum suppression in the direct detection rate at low recoil energies. Ensuring the dominance of such interactions, however, requires suppressing the operators in \Eqm{SI 1}{SD}, which
may indicate a tuning of UV parameters, or that the 
pseudoscalar is much lighter than any scalar or vector particle that can mediate the processes in \Eqm{SI 1}{SD}. If the scattering is mediated by either the (dark) photon or a pseudoscalar, the associated nonstandard nuclear responses could upend notions of which target element is most sensitive to scattering. One purpose of this work is to explore what sets of target nuclei are most effective in distinguishing models when both standard and nonstandard interactions are considered. 

To evaluate prospects for direct detection experiments to distinguish different models of DM--nucleus scattering, we first choose competing hypotheses. The full list of models considered in this work is given in Table~\ref{tab:operators}.  It represents a balanced subset of the allowed interactions, involving a compromise between exhausting the set of operators allowed by the symmetries of the theory and restricting to the set of operators produced by relatively simple UV completions.  The set of operators mostly follows the discussion of \cite{Gresham:2014vja}. 
In addition to representing a diversity of models in terms of UV completions, the operators chosen in Table~\ref{tab:operators} form a representative (if not exhaustive) class of models in terms of momentum and velocity dependence, and triggered nuclear response. In this Section, we sketch examples of each of these categories of models and demonstrate how they might fall out of a UV complete theory.  
\begin{table}[!t]
\begin{centering}
\renewcommand{\arraystretch}{1.4}
\begin{tabular}{c |>{$}l<{$} >{$}c<{$} >{$}c<{$} c }
\hline
 Model name & {\rm Lagrangian} & \text{$\vec q$, $v$ Dependence} &  {\rm Response}  & $f_n/f_p$
\\
\hline 
 SI & \bar \chi \chi \bar N N & 1 & M & +1 \\ 
\hline SD & \bar \chi \gamma^\mu \gamma_5 \chi \bar N \gamma_\mu \gamma_5 N  & 1 & \Sigma' + \Sigma'' & --1.1
\\
\hline 
 \multirow{2}{*}{Anapole} & \multirow{2}{*}{$\chibar \gamma^\mu \gamma_5 \chi \partial^\nu F_{\mu \nu}$} & v^{\perp 2} & M & \multirow{2}{*}{photon--like} \\  
 & & \qsq/m_N^2 & \Delta	+ \Sigma' \\
\hline
 Millicharge & \chibar \gamma^\mu \chi A_\mu & m_N^2m_\chi^2/\vec q^{\,4} &  M & photon--like \\ 
\hline
 \multirow{2}{*}{\pbox{20cm}{MD (light med.)}} & \multirow{2}{*}{$\chibar \sigma^{\mu \nu} \chi F_{\mu \nu}$} & 1+ \frac{v^{\perp 2} m_N^2}{\qsq} & M & \multirow{2}{*}{photon--like} \\
 & & 1 & \Delta + \Sigma'\\
\hline
 ED (light med.) & \chibar \sigma^{\mu \nu} \gamma_5 \chi F_{\mu \nu}  & m_N^2/\qsq & M & photon--like \\
\hline 
 \multirow{2}{*}{\pbox{20cm}{MD (heavy med.)}} & \multirow{2}{*}{$\chibar \sigma^{\mu \nu} \partial_\mu \chi \partial^\alpha F_{\alpha \nu}$} & \frac{\vec q^{\,4}}{\Lambda^4}+ \frac{v^{\perp 2} m_N^2 \qsq }{\Lambda^4} & M & \multirow{2}{*}{photon--like} \\
 & & \vec q^{\,4}/\Lambda^4 & \Delta + \Sigma' \\
\hline
 ED (heavy med.) & \chibar \sigma^{\mu \nu} \gamma_5 \partial_\mu \chi \partial^\alpha F_{\alpha \nu}  & \qsq m_N^2/\Lambda^4 & M & photon--like \\
\hline
 SI$_{q^2}$ & i\chibar \gamma_5 \chi \Nbar N & \qsq/m_\chi^2 & M & +1 \\ \hline
 SD$_{q^2}$ (Higgs-like/flavor--univ.) & i\chibar \chi \Nbar \gamma_5 N & \qsq/m_N^2 & \Sigma''  &$+1/-0.05$ \\  \hline
 SD$_{q^4}$ (Higgs-like/flavor--univ.) &  \chibar \gamma_5 \chi \Nbar \gamma_5 N & \vec q^{\, 4}/m_\chi^2 m_N^2 & \Sigma'' & $+1/-0.05$ \\\hline
 \multirow{3}{*}{$\vec{L}\cdot\vec{S}$-like} & \multirow{3}{*}{\pbox{20cm}{$\bar{\chi} \gamma_\mu \chi {{\partial^2} \Nbar  \gamma^\mu N  \over m_N^2} +$ \\ $+ \bar{\chi} \gamma_\mu \chi {{\partial_\nu } \Nbar \sigma^{\mu \nu} N \over 2 m_N} $} } & \vec q^{\, 4}/m_N^4 & M & \multirow{3}{*}{+1} \\
 & & \vec q^{\, 4}/m_N^4 & \Phi'' & \\
 & & \frac{\qsq v^{\perp 2}}{m_N^2} + \frac{\vec q^{\, 4}}{m_\chi^2m_N^2} & \Sigma' \\
\hline 
\end{tabular}
\caption{Scattering models and corresponding relativistic operators considered in this work are listed here. In model names, ``MD'' stands for ``magnetic dipole'', and ``ED'' for ``electric dipole''; ``heavy med.'' and ``light med.'' refers to the mediator mass, as compared to the characteristic value of momentum transfer. SI$_{q^2}$, SD$_{q^2}$, and SD$_{q^4}$ represent the pseudoscalar--mediated models; note that they do {\it not} simply correspond to $q^2(q^4) \times$ SI(SD); also note that SD$_{q^2}$ and SD$_{q^4}$ taken with two different values of $f_n/f_p$ (denoted as Higgs--like and flavor--universal; values are listed in the last column) are treated as two separate models each for the purposes of our analysis in later sections.
$N$ is a nucleon field; $A_\mu$ is the photon field; $F_{\mu \nu}$ is the EM field strength; $v^\perp$ is the transverse velocity; $\qsq=2m_TE_R$ is the three--momentum transfer; and $\Lambda$ is a heavy--mass or compositeness scale appearing in the dipole models with a heavy mediator. The leading momentum and velocity dependence of the response corresponding to the given operator is listed (schematically) in the third column, and the corresponding response functions in the fourth column. The last column contains a list of benchmark values for $f_n/f_p$ used for simulations in this work. ``Photon-like'' refers to the fact that ratios analogous to $f_n/f_p$ are completely determined by the EM properties of the target nucleus when the mediator is a photon.}
\label{tab:operators} 
\end{centering}
\end{table}

The models we list in Table~\ref{tab:operators} are motivated by their UV completions. We write them explicitly in terms of Lorentz--invariant operators of a relativistic quantum field theory.  To map onto low--energy (nuclear) physics appropriately, we need to relate these relativistic operators to the appropriate non--relativistic expansion.  Completing this map allows one to apply the correct nuclear response, which is encoded in a form factor for each type of response.  In the case of standard SI or SD scattering, this is normally written
\alg{ \label{eq:basic SI and SD}
\sigma_T^{\rm SI} &= \frac{\mu_T^2}{\mu_p^2} \sigma_p^{\rm SI} \bL Z+\frac{f_n}{f_p}(A-Z) \bR^2 F_{{\rm SI},T}^2(E_R),  
\\ 
\sigma_T^{\rm SD} &= \frac{\mu_T^2}{\mu_p^2} \sigma_p^{\rm SD} \frac43 \frac{J+1}{J} \pL \langle S_p \rangle + \frac{f_n}{f_p}\langle S_n \rangle \pR^2 F_{{\rm SD},T}^2(E_R),
}
where $Z$, $A$, and $J$ are the atomic number, atomic mass, and spin of the nucleus, respectively; $\mu_T$ and $\mu_p$ are the reduced DM-target and DM-proton masses, respectively; $f_n/f_p$ is the ratio of DM-neutron and DM-proton coupling strengths; $\langle S_{p} \rangle$ and $\langle S_{n} \rangle$ are the proton and neutron spin content of the nucleus, respectively; and $\sigma_p$ is the cross section for scattering off a proton. The form factors $F$ describe the dependence of the scattering on energy, defined such that $F_{{\rm SI},T}^2(0)=F_{{\rm SD},T}^2(0)=1$. While $F_{{\rm SI,}T}^2(E_R)$ could in principle strongly depend on the nucleus in question, a Helm form factor \cite{Helm:1956zz}
\beq
F_{{\rm SI},T}(q) = F_{\rm SI}(q) = \frac{3~ j_1\pL \sqrt{\vec q^{\, 2} r_T^2} \pR}{\sqrt{ \vec q^{\, 2}r_T^2} } \exp \pL- \vec q^{\, 2} s^2 \pR ,
\eeq
is found to be a good approximation across many nuclei \cite{Freese:2012xd}. Here, $j_1$ is a spherical Bessel function of the first kind; $\vec q^{\, 2}=2m_T E_R$; $s\simeq 0.9$~fm; and $r_T^2 = c^2 + \frac{7}{3} \pi^2 a^2 - 5s^2$ is an effective nuclear radius, with $a \simeq 0.52$~fm and $c\simeq 1.23 A^{1/3} - 0.60$~fm. No such universal form suffices for $F_{{\rm SD},T}$, although measured values for these functions are available \cite{Bednyakov:2006ux}. 

For more general interactions, universal form factors are insufficient, but a suitable basis that can describe all nuclear responses compatible with the low--energy symmetries has recently been found \cite{Fitzpatrick:2012ix}. Instead of two responses ($M$ for SI and $\Sigma' + \Sigma''$ for SD), the generalized scattering can trigger five responses ($M,\Sigma',\Sigma'',\Delta,\Phi''$) along with the relevant interference terms. 
Refs.~\cite{Fitzpatrick:2012ix,Anand:2013yka} parameterized the generalized responses as follows\footnote{Note that Eq.~3 of \cite{Anand:2013yka} is schematic. Compare the following to Eq.~40 of \cite{Anand:2013yka}.} 
\alg{{d\sigma_T \over dE_R}(E_R, v)  = {m_T \over 2 \pi v^2} &\sum_{(N,N')} \bigg[ \sum_{X=M,\Sigma',\Sigma''} R_X\left(E_R,v, c_i^{(N)}, c_j^{(N')}\right) \W_X^{(N,N')}(y)\\ &+{2m_TE_R \over m_N^2} \sum_{X=\Phi'', \Delta, M \Phi'', \Delta \Sigma'} R_X\left(E_R,v, c_i^{(N)}, c_j^{(N')}\right) \W_X^{(N,N')}(y) \bigg],
\label{eq:EFT cross section}
}
where $X\in \{ M,\Sigma',\Sigma'',\Phi'', \Delta, M \Phi'', \Delta \Sigma'\}$, and $(N,N')\in \{ (p,p), (n,n), (p,n), (n,p)\}$.  The functions $R_X$ in \eq{eq:EFT cross section} have mass dimension negative four since they arise from dimension--six operators in the Lagrangian. The $R_X$ encode the momentum and velocity dependence arising from the interaction(s) between DM and the SM. Through the functions $R_X$, the particle physics content is carried through in a target--independent way. Explicit forms of the $R_X$ in terms of the low--energy DM--nucleon operator coefficients $c_i$ are given by Eq.~(38) of Ref.~\cite{Anand:2013yka}. These $c_i$ are mass dimension negative two parameters that control which non--relativistic operators $\mathcal O_i$ enter the cross section. The target--specific information is carried by the $\widetilde{W}_X^{(N,N')} (y)$, which we call the nuclear response functions. These can be obtained from Ref.~\cite{Anand:2013yka}\footnote{For convenience, we define $\W_X^{(N,N')} (y)\equiv 4 \pi W_X^{(N,N')} (y)/ (2 J + 1)$, where $J$ is nuclear spin. The $W_X$ are given in Eq.~(41) of Ref.~\cite{Anand:2013yka}.}; we summarize some low--energy limits of these responses in Table \ref{tab:responses}, which we have reproduced from Ref.~\cite{Gresham:2014vja}.  The dimensionless variable $y \equiv m_T E_R b^2 / 2$ stands in for the nuclear recoil, where $b \equiv \sqrt{41.467/(45 A^{-1/3} - 25 A^{-2/3})} \fm$ is the harmonic oscillator parameter for an atom with mass number $A$. Of the five different responses that can arise in elastic scattering mediated by the exchange of spin--0 or spin--1 particles ($M,\Sigma',\Sigma'', \Phi''$, and $\Delta$) only the pairs $M,\Phi''$ and $\Sigma',\Delta$ can interfere. The response entering into ``standard'' SI scattering is $M$, while that entering into ``standard'' SD scattering is $\Sigma'' + \Sigma'$. The $\Phi''$ and $\Delta$ responses (and their interference terms with the others) are ``novel'' in the sense that they do not appear in the conventional scenarios. 

We now break \eq{eq:EFT cross section} into expressions for each of the operators from Table~\ref{tab:operators}.
\begin{table}
\begin{centering}
\begin{tabular}{c l >{$}l<{$}}
Response & Colloquial name &  \widetilde{W}_X^{(p,p)}(E_R\rightarrow 0)\\
\hline
$M$	& spin--independent & Z^2 \\
$\Sigma''$		& spin--dependent (longitudinal)  & 4 {J+1 \over 3 J} \langle S_p \rangle^2 \\
$\Sigma'$		& spin--dependent (transverse) 	 & 8 {J+1 \over 3 J} \langle S_p \rangle^2\\
$\Delta$		& angular--momentum--dependent & {1 \over 2} {J+1 \over 3 J} \langle L_p \rangle^2 \\
$\Phi''$		& angular--momentum--and--spin--dependent  & \sim \langle \vec{S}_p \cdot \vec{L}_p \rangle^2\\ 
\hline
\end{tabular}
\caption{Some nuclear response functions relevant for DM direct detection. The $E_R\rightarrow0$ limit for $(N,N')=(p,p)$ is shown in the right-most column; in this limit, we also have ${4 \pi \over 2 J + 1}W_{\Delta \Sigma'}^{(N,N')} \rightarrow -2 {J+1 \over 3 J}\langle L_N\rangle \langle S_{N'}\rangle$. }\label{tab:responses}
\end{centering}
\end{table}
\subsection{Standard spin--independent and spin--dependent scattering}
\label{sec:SI_SD_models}

Given heavy mediators and order one coefficients for all Lorentz--invariant UV operators, standard SI or SD scattering as presented in \eq{eq:basic SI and SD} will dominate the behavior of the scattering cross section at low energies. This happens because these are the only terms not suppressed by the DM velocity or momentum. At the level of the DM--nucleon interaction, these standard scattering cross sections are produced by the Lagrangians\footnote{The interaction $\bar \chi \gamma^\mu \chi \bar N \gamma_\mu N$ also includes a momentum--unsuppressed SI scattering cross section, and for the purposes of this discussion is equivalent to $\bar \chi \chi \bar N N$.}
\alg{
\cL_{\rm leading} = \cbL \begin{array}{ll} \sum_N f^{\rm SI}_N \bar \chi \chi \bar N N/M^2 \qquad  &{\rm ~~(SI)}
\\ \sum_N f^{\rm SD}_N \bar \chi \gamma^\mu \gamma_5 \chi \bar N \gamma_\mu \gamma_5 N /M^2 \qquad &{\rm ~~(SD)}
\end{array} \right. , }
where we assume that the particle mediating the interaction has mass $m_{\rm med} \sim M/\sqrt{f_N}$, and we reserve the symbol $N=n,p$ to represent the nucleon involved in the interaction. The corresponding cross sections for scattering off a nucleus are
\beq
\sigma^{\rm SI}_T=\frac{\mu_T^2}{\pi M^4} \sum_{NN'} f^{\rm SI}_N f^{\rm SI}_{N'} \W_M^{(N,N')}, \qquad \sigma^{\rm SD}_T=\frac{\mu_T^2}{\pi M^4} \sum_{NN'} f^{\rm SD}_N f^{\rm SD}_{N'} \bL \W_{\Sigma'}^{(N,N')} + \W_{\Sigma''}^{(N,N')} \bR \label{eq: SI SD} .
\eeq
The SI scattering acts coherently on the entire nucleus, and consequently scales quadratically with nucleon number $A$, such that $\sum_{NN'} \W_M^{(N,N')}|_{y=0} \propto A^2$, as in Eq.~(\ref{eq:basic SI and SD}). In contrast, the SD scattering response depends primarily on the average spin of unpaired nucleons, so the resulting cross section does not scale with nuclear size. Assuming heavy mediators, if any Lagrangian gives rise to one of these responses (without momentum suppression), the cross section will be well approximated by \eq{eq: SI SD} at low momentum transfer. 

When reporting limits on the strength of these standard interactions, and performing corresponding model fits, we refer to the cross section off the proton, defined as
\beq
\sigma^{\rm SI}_p = \frac{\mu_p^2}\pi \pL \frac{f^{\rm SI}_p}{M^2} \pR^2, \qquad \sigma^{\rm SD}_p = \frac{3\mu_p^2}\pi \pL \frac{f^{\rm SD}_p}{M^2} \pR^2, 
\label{eq:sigma_p_SI_SD}
\eeq
where $\mu_p = m_p m_\chi/(m_p + m_\chi)$ is the reduced mass of the proton--DM system; the factor of 3 in $\sigma^{\rm SD}_p$ is included to agree with conventions.  Instead of parametrizing cross sections in terms of $f_p$, $f_n$, and $M$, we rescale the DM--neutron cross section to the DM--proton cross section.  This leaves two free parameters to describe the scattering strength: $\sigma^{\rm SI}_p$ and $f_n/f_p$. 
\subsection{Photon-mediated scattering}
\label{sec:photon_models}

Some of the best--motivated models that produce nonstandard scattering are those where the mediator of the scattering is the SM photon. Electrically neutral DM with an anapole, electric dipole, or magnetic dipole moment can arise in composite DM models (where the constituents carry electromagnetic charges) \cite{Bagnasco:1993st}, models with EW--neutral DM that couples to heavy, charged messengers \cite{Weiner:2012gm}, or from models where the photon kinetically mixes with a dark photon \cite{Fitzpatrick:2010br, ArkaniHamed:2008qn}. DM with a small electric charge can arise when the SM photon kinetically mixes with a heavy $U(1)$ gauge boson \cite{Holdom:1985ag} and its mass is stabilized by the St\"uckelberg mechansim \cite{Kors:2004dx}.  For each of these cases, we recap simple models and their resulting momentum and nuclear response dependence. In all of these cases, the couplings to the SM nucleons are entirely fixed. This leaves no freedom to fine--tune operators against each other, but it still allows non-standard nuclear responses (for which the relative sensitivities between elements differ from the standard cases) to dominate the recoil spectrum.

All photon-mediated scattering involves an interaction with the nucleon-level electromagnetic current,
\beq
J_\mu^{\rm EM} \equiv e \sum_{N=p,n} \bar N\left( Q_N \frac{K_\mu}{2 m_N} - \tilde{\mu}_N \frac{i \sigma_{\mu \nu} q^\nu}{2 m_N} \right) N,
\eeq
where we explicitly pull out the elementary charge $e>0$, $\tilde{\mu}_N = {\text{magnetic moment} \over \text{nuclear magneton}}$ is the dimensionless magnetic moment, and $K_\mu$ is the sum of incoming and outgoing nucleon momentum. The scattering behavior is determined by the DM bilinear. For neutral DM, the bilinears are restricted by gauge invariance to be
\begin{eqnarray}
{\cal O}_{\rm \chi, Anapole}^\mu & = & g^{\rm Anapole}\bar \chi \gamma^\mu \gamma_5 \chi, \\
{\cal O}_{\rm \chi, MD}^\mu & = & \frac{g^{\rm MD}}{\Lambda}\bar \chi i \sigma^{\mu \nu} q_\nu \chi ,\\
{\cal O}_{\rm \chi, ED}^\mu & = & \frac{g^{\rm ED}}{\Lambda} \bar \chi i \sigma^{\mu \nu} \gamma_5 q_\nu \chi.
\end{eqnarray}
The magnetic and electric dipoles can arise when the DM is a Dirac fermion which is a singlet under the SM gauge group but which couples to the SM through heavy fermion and scalar messenger states \cite{Weiner:2012gm}; they may also arise if DM is a fermionic SM singlet composite particle \cite{Bagnasco:1993st}. The scale of the heavy fermion or scalar messenger states, or alternately the confinement scale, is $\Lambda$. If CP is nearly conserved in the dark sector, $\Lambda_\text{ED} \gg \Lambda_\text{MD}$. If certain conditions on the mass and kinetic mixing of the gauge bosons hold, the DM may pick up an electric charge \cite{Holdom:1985ag, Kors:2004dx}, typically parametrized by a small number multiplied by the electron charge, $q_\chi = \epsilon e$. Such millicharged DM interacts with $J_\mu^{\rm EM}$ via the bilinear
\beq 
{\cal O}_{\rm \chi, Millicharge}^\mu = \epsilon e \bar \chi \gamma^\mu \chi,
\eeq
and is necessarily coupled through the SM photon. In these cases, the mediator is naturally the photon, which is exactly massless. The contact operator for direct detection scattering is
\beq
{\cal O}_{\rm [MD, ED, Millicharge]} = \frac1{\qsq} {\cal O}_{\rm \chi,[MD, ED, Millicharge]}^\mu J_\mu^{\rm EM},
\label{Op:MdEd}
\eeq 
where $|\vec{q}| = \sqrt{-q^\mu q_\mu}$ is the magnitude of the momentum transfer in the event.  The corresponding scattering is dominated by a SI term that is {\it enhanced} rather than suppressed at low momentum transfer. In general, other photon--mediated interactions of DM (such as through the nucleon charge radius \cite{Bagnasco:1993st,Weiner:2012gm} or polarizability \cite{Kribs:2009fy, Xu:2015wha}) can be important, but we neglect them in this work since they arise from higher--dimension operators.
 
If the DM is a Majorana fermion, the anapole is the only operator that can give rise to a spin--one--mediated interaction with SM matter. In this case, kinetic mixing with the gauge boson of a broken $U(1)$ plays an irreducible role. A gauge boson of mass $M$ that kinetically mixes with the photon gives rise to an operator
\beq
{\cal O}_{\rm Anapole} = \frac1{\qsq + M^2} {\cal O}_{\rm\chi, Anapole}^\mu J_\mu^{\rm EM}.
\eeq
The gauge boson may in principle have a very small mass, $M^2\ll \qsq$, in which case the scattering would appear to be through a light mediator (compared to the momentum transfer). However, masses this small are constrained by astrophysical and collider searches \cite{Essig:2013lka}, and we focus on the $M^2\gg \qsq$ regime instead. Of course, kinetic mixing with a heavy mediator may also play a role in the preceding cases; in general, $\vqsq$ in Eq.~\ref{Op:MdEd} should be replaced with $\vqsq + M^2$.

The scattering cross sections for these operators are \cite{Fitzpatrick:2012ix,Gresham:2014vja}
\begin{align}
\sigma_T^{\rm Anapole} &= \frac{\mu_T^2}{\pi} \left(\frac{e g^{\rm Anapole}}{\qsq+M^2}\right)^2 \Bigg\{ {{\vec v_T}^{\perp 2}} \W_M^{(p,p)} +  \frac{\vqsq}{m_N^2} \Bigg[ \W_\Delta^{(p,p)} - \tilde \mu_n \W_{\Delta \Sigma'}^{(p,n)} - \tilde \mu_n \W_{\Delta \Sigma'}^{(p,p)} + \nonumber  
\\ & \qquad \qquad \qquad\qquad\qquad  + \frac{1}{4} \left( \tilde \mu_p^2 \W_{\Sigma'}^{(p,p)} +  \tilde \mu_p \tilde \mu_n \W_{\Sigma'}^{(p,n)} +  \tilde \mu_n^2 \W_{\Sigma'}^{(p,n)} \right)\Bigg]\Bigg\},  \label{ana sigma}
\\ \sigma_T^\text{MD}  &= {{\mu_T^2} \over \pi} \left(\frac{e g^\text{MD}}{\vec q^2 + M^2}\right)^2  \frac{\vqsq}{\Lambda^2} \bigg\{ \left[ \vec{v}^{\perp 2}_T + {\vqsq \over 4  m_\chi^2} \right]\W_M^{(p,p)} + {\vqsq \over m_N^2} \bigg[\W_\Delta^{(p,p)}-\nonumber
\\ & - \mag_n \W_{\Delta \Sigma'}^{(p,n)}-\mag_p \W_{\Delta\Sigma'}^{(p,p)}  
+ {1 \over 4} \left( \mag_p^2 \W_{\Sigma'}^{(p,p)}+2 \mag_n \mag_p \W_{\Sigma'}^{(p,n)}+\mag_n^2 \W_{\Sigma'}^{(n,n)}\right) \bigg] \bigg\},
\label{eq: mag dip rate}   
\\ \sigma_T^\text{ED}  &= {{\mu_T^2} \over \pi}\left({e g^\text{ED} \over \vqsq +M^2}\right)^2 {\vqsq \over \Lambda^2} \left[  \W_M^{(p,p)} + \left(\text{terms of order }{\vec{q}^{\,2} \over m_N^2} \right) +\dots \right],
\\ \sigma_T^\text{Millicharge}  &= {{\mu_T^2} \over \pi}\left({\epsilon e^2 \over \vqsq }\right)^2 \left[ \W_M^{(p,p)} + \left(\text{terms of order }{\vec{q}^{\,2} \over m_N^2} \right) +\dots \right],
\end{align}
where the transverse velocity $v_T^{\perp 2}=v^2-\vqsq/4\mu_T^2$ generates nonstandard velocity dependence. The difference between a velocity-- and momentum--dependent rate is seen in the differential energy spectrum: a velocity--dependent rate has a suppressed normalization but a relatively unchanged differential energy spectrum compared to an unsuppressed rate. In contrast, a momentum--dependent rate will contain the novel feature of a characteristic energy below which the event rate rises with $E_R$. The various rates are compared in Figure \ref{fig:models_set1} below.

\begin{figure}
\includegraphics[width=\textwidth]{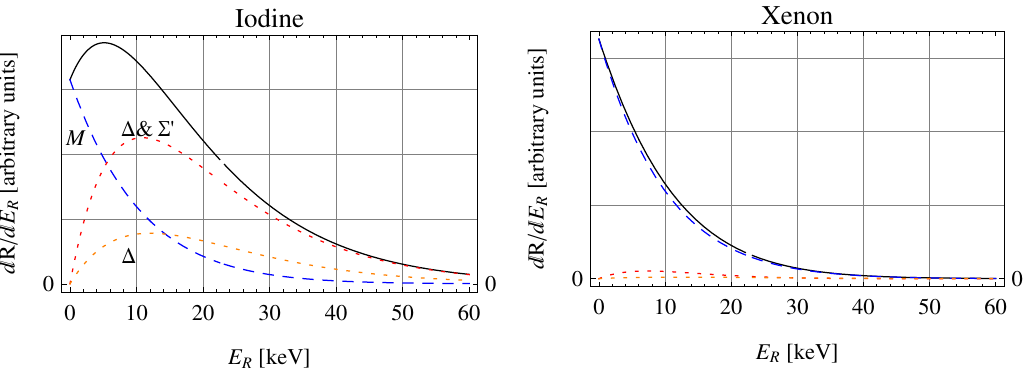}
\caption{Rate for anapole DM to scatter off of iodine or xenon, decomposed in terms of the SI (dashed blue), and magnetic--moment--dependent ($\Delta$\&$\Sigma'$, dotted red) contributions to \eq{ana sigma}. The orbital--angular--momentum contribution ($\Delta$, dotted orange) is also shown separately. \label{SIvsSD}}
\end{figure}

In addition to generating momentum and velocity dependence in the cross section, these interactions stimulate different nuclear responses than the ones probed by standard scattering. The average orbital angular momentum, $L$, of the constituent nucleons (via the $\Delta$ response) is important for photon--mediated interactions.  In Figure \ref{SIvsSD}, we illustrate the relative size of the standard SI versus SD responses for the case of DM interacting through the anapole interaction. To highlight the fact that the responses diverge from the standard case (and, crucially, diverge in different ways for different detector elements) we show the rates on xenon and iodine targets. These elements have isotopes with an unpaired neutron and proton, respectively. The SI contribution dominates the rate for xenon because of its unpaired neutron: since the angular momentum coupling is dominantly through the proton, the overall angular momentum response of xenon is small. In contrast, iodine has an unpaired proton and so has a significant angular momentum response. This simple, well--motivated example illustrates the importance of considering non--standard interactions at direct detection experiments. No tuning was necessary to bring the novel nuclear response to the forefront. On the other hand, the novel nuclear responses play an important role only for certain targets. In particular, a xenon target experiment is still rather sensitive to the underlying interaction due to its large $M$ response.

As in the case of standard interactions, for the purposes of our numerical analysis, we define quantities analogous to the cross section for scattering off the proton,
\beq
\sigma_p^{\rm Anapole}= {  \mu_p^2 \over \pi} \pL {e g^{\rm Anapole} \over M^2} \pR^2 , ~~\sigma_p^{\rm MD, heavy}={ \mu_p^2 \over \pi} \pL {e g^{\rm MD} \over M^2} \pR^2 {\vec q_{\rm ref}^{\,2} \over \Lambda^2} , ~~\text{and}~~\sigma_p^{\rm ED, heavy}={ \mu_p^2 \over \pi} \pL {e g^{\rm ED} \over M^2}\pR^2 {\vec q_{\rm ref}^{\,2} \over \Lambda^2},
\label{eq:sigma_p_photon_heavy}
\eeq
for the heavy--mediator cases; analogously, for the light--mediator cases,
\beq
\sigma_p^{\rm MD, light} = { \mu_p^2 \over \pi} \pL {e g^{\rm MD} \over \Lambda |\vec q_\text{ref}| } \pR^2 , ~~\sigma_p^{\rm ED, light}={ \mu_p^2 \over \pi} \pL {e g^{\rm ED} \over \Lambda |\vec q_\text{ref}| } \pR^2, ~~\text{and}~~\sigma_p^{\rm Millicharge}={ \mu_p^2 \over \pi} \pL {\epsilon e^2 \over \vec q_{\rm ref}^{\,2}}\pR^2.
\label{eq:sigma_p_photon_light}
\eeq
We choose a reference momentum typical of direct detection experiments, $\vec q_{\rm ref}^{\,2}=\pL 100 \mev\pR^2$. In terms of $\vec q_{\rm ref}$, the characteristic turn-over energy described above is $E_{\rm turn}=\vec q_{\rm ref}^{\,2}/2m_T$.
\subsection{Pseudoscalar--mediated DM}

We next consider interactions mediated by the exchange of a spin--0 particle with at least one CP--odd vertex. There is more parameter freedom in these pseudoscalar--mediated models than in the photon--mediated examples, which allows us to isolate additional phenomenologically nontrivial responses. However, this freedom comes at the cost of having a somewhat fine--tuned UV theory. The symmetries that protect the novel responses in the photon--mediated case are absent here and some sort of tuning, or hierarchy of mass scales between scalars and pseudoscalars (with the pseudoscalars being much lighter), is required so the novel responses are not dominated by the standard responses.

We begin with the Lagrangian
\beq \label{pseudo lint}
\lagint^\text{pseudoscalar}={1 \over M^2} \sum_{N=n,p} \left( f^{{\rm SI}q^2}_N i \chibar \gamma_5 \chi  \Nbar N + f^{{\rm SD}q^2}_N i \chibar \chi  \Nbar \gamma_5 N+ f^{{\rm SD}q^4}_N \chibar \gamma_5 \chi  \Nbar \gamma_5 N \right).
\eeq
The terms involving $\bar N \gamma_5 N$ give rise to a longitudinal SD ($\Sigma''$) response; depending on the CP properties of the DM vertex, one additionally expects either a $q^2$ or $q^4$ suppression in the rate. The cross section for scattering with at least one CP-odd  vertex is given by \cite{Fitzpatrick:2012ix,Gresham:2014vja}
\alg{
&\sigma_T^\text{pseudoscalar}  = {{\mu_T^2} \over \pi}\left({1 \over M^2}\right)^2 \times \sum_{N,N'} \Bigg[ {\vec{q}^{\,2} \over 4 m_\chi^2}  f^{{\rm SI}q^2}_N f^{{\rm SI}q^2}_{N'} \W_M^{(N,N')} + \\ &\qquad\qquad\qquad\qquad 
 + \pL {\vec{q}^{\,2} \over 4 m_N^2}f^{{\rm SD}q^2}_N f^{{\rm SD}q^2}_{N'} + {\vec{q}^{\,4} \over 16 m_N^2 m_\chi^2} f^{{\rm SD}q^4}_N f^{{\rm SD}q^4}_{N'} \pR \W_{\Sigma''}^{(N,N')} \Bigg] \label{eq: pseudoscalar rate}  
.}
The SD part of the interaction depends on only the longitudinal SD response rather than the longitudinal plus transverse SD response that is conventionally considered for SD DM interactions. For this reason, it is inappropriate to treat the cross section as $(\qsq / \vec q_{\rm ref}^{\,2})^n \times \sigma_T^\text{SD}$. Nonetheless, we will use the (potentially misleading) nomenclature ${\rm SD}_{q^2}$ and ${\rm SD}_{q^4}$ when referring to these models in the context of model selection. 

From the model building perspective, there is an immediate challenge with models that rely on pseudoscalar mediators to suppress a SI $\bar \chi \chi \bar N N$ term: generally, pseudoscalars are accompanied by scalars in a complete theory, and CP violation at either vertex can mimic this effect.  If present, a scalar interaction (from a CP--even particle or from CP violation at a vertex) will mediate a purely SI, non--momentum--suppressed interaction with nuclei, which will dominate the scattering rate. 
An explicit example of a model in which the $f^{{\rm SD}q^2}$ term can naturally compete with a SI term is given in Sec.~6 of \cite{Fitzpatrick:2012ix}. 

Analogous to the case of standard interactions and photon--mediated models, when discussing current constraints and fitting pseudoscalar--mediated models to simulated data in later sections, we parameterize coupling strength via the cross section for scattering off a proton, defined as
\beq
\sigma^{{\rm SI}q^2}_p = {\mu_p^2 \over \pi} {\vec q_{\rm ref}^{\,2} \over 4 m_\chi^2} {(f^{{\rm SI}q^2}_p)^2 \over M^4},~~~~ \sigma^{{\rm SD}q^2}_p = {\mu_p^2 \over \pi} {\vec q_{\rm ref}^{\,2} \over 4 m_p^2} {(f^{{\rm SD}q^2}_p)^2 \over M^4}, ~~~~\sigma^{{\rm SD}q^4}_p = {\mu_p^2 \over \pi}  {\vec q_\text{ref}^{\,4} \over 16 m_p^2 m_\chi^2} {(f^{{\rm SD}q^4}_p)^2 \over M^4},
\label{eq:sigma_p_PS}
\eeq
with $\vec q_{\rm ref}^{\,2}=\pL100 \mev\pR^2$ depending on which interaction dominates. To make contact with earlier literature we write the low--momentum--transfer expansion for the $\Sigma''$ response,
\beq
\sum_{N,N'} f^{{\rm SD}q^{2,4}}_N f^{{\rm SD}q^{2,4}}_{N'}\W_{\Sigma''}^{(N,N')}(0)={4 \over 3}{J+1 \over J}\left( f^{{\rm SD}q^{2,4}}_p \langle S_p \rangle + f^{{\rm SD}q^{2,4}}_n \langle S_n \rangle \right)^2.
\eeq
The corresponding cross section $\sigma_p$ and $f^i_n/f^i_p$ are, in principle, the two free parameters of each of the pseudoscalar--mediated scattering models.

The values for $f^{{\rm SD}q^{2,4}}_n/f^{{\rm SD}q^{2,4}}_p$ required to generate events for our mock simulations arise as a consequence of the couplings to quarks. The quark couplings, $f^{{\rm SD}q^{2,4}}_q$, are the free parameters. Starting with a UV theory that describes these interactions, we then construct the couplings to nucleons. We define the nucleon coupling to be
\beq \label{nucleon PS couplings}
f^{{\rm SD}q^{2,4}}_N=\sum_q \langle N |  i f^{{\rm SD}q^{2,4}}_q \bar q \gamma_5 q | N \rangle.
\eeq
The pseudoscalar content of the nucleons is defined by $f_5^{N(q,0)} \equiv \langle N | i m_q \bar q \gamma_5 q | N \rangle $, for which we use quark masses $m_q$ from current measurements. Finally, we calculate the ratio of the nucleon couplings as \cite{Hill:2014yxa}
\beq \label{PS coupling ratio}
\frac{f^{{\rm SD}q^{2,4}}_n}{f^{{\rm SD}q^{2,4}}_p}= \frac{m_n}{m_p}  \frac{ \sum_q  f_5^{n(q,0)} f^{{\rm SD}q^{2,4}}_q/m_q}{\sum_q  f_5^{p(q,0)} f^{{\rm SD}q^{2,4}}_q/m_q}.
\eeq
For a given set of quark couplings $f^{{\rm SD}q^{2,4}}_q$, the ratio $f^{{\rm SD}q^{2,4}}_n/f^{{\rm SD}q^{2,4}}_p$ is fixed by nuclear measurements \cite{Cheng:2012qr}. Unfortunately, such measurements are beset by large uncertainties \cite{Hill:2014yxa}. We emphasize that for some of the best--motivated choices of $f^{{\rm SD}q^{2,4}}_q$, the central values of \eq{PS coupling ratio} are close to cancellation points and the ratio is very uncertain. For this reason, selecting central values of $f_5^{N(q,0)}$ as inputs to evaluate $f^{{\rm SD}q^{2,4}}_n/f^{{\rm SD}q^{2,4}}_p$ for Higgs--like (flavor--universal) scenarios should be understood to carry significant error bars. With couplings to heavier quarks or the inclusion of two Higgs doublets, the bounds on this ratio become even less constrained. In the Type-II two Higgs doublets Models \cite{Djouadi:2005gj}, for example, couplings to leptons and down--type quarks are enhanced by $\tan \beta$ while couplings to up-type quarks are suppressed by $\cot \beta$, where $\tan \beta = v_u / v_d > 1$ is a free parameter in the model. Thus, while $|f_p| \gtrsim |f_n|$ is expected at the effective operator level \cite{Arina:2014yna}, it is clear that $|f_n| \gtrsim |f_p|$ is equally well motivated in UV--complete models. As a choice of representative benchmark values for the purposes of simulating data under these models in \S\ref{sec:simulations}, we use:
\begin{itemize}
\item $f^{{\rm SD}q^{2,4}}_n/f^{{\rm SD}q^{2,4}}_p=+1$, motivated by a high--scale MFV--like model in which all quarks couple to the mediator proportionally to their mass
\item $f^{{\rm SD}q^{2,4}}_n/f^{{\rm SD}q^{2,4}}_p=-0.05$, motivated by a flavor--universal model as in \cite{Arina:2014yna}.
\end{itemize} 
\subsection{$\vec L \cdot \vec S$ Scattering}
\label{sec:LS}

In some models, the interaction
\begin{align}
\lagint^\text{LS} &= {g^\text{LS} \over \Lambda^2} \bar{\chi} \gamma_\mu \chi \sum_{N=n,p}  \left( {f_1^N q_\alpha q^\alpha \over m_N^2} \Nbar \gamma^\mu N + \frac{f_2^N}{2m_N} \Nbar i \sigma^{\mu \nu} q_\nu N \right) \label{LS operator}
\end{align}
arises, which leads to a DM--nucleus interaction in which the $\Phi''$ response competes with the $M$ response \cite{Fitzpatrick:2012ix}.  For computational tractability, we reduce the number of free parameters by making the arbitrary (though, in this case, untuned) choice $f_1^N= f_2^N/2 \equiv 2f_N$. The scattering rate is then given by \cite{Fitzpatrick:2012ix,Gresham:2014vja}
\alg{
\sigma_T^\text{LS} & = {{\mu_T^2} \over \pi}\left({g^{\rm LS} \over \Lambda^2}\right)^2 \sum_{N,N'} \frac\vqsq{m_N^2}  f_N f_{N'}  \Bigg\{ \frac\vqsq{m_N^2}  \bigg[ \W_M^{(N,N')} + 4\W_{\Phi''M}^{(N,N')} + 4\W_{\Phi''}^{(N,N')} +\\ &\qquad\qquad\qquad\qquad\qquad\qquad\qquad\qquad\qquad\qquad+ {m_N^2 \over m_\chi^2} \W_{\Sigma'}^{(N,N')}   \bigg]  + 2\vec{v}_T^{\perp \,2} \W_{\Sigma'}^{(N,N')} \Bigg\}.
\label{eq: LS rate}  
}
To parameterize the overall coupling strength we will use
\beq
\sigma^{\rm LS}_p = \frac{(g^\text{LS})^2 \mu_p^2 f_p^2}{\pi \Lambda^4}.
\label{eq:sigma_p_LS}
\eeq
Since the $\Phi''$ response is related to products of orbital and spin angular momenta of constituent nucleons (see \Tab{tab:responses}), we refer to this model as ``$\vec L \cdot \vec S$--like'' and note that it is another example of a model in which a novel nuclear response contributes significantly to the scattering rate.
\section{Simulations}
\label{sec:simulations}

We have described all theory inputs necessary to compute scattering rates and simulate nuclear recoil--energy spectra for a wide range of scattering theories on a variety of nuclear targets. To simulate future data, we use a total of 14 scattering models, listed in Table \ref{tab:operators}. Each scattering model in this Table is defined by the choice of the interaction operator (corresponding to specific nuclear responses), and the choice of $f_n/f_p$ (note that some models in this list only differ by their value of $f_n/f_p$), leaving us with two parameters that can be freely chosen for each model: the cross section for scattering off protons $\sigma_p$ (defined for different models in Eqs.~(\ref{eq:sigma_p_SI_SD}), (\ref{eq:sigma_p_photon_heavy}), (\ref{eq:sigma_p_photon_light}), (\ref{eq:sigma_p_PS}), and (\ref{eq:sigma_p_LS})), and the DM particle mass $m_\chi$. Our benchmark values for $m_\chi$ used for most simulations are 15 GeV, 50 GeV, and 500 GeV, but we also explore a 7 GeV DM in a small subset of simulations in \S\ref{sec:othertargets}.
The choices of $\sigma_p$ values used to simulate future experiments are detailed in \S\ref{sec:exclusions}; overall, they cover both the optimistic scenarios where the DM signal is close to the current upper limit, and the scenarios where it is just below the reach of the G2 experiments.

\begin{figure*}[h]
\centering
\includegraphics[width=.31\textwidth,keepaspectratio=true]{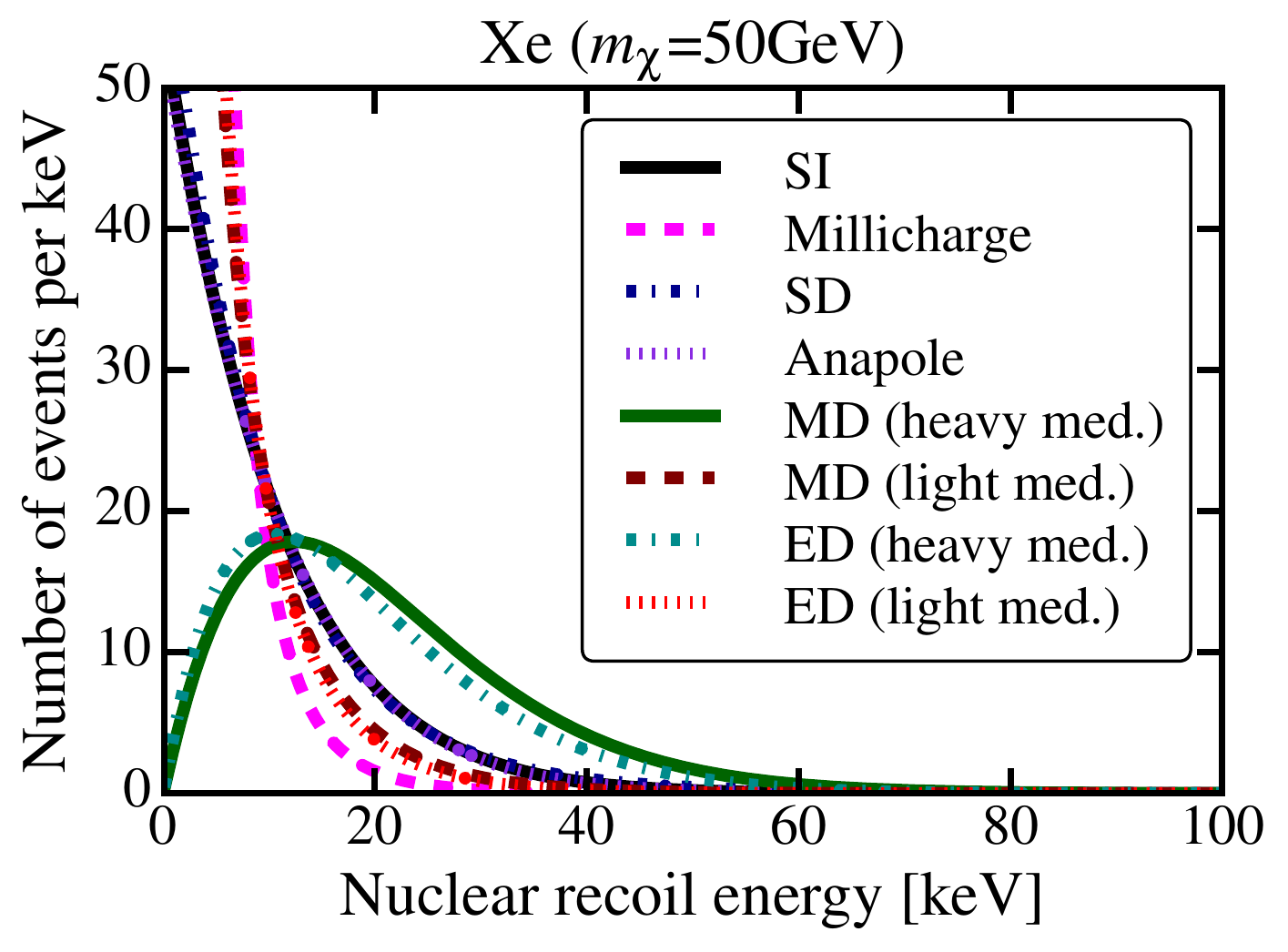}
\includegraphics[width=.31\textwidth,keepaspectratio=true]{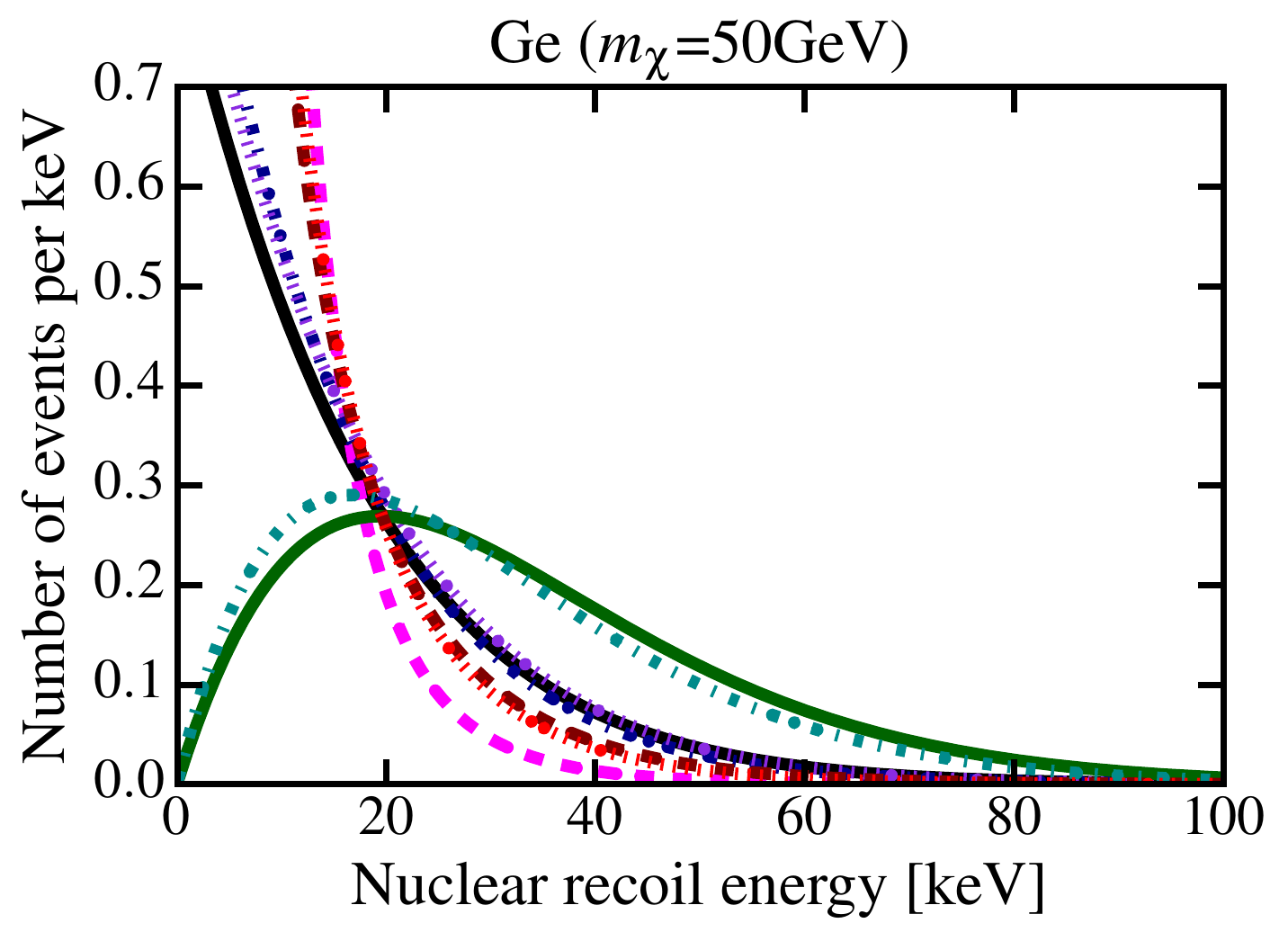}
\includegraphics[width=.31\textwidth,keepaspectratio=true]{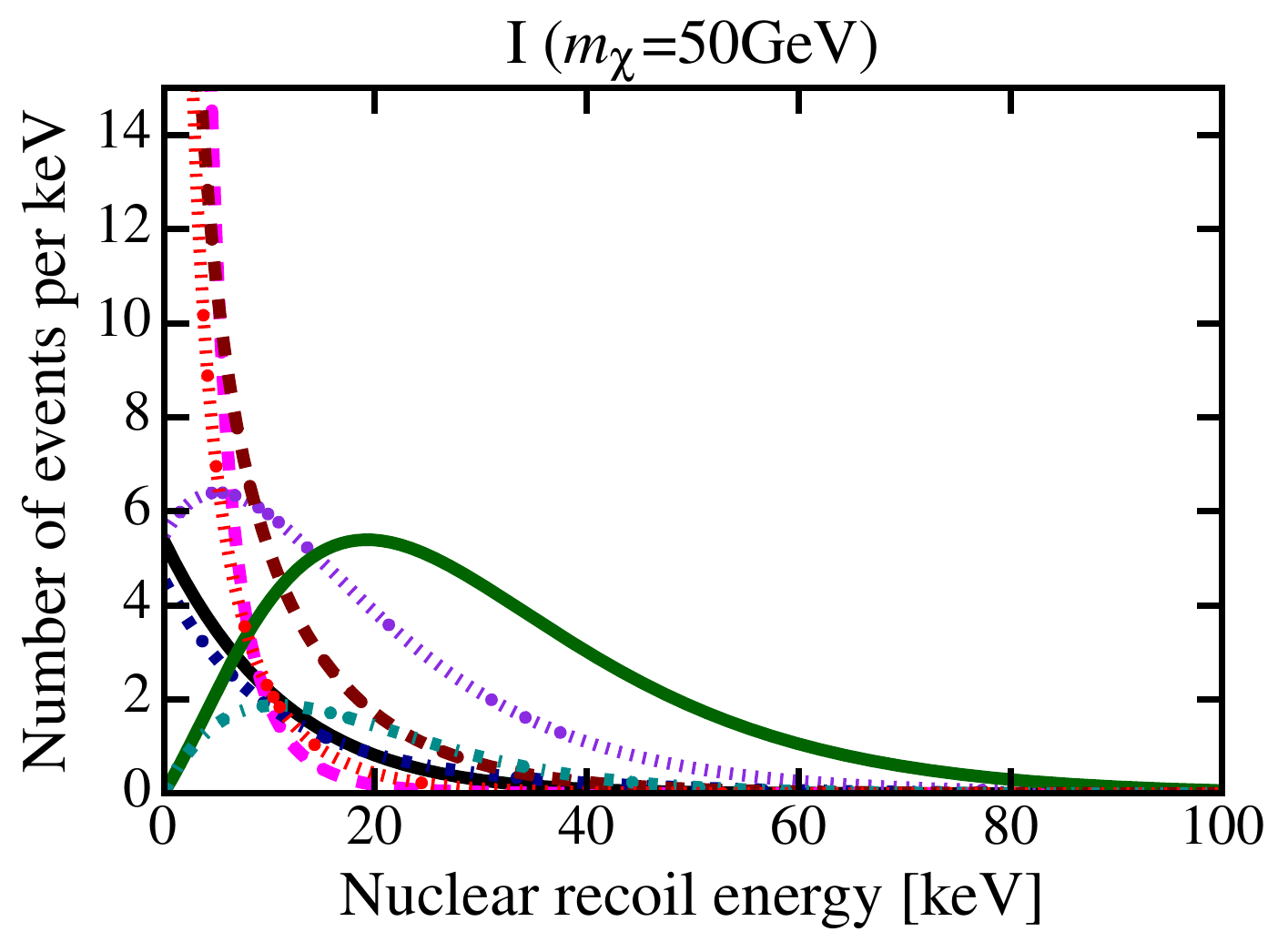}
\includegraphics[width=.31\textwidth,keepaspectratio=true]{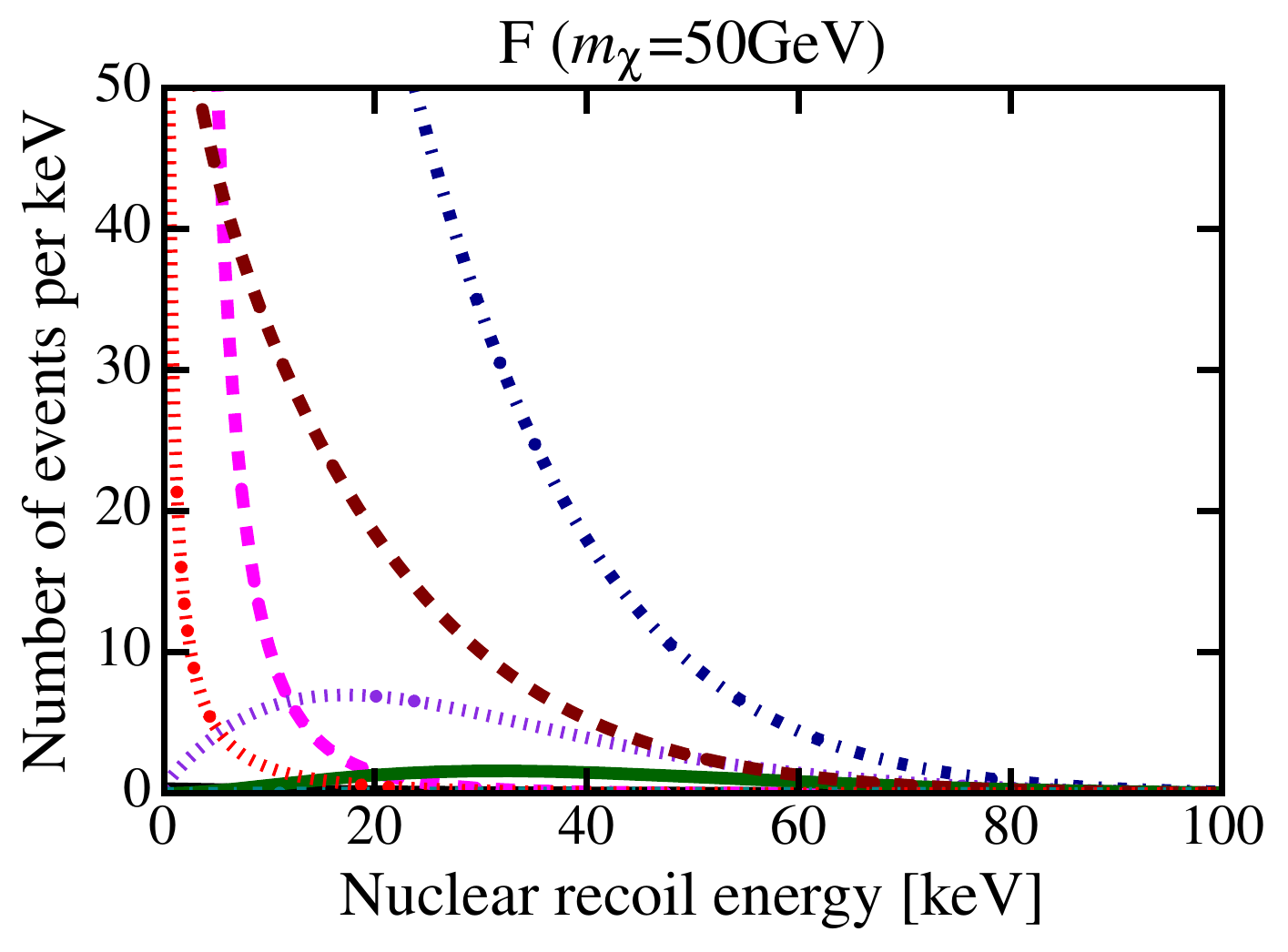}
\includegraphics[width=.31\textwidth,keepaspectratio=true]{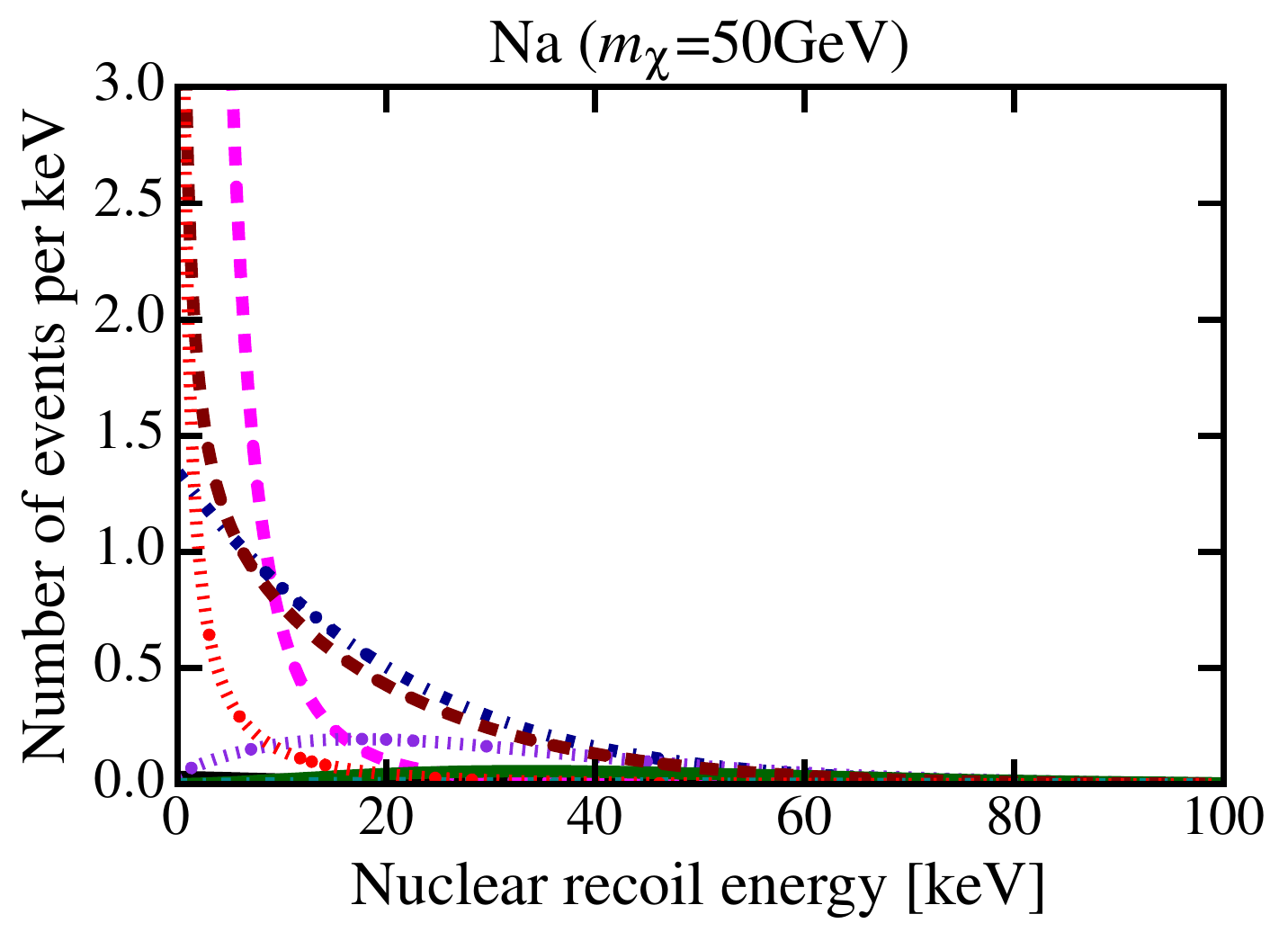}
\includegraphics[width=.31\textwidth,keepaspectratio=true]{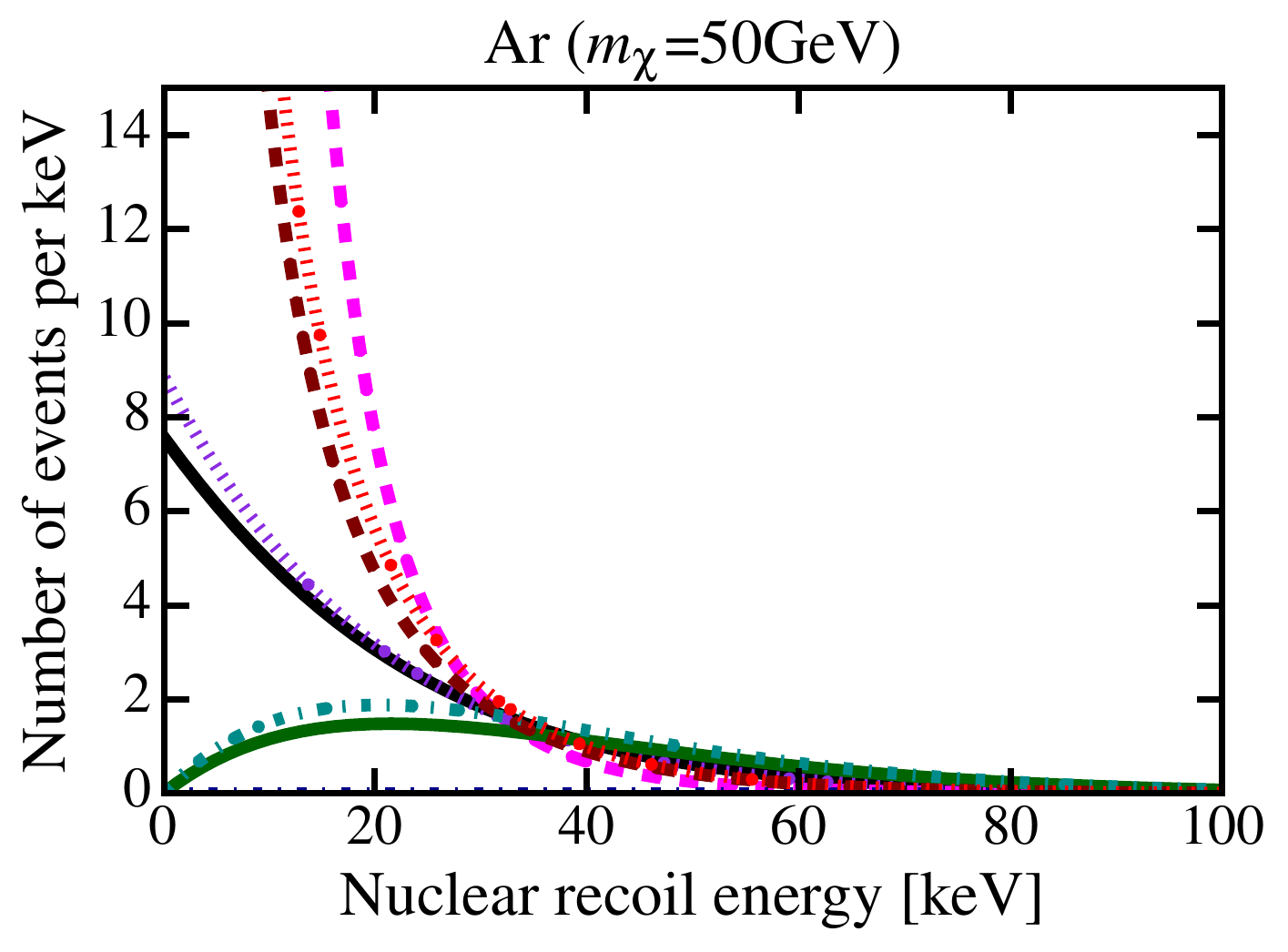}
\includegraphics[width=.31\textwidth,keepaspectratio=true]{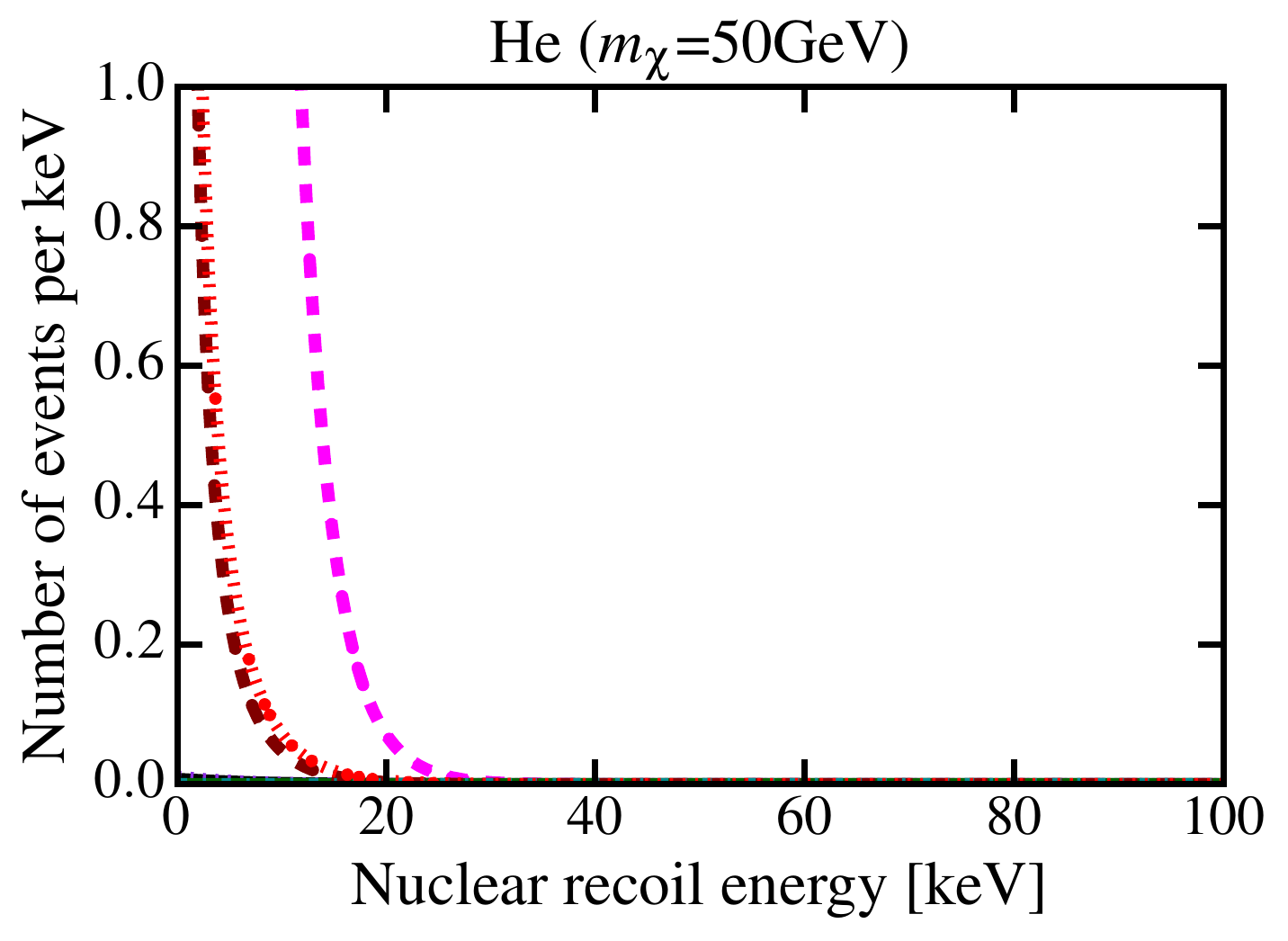}
\caption{Theoretical nuclear recoil energy spectra for the first 8 models  of Table \ref{tab:operators} (``set--I'' models used for our baseline analysis; see also discussion of models in \S\ref{sec:analysis}), shown for a variety of G2--like experiments (described in Table \ref{tab:experiments}). Spectra are calculated for a 50 GeV DM particle, with scattering cross sections set close to their current upper limits. \label{fig:models_set1}}
\end{figure*}

\begin{figure*}
\centering
\includegraphics[width=.31\textwidth,keepaspectratio=true]{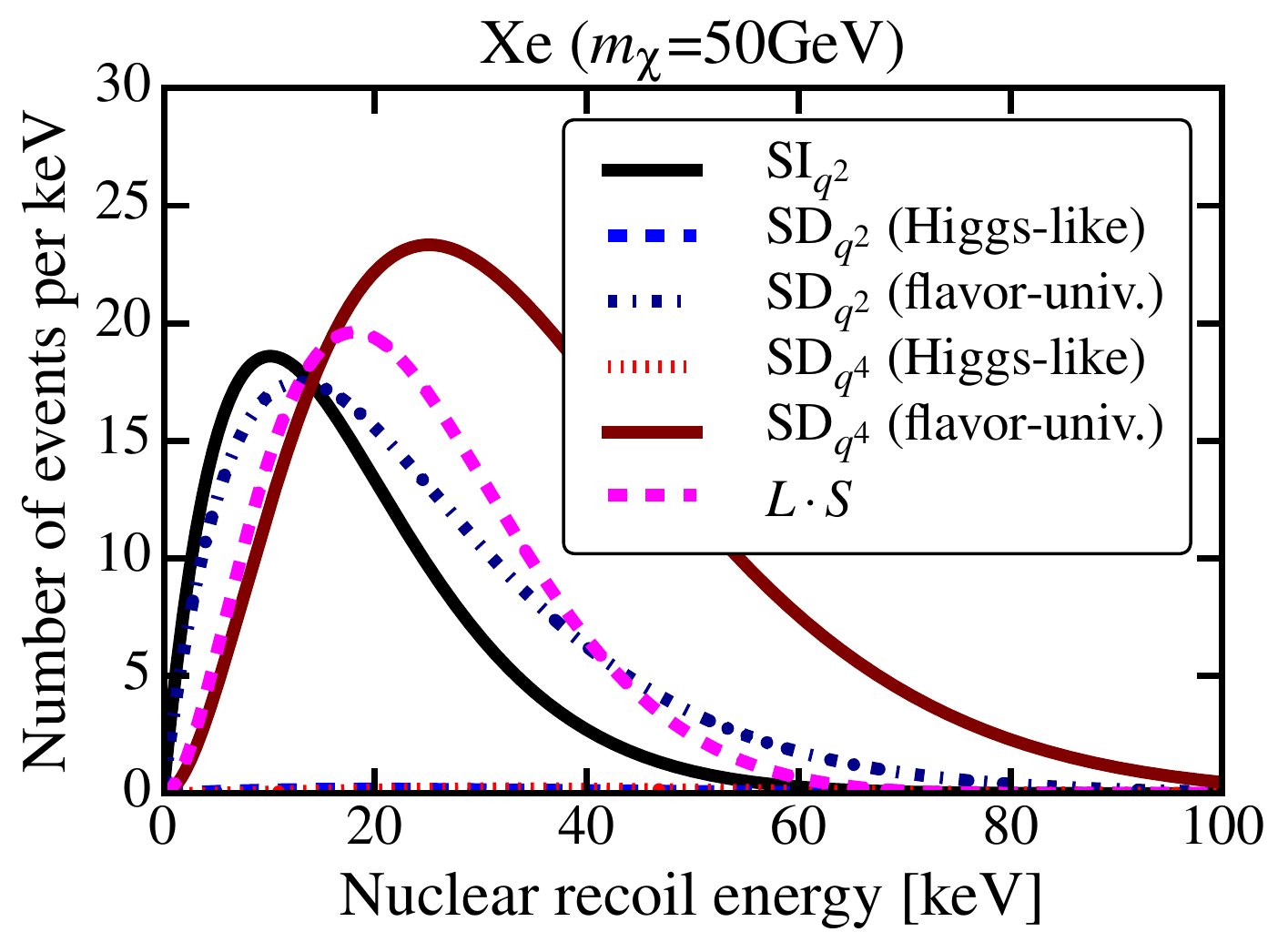}
\includegraphics[width=.31\textwidth,keepaspectratio=true]{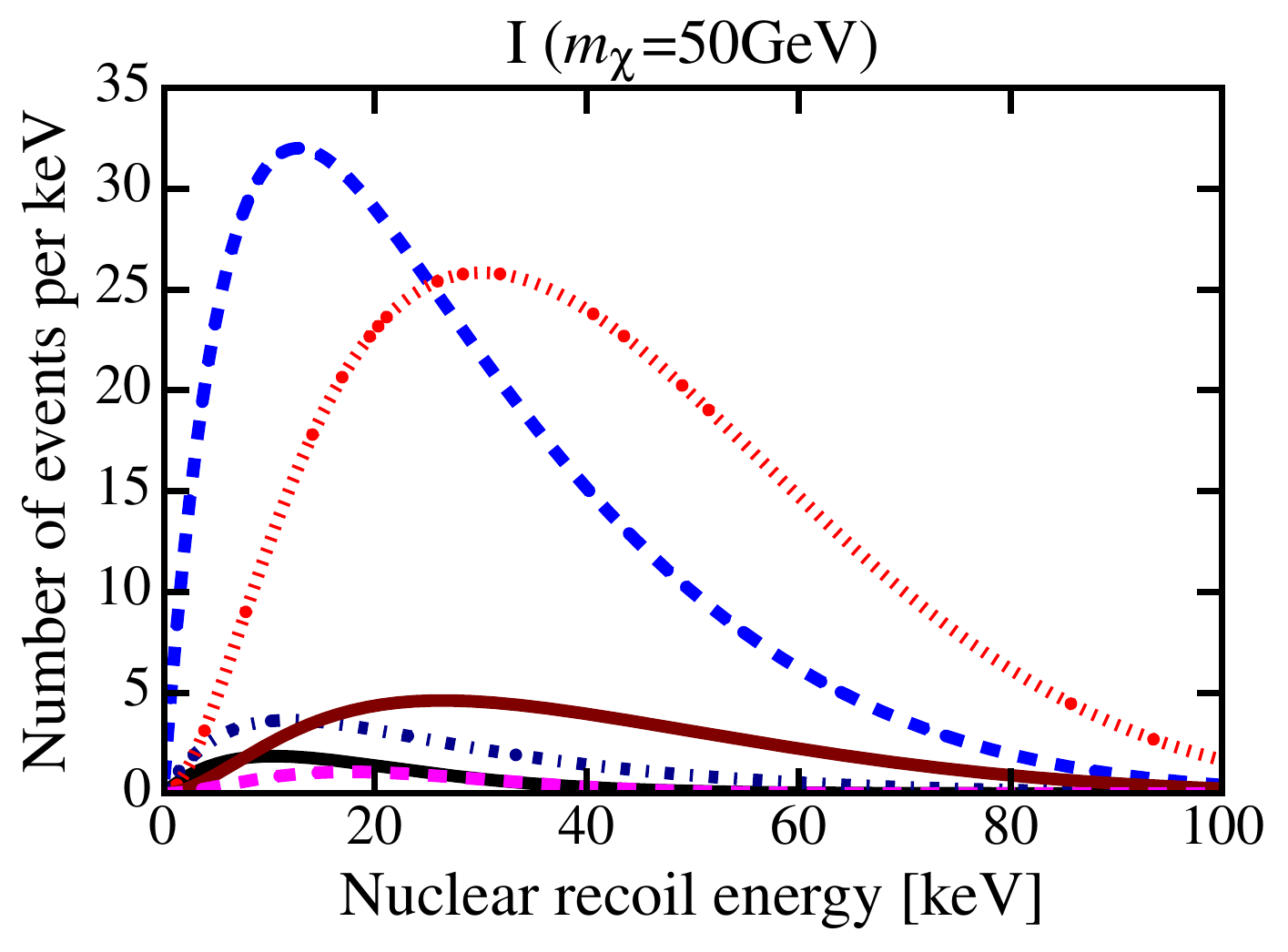}
\includegraphics[width=.31\textwidth,keepaspectratio=true]{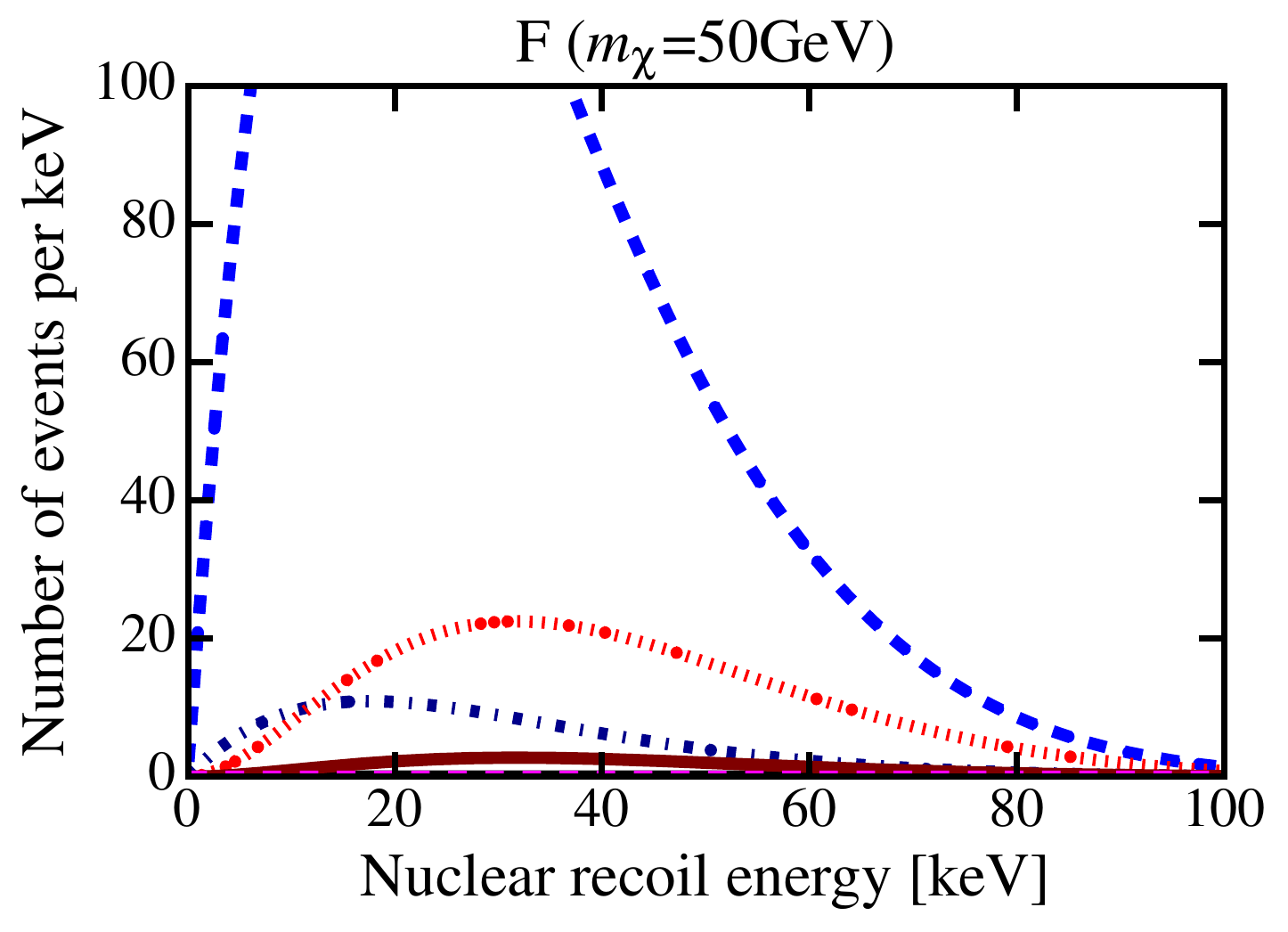}
\caption{Same as Figure \ref{fig:models_set1}, but for models of ``set II'' discussed in \S\ref{sec:analysis} (note that the MD and ED with heavy mediators are omitted from this Figure, for clarity, even though they are treated as members of set II). All of the models here have a suppressed scattering rate at the lowest nuclear--recoil energies, and a corresponding turnover in the event--energy spectrum.\label{fig:models_set2}}
\end{figure*}

The nuclear recoil--energy spectra produced by our scattering models are shown in Figures \ref{fig:models_set1} and \ref{fig:models_set2}.  
From these Figures, it is apparent how the momentum dependence of the leading response for each model gives rise to a distinct phenomenology. 
In particular, for germanium and xenon targets, a positive power of momentum transfer in the cross section ({\em e.g.}~$q^2$ and $q^4$ dependence for ED and MD models with heavy mediators, respectively) produces a suppression (turnover) of the rate going to smaller nuclear--recoil energy.   On the other hand, models with a negative power of momentum transfer (such as those with a mediator much lighter than momentum transfer, i.e.~the Millicharge, and the ED and MD with light mediators) show a sharp rise in the event rate for the lowest--energy recoils. Another striking feature is the target dependence of the spectra: as emphasized in previous literature \cite{McDermott:2011hx,Peter:2013aha,Gresham:2014vja,Gluscevic:2014vga}, the nuclear--target dependence of DM spectra will be essential for breaking degeneracies between different interaction models. We quantify this statement in detail in \S\ref{sec:results}. 

We consider a wide range of experimental setups, with a particular focus on G2 experiments. For the baseline analysis (results shown in \S\ref{sec:model_selection}), we focus on xenon (labeled as ``Xe''), germanium (``Ge''), iodine (``I''), and fluorine (``F'') targets, with an outlook towards some existing and proposed experiments (LZ \cite{LZLBL}, SuperCDMS Snolab \cite{SCDMSSnolab}, sodium--iodide experiments \cite{Amare:2014aka}, and fluorine--based bubble--chamber experiments \cite{PICO250L}, respectively). Later in \S\ref{sec:results}, we explore prospects for argon (``Ar''), sodium (``Na'')\footnote{For I and Na, as guidance, we adopt parameters of the proposed ANAIS--250 experiment \cite{2015PhPro..61..157A}, with appropriate weighting for sodium and iodine components of the target, and scale the proposed energy window (given in units of keVee) using quenching factors of Ref.~\cite{Xu:2015wha}.}, and helium (``He'')\footnote{For the He experiment we assume exposure larger than in the original experiment proposals: we assume availability of 100 kilograms of target and 3 years of exposure time. We make this choice in order to explore model--selection prospects for very low DM masses in \S\ref{sec:othertargets} with a helium target, and are interested only in its complementarity to other targets that are already available or in advanced planning stages.} targets (inspired by proposals of Refs.~\cite{Cushman:2013zza}, \cite{2015PhPro..61..157A,Amare:2014aka}, and \cite{2013PhRvD..87k5001G}, respectively). 
Furthermore, we consider somewhat modified versions of Xe and I, to quantify the impact of changes to the nuclear recoil--energy window on the results of model selection: ``Xe(lo)'' is equivalent to Xe with a lower energy threshold; Xe(hi) is equivalent to Xe with a higher upper end of the energy window; ``Xe(wide)'' is Xe with a wider energy coverage; and ``I(lo)'' is equivalent to I with a lower threshold. And finally, we consider a next--stage xenon experiment we name ``XeG3'' with exposure that reaches the irreducible neutrino background, and explore futuristic fluorine-- and iodine--based experiments ``F+'' and ``I+'' (with larger energy windows and exposures than those of I and F), in order to gain a sense of the ultimate reach of direct detection.
   
We assume perfect energy resolution for all experiments, except for F, for which we assume no resolution within the analysis window. We leave careful consideration of efficiency and experimental backgrounds for future analysis, and assume that the signal is entirely DM--dominated. We thus adjust the exposures of Xe and Ge to reproduce the exclusion curves presented in Ref.~\cite{Cushman:2013zza}, assuming zero background events. 
\begin{table*}[tbp]
  \setlength{\extrarowheight}{3pt}
  \setlength{\tabcolsep}{10pt}
  \begin{center}
	\begin{tabular}{c|m{2.3cm}m{4.2cm}m{2.8cm}}  
	Label & A (Z) & Energy window [keVnr] & Exposure [kg-yr] \\
	\hline
	Xe & 131 (54) & 5-40 & 2000 \\
	Ge & 73 (32) & 0.3-100 & 100  \\
	I & 127 (53) & 22.2-600 & 212 \\
	F &  19 (9) & 3-100 & 606 \\
	Na & 23 (11) & 6.7-200 & 38 \\
	Ar & 40 (18) & 25-200 & 3000 \\
	He & 4 (2) & 3-100 & 300 \\
	\hline
	Xe(lo) & 131 (54) & 1-40 & 2000 \\
	Xe(hi) & 131 (54) & 5-100 & 2000 \\
	Xe(wide) & 131 (54) & 1-100 & 2000 \\
	I(lo) & 127 (53) & 1-600 & 212 \\
	\hline
	XeG3 & 131 (54) & 5-40 & 40 000 \\
	I+ & 127 (53) & 1-600 & 424 \\
	F+ &  19 (9) & 3-100 & 1200 \\
	\end{tabular}
  \end{center}
\caption{Mock experiments considered in this work. The efficiency and the fiducialization of the target mass are included in the exposure. The first group of experiments is used for most of the simulations in this work and is chosen such to be representative of the reach of G2 experiments for Xe, Ge, I, and F. The exposure for Xe and Ge is chosen to agree with the projected exclusion curves for LZ and SuperCDMS presented in Ref.~\cite{Cushman:2013zza}. The second group of experiments is used to test impact of the energy window on prospects for model selection; note that only the energy window differs from the corresponding experiments of the first group. The last group represents futuristic experiments, where XeG3 reaches the level of atmospheric neutrino backgrounds. }
\label{tab:experiments}
\end{table*}

\begin{figure*}    
\centering 
\includegraphics[width=.3\textwidth,keepaspectratio=true]{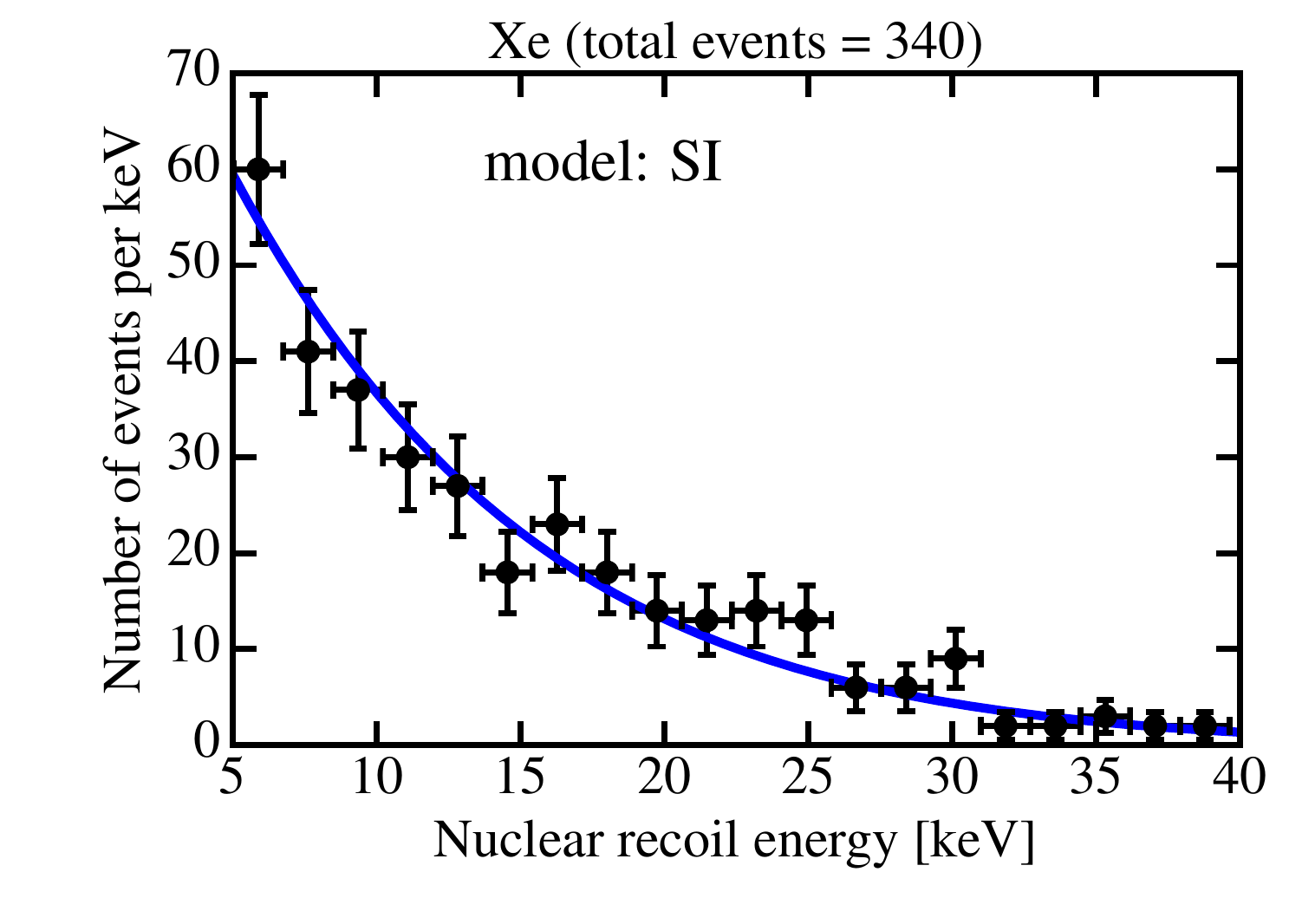}
\includegraphics[width=.3\textwidth,keepaspectratio=true]{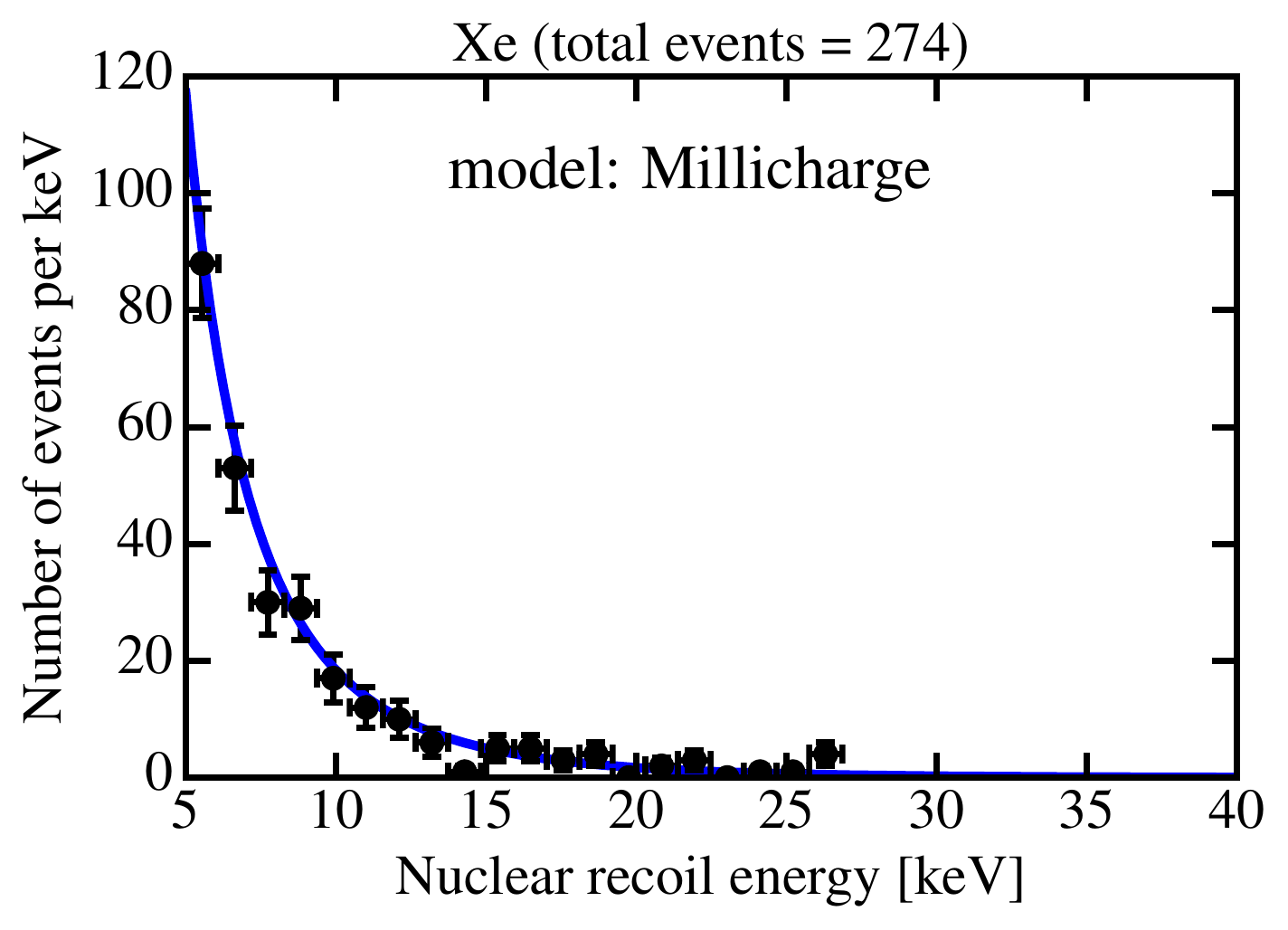}
\includegraphics[width=.3\textwidth,keepaspectratio=true]{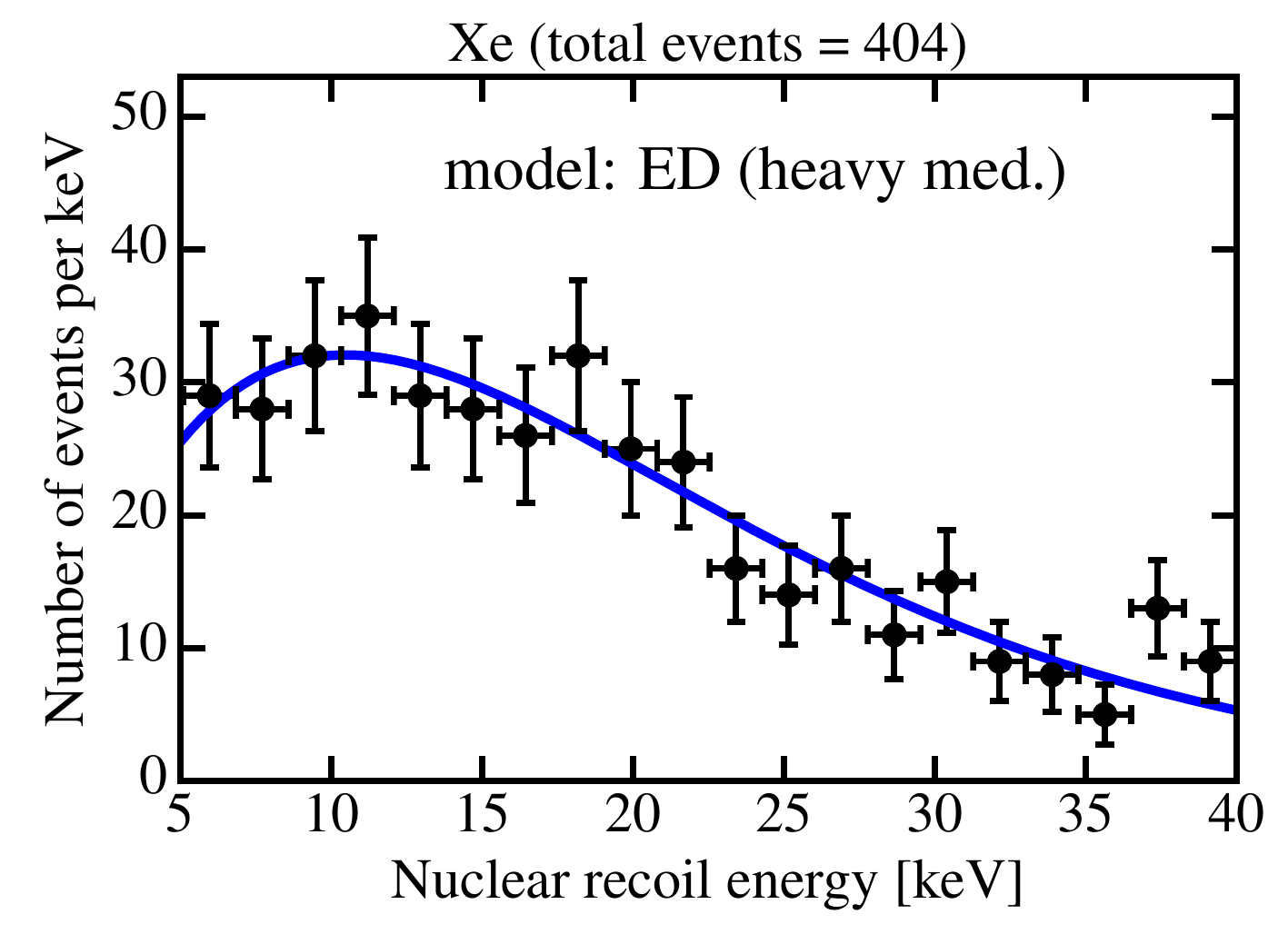}
\caption{Examples of simulated nuclear recoil energy spectra, for three different models from Table \ref{tab:operators}, on a Xe experiment described in Table \ref{tab:experiments}, for a 50 GeV DM particle, with a cross section set to current upper limit, calculated in \S\ref{sec:exclusions}. Error bars include only the Poisson noise. For this work, we create a large number of simulated spectra such as the ones shown here. For illustration purposes only, we bin the events according to their energy; we perform all analyses on unbinned data. \label{fig:spectra_examples}}
\end{figure*}

In order to statistically capture the impact of Poisson noise on future data analyses, we create a large number of simulations under each of the scattering hypotheses (allowing for independent realizations of the noise), following Ref.~\cite{Gluscevic:2014vga}. We repeat this procedure for all experiments in Table \ref{tab:experiments}, for all benchmark DM particle masses and $\sigma_p$ values (set to the current upper limit for our baseline analysis).  Examples of simulated spectra for three scattering models are shown in Figure \ref{fig:spectra_examples} to get a visual sense for the relevant level of noise.

To simulate a recoil--energy spectrum observed with a single experiment under a chosen scattering model $\mathcal{M}$, given a set of its parameter values $\Theta$ ($m_\chi$, $\sigma_p$, and $f_n/f_p$), we use the following procedure. For each simulation, we first draw a number $N$ from a Poisson distribution
\beq \label{eq:Poisson probability}
P(N|\langle N \rangle) = \frac{\left<N\right>^N e^{-\left<N\right>}}{N!},
\eeq
whose mean is the expected number of events $\left<N\right>$.
We then assign energies to each of these $N$ events following the probability distribution for observing a single event at a given energy,
\begin{equation}
P_1(E_R | \Theta, \mathcal{M}) = \frac{dR/dE_R}{R}.
\label{eq:P1}
\end{equation}
This way, a single ``data set'' (list of nuclear--recoil energies $\{E_R\}$) is obtained, including Poisson noise. Repeating this procedure, we create a large number of simulations under scattering model $\mathcal{M}$ (defined by the choice of ${dR/dE_R}$), for a fixed choice of its parameter values $\Theta$, and thus obtain a statistical representation of possible outcomes of future observations. 

Before moving on to describing the analysis of the simulated data sets, we again emphasize the importance of this statistical approach: given the absence of a confirmed DM signal from present--day experiments, we already know that the limitations from Poisson noise may severely impact the prospects for identifying the right underlying interaction \cite{Gluscevic:2014vga}. Making a statement about probable outcomes of future analyses from a single data realization is thus not informative; it is an essential feature of our approach to take this into account and evaluate possible experimental outcomes in a probabilistic sense.
\section{Analysis}
\label{sec:analysis}

The choice of analysis framework adopted here is Bayesian inference, and in particular, Bayesian model selection. We analyze each simulated energy spectrum either by itself or in combination with spectra from other experiments. The main computational step is evaluation of the posterior probability $\mathcal{P}(\Theta | \{E_R\},M)$, as a function of parameters $\Theta$ of the model (hypothesis) $\mathcal{M}$, given simulated nuclear recoil energies $\{E_R\}$,
\begin{equation}
\mathcal{P}(\Theta | \{E_R\},\mathcal{M}) = \frac{\mathcal{L}(\{E_R\} | \Theta, \mathcal{M})p(\Theta  | \mathcal{M})}{\mathcal{E}(\{E_R\} | \mathcal{M})}.
\label{eq:bayes_theorem}
\end{equation}
In our case, $\Theta$ is a vector of the model parameters: $m_\chi$, $\sigma_p$, and, optionally, $f_n/f_p$; note that $f_n/f_p$ is held fixed in most of our analyses, unless otherwise noted (specifically, it is only a free parameter for the purposes of \S\ref{sec:fnfp}). The nuclear response functions are fixed.  All astrophysical parameters are fixed to the values in \S\ref{sec:scattering}.\footnote{In Appendix \ref{app:nuclear} and \ref{app:astrophysical}, we discuss model selection in the context of existing nuclear and astrophysical uncertainties.} The posterior is reconstructed as a product of the likelihood $\mathcal{L}(\{E_R\} | \Theta, \mathcal{M})$ (the probability of data\footnote{If experiments are ``jointly analyzed'', a combined likelihood for the corresponding set of several energy spectra is evaluated.}, given theory) and the prior $p(\Theta  | \mathcal{M})$. We choose wide priors on all parameters, in order to remain as agnostic as possible\footnote{For $m_\chi$ and $\sigma_p$ our prior is spaced logarithmically between 1--1000 GeV and within 8 orders of magnitude around the relevant normalization, respectively. When $f_n/f_p$ is a free parameter, we assign it a large flat prior, discussed in detail in \S\ref{sec:fnfp}.}.  The normalization in \eq{eq:bayes_theorem} is the evidence of model $\mathcal{M}$, i.e.~the integral of the likelihood over the entire prior parameter space,
\begin{equation}
\mathcal{E}(\{E_R\} |\mathcal{M})=\int d\Theta \mathcal{L}(\{E_R\} | \Theta, \mathcal{M})p(\Theta | \mathcal{M}),
\label{eq:evidence}
\end{equation}
which can be thought of as a measure of how well model $\mathcal{M}$ fits the data overall. The likelihood in \eq{eq:bayes_theorem} takes into account the energy distribution of all recoil events,
\begin{equation}
\mathcal{L}(\{E_R\} | \Theta, \mathcal{M})=P(N|\Theta, \mathcal{M})\prod_{i=1}^{N} P_1(E^i_R | \Theta, \mathcal{M}),
\end{equation} 
where $P$ and $P_1$ are defined in Eqs.~(\ref{eq:Poisson probability}) and (\ref{eq:P1}), and the product is over all observed events. (In the case of the F experiment that has no energy resolution, the likelihood is taken to be only the Poisson factor.) To reconstruct the posterior probability on the relevant parameter space, we use \texttt{PyMultiNest} \cite{Buchner:2014nha} (a Python wrapper of the \texttt{MultiNest} multi--modal nested sampler \cite{2009MNRAS.398.1601F}), with the following parameters: \texttt{efr}$=0.3$, \texttt{tol}$=0.01$, and $2000$ live points.

Within the Bayesian framework, the probability the data $\{E_R\}$ assigns to the model $\mathcal{M}_j$ (where $\mathcal{M}_j$ is one of the models listed in Table \ref{tab:operators}) is given by the ratio of its evidence to the sum of evidences of all the competing hypotheses,
\begin{equation}
\text{Pr}(\mathcal{M}_j) = \frac{\mathcal{E}(\{E_R\} |\mathcal{M}_j)}{\sum\limits_{i}^\mathrm{}\mathcal{E}(\{E_R\} |\mathcal{M}_i)}.
\label{eq:E_ratio}
\end{equation}
Data are said to ``prefer'' a given model if the corresponding evidence ratio evaluated from Eq.~(\ref{eq:E_ratio}) is high.  To summarize our results in a succinct manner in the following Section, we introduce a (somewhat arbitrary) label of ``successful model selection'' for the case where the right underlying model is assigned more than $90\%$ probability; in this case, we say that the right model was ``confidently selected.'' Conversely, data is ``indecisive'' if it assigns about equal probability to several competing models.

For the purposes of model selection, we divide the 14 scattering models listed in Table \ref{tab:operators} into two groups of competing hypotheses, ``set I'': SI, SD, Anapole, Millicharge, MD (light med.), MD (heavy med.), ED (light med.), ED (heavy med.); and ``set II'': SI$_{q^2}$, SD$_{q^2}$ (Higgs-like), SD$_{q^2}$ (flavor univ.), SD$_{q^4}$ (Higgs-like), SD$_{q^4}$ (flavor univ.), $\vec L \cdot \vec S$-like, MD (heavy med.), and ED (heavy med.). We compare evidences of models in set I against each other (for nuclear recoil spectra simulated under those models), and do the same separately for set II. This division of models is physically motivated: models of set I include the two standard interactions, supplemented by six photon--mediated scattering scenarios, and are representative of the theoretically best--motivated spectral features that may arise in recoil spectra (driven by specific momentum dependence of the scattering interaction at hand). Moreover, set I comprises a smaller number of hypotheses than the entire Table \ref{tab:operators}, and its statistical analysis is more computationally tractable. Most of the model--selection results presented in \S\ref{sec:results} focus on set I. 
On the other hand, models of set II incorporate all models of Table \ref{tab:operators} that show a turnover in recoil rate at low recoil energies, and additionally isolate nuclear responses that are subdominant or present in different linear combinations in set I. In addition to MD (heavy med.)~and ED (heavy med.)~which overlap with set I, they also include the pseudoscalar--mediated and $\vec L \cdot \vec S$--like scattering. This subset expands the coverage of modeling space and lets us investigate distinguishability amongst a pool of models that give rise to qualitatively similar spectral features. 
\section{Results}
\label{sec:results}

As a first step, in \S\ref{sec:exclusions} we calculate current upper limits of the scattering cross section for each model considered in this work, and make predictions for the maximum number of recoil events G2 experiments can hope to see for each model. We then present the results of model selection in \S\ref{sec:model_selection}. In \S\ref{sec:mass}, we examine the projected quality of DM mass measurements under different models. In \S\ref{sec:fnfp}, we discuss the impact of uncertainties on the $f_n/f_p$ parameter on both mass reconstruction and on the success of model selection. In order to provide guidance for future efforts, we investigate the dependence of the recoil signal and the prospects for model selection on the choice of recoil energy window used for data analysis, in \S\ref{sec:experimental_params}. 
\subsection{Exclusions and expectations}
\label{sec:exclusions}

\begin{figure}[h]
\centering
\begin{tabular}{c c}
\includegraphics[width=.49\textwidth,keepaspectratio=true]{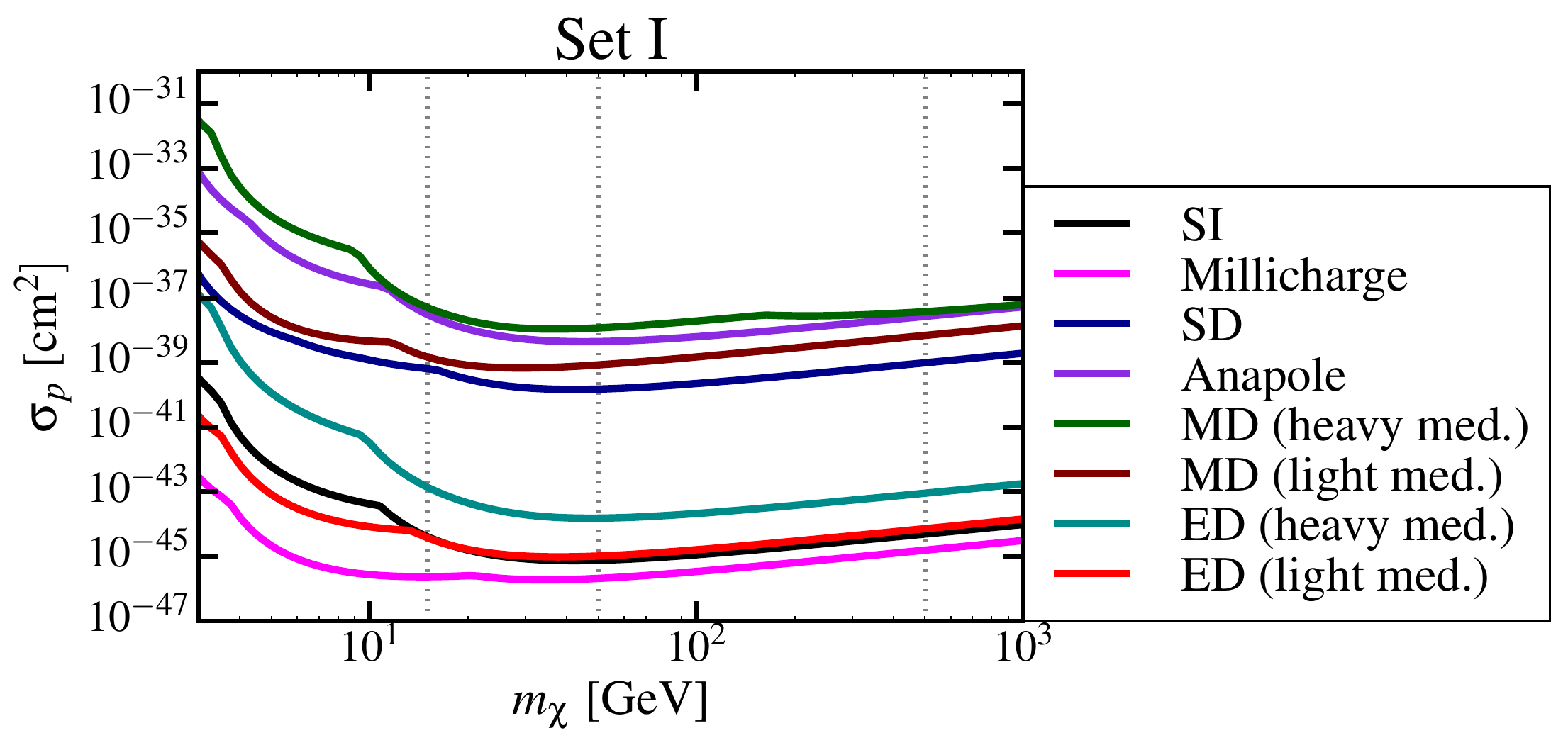} &
\includegraphics[width=.49\textwidth,keepaspectratio=true]{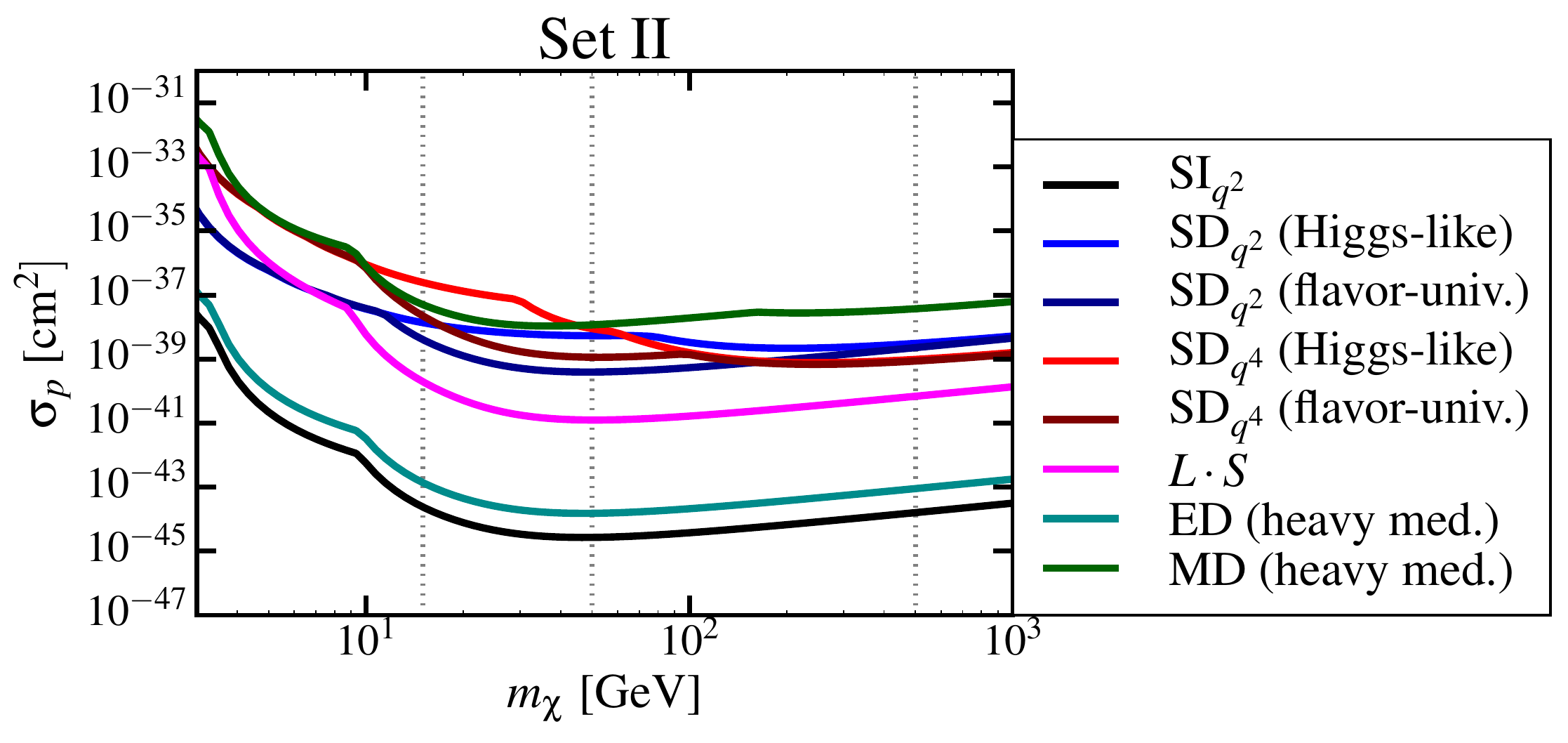} 
\end{tabular}
\caption{Current upper limits on $\sigma_p$ for all DM--nucleon scattering models considered in this work; see Eqs.~(\ref{eq:sigma_p_SI_SD}), (\ref{eq:sigma_p_photon_heavy}), (\ref{eq:sigma_p_photon_light}), (\ref{eq:sigma_p_PS}), and (\ref{eq:sigma_p_LS}) for definitions of $\sigma_p$ for different models. We calculate the upper limits assuming the absence of DM signal on the LUX, SuperCDMS, KIMS, and PICO experiments, and taking $f_n/f_p$ values given in Table \ref{tab:operators}. Models in the left panel constitute ``set I'' discussed in \S\ref{sec:analysis} (top to bottom, on the far left:  MD (heavy med.), Anapole, MD (light med.), SD, ED (heavy med.), SI, ED (light med.), and Millicharge); models on the right, ``set II'', are those that show a turnover in the recoil spectrum (from top to bottom, on the far right: MD (heavy med.), SD$_{q^2}$ (Higgs--like) and SD$_{q^2}$ (flavor--univ.) (these two overlap at this point, but SD$_{q^2}$ (Higgs--like) is the higher line at lower masses), SD$_{q^4}$ (Higgs--like) and  SD$_{q^4}$ (flavor--univ.) (the last two overlap at this point, but SD$_{q^4}$ (flavor--univ.) is the lower line at lower masses), $\vec L\cdot \vec S$--like, ED (heavy med.), and SI$_{q^2}$). Vertical dotted lines mark the  three main DM--mass benchmarks used for the simulations of our baseline model--selection analysis presented in \S\ref{sec:baseline}. \label{fig:exclusions}}
\end{figure}

In order to simulate DM signals just beyond the reach of current experiments, we first estimate current upper limits of the scattering cross section for each model considered in Table \ref{tab:operators}. The corresponding exclusion curves are shown in Figure \ref{fig:exclusions}. They are calculated under the assumption that the  number of DM--induced events has not exceeded the number of expected background events in LUX \cite{Akerib:2013tjd}, SuperCDMS \cite{Anderson:2014fka}, CDMSlite \cite{Agnese:2013jaa}, PICO \cite{Amole:2015lsj}, and KIMS\footnote{For KIMS, which is not near--background--free like the other experiments, we obtain the ``exclusions'' by setting the number of DM--induced recoils to be less than $\sqrt{N_\text{observed}}$, i.e.~we assume the signal is not larger than the Poisson noise of the background events.} \cite{Lee:2014zsa}. In Figure \ref{fig:exclusions}, exclusion limits from individual experiments are combined into a single curve by setting the upper limit equal to the most stringent constraint from the individual experiments. For the models where limits are available, we have checked that our limits qualitatively reproduce the more exact ones available from individual collaborations. However, even a rough estimation of upper limits suffices for our purposes: we simply use it to inform a choice of benchmark cross sections for simulating the data that upcoming and future experiments might hope to see.

\begin{table*}[t]
\setlength{\extrarowheight}{1pt}
\setlength{\tabcolsep}{3pt}
\begin{center}
\begin{tabular}{l|lll|lll|lll|lll}
Model & Xe&&& Ge&&& I&&& F&&\\
\hline 
$m_\chi$ [GeV]  & 15 & 50 & 500 & 15 & 50 & 500 & 15 & 50 & 500 & 15 & 50 & 500 \\ \hline
\hline
{SI}  & (290, &331, &383)& (61, &14, &17)& (0, &6, &16)& (32, &5, &6)\\
{SD}  & (239, &334, &390)& (59, &15, &18)& (0, &11, &86)& (12922, &2724, &2975)\\
{Anapole} & (290, &330, &386)& (79, &15, &19)& (0, &52, &382)& (1057, &283, &420)\\
{Millicharge} & (116, &294, &301)& (2591, &741, &554)& (0, &1, &2)& (916, &363, &303)\\
{MD (heavy med.)}& (290, &428, &227)& (14, &12, &11)& (0, &129, &1449)& (77, &79, &92)\\
{MD (light med.)}& (290, &309, &325)& (217, &49, &44)& (0, &14, &65)& (1635, &819, &1110)\\
{ED (heavy med.)}& (290, &402, &529)& (22, &12, &25)& (0, &17, &51)& (8, &2, &2)\\
{ED (light med.)}& (290, &305, &325)& (466, &57, &46)& (0, &2, &6)& (277, &43, &38)\\
SI$_{q^2}$& (290, &395, &513)& (19, &11, &22)& (0, &15, &45)& (6, &1, &2)\\
SD$_{q^2}$ (Higgs-like)& (1, &6, &1)& (0, &0, &0)& (0, &676, &1195)& (8558, &6285, &899)\\
SD$_{q^2}$ (flavor-univ.)& (290, &458, &652)& (18, &9, &19)& (0, &65, &1009)& (2432, &444, &640)\\
SD$_{q^4}$ (Higgs-like)& (2, &5, &0)& (0, &0, &0)& (0, &1124, &1300)& (6817, &1109, &44)\\
SD$_{q^4}$ (flavor-univ.)& (290, &655, &148)& (9, &10, &5)& (0, &182, &1277)& (591, &137, &38)\\
$L \cdot S$& (290, &500, &714)& (11, &13, &39)& (0, &14, &110)& (5, &2, &3)\\
\end{tabular}
\end{center}
\caption{Expected number of nuclear--recoil events for our mock G2 experiments (as defined in Table \ref{tab:experiments}), for each of the scattering models discussed in this work. Cross sections for scattering are set to their current upper limits, presented in Figure \ref{fig:exclusions}. The three entries in parentheses correspond to 15, 50, and 500 GeV DM masses, in order. }
\label{tab:Nexp}
\end{table*}

\begin{figure*}[h]
\includegraphics[width=.25\textwidth,keepaspectratio=true]{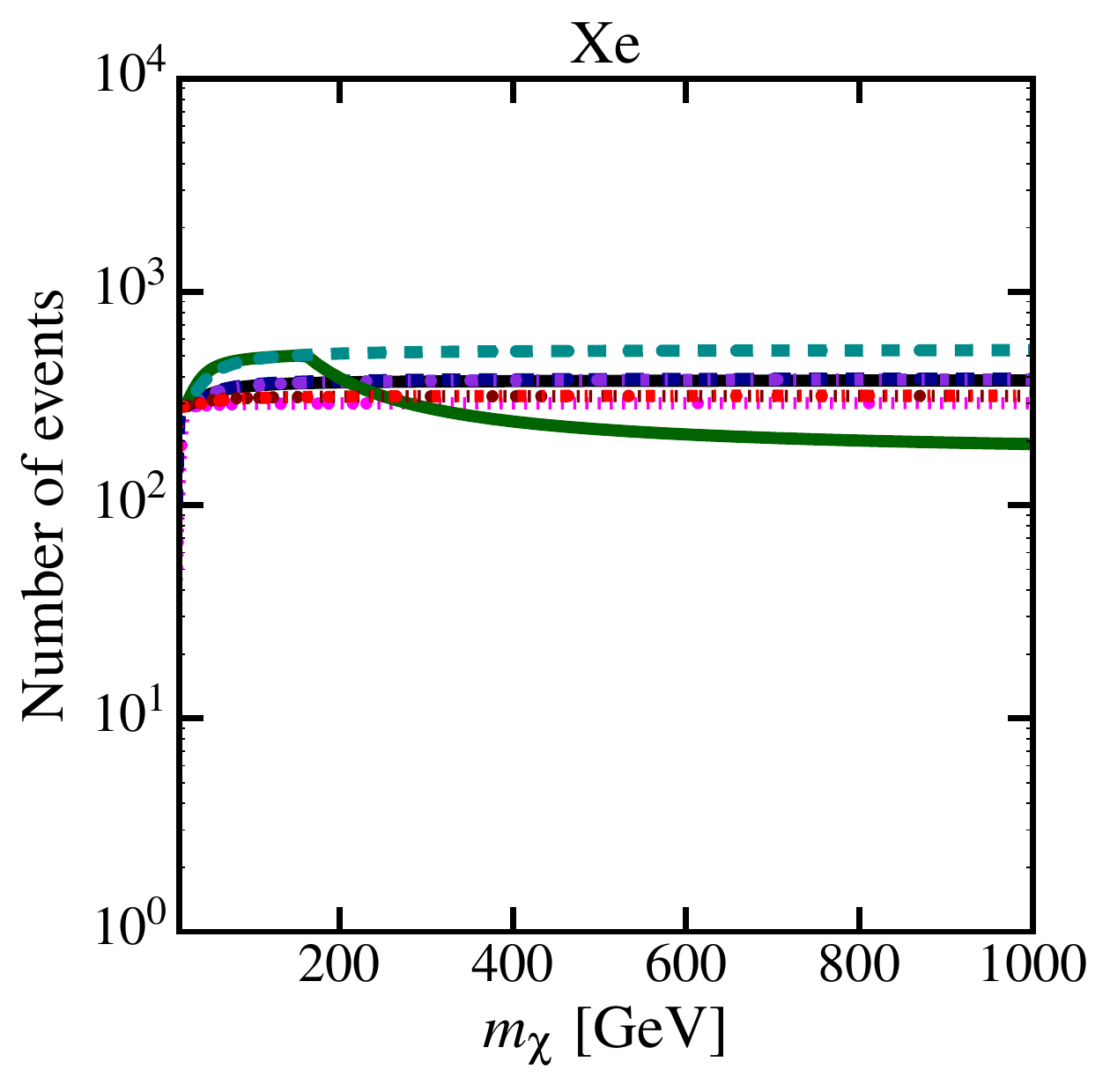} 
\includegraphics[width=.25\textwidth,keepaspectratio=true]{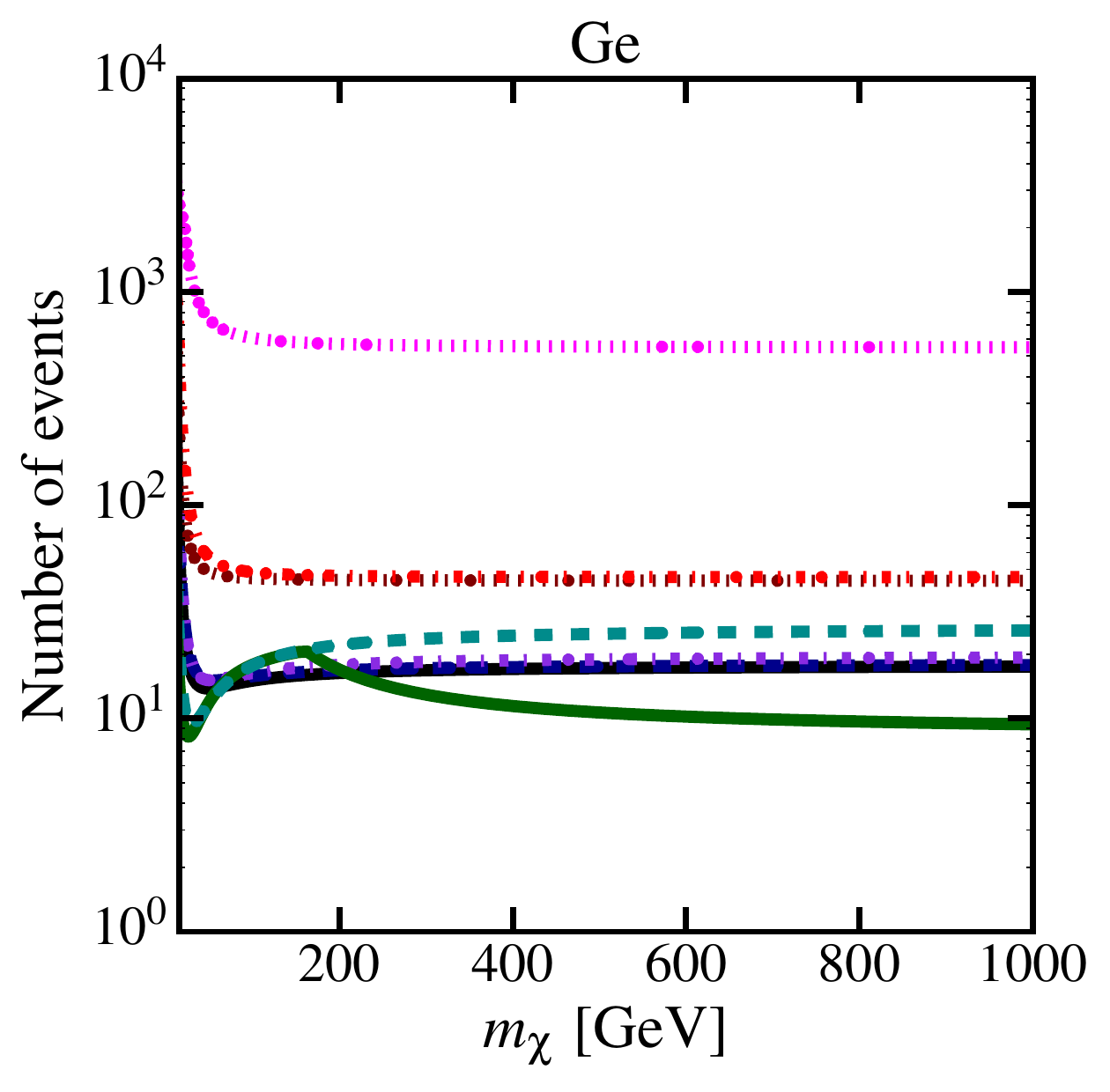}
\includegraphics[width=.39\textwidth,keepaspectratio=true]{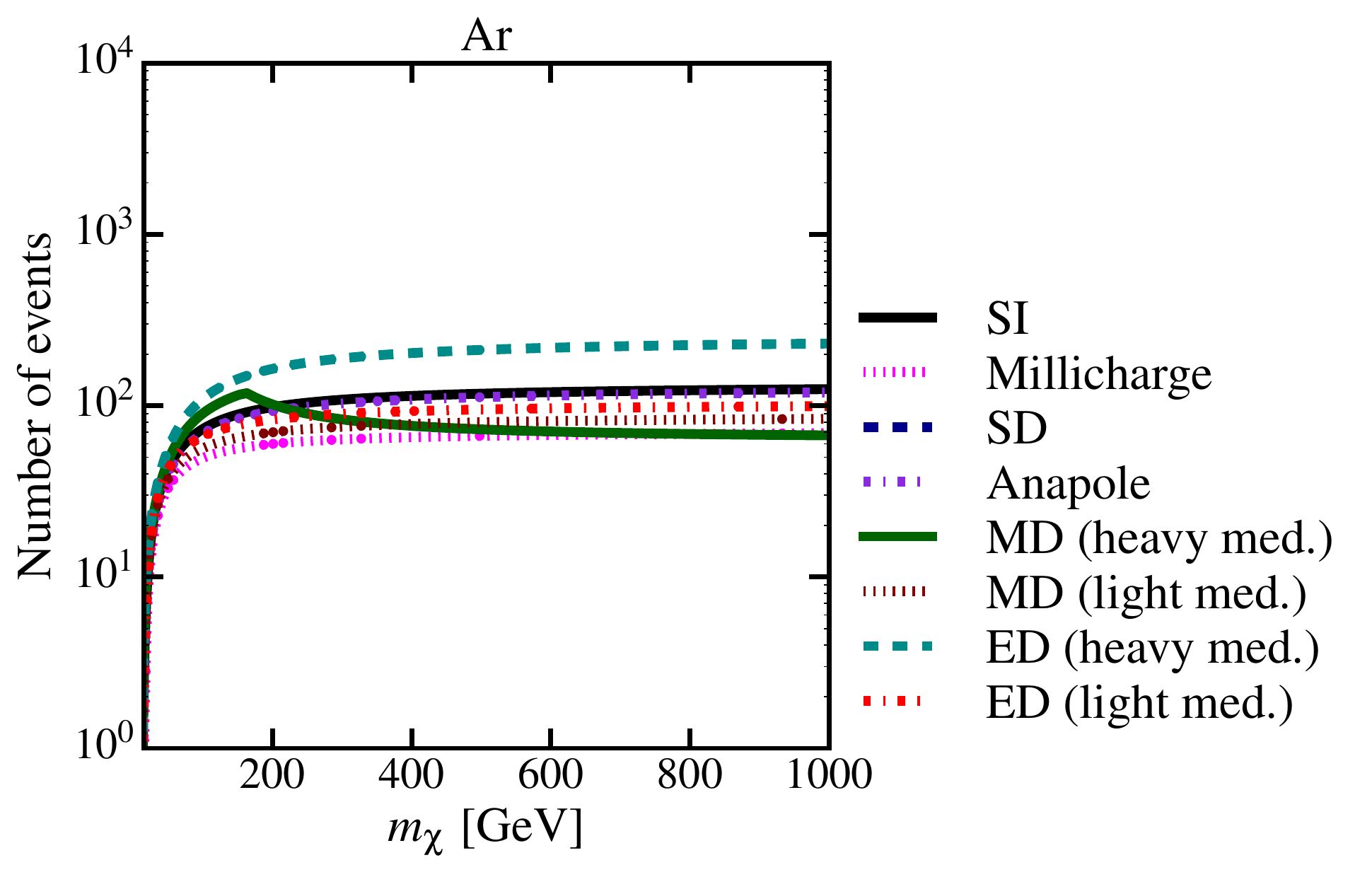}\\
\includegraphics[width=.25\textwidth,keepaspectratio=true]{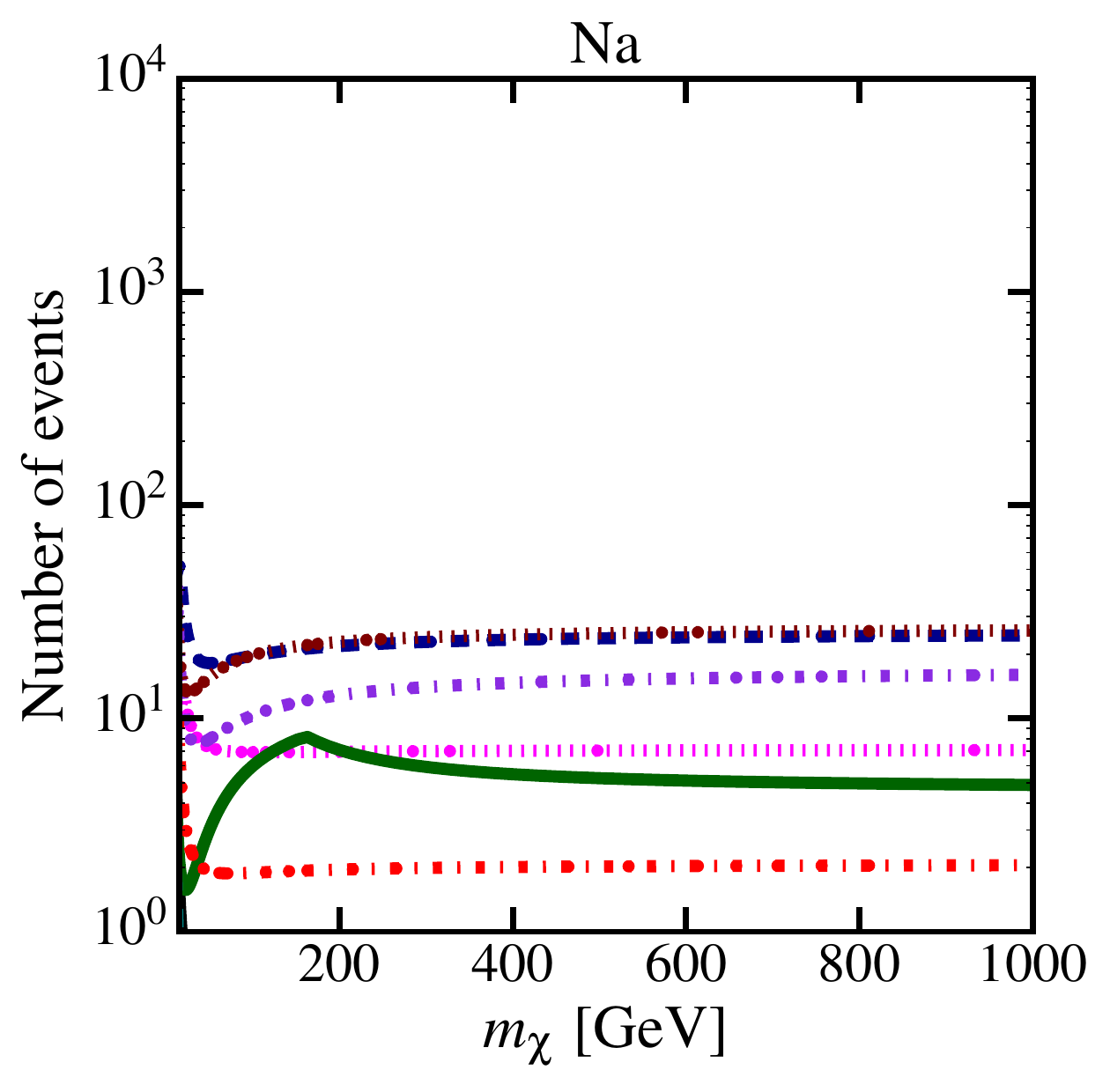}
\includegraphics[width=.25\textwidth,keepaspectratio=true]{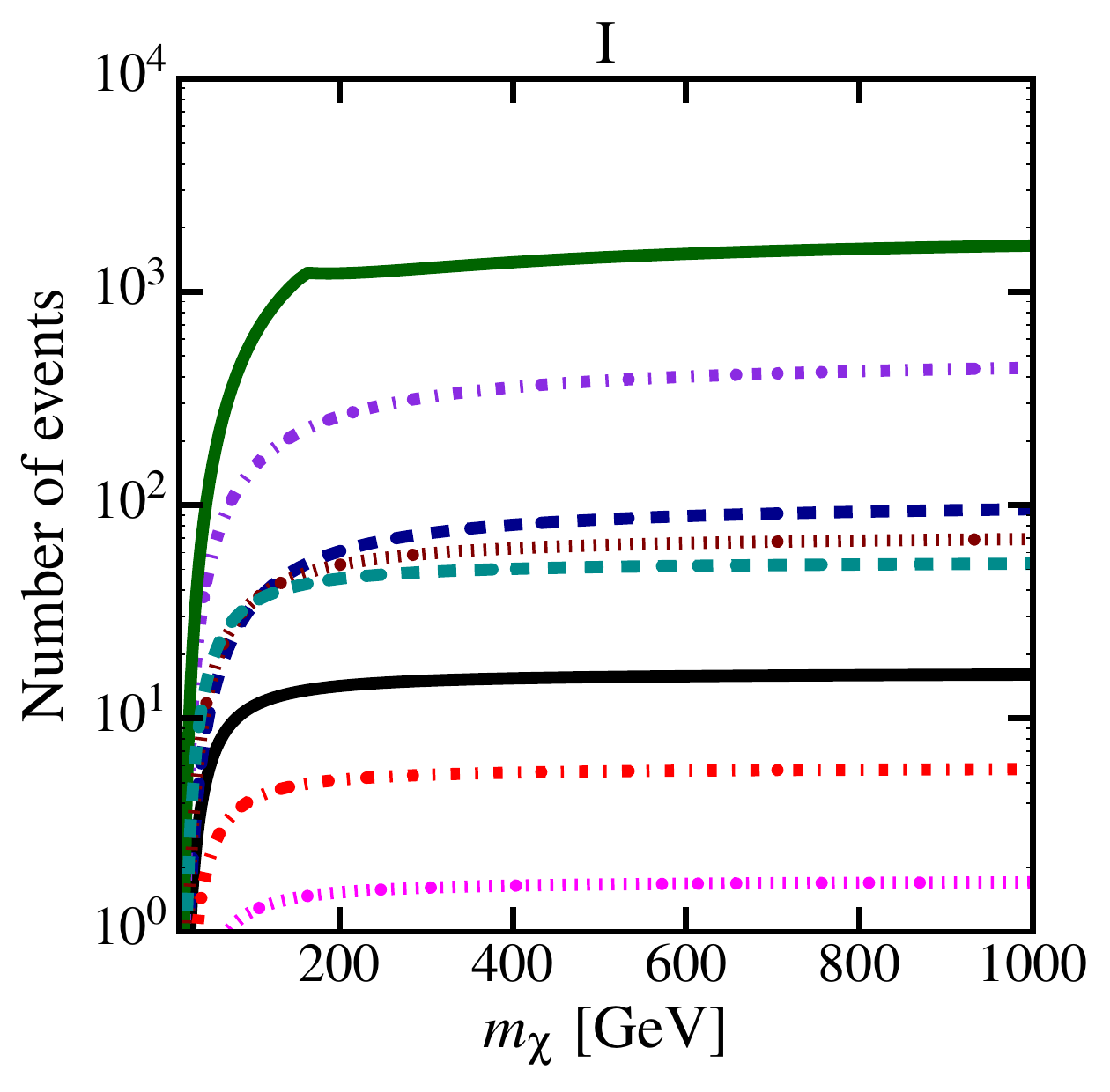}
\includegraphics[width=.25\textwidth,keepaspectratio=true]{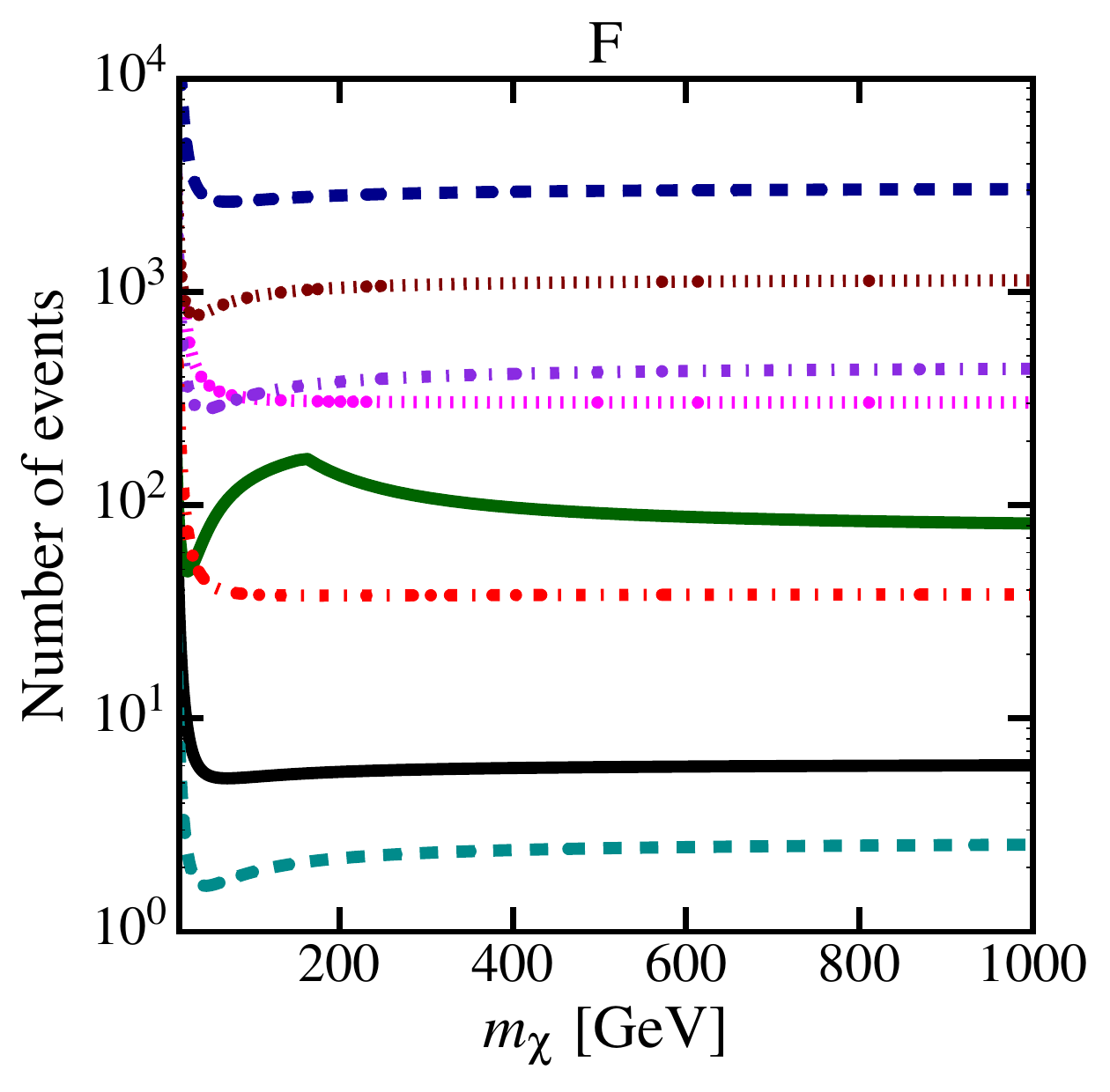}
\caption{The number of expected events as a function of DM mass for each  DM--nucleon scattering model of ``set I'' (discussed in \S\ref{sec:analysis} and summarized in Table \ref{tab:operators}) on a variety of G2 targets with experimental parameters listed in Table \ref{tab:experiments}. Cross sections are set to their respective upper limits; see Figure \ref{fig:exclusions}.\label{fig:Nexps1}}
\end{figure*}

\begin{figure*}[h]
\centering
\includegraphics[width=.25\textwidth,keepaspectratio=true]{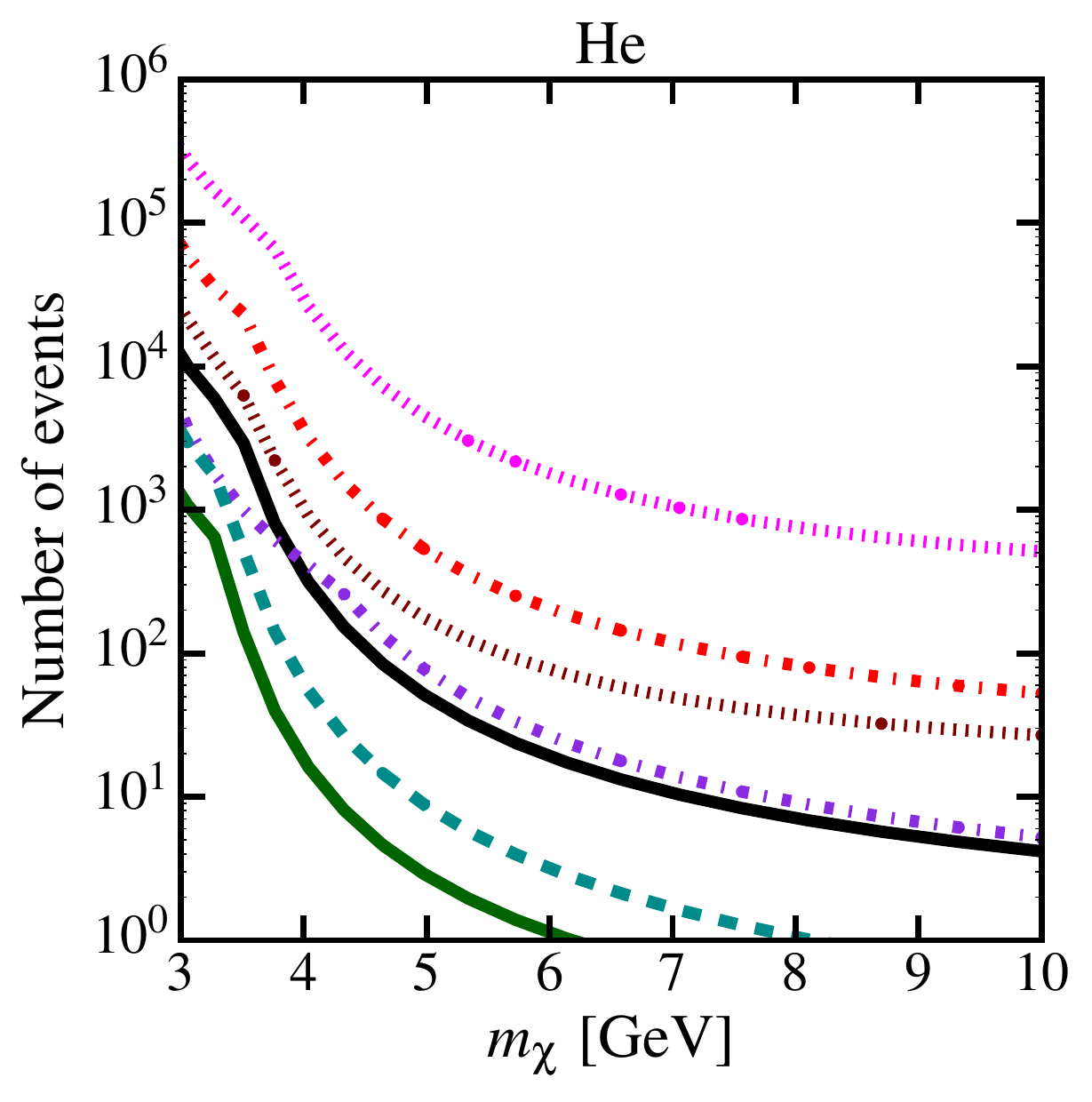}
\includegraphics[width=.25\textwidth,keepaspectratio=true]{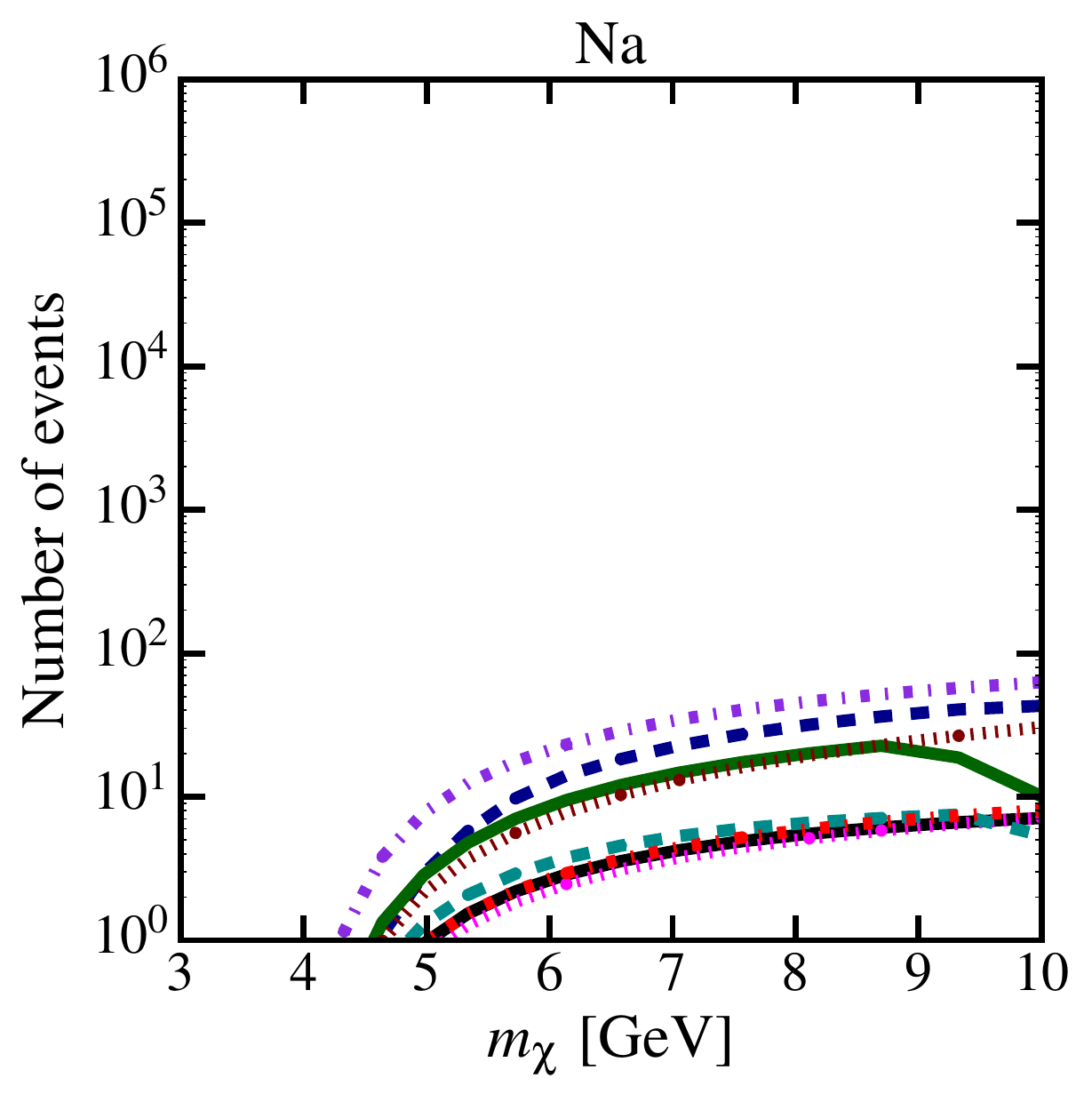}
\includegraphics[width=.39\textwidth,keepaspectratio=true]{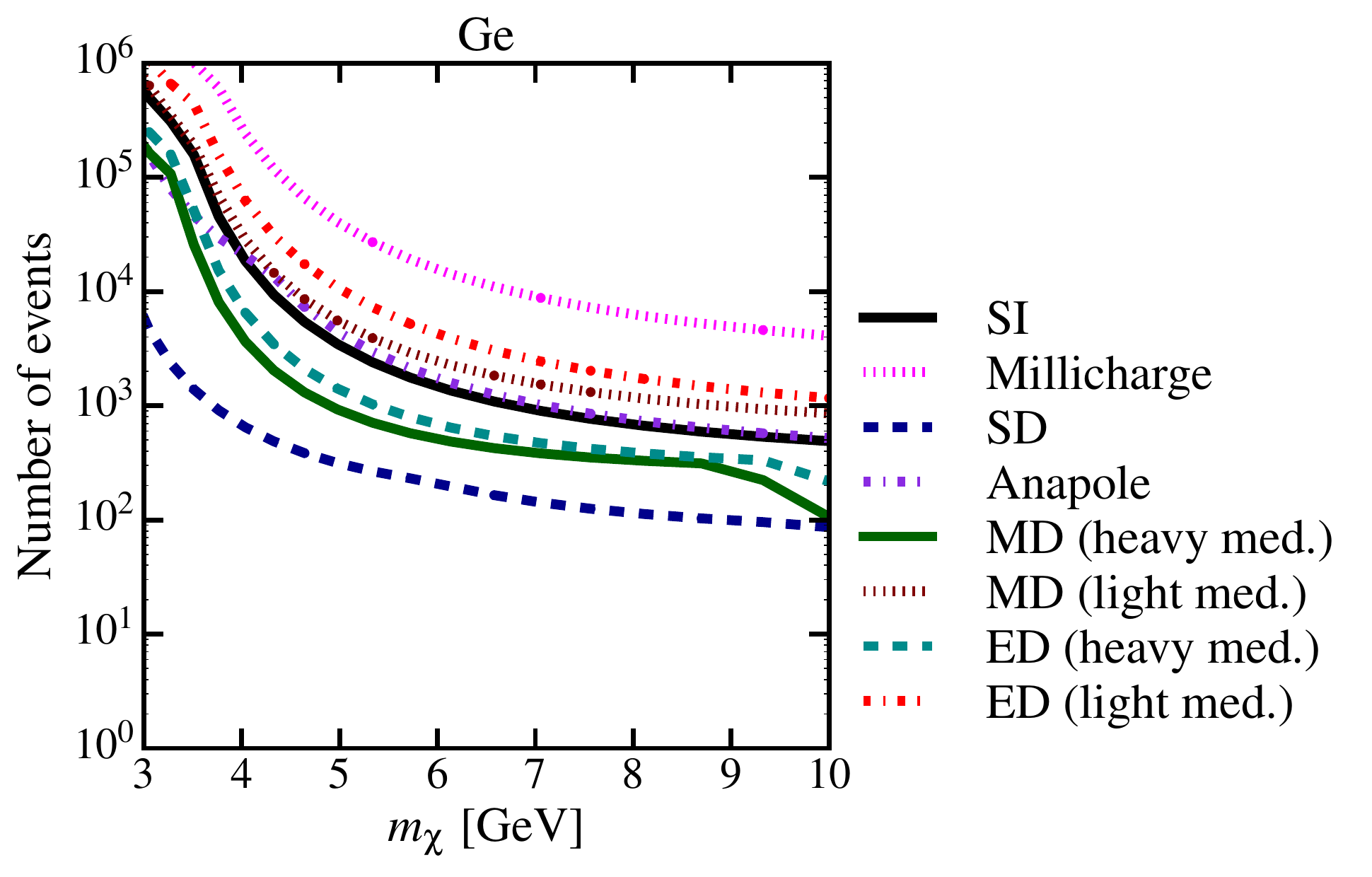}
\caption{Same as Figure \ref{fig:Nexps1}, but for low--mass DM, and only for targets that are kinematically favorable for low--mass DM detection.\label{fig:Nexps_lowm}}
\end{figure*}

We use the upper limits for $\sigma_p$ at the three benchmark DM masses denoted with dotted lines in Figure \ref{fig:exclusions} (15 GeV, 50 GeV, and 500 GeV) to create simulations used for our baseline model--selection analysis in \S\ref{sec:baseline}; the same benchmarks are also used when exploring the accuracy of mass reconstruction in \S\ref{sec:mass}, and the uncertainty on $f_n/f_p$ in \S\ref{sec:fnfp}. In \S\ref{sec:othertargets} where we explore targets that favor detection of low--mass DM particles, we introduce another benchmark of a 7 GeV DM particle mass, and use appropriate upper--limit values for $\sigma_p$. In \S\ref{sec:G3}, we analogously compute projected upper limits assuming that G2 Xe, Ge, I, and F experiments do not see a signal, and simulate data using those limits in order to investigate the ultimate reach of futuristic experiments.
To illustrate the statistical sample that represents the most optimistic G2 output under a particular model, we list the number of expected events for our main mock experiments in Table \ref{tab:Nexp}; this is representative of the typical numbers of events in simulations used for the baseline model--selection analysis of \S\ref{sec:baseline}. 
To show that the three main benchmark DM masses we choose for our baseline simulations are representative of the statistical sample that might arise for a wide range of DM particle masses, we show the total number of events each one of the experiments in Table \ref{tab:experiments} would see for models in set I, as a function of DM mass, in Figures \ref{fig:Nexps1} and \ref{fig:Nexps_lowm}. These Figures also provide a rough visualization of the sensitivity of different experiments to a wide range of DM masses, under different scattering models; when interpreting results of \S\ref{sec:baseline} and \S\ref{sec:othertargets}, looking back at these Figures will be particularly useful to qualitatively understand how different targets break degeneracy between some of the models.   
\subsection{Model selection}
\label{sec:model_selection}

In this Section, we present the key results of this work: we address the question of how likely it is that direct detection experiments will correctly identify the right underlying scattering theory from a wide variety of competing hypotheses. In addition, we try to identify the particle physics information that we can robustly extract from the same data. 

This Section is organized as follows. In \S\ref{sec:baseline}, we consider Xe, Ge, I, and F experiments from Table \ref{tab:experiments} for our baseline model selection. This choice covers all major direct detection technologies and considers target elements that would contribute the majority of exposure in a given experiment\footnote{Note that, for example, the exposure on silicon in SuperCDMS is one tenth of the exposure on germanium; similarly, in this sense sodium is a subdominant component to iodine for sodium--iodide crystals.}. In \S\ref{sec:othertargets}, we additionally evaluate model selection prospects for other targets: Ar, Na, and He, for a specific subset of scenarios. Finally, in \S\ref{sec:G3}, we explore the reach of futuristic experiments, some of which are projected to tap into the irreducible neutrino background. In all simulations used in \S\ref{sec:baseline} and \S\ref{sec:othertargets} the signal is set to the current upper limit (cross sections are chosen to match exclusion values of Figure \ref{fig:exclusions}), and those in \S\ref{sec:G3} have a signal at the projected exclusion limit derived assuming a non--detection on G2 Xe and Ge.
\subsubsection{Baseline model selection}
\label{sec:baseline}

\begin{figure*}[t]
\centering
\includegraphics[width=.45\textwidth,keepaspectratio=true]{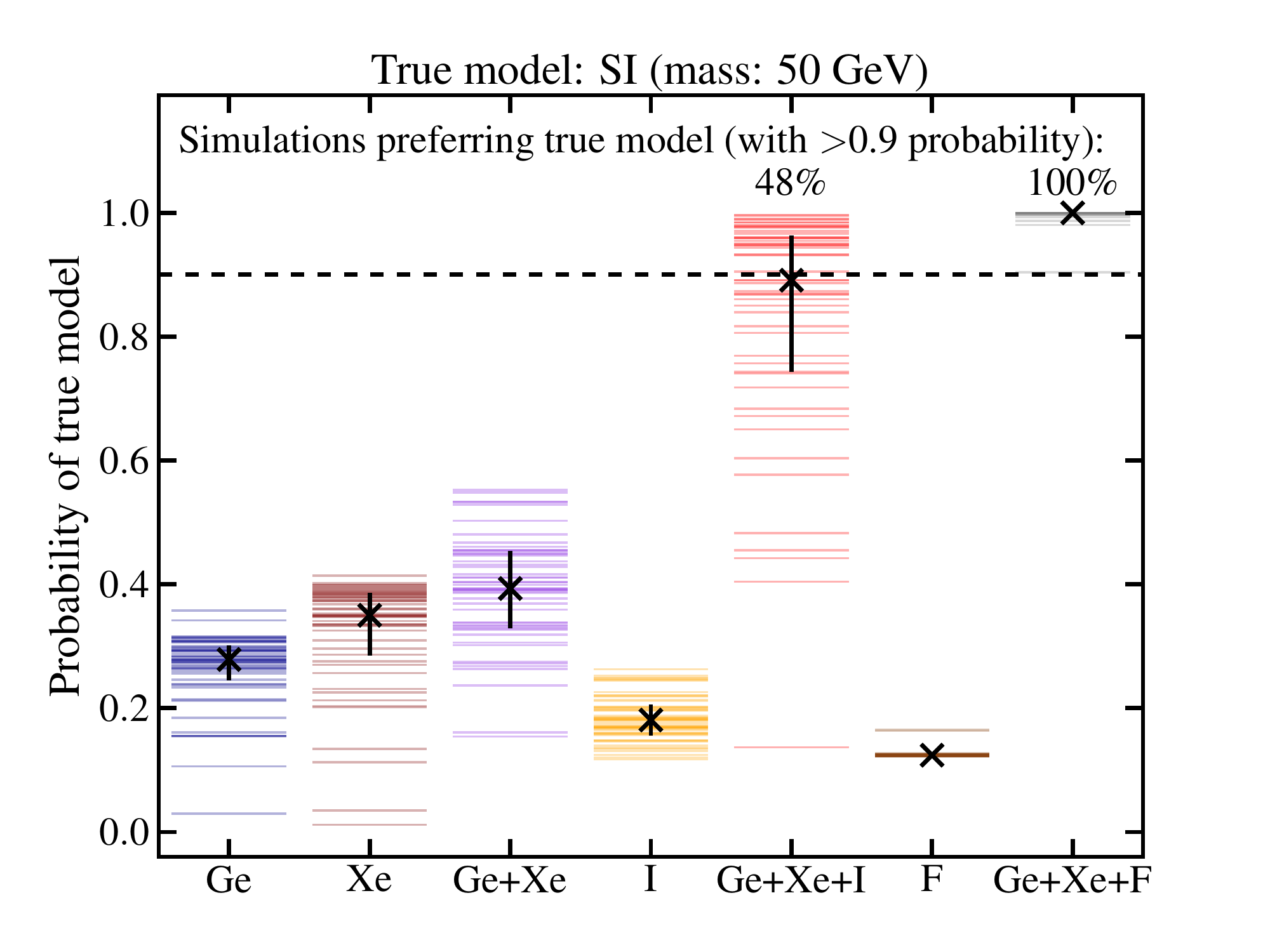}
\includegraphics[width=.45\textwidth,keepaspectratio=true]{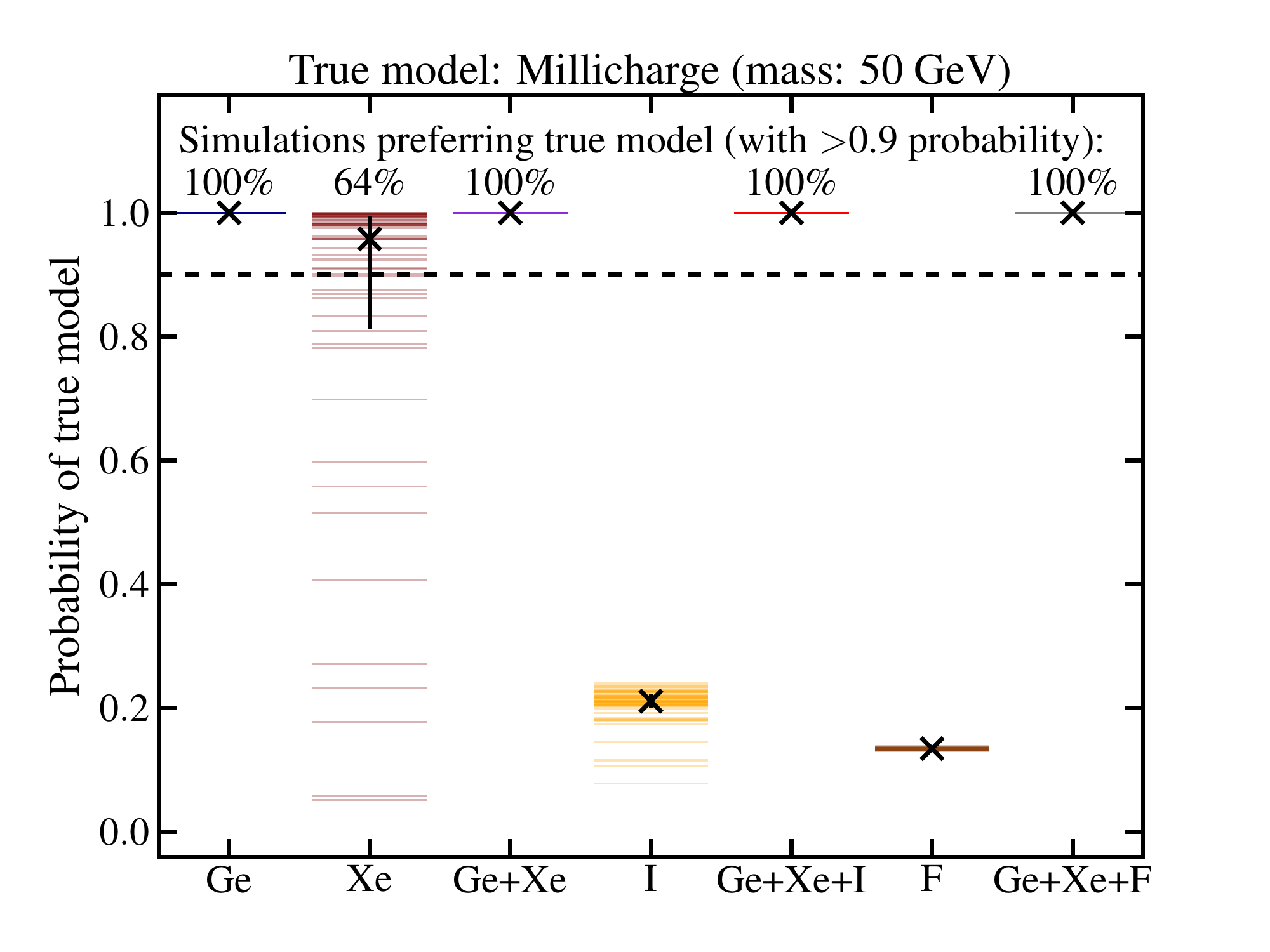}
\includegraphics[width=.45\textwidth,keepaspectratio=true]{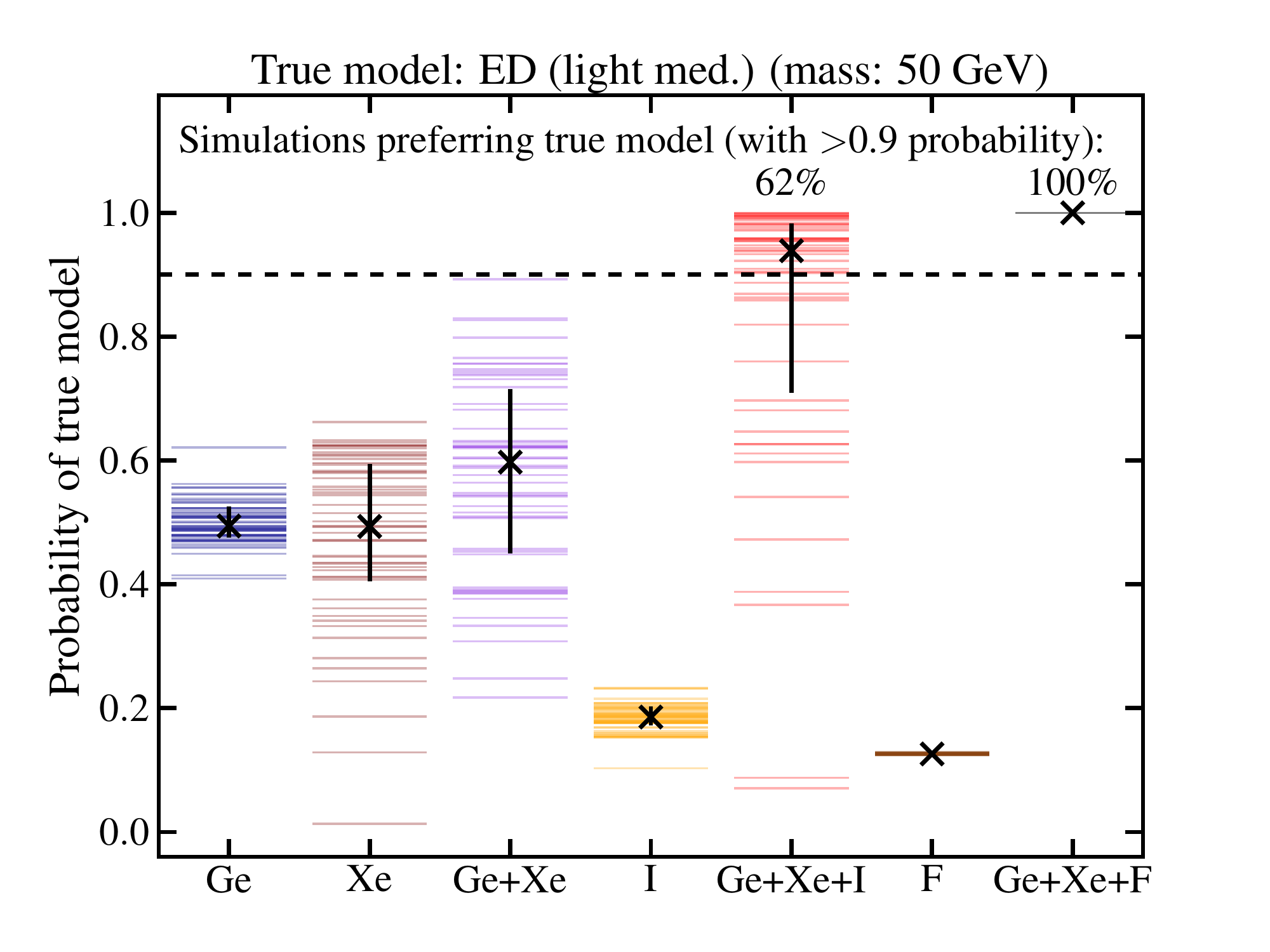}
\includegraphics[width=.45\textwidth,keepaspectratio=true]{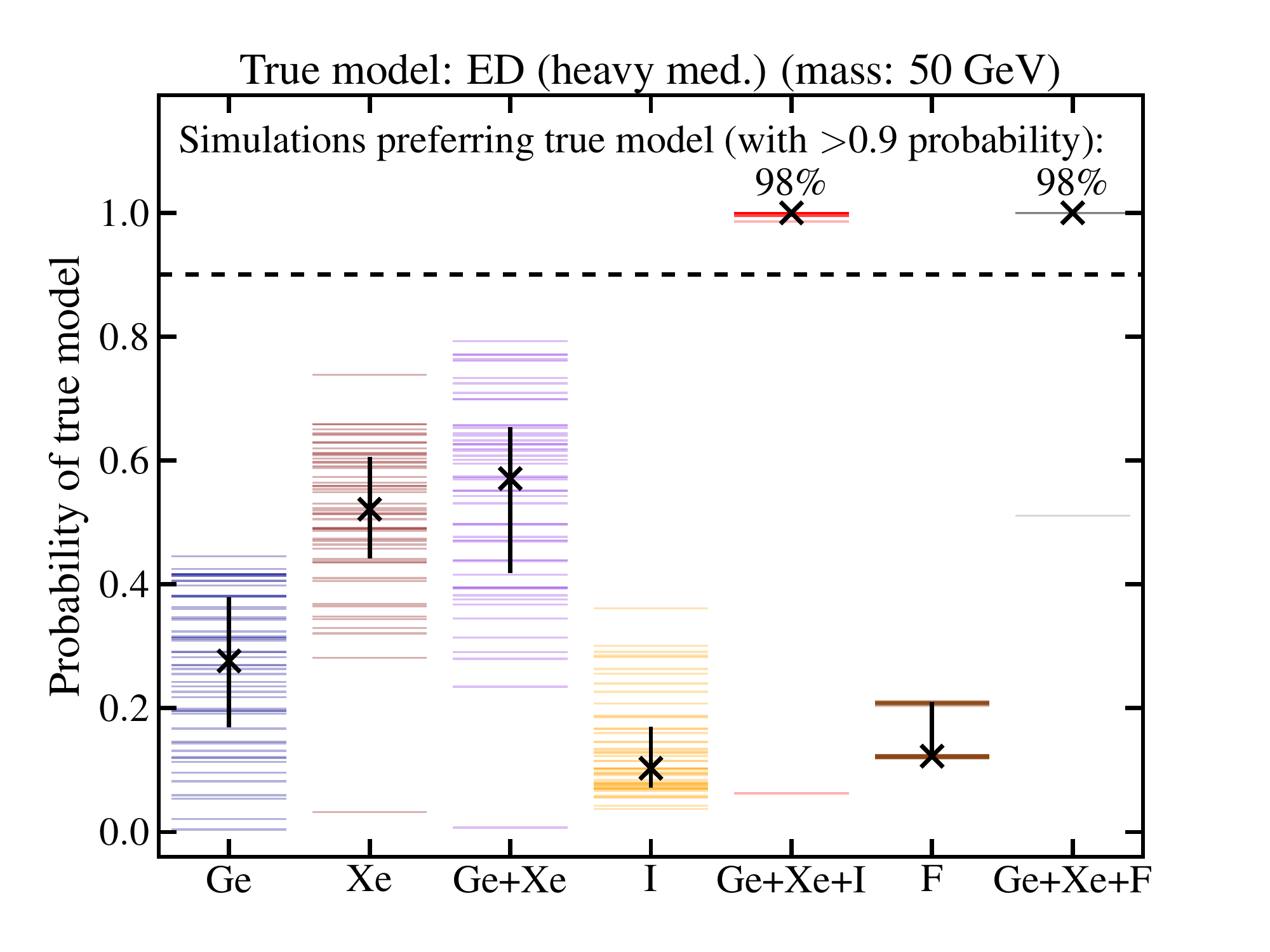}
\caption{The key results of set--I model selection are shown in this Figure. Each of the columns represents analysis of simulations from either a single experiment, or a joint analysis of several experiments (denoted on the $x$--axis). Each panel corresponds to simulations for a single underlying scattering model (indicated in the header) for 50 GeV DM mass; $\sigma_p$ is set to its current upper limit. Each horizontal colored line represents a single realization of the data under that model, and the spread of the lines on the $y$--axis is due to the Poisson noise. Evidences for all 8 models from set I are compared to compute the probability of the right underlying model, shown on the $y$--axis. A pile--up of many simulations at more than $90 \%$ probability signifies a high chance of selecting the correct underlying model. This chance is generally low if only Xe and Ge are considered, but greatly improved if F or I targets of modest exposure are added to the analysis. See Figures \ref{fig:model_selection_gexe_50gev}, \ref{fig:model_selection_gexe_15gev}, and \ref{fig:model_selection_gexe_500gev} for related results for other models of set I and for other DM particle masses. \label{fig:model_selection_gexe_50gev_select}}
\end{figure*}

\begin{figure*}[t]
\centering
\includegraphics[width=.45\textwidth,keepaspectratio=true]{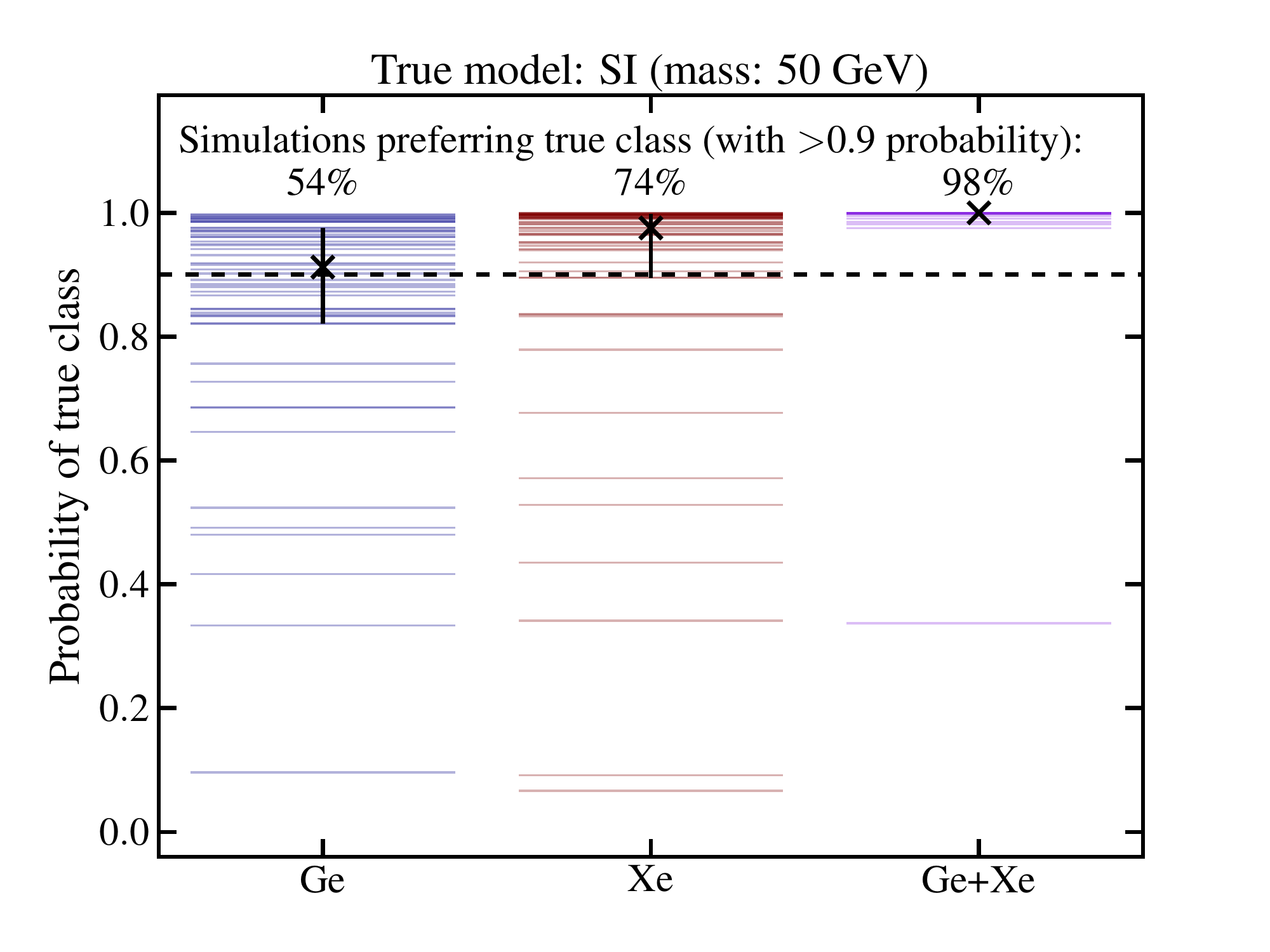}
\includegraphics[width=.45\textwidth,keepaspectratio=true]{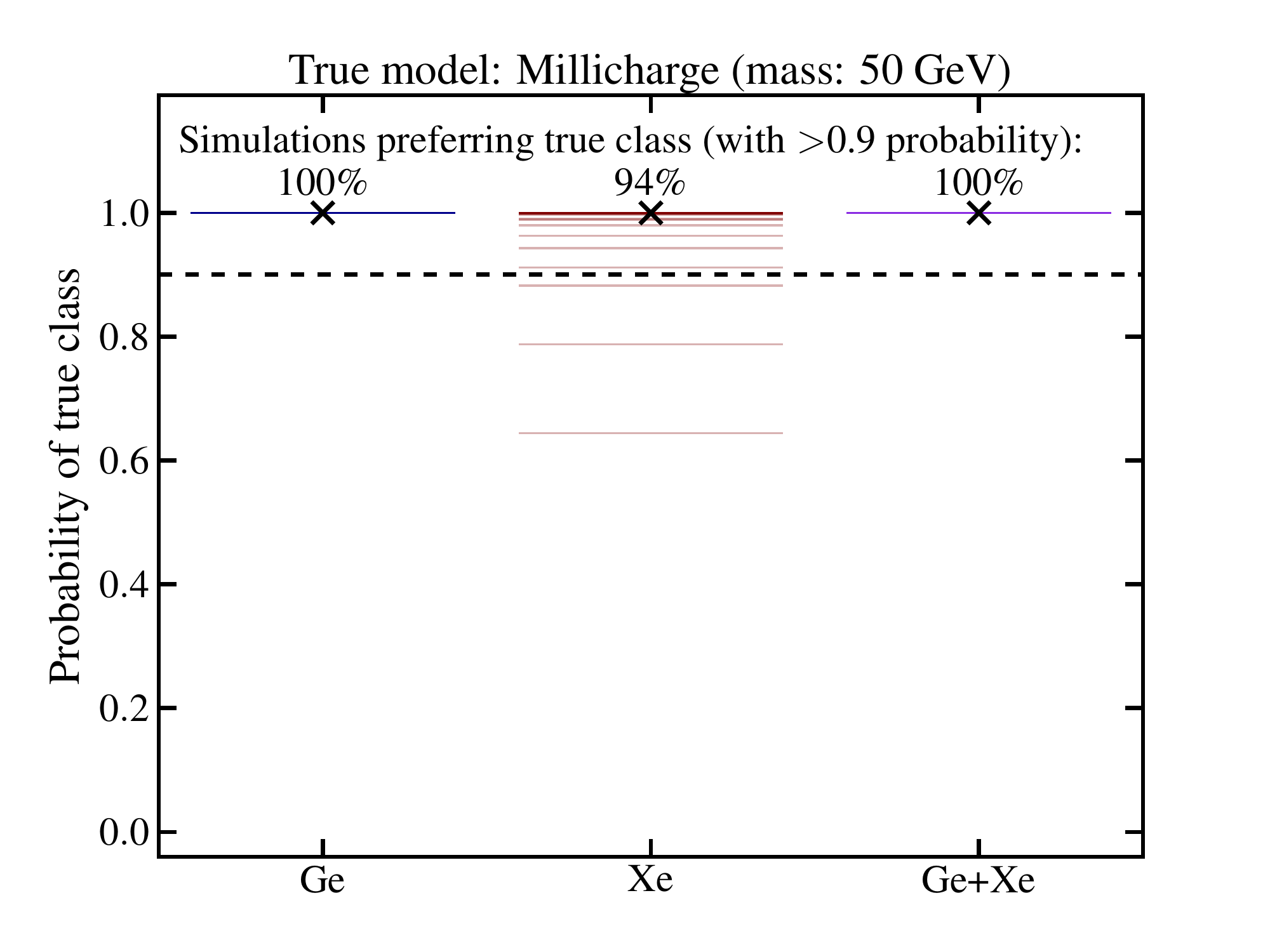}
\includegraphics[width=.45\textwidth,keepaspectratio=true]{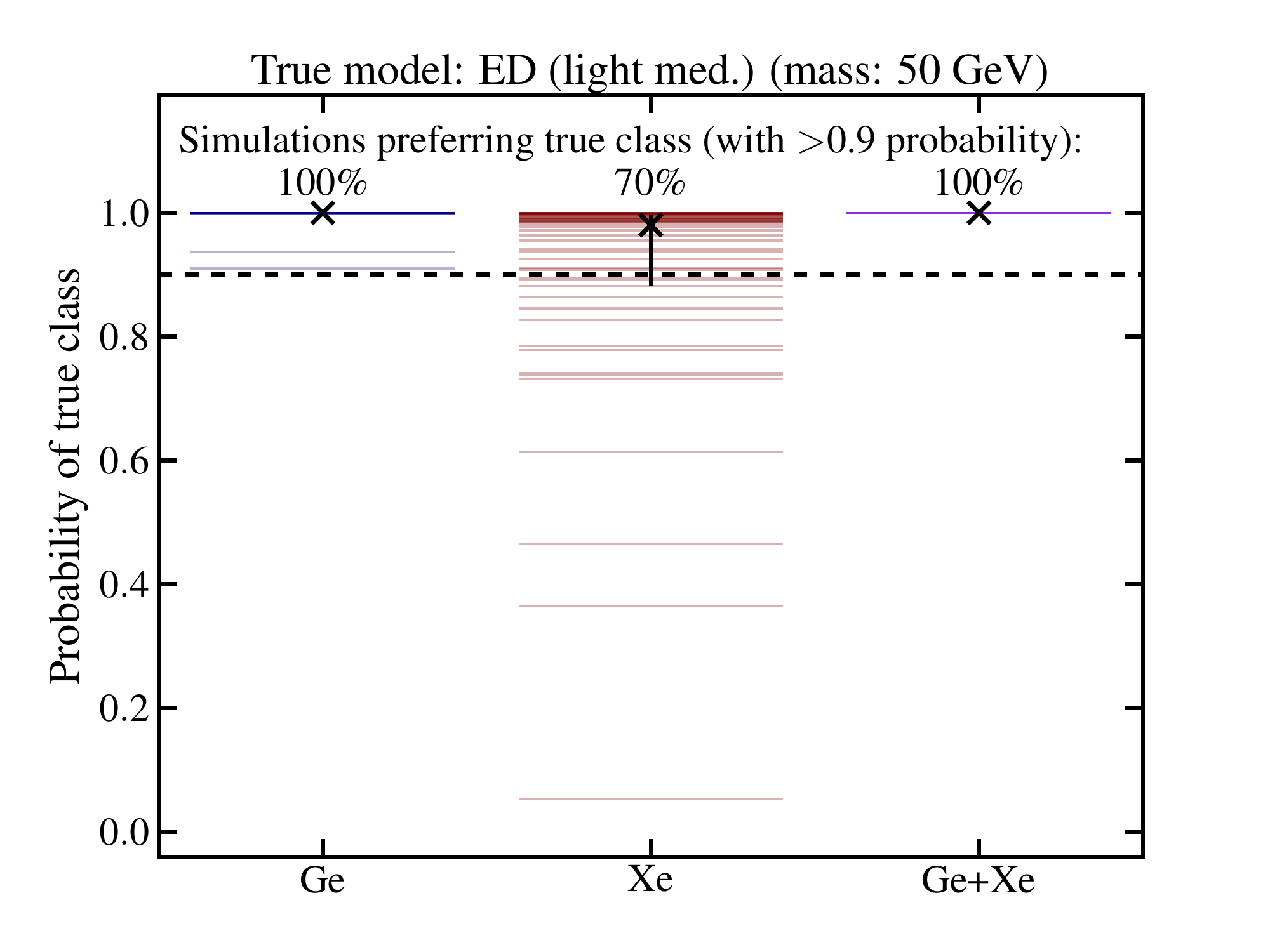}
\includegraphics[width=.45\textwidth,keepaspectratio=true]{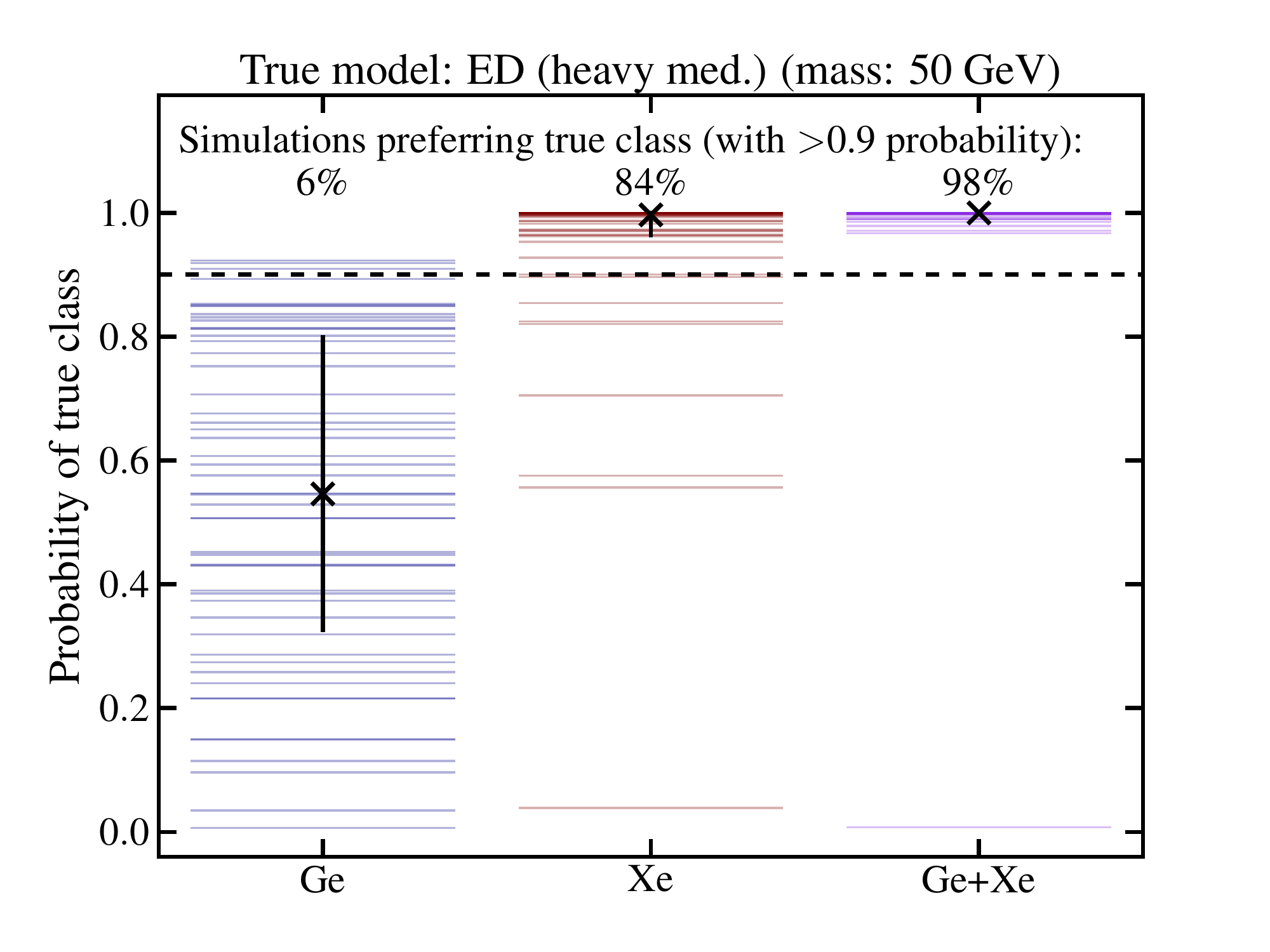}
\caption{Same as Figure \ref{fig:model_selection_gexe_50gev_select}, except that the vertical axis shows the sum of probabilities for all models that have the same momentum dependence. In contrast with prospects for selecting the right underlying model (shown in Figure \ref{fig:model_selection_gexe_50gev_select}), Ge and Xe targets are able to confidently identify momentum dependence of the underlying interaction. The simulations used here are for DM particle mass of 50 GeV.\label{fig:class_selection_gexe_50gev_select}}
\end{figure*}

\begin{figure*}[t]
\centering
\includegraphics[width=.45\textwidth,keepaspectratio=true]{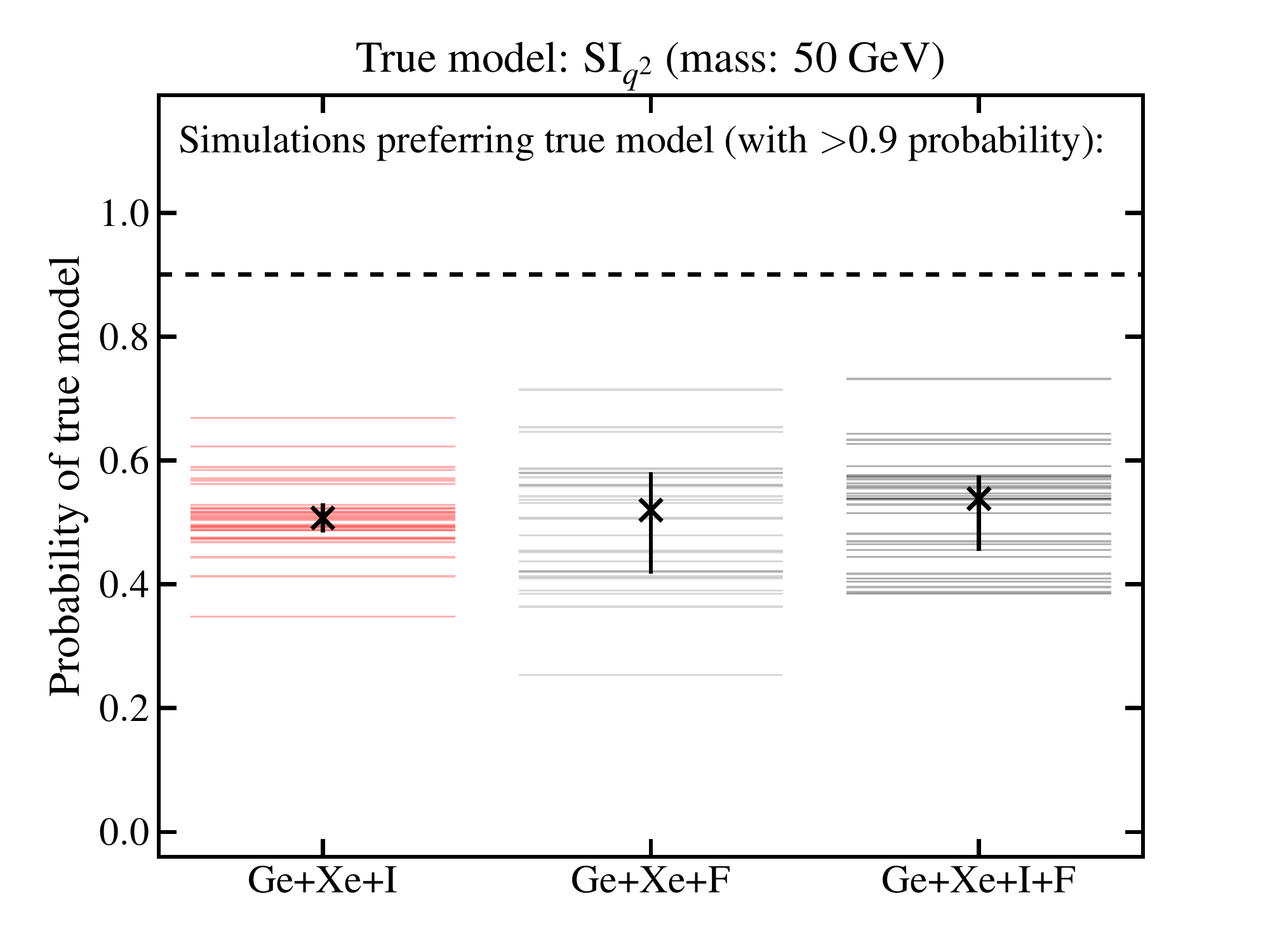}
\includegraphics[width=.45\textwidth,keepaspectratio=true]{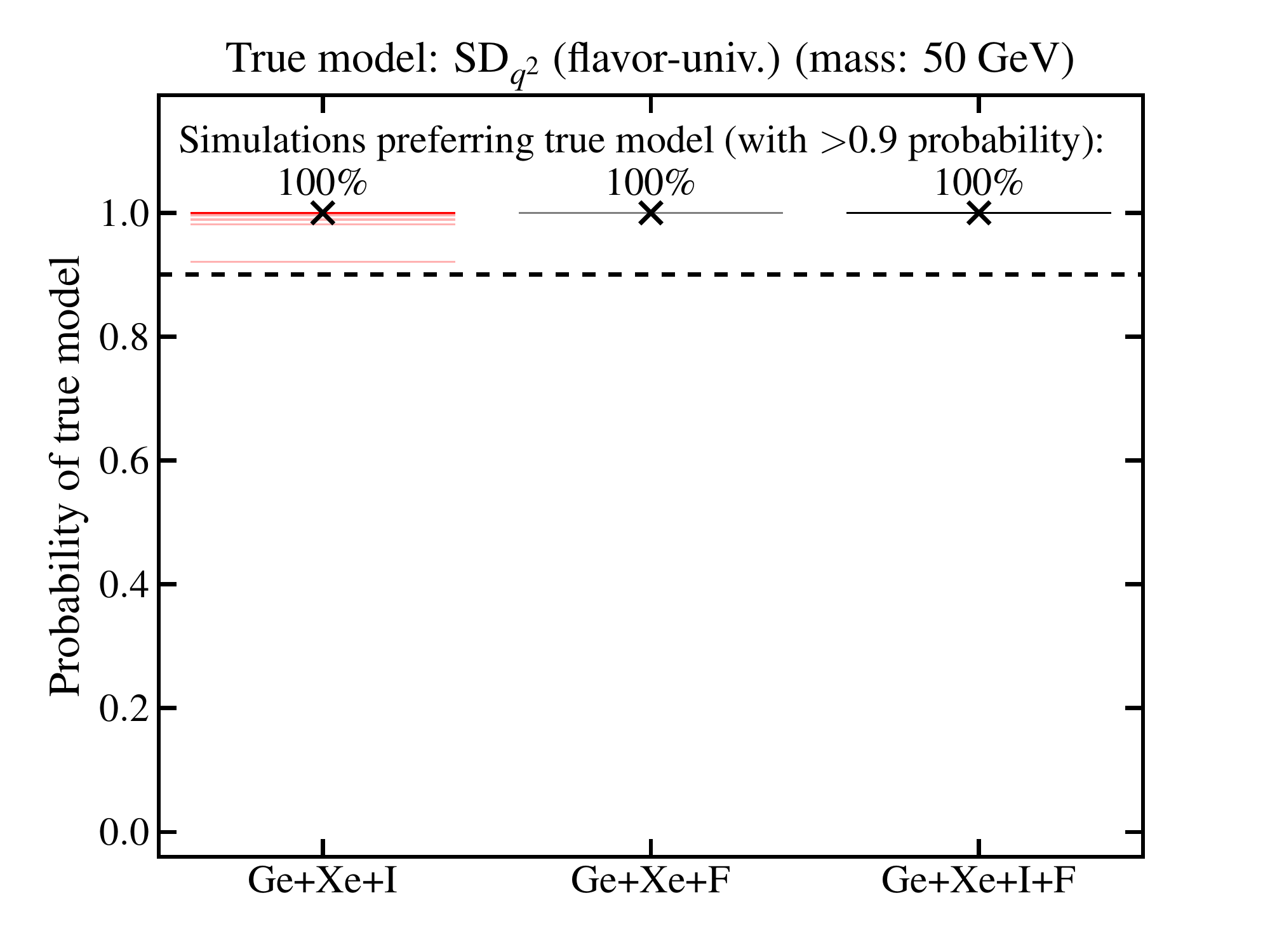}
\includegraphics[width=.45\textwidth,keepaspectratio=true]{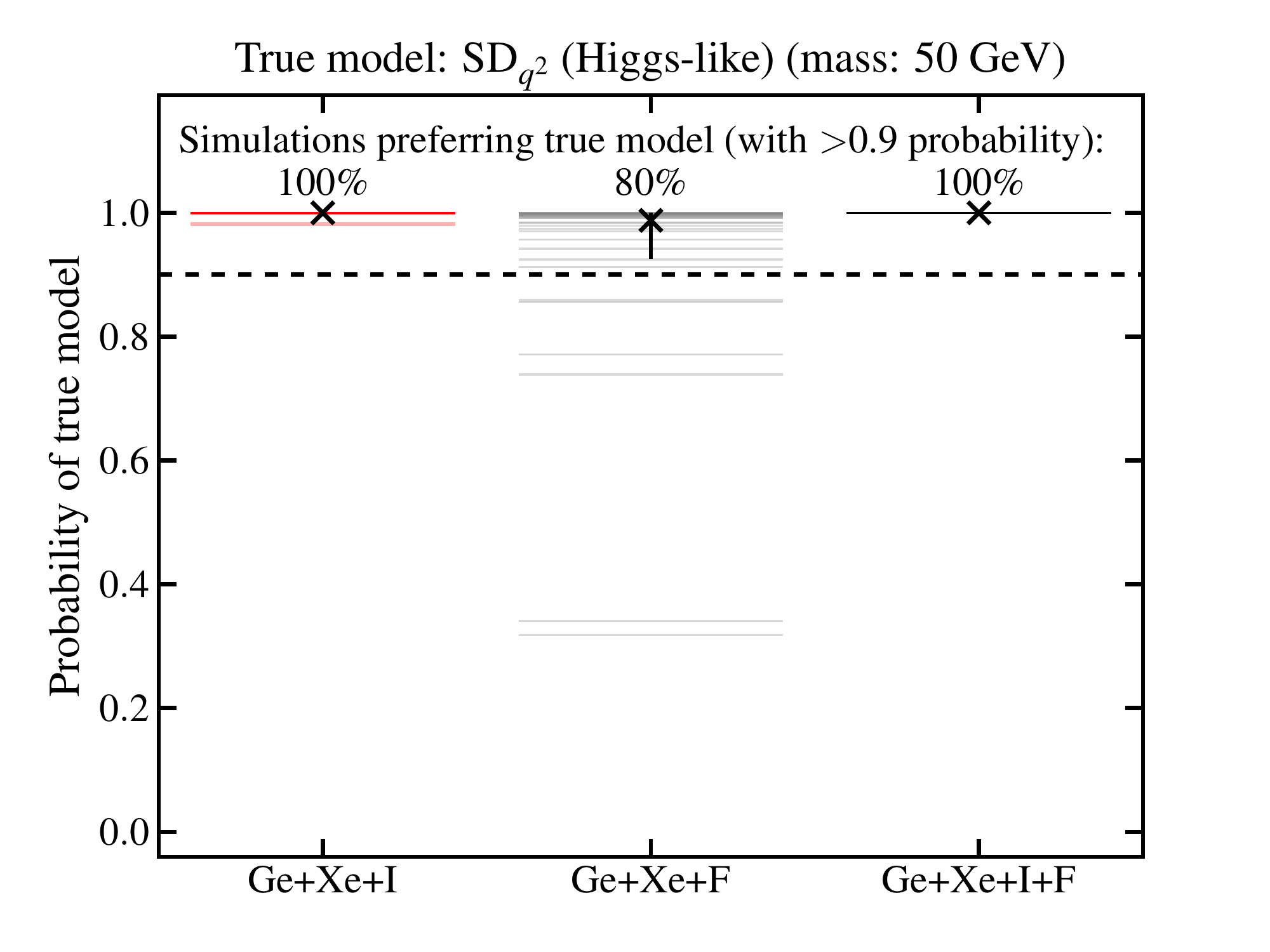}
\includegraphics[width=.45\textwidth,keepaspectratio=true]{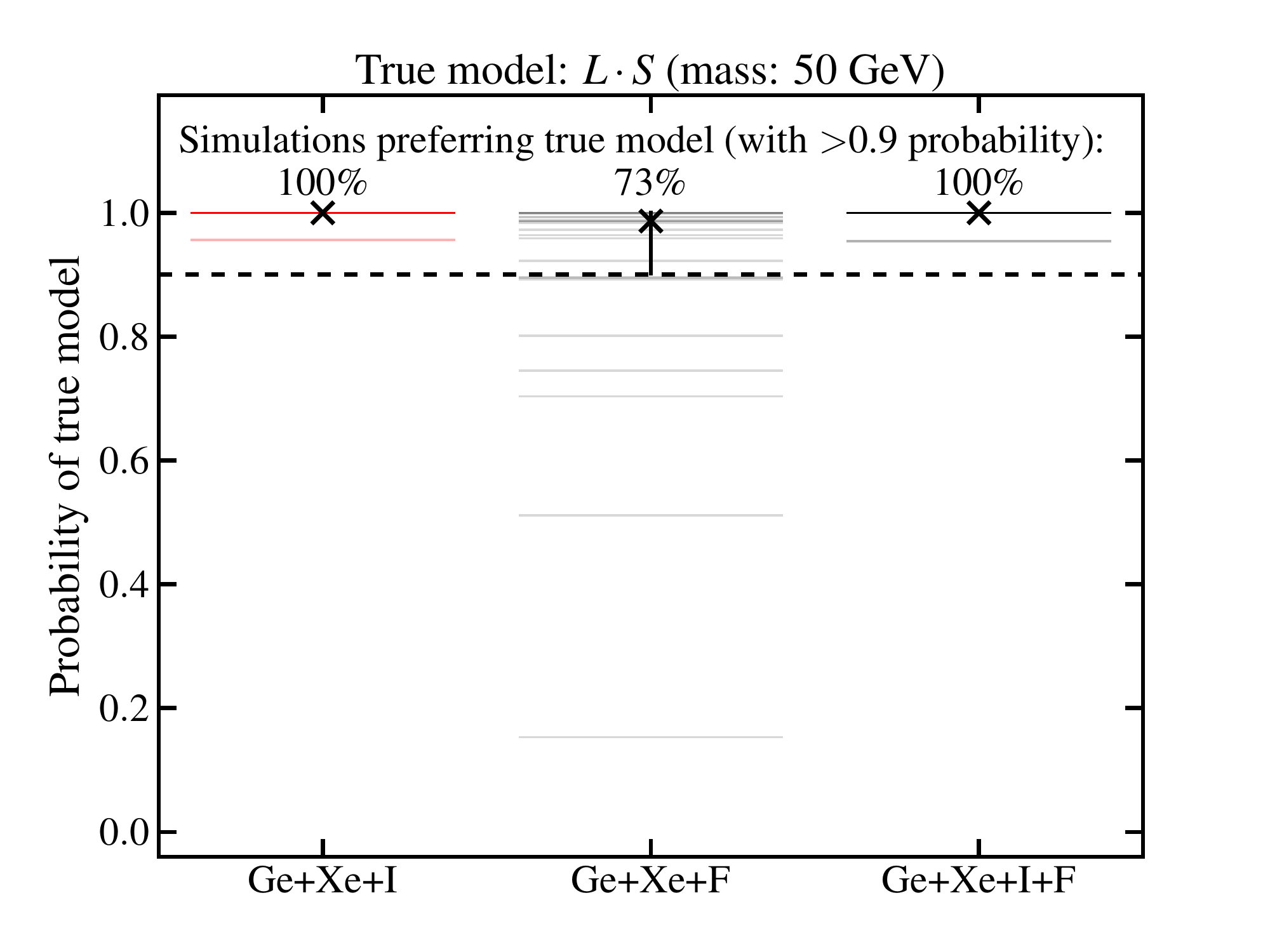}
\caption{Same as Figure \ref{fig:model_selection_gexe_50gev_select}, but for models of set II (discussed in  \S\ref{sec:analysis}) compared against each other. Model selection prospects for identifying the right interaction from amongst many alternatives that all produce a turnover in the recoil energy spectrum are excellent, for a signal just below the current limit, when G2 xenon and germanium targets are combined with a modest exposure on fluorine or iodine. See Figure \ref{fig:model_selection_2_50gev} of Appendix \ref{app:model_selection} for the rest of the related plots. \label{fig:model_selection_2_50gev_select}}
\end{figure*}

First, we perform model selection taking simulations created under set I models (discussed in \S\ref{sec:analysis}) and treating all models from that set as competing hypotheses. After that, we perform analogous hypothesis testing on set II. In order to gain a sense for the complementarity of different targets, we analyze simulated data corresponding to individual experiments followed by combined analyses of several experiments at a time. 
The main results of this Section are shown in Figures \ref{fig:model_selection_gexe_50gev_select}, \ref{fig:class_selection_gexe_50gev_select}, and \ref{fig:model_selection_2_50gev_select}. These Figures only show the most representative cases, for a 50 GeV DM particle; a more complete set of baseline model--selection plots can be found in Appendix \ref{app:model_selection}. 

Each panel in Figure \ref{fig:model_selection_gexe_50gev_select} corresponds to simulations under a single underlying scattering model (where $\sigma_p$ is set to its current upper limit for a chosen DM mass). Each column represents an experiment or a combination of experiments (denoted on the $x$--axis). Each horizontal colored line represents a single realization of the data under that same model (created for the same value of mass, cross section, and $f_n/f_p$). The posterior probability distribution and the corresponding evidence was reconstructed eight times for each simulation (once for each of the competing hypotheses in a given model set); these results are then used to calculate the probability of the right underlying model by evaluating Eq.~(\ref{eq:E_ratio}). The probability the data set assigns  to the right underlying model (given a pool of 8 alternatives) is shown on the $y$--axis; the spread along the $y$--axis is due solely to Poisson noise.
In Figure \ref{fig:model_selection_gexe_50gev_select}, we show only the results obtained from simulations of standard SI scattering, millicharged DM, and electric--dipole scattering through a heavy and a light mediator. These constitute distinct cases in the following sense: anapole (on a target with low intrinsic spin) and SD spectra are similar to SI, while the dipole models with a mediator of the same mass have a similar nuclear recoil spectrum.  Millicharge is the only model that stands alone in a phenomenological sense, since it has the steepest recoil--energy dependence. 

By examining these plots (and the corresponding model probabilities), we can immediately see that the probability for successful model selection is not a monotonic function of the total number of observed events. Instead, model selection depends on a complex interplay of several factors: the type of underlying model, DM mass, and the complementarity of targets used to measure the scattering signal. For example, the only scattering model that is confidently identified regardless of the DM mass, when any two or more experiments are jointly analyzed, is millicharged DM. For all others, the outcome can be significantly different. Similarly, different targets show a varied level of success for different models and masses.

We make two additional observations from this Figure. Firstly, by considering the 3 leftmost columns in each panel of Figure \ref{fig:model_selection_gexe_50gev_select}, we see that Xe and Ge alone (as realized in experiments like the LZ, Xenon1T, and SuperCDMS) are generically {\it not} able to pick out the right underlying scattering process when analyzing data in an agnostic way (i.e.~assuming equal prior probabilities on each scattering model); this statement holds for intermediate and high DM masses (see also Figure \ref{fig:model_selection_gexe_500gev} of Appendix \ref{app:model_selection}). For lighter DM particles, prospects are more optimistic only for ED and MD models, especially with a light mediator (Figure \ref{fig:model_selection_gexe_500gev}). Even in these cases, however, it is only when Ge and Xe data are jointly analyzed that the probability of successful model selection becomes high. 
Secondly, and crucially, addition of either an iodine or a fluorine target with relatively modest exposure (corresponding to about 200 kilogram--years for our I experiment, and about 90 liter--years on $C_3F_8$ for our F experiment) improves the results dramatically, even despite the absence of energy resolution on fluorine. Fluorine shows poor chances for model selection on its own, and iodine alone is only successful for some of the models (only in the case of large DM mass; see Figure \ref{fig:model_selection_gexe_500gev} of Appendix \ref{app:model_selection}). However, these targets are {\it highly complementary} to Xe and Ge for the purposes of model selection---the probability of selecting the correct model when data from Xe, Ge, and either I or F is analyzed jointly is much greater than when Xe and Ge are considered on their own. In other words, even if $\mathcal{E}(\{E_R^{\rm Ge},E_R^{\rm Xe}\}  | \mathcal{M}_{i}) \simeq \mathcal{E}(\{E_R^{\rm Ge},E_R^{\rm Xe}\}  | \mathcal{M}_{j})$, such that model $i$ and $j$ are indistinguishable using only Xe and Ge, we find that including a F or I experiment can break the degeneracy, so that $\mathcal{E}(\{E_R^{\rm Ge},E_R^{\rm Xe},E_R^{\rm F(I)}\}  | \mathcal{M}_{i}) \neq \mathcal{E}(\{E_R^{\rm Ge},E_R^{\rm Xe},E_R^{\rm F(I)}\}  | \mathcal{M}_{j})$.

We now discuss the first conclusion in more detail. If we look at the shapes of the recoil spectra for set I, as shown in Figure \ref{fig:models_set1}, we see that certain subsets of models look similar on these targets within the relevant energy windows. These are the models with a similar energy (momentum) dependence of the dominant scattering response $R(E_R,v)$ of Eq.~(\ref{eq:EFT cross section}) (see also the middle column of Table \ref{tab:operators}). For SI, SD, and Anapole, the dominant response on a target with a small net spin (such as Xe and Ge) does not have any additional energy dependence, so $R$ does not vary strongly with $E_R$. For the light--mediator dipole models, there is an additional $\sim 1/E_R$ enhancement of the recoil spectrum at low energies. For the heavy--mediator dipole models, there is a turnover feature due to an additional $E_R$ suppression. Millicharge DM has a steep enhancement that goes like $\sim 1/E_R^2$. These dependences dictate the level of model degeneracy we observe in results for Xe and Ge. Combining the data from these two experiments does not alleviate this degeneracy. For example, the probabilities of the right model pile up around $33\%$ for SI, SD, and Anapole (as expected from this three--fold degeneracy), and around $50\%$ for the dipole models with the light mediator, and for the dipole models with a heavy mediator. When we examine the probabilities of individual models, we confirm this picture; for example, electric-- and magnetic--dipole models with a light mediator always appear to be ``false positives'' to each other in the sense that the model probability in simulations created under one of the two models is roughly evenly distributed between the correct model and its counterpart. 

These observations motivate a slightly different presentation of the results. In Figure \ref{fig:class_selection_gexe_50gev_select}, the simulations and fits are the same as those used for Ge and Xe in  Figure \ref{fig:model_selection_gexe_50gev_select}, with one difference in the presentation: the $y$--axis now represents the {\it sum of probabilities} for the several models that share the same energy dependence of the scattering rate. In other words, we sum the probability for SI + SD + Anapole, and also for light--mediator models (including dipoles and the Millicharge), and finally for the heavy--mediator dipoles. This Figure shows that Xe and Ge with G2 capability are able to confidently select the right momentum dependence ``class'' of the model at hand, for a direct detection signal close to the current upper limit. If data from the two experiments are jointly analyzed, this outcome is close to 100$\%$ likely, despite appreciable Poisson noise in the data. Therefore, even without considering iodine or fluorine targets, it is likely that Xe and Ge experiments will correctly select the mediator mass and rule out models with ``heavy'' or ``light'' mediators, depending on the results of model selection, if the signal is confidently detected on both targets.  This would allow us to narrow the set of operators under consideration, but would not provide sufficient information to identify the correct UV completion.  

Let us now examine the ``complementarity'' of I and F experiments with Ge and Xe in more detail. Ge and Xe can provide a handle on the momentum dependence of the underlying model owing to their energy resolution. However, germanium and xenon are dominated by isotopes with relatively simple nuclear structure, and signals arising from phenomenologically similar models (for instance, SI, SD, and Anapole) are usually indistinguishable on these targets.
However, both fluorine and iodine are dominated by isotopes with large spin. As a result, they yield widely different (total) number counts for models that appear to have the same spectral shapes on xenon and germanium (see, for instance, SI, SD, and Anapole in Table \ref{tab:Nexp}, and in Figure \ref{fig:Nexps1}), breaking degeneracies between individual models within the same momentum--dependence ``class''. In some cases, even a small event count (or a null result) on these targets provides enough complementary information to result in highly successful model selection, when considered in combination with other data.

Finally, since we have concluded that it is likely that the energy dependence of the response can be deduced in a robust manner using future data sets, we now examine a set of models that all display a similar feature in the recoil spectrum. In particular, we examine the 8 models with a turnover in the spectrum at low recoil energies; such models comprise set II, as discussed in \S\ref{sec:analysis}. The results of this analysis are shown in  Figure \ref{fig:model_selection_2_50gev_select}. For this Figure, all the simulations are created under models of set II, and only models of that set are compared against each other. The content of that Figure is completely analogous to Figure \ref{fig:model_selection_gexe_50gev_select}, except that we now only present combined analysis for Ge+Xe plus either one, or both, F and I. 
From this Figure (and related Figure \ref{fig:model_selection_2_50gev} of Appendix \ref{app:model_selection}), we see that the prospects for selecting out the correct model from amongst 8 possibilities with the same spectral feature is excellent, given a signal close to the current upper limit, and if data from germanium, xenon, and either fluorine, or iodine targets are jointly analyzed. The only exceptions are the SI$_{q^2}$ and the electric--dipole scattering through a heavy mediator, which remain degenerate with each other. Since momentum suppression is a common feature of a wide class of nonstandard models (indeed, the momentum suppression is typically what ensures their subdominance compared to the standard models), being able to identify the single correct model from an agnostic analysis of many alternatives is an impressive and valuable capability of G2--like experiments.
\subsubsection{Other targets}
\label{sec:othertargets}

\begin{figure*}
\centering
\includegraphics[width=.31\textwidth,keepaspectratio=true]{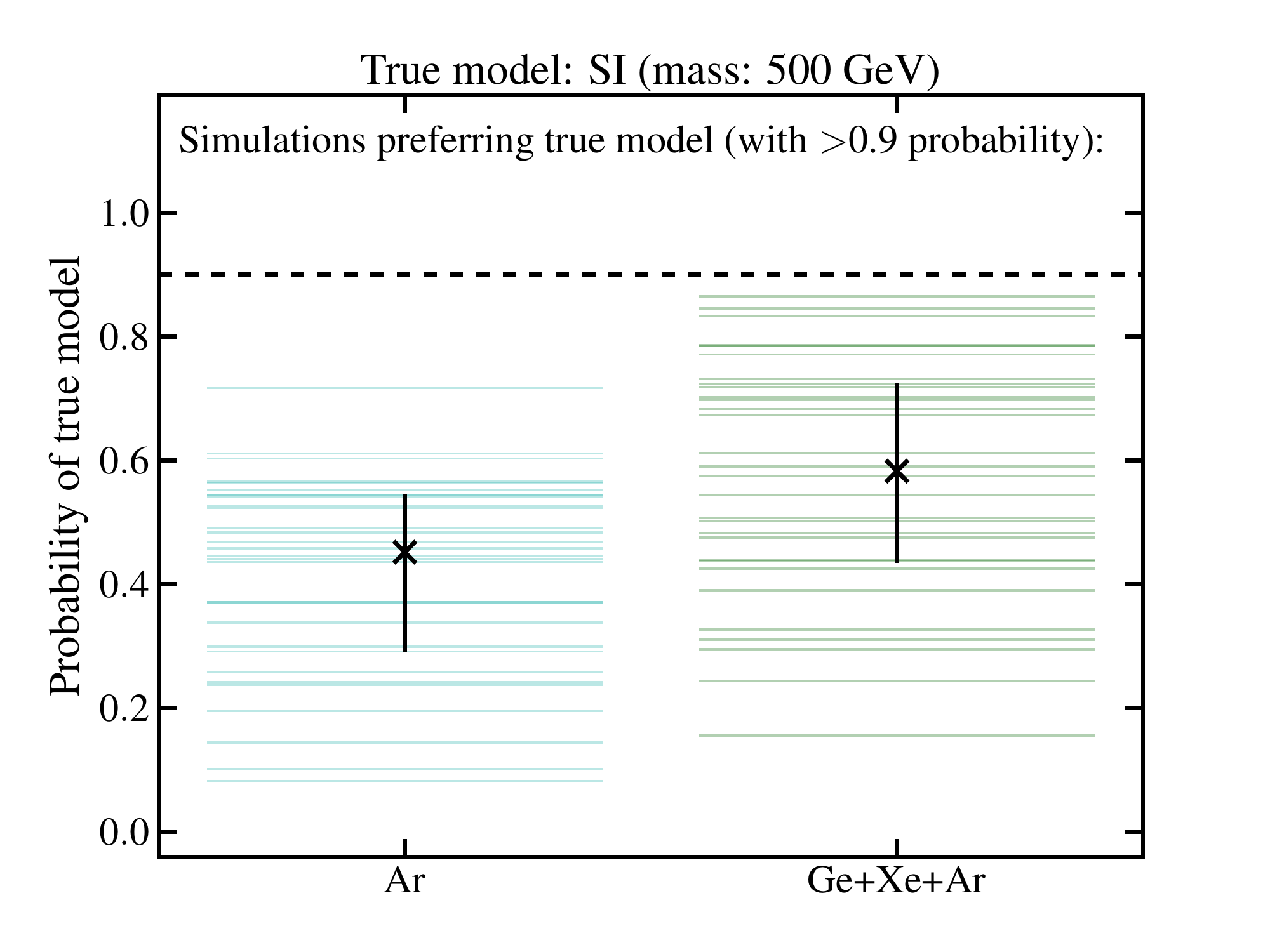}
\includegraphics[width=.31\textwidth,keepaspectratio=true]{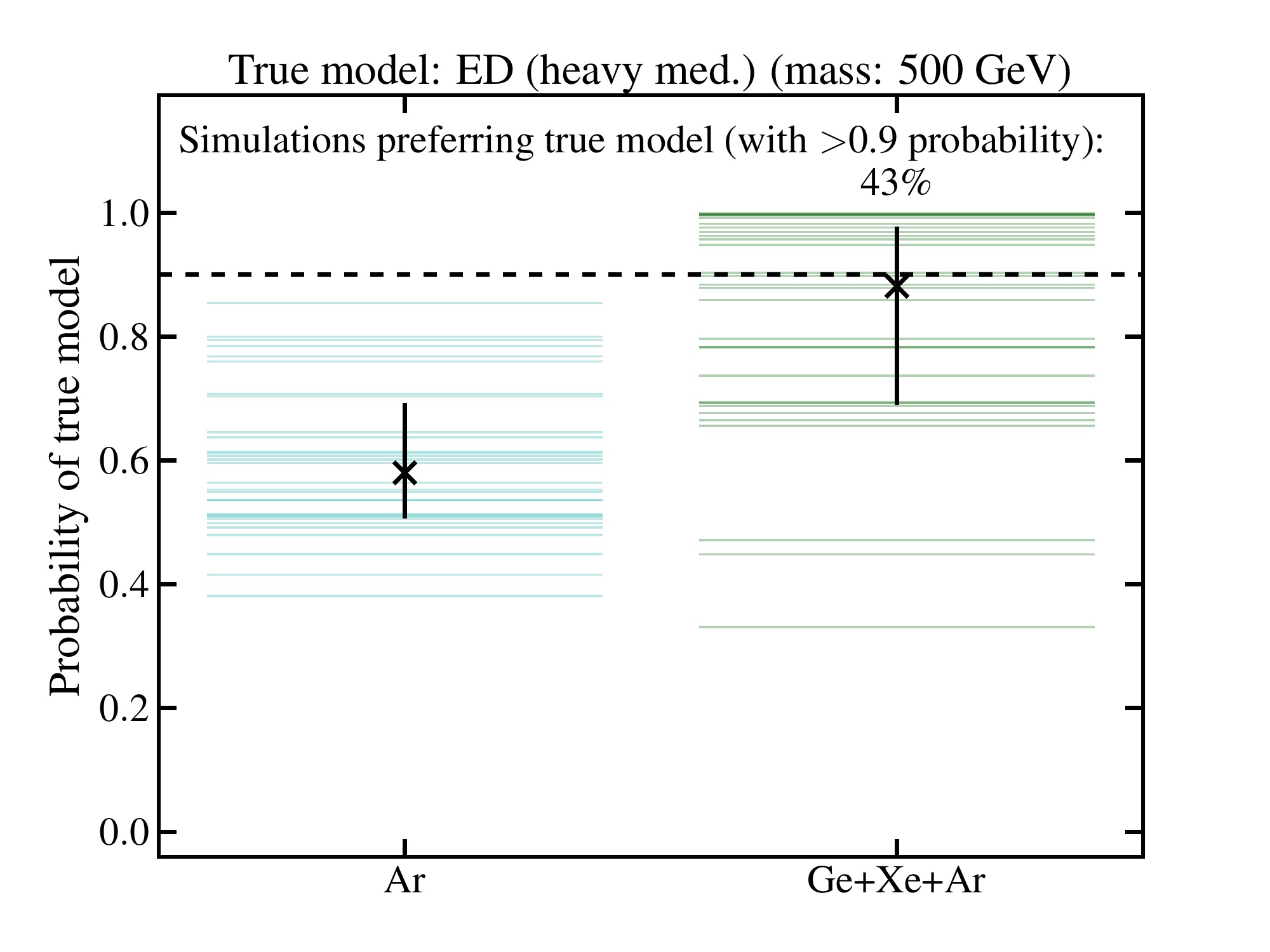}
\includegraphics[width=.31\textwidth,keepaspectratio=true]{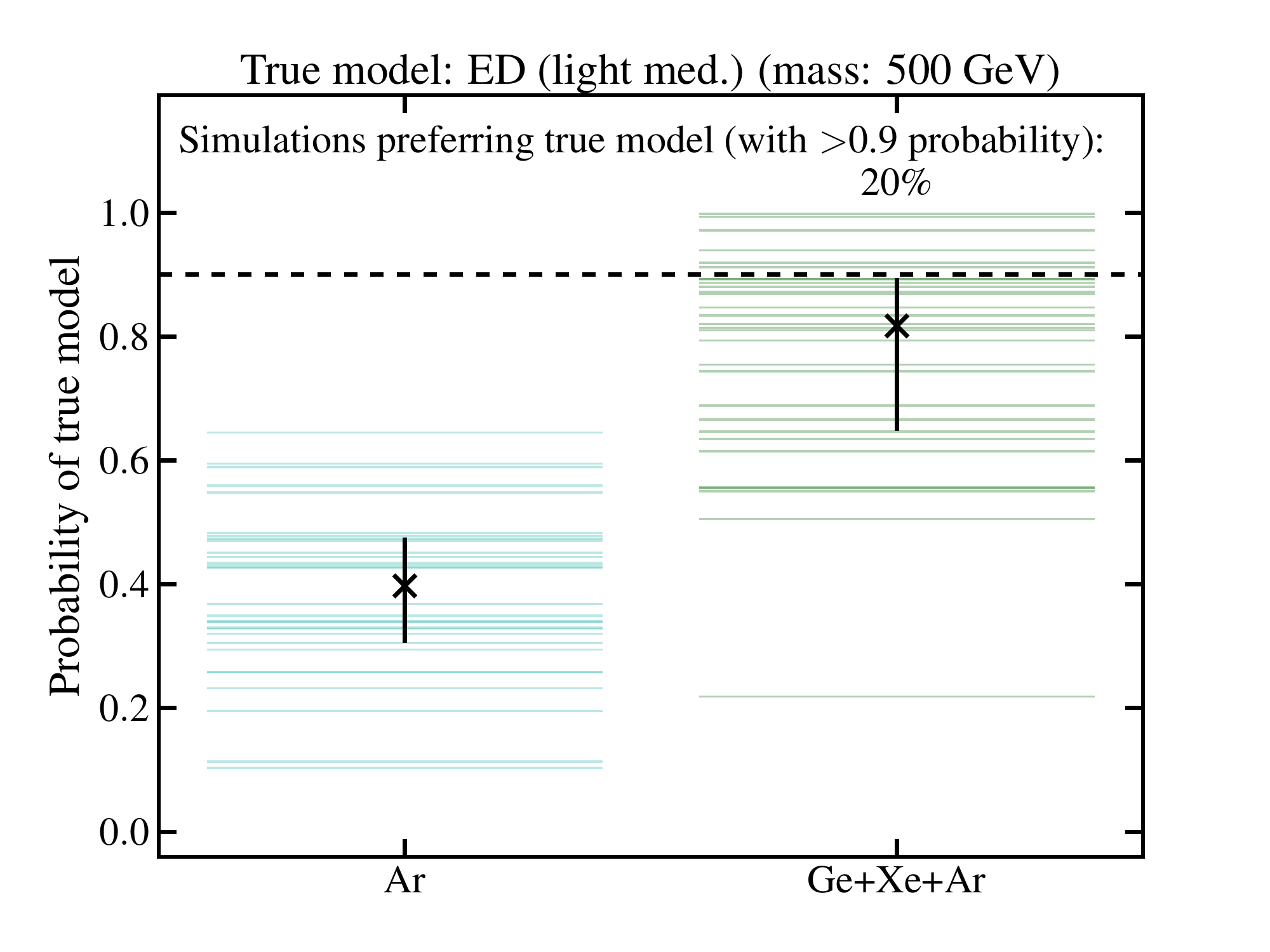}
\caption{Results of model selection preformed amongst set--I models, analogous to Figure \ref{fig:model_selection_gexe_50gev_select}, but for argon--based G2 experiment defined in Table \ref{tab:experiments}, for a 500 GeV DM particle. \label{fig:model_selection_Ar}}
\end{figure*}
In this Section, we briefly consider the model selection capability of additional nuclear targets. We start by examining an argon target with G2 capability, as defined in Table \ref{tab:experiments}. Due to its high energy threshold as compared to most other experiments (30 keV), Ar would not capture most of the telltale spectral features that would give a unique handle on the underlying interaction. However, because of the wide energy window such a target can achieve (up to 200 keV or more), Ar may contribute statistical information about the high--energy end of the recoil spectrum for large DM masses. For this reason, we examine a mock version of this experiment on simulations for 500 GeV DM, and only perform model selection amongst set--I models, in analogy with our baseline results in  \S\ref{sec:baseline}. The results forming a representative set of simulations are shown in Figure \ref{fig:model_selection_Ar}. From this Figure we see that Ar produces a moderate improvement when combined with Xe and Ge. On its own, it has model--selection power comparable to that of Xe, with slightly better prospects for heavy--mediator dipole model, and slightly worse for the light--mediator model. As expected, Ar performs best for models with a turnover feature (like the heavy--mediator model shown in the middle panel of this Figure), where the number of expected events at high energies is maximal. In these cases, the high--energy lever arm compensates for the high energy threshold. However, the complementarity of Ar with Xe and Ge, despite its superior exposure and high--energy capabilities, does not match that of F or I (see Figure \ref{fig:model_selection_gexe_500gev} in Appendix A for comparison).

\begin{figure*}
\centering
\includegraphics[width=.31\textwidth,keepaspectratio=true]{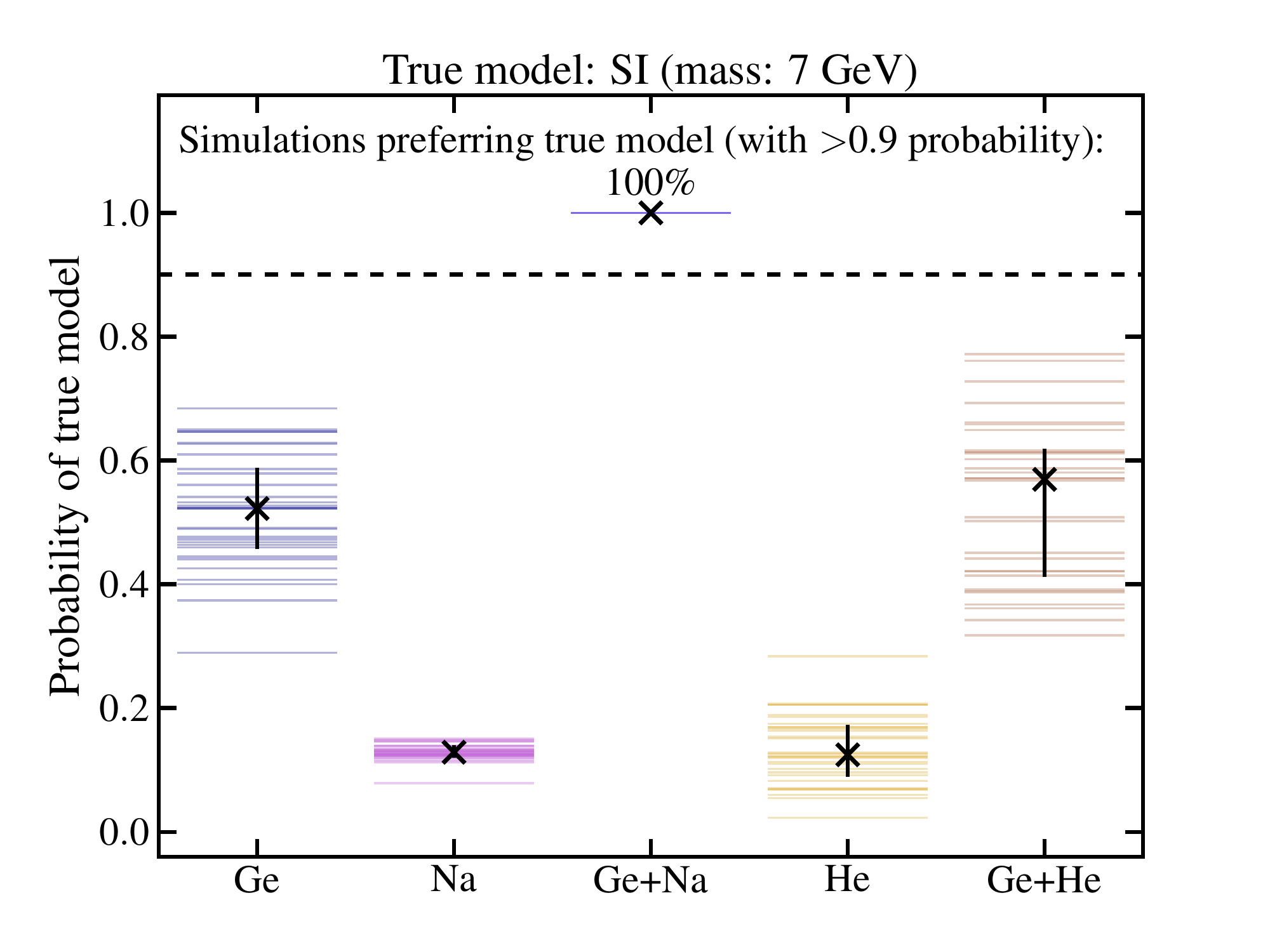}
\includegraphics[width=.31\textwidth,keepaspectratio=true]{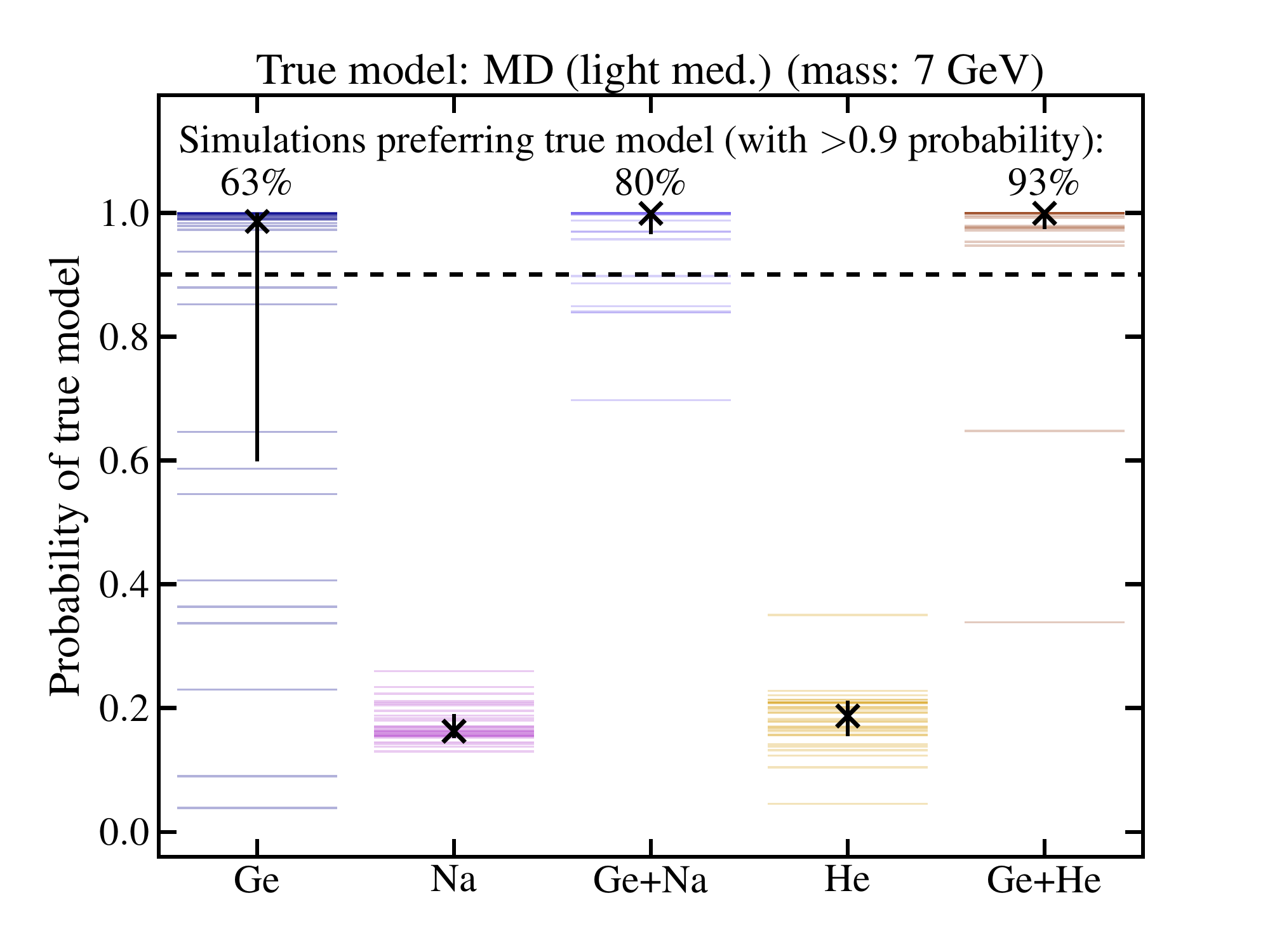}
\includegraphics[width=.31\textwidth,keepaspectratio=true]{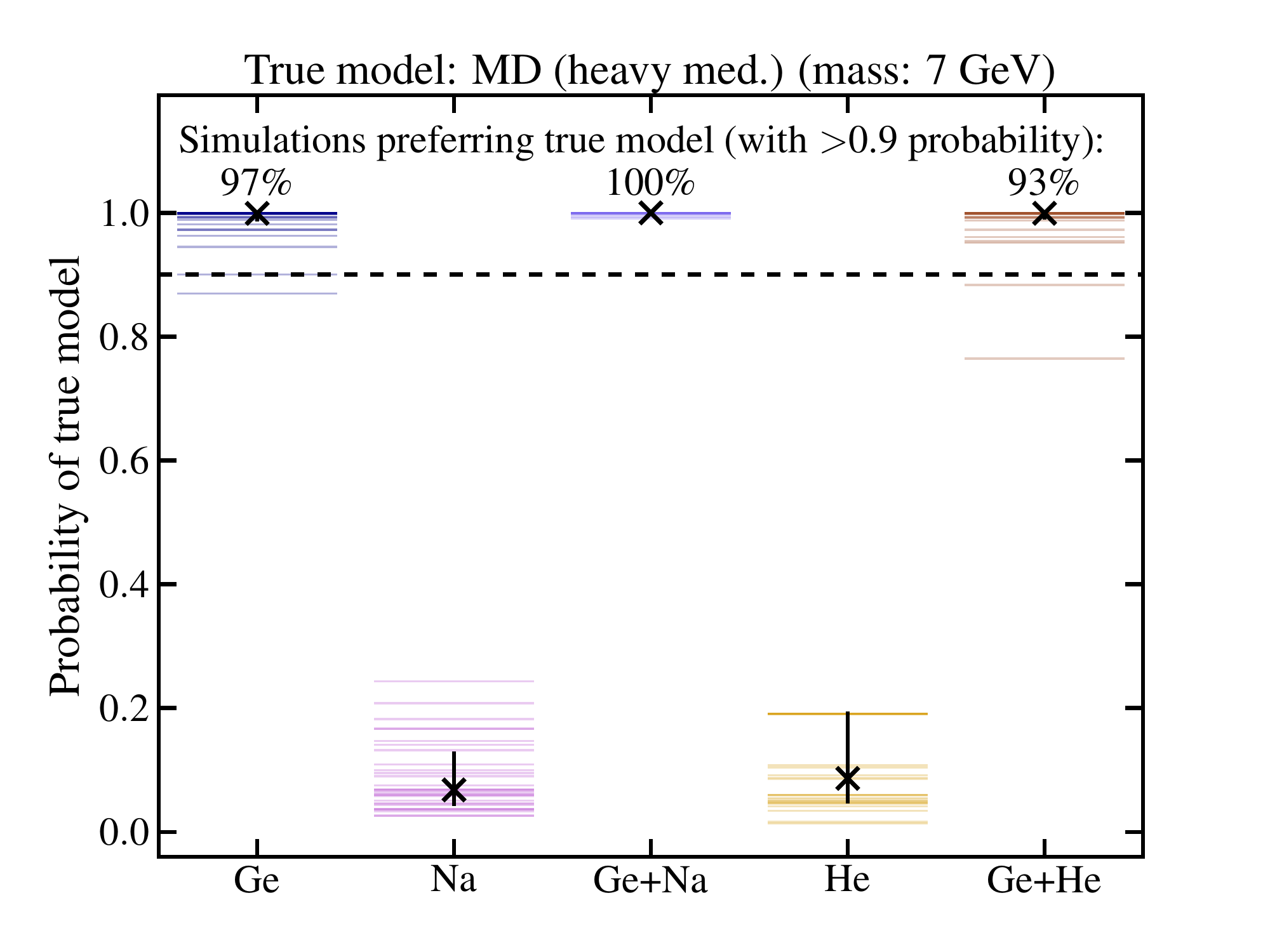}
\caption{Results of model selection preformed amongst set--I models, analogous to Figure \ref{fig:model_selection_gexe_50gev_select}, but for He, Ge, and Na experiment defined in Table \ref{tab:experiments}, for a 7 GeV DM particle. \label{fig:model_selection_lowm}}
\end{figure*}
We also examine the model--selection capability of targets that are kinematically favorable for detection of very low--mass DM particles that are otherwise below the sensitivity of most targets. For this purpose, we create simulations for a 7 GeV DM particle, the lowest DM mass considered in this work. We simulate spectra on helium, sodium, and germanium targets as represented in Table \ref{tab:experiments}, and we once again repeat the model selection exercise using set--I models as competing hypotheses. Results for a representative subset of simulations is shown in Figure \ref{fig:model_selection_lowm}. We also draw attention to Figure \ref{fig:Nexps_lowm}, showing the number of expected events for low--mass DM particles on these targets. This gives a sense of the statistical sample on hand for a single simulated data set. Figure \ref{fig:model_selection_lowm} shows that the single most successful experiment is Ge, partly owing to its low energy threshold. He and Na do not seem to have much model selection power by themselves for the model simulations examined in this Figure. However, there is once again a degree of complementarity amongst these three targets that shows as a visible improvement in model selection success probability when several data sets are combined.
\subsubsection{Ultimate prospects}
\label{sec:G3}

\begin{figure*}
\centering
\includegraphics[width=.3\textwidth,keepaspectratio=true]{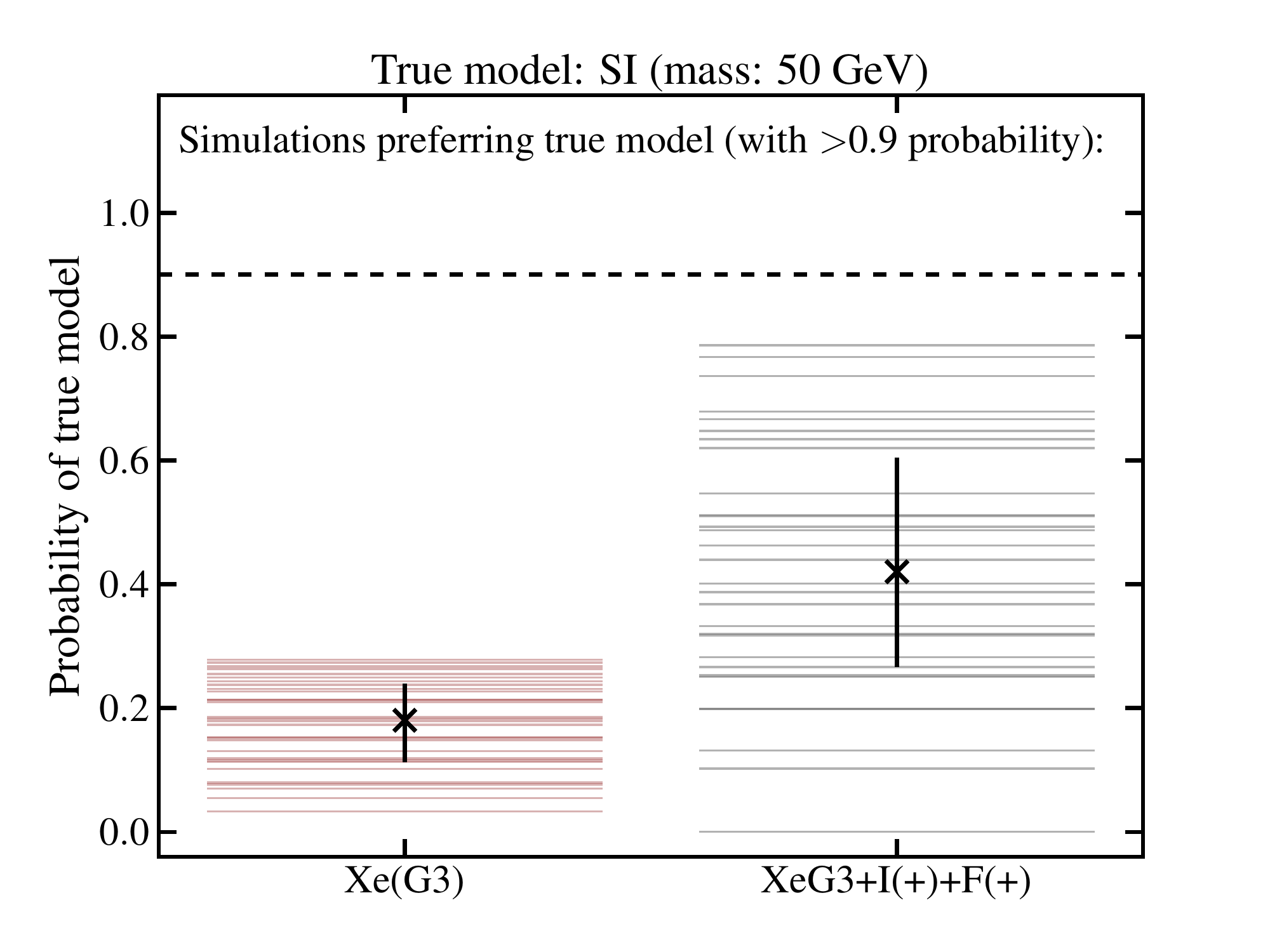}
\includegraphics[width=.3\textwidth,keepaspectratio=true]{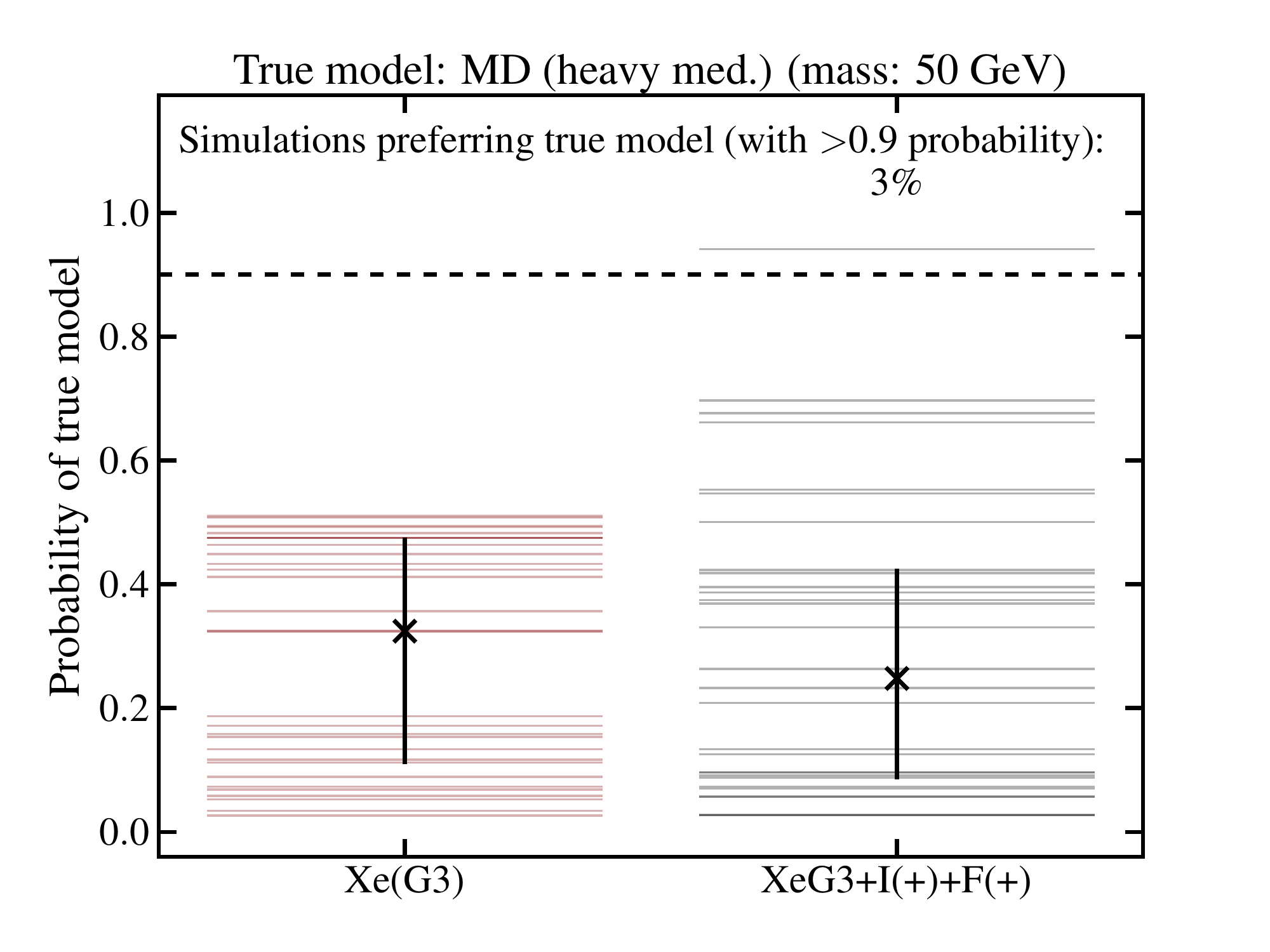}
\includegraphics[width=.3\textwidth,keepaspectratio=true]{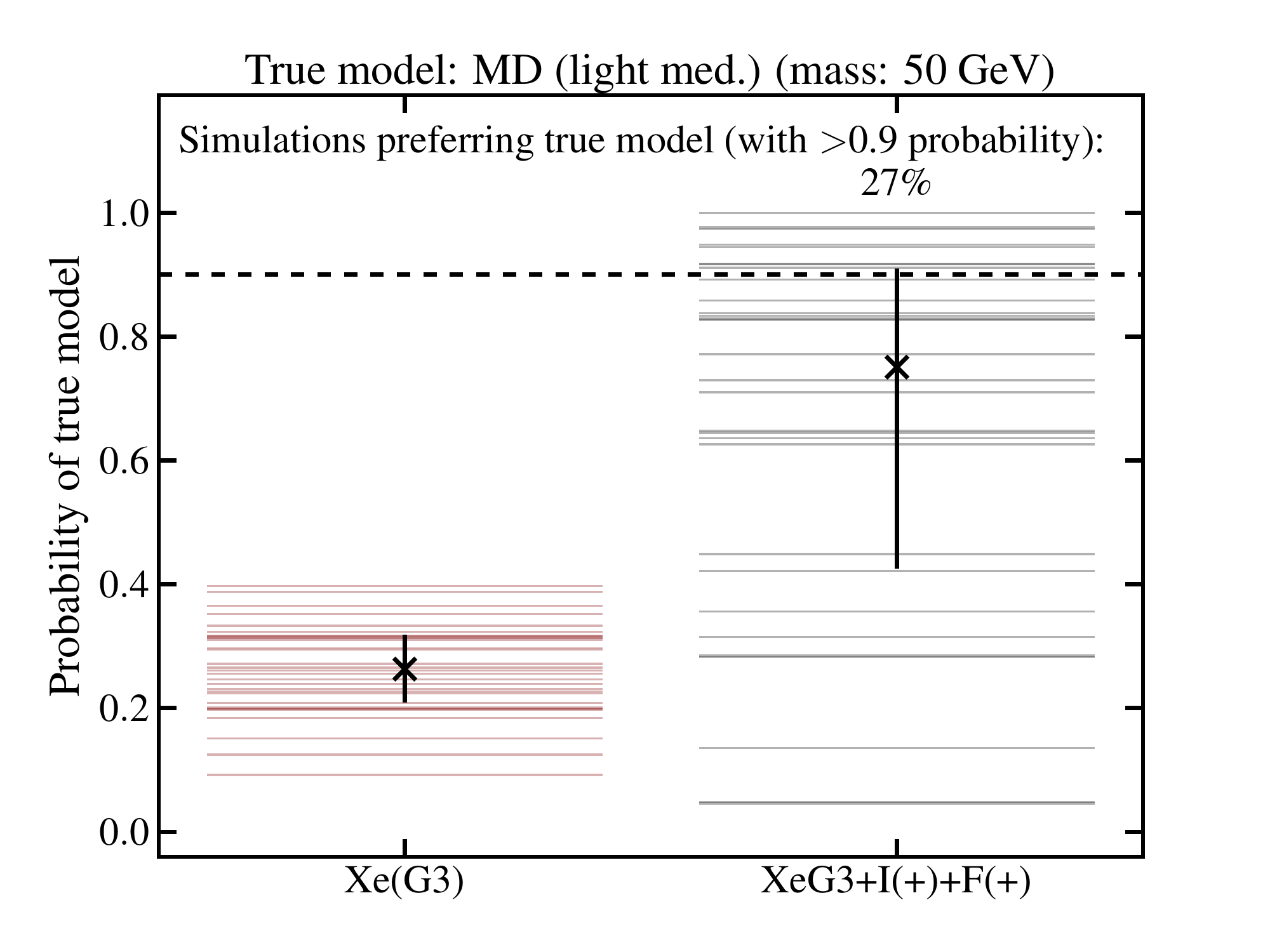}
\caption{This Figure illustrates prospects for selecting the right underlying interaction with futuristic direct detection experiments, if the signal is just below the reach of G2 experiments described in Table \ref{tab:experiments}. The content of this Figure is analogous to Figure \ref{fig:model_selection_gexe_50gev_select}, where only set--I interaction models are compared as competing hypotheses, in light of simulated data from futuristic mock experiments (denoted on the x--axis). Results for a representative subset of simulations are shown here.\label{fig:model_selection_G3}}
\end{figure*}
To gain a sense for the ultimate model--selection capability of direct detection, we generate a new set of simulations with experiments that reach the irreducible background of atmospheric neutrinos. For these simulations, the signal is assumed to be just below the thresholds of G2 Ge and Xe (described in Table \ref{tab:experiments}). The following experiments are considered: a next--generation xenon--based experiment ``XeG3'', a large--exposure fluorine experiment ``F+'', and a low--threshold next--generation iodine experiment ``I+'' (all described in Table \ref{tab:experiments}). We analyze these simulations in the same way as for our baseline analysis of \S\ref{sec:baseline}, for set--I scattering models only.  The results are shown in Figure \ref{fig:model_selection_G3}. The takeaway point from this Figure confirms results reported in previous studies \cite{Gluscevic:2014vga}: if a signal is not seen at G2 experiments, prospects for model selection with future experiments are slim. Addition of fluorine and iodine targets helps, but a large exposure on these targets, combined with data from liquid--noble (and other) experiments might be necessary to have a chance of identifying the interaction producing any putative recoil signals.
\subsection{Reconstruction of DM particle mass}
\label{sec:mass}

\begin{figure*}[t]
\centering
\includegraphics[width=.3\textwidth,keepaspectratio=true]{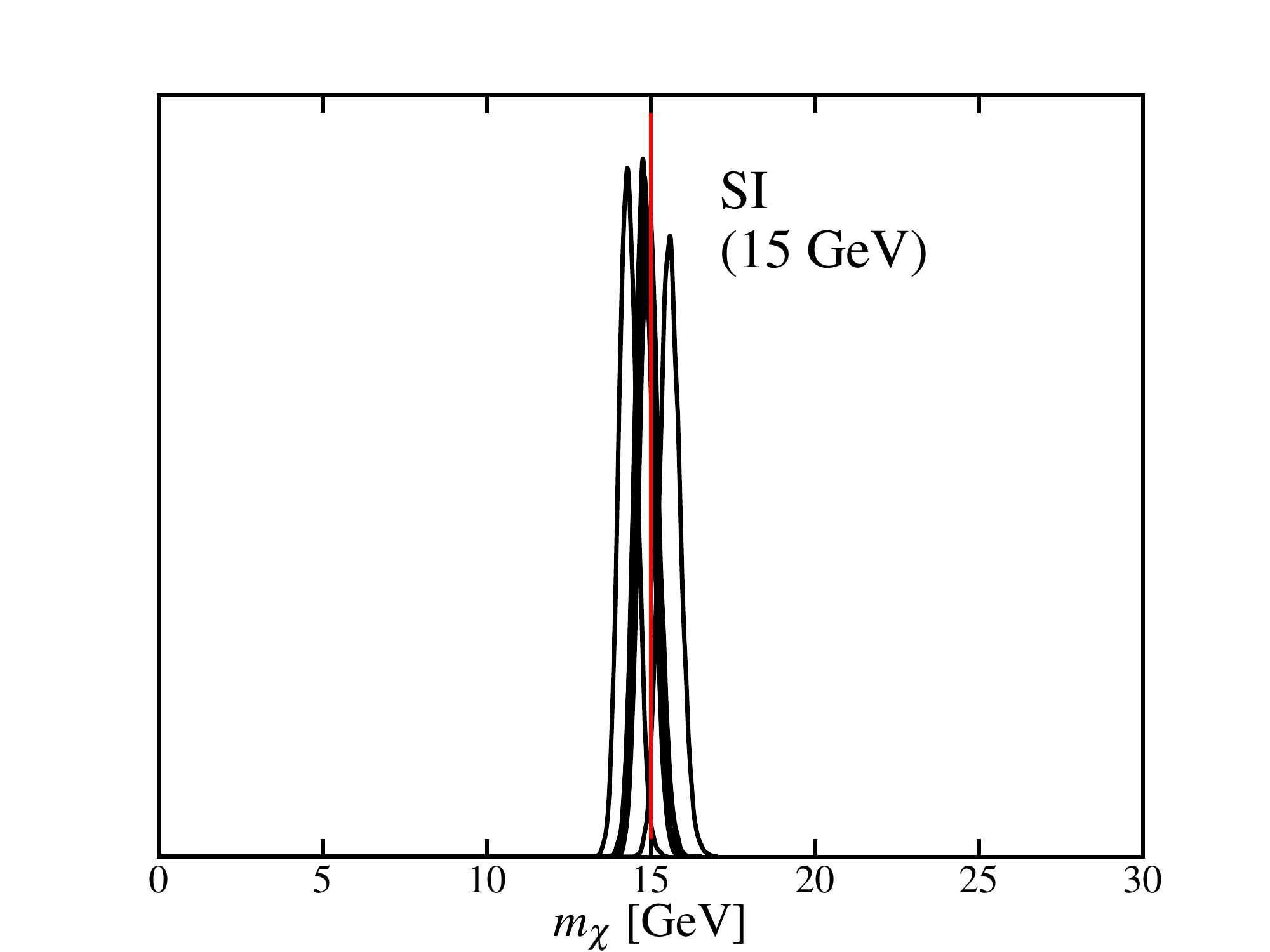}
\includegraphics[width=.3\textwidth,keepaspectratio=true]{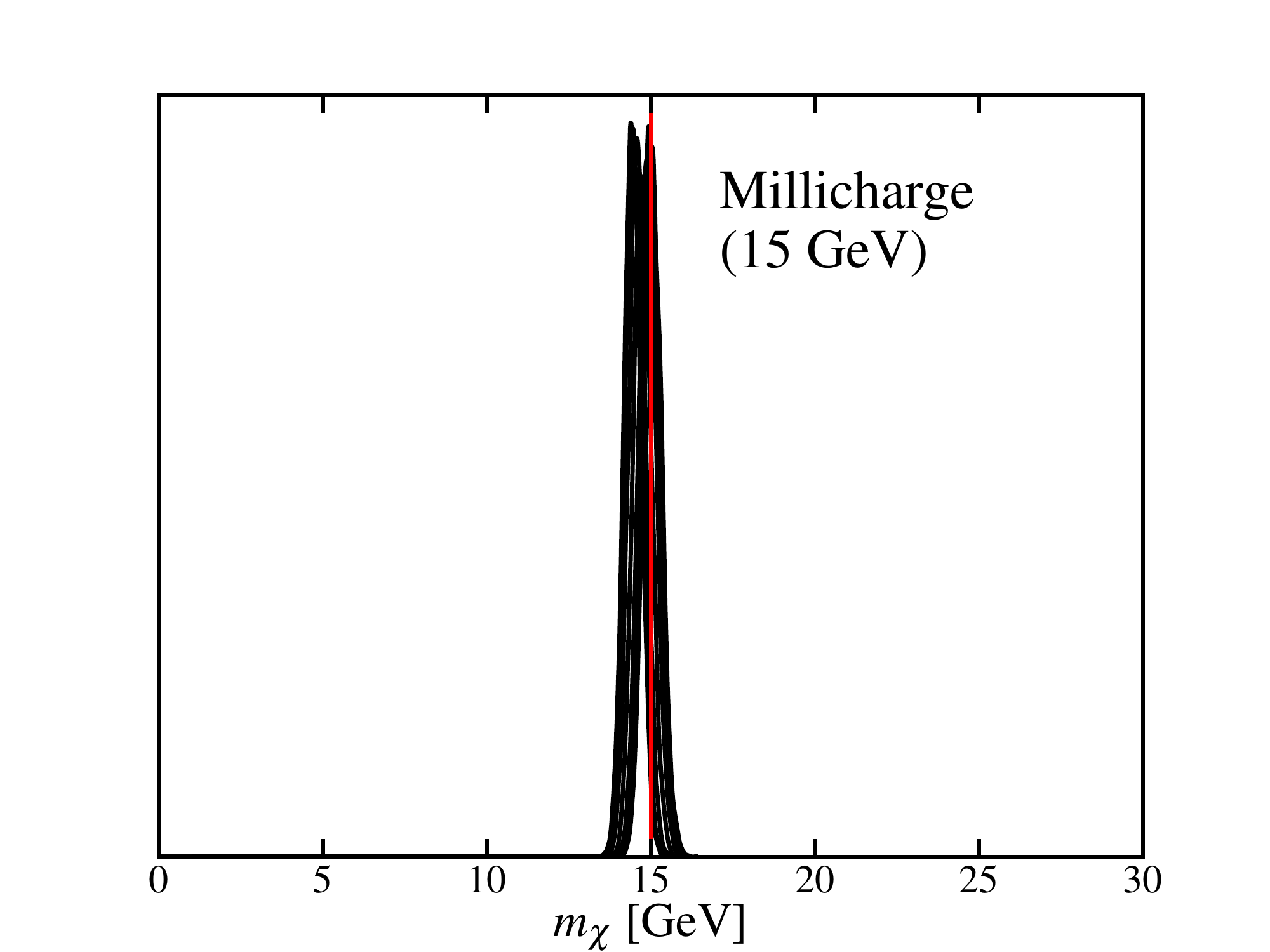}
\includegraphics[width=.3\textwidth,keepaspectratio=true]{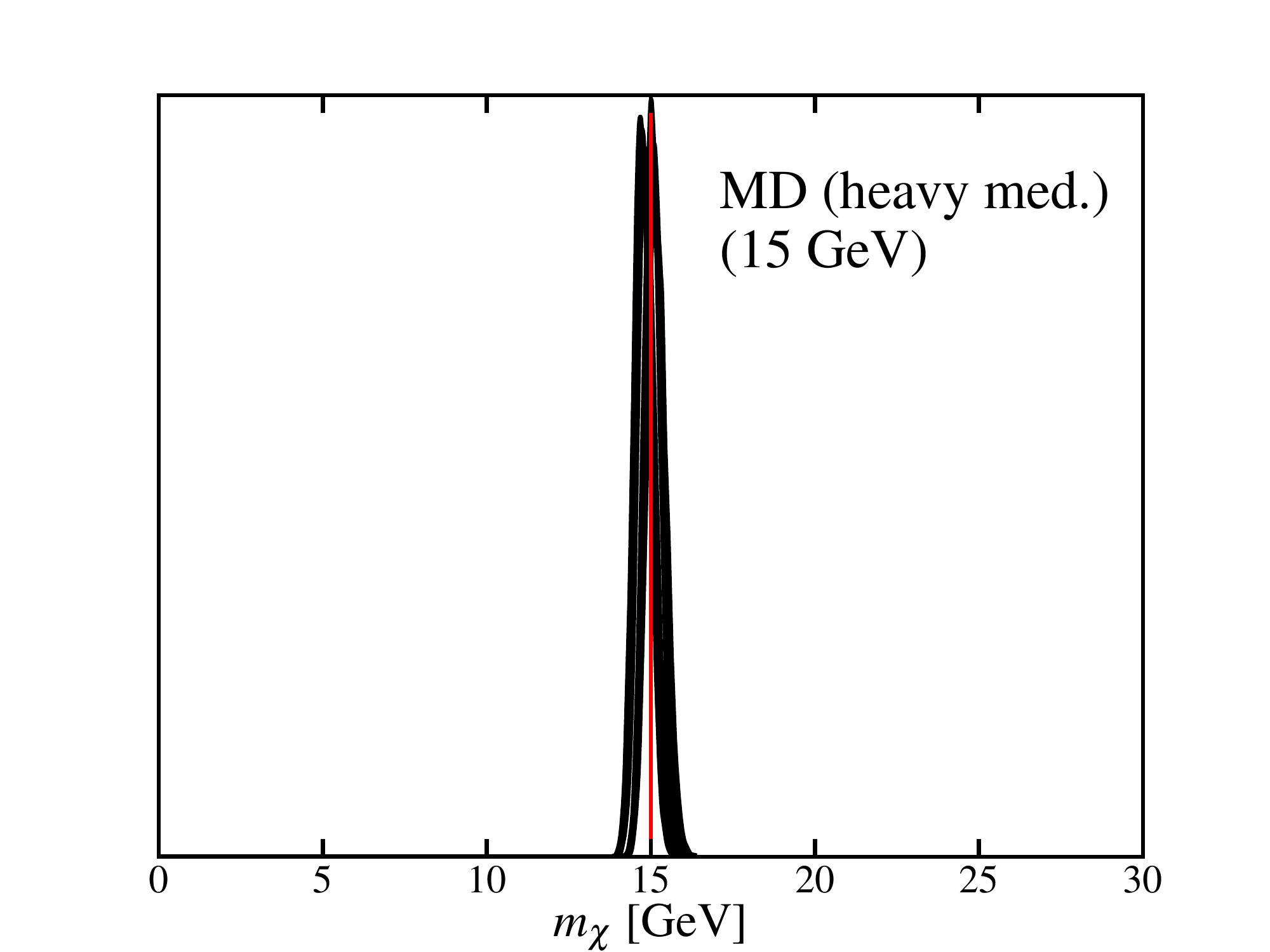}
\includegraphics[width=.3\textwidth,keepaspectratio=true]{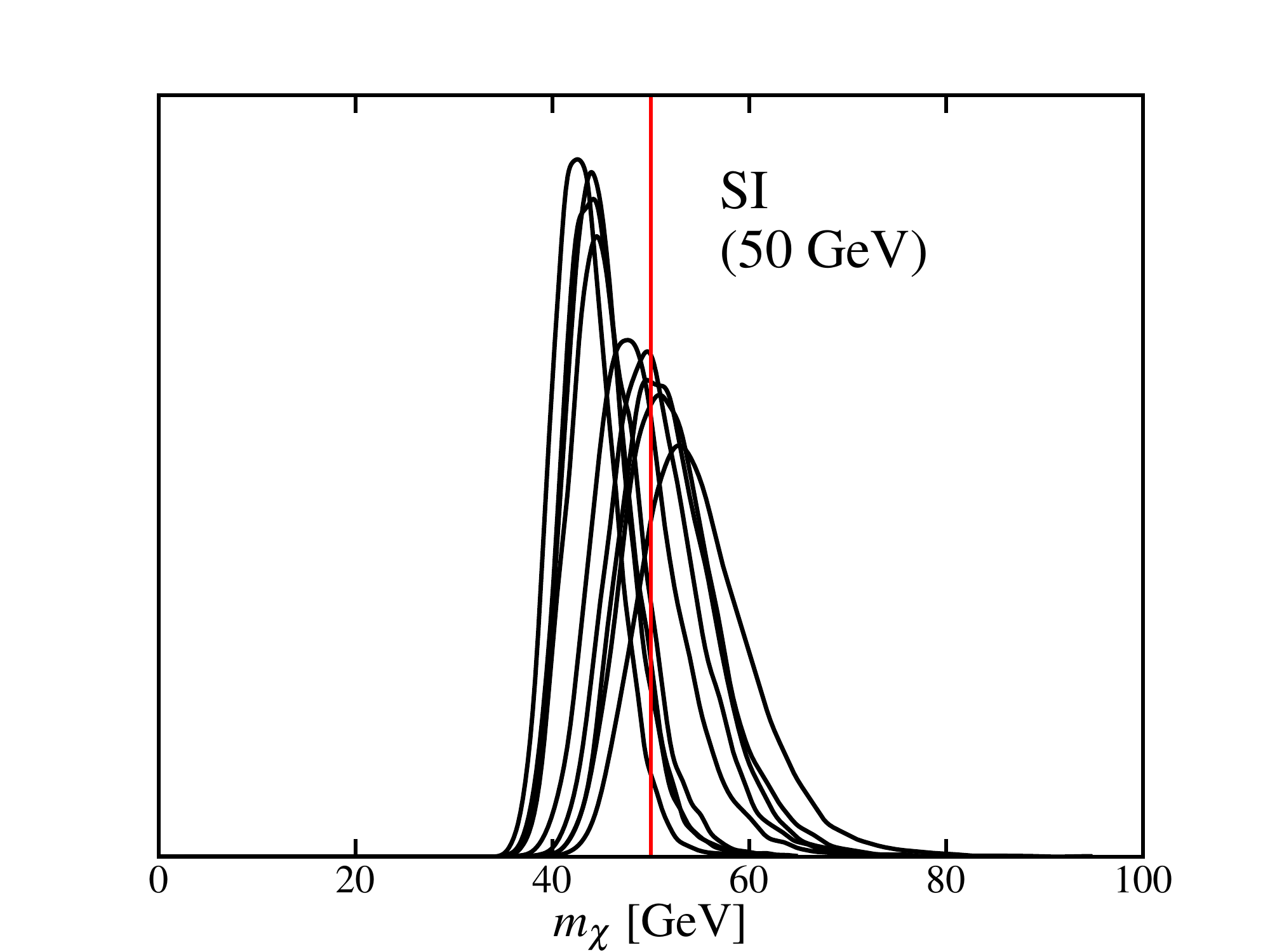}
\includegraphics[width=.3\textwidth,keepaspectratio=true]{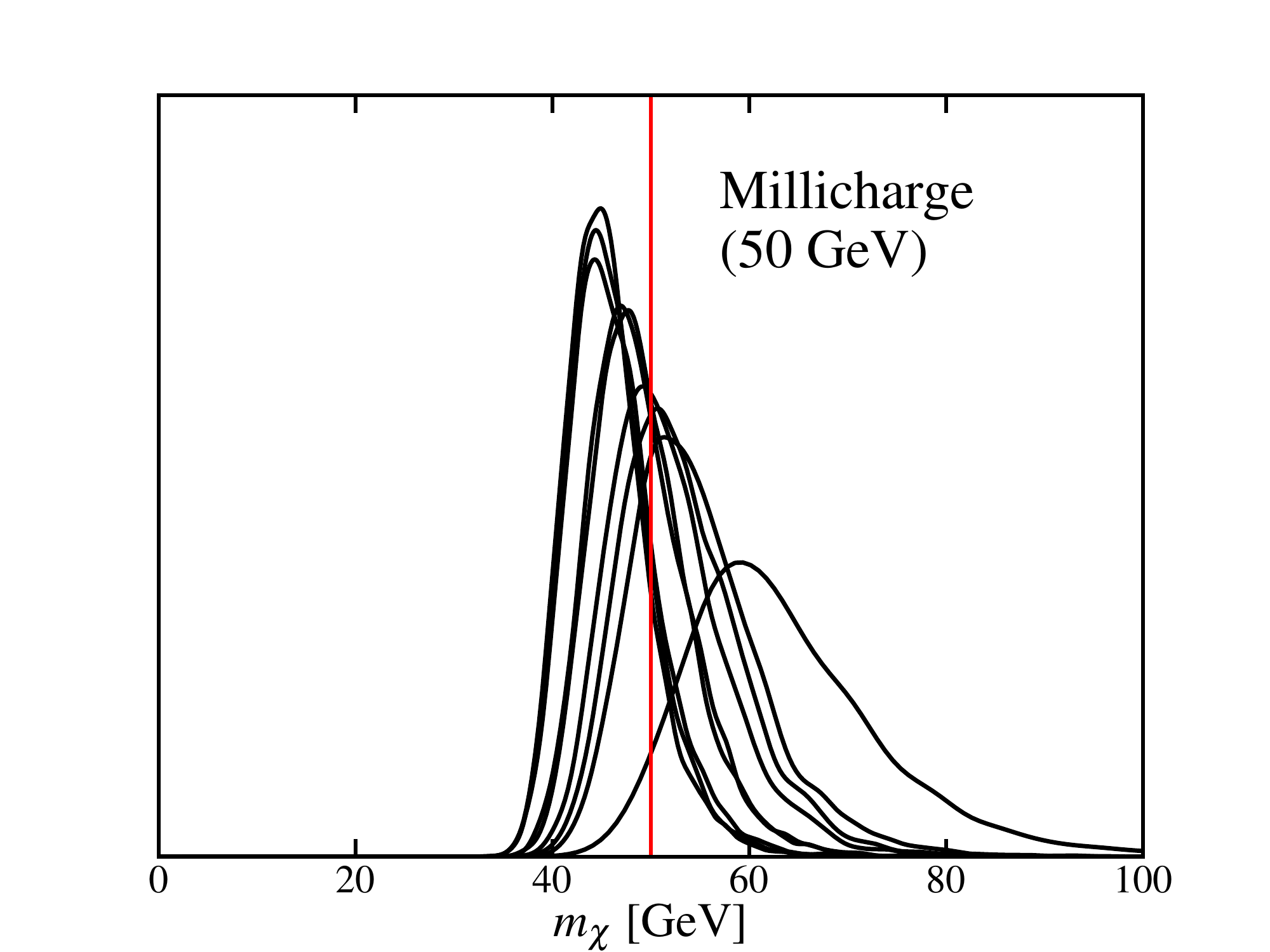}
\includegraphics[width=.3\textwidth,keepaspectratio=true]{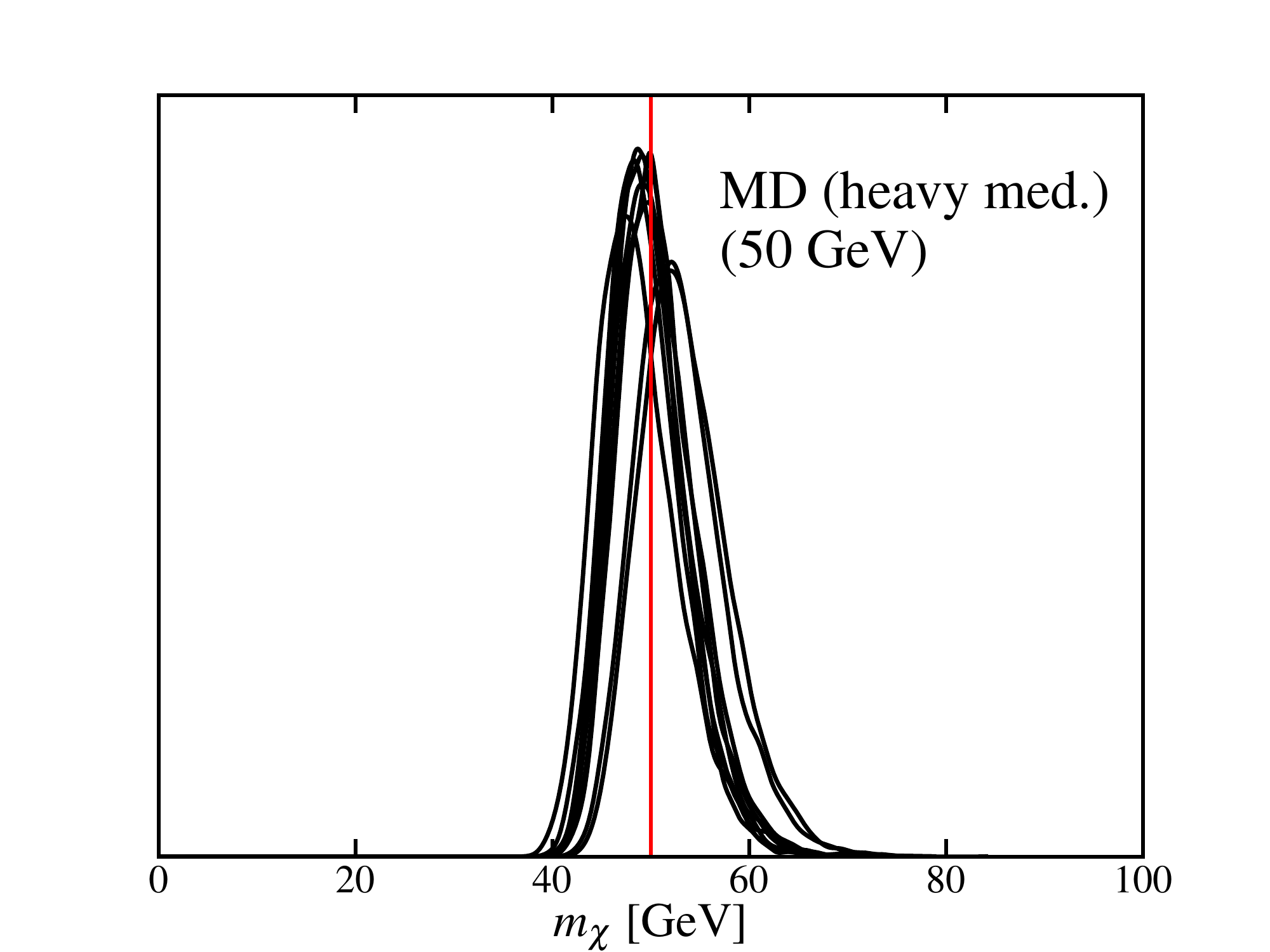}
\includegraphics[width=.3\textwidth,keepaspectratio=true]{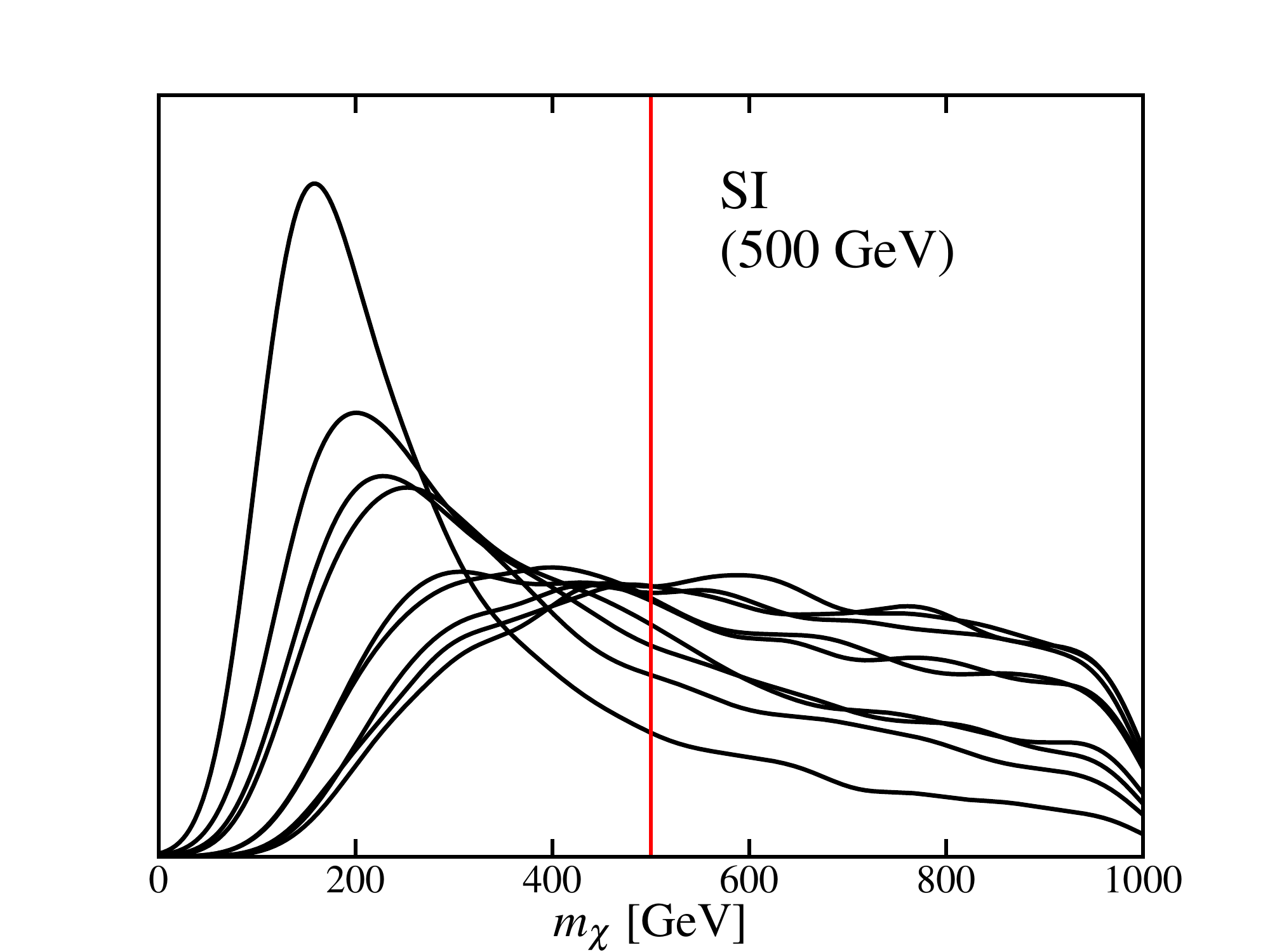}
\includegraphics[width=.3\textwidth,keepaspectratio=true]{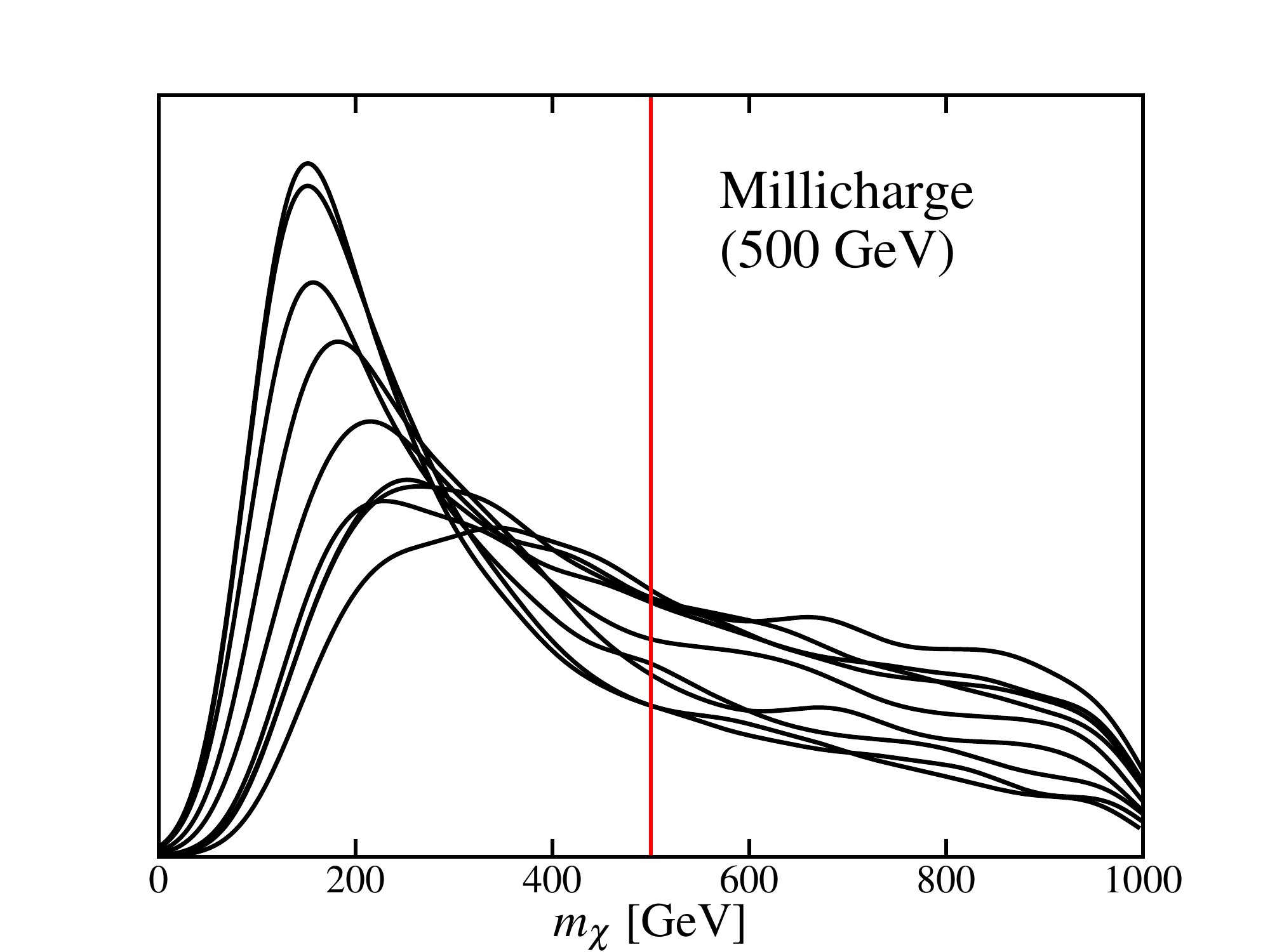}
\includegraphics[width=.3\textwidth,keepaspectratio=true]{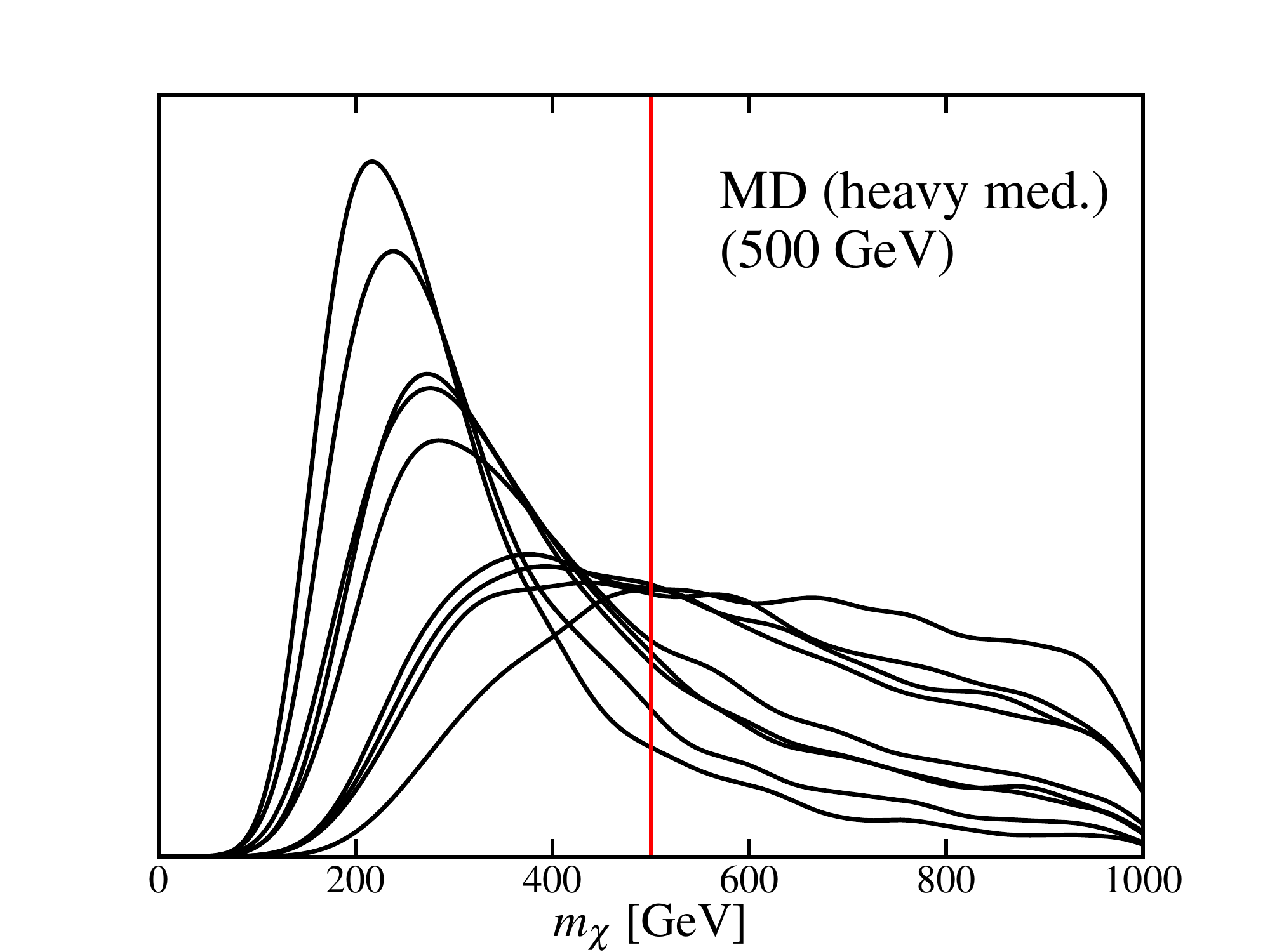}
\caption{Examples of marginalized posteriors for the DM mass are shown to demonstrate the quality of projected mass estimation. A representative subset of baseline simulations described in \ref{sec:baseline} is used, and the right scattering model is fit to simulated data. Each posterior line is reconstructed from a joint analysis of Ge, Xe, and F data. The input DM mass is (top to bottom): 15, 50, and 500 GeV, shown with a vertical red line. As usual, the accuracy of mass reconstruction is degraded for higher masses; however, it does not have a strong dependence on the underlying model.\label{fig:mass_marginals}} 
\end{figure*}

Since the DM particle mass is one of the key DM properties of interest, and likely to be the first measurement compared amongst different DM searches, we examine in more detail the problem of mass reconstruction under different underlying scattering models. First, we analyze the accuracy of mass measurements in an optimistic scenario where a signal is just below the current limit, assuming data from germanium, xenon, and fluorine targets are jointly analyzed. We show marginalized posterior probability distributions for the mass in Figure \ref{fig:mass_marginals}, derived from a joint analysis of data from Xe, Ge, and F experiments of Table \ref{tab:experiments}. The simulations used for this Figure constitute a phenomenologically representative subset of the simulations of our baseline analysis discussed in \S\ref{sec:model_selection}. We show the posteriors obtained assuming the correct scattering hypothesis only.
From this Figure, we see that the mass reconstruction accuracy for all models we consider is similar to that expected from SI and SD interactions, if the correct model is fit to the data. The accuracy is degraded for higher masses, as usual, due to the degeneracy between mass and cross section in this regime.

On the other hand, we confirm results from Ref.~\cite{Gluscevic:2014vga}: if a {\it wrong} model is fit to data, mass estimates can be severely biased, inconsistent amongst different experiments, and have poor accuracy. To illustrate this, we examine the scenario where the standard SI model is fit to recoil--energy spectra generated by another (``true'') model.  This reflects a likely choice for parameter estimation in the DM direct detection discovery phase, that data will be fit with SI scattering; however, the key points we make based on this scenario are more general in scope.
\begin{figure}
\centering
\Large{Xe} \hskip 4cm \Large{Ge} \\
\includegraphics[width=.3\textwidth,keepaspectratio=true]{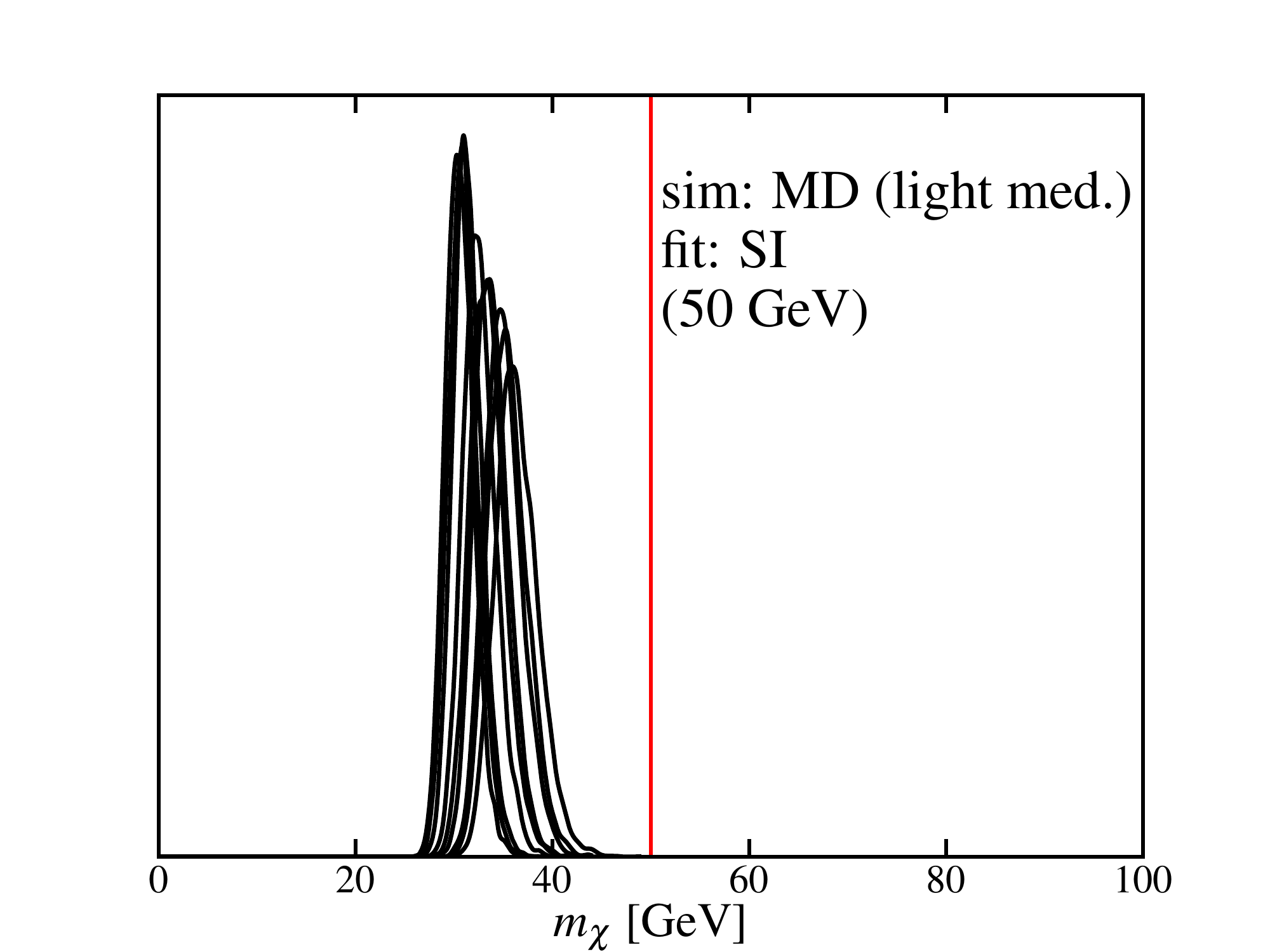}
\includegraphics[width=.3\textwidth,keepaspectratio=true]{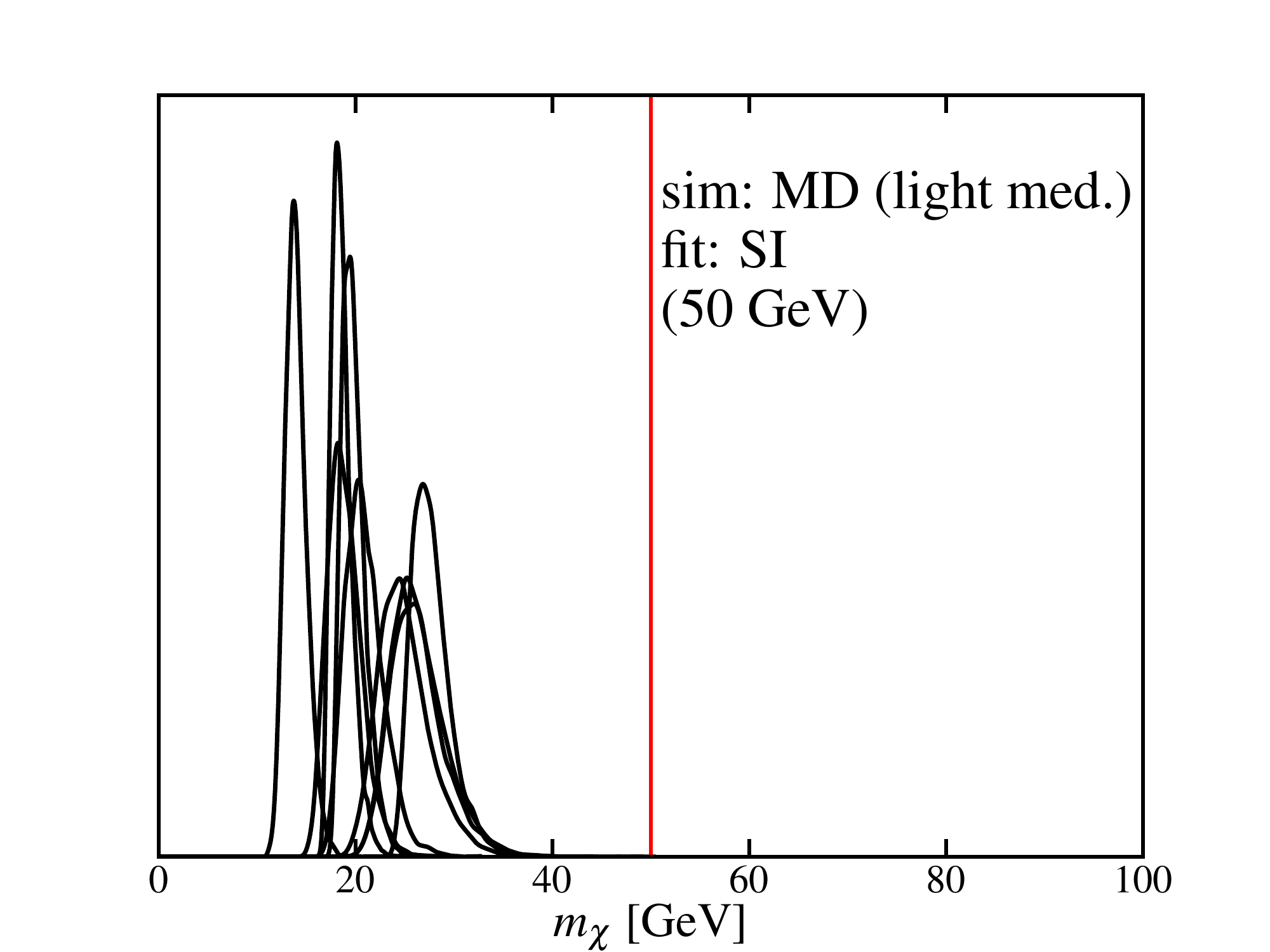}\\
\includegraphics[width=.3\textwidth,keepaspectratio=true]{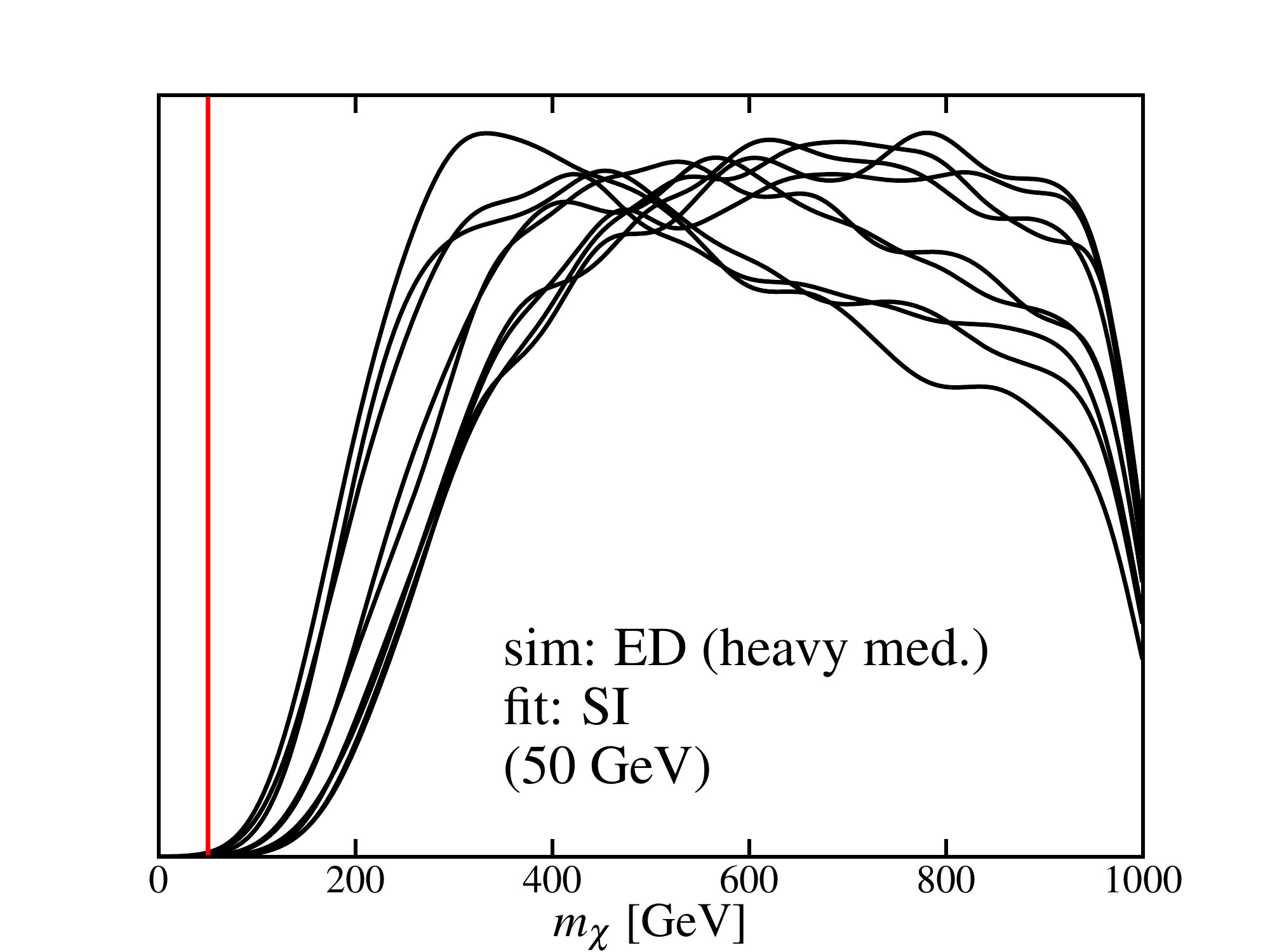}
\includegraphics[width=.3\textwidth,keepaspectratio=true]{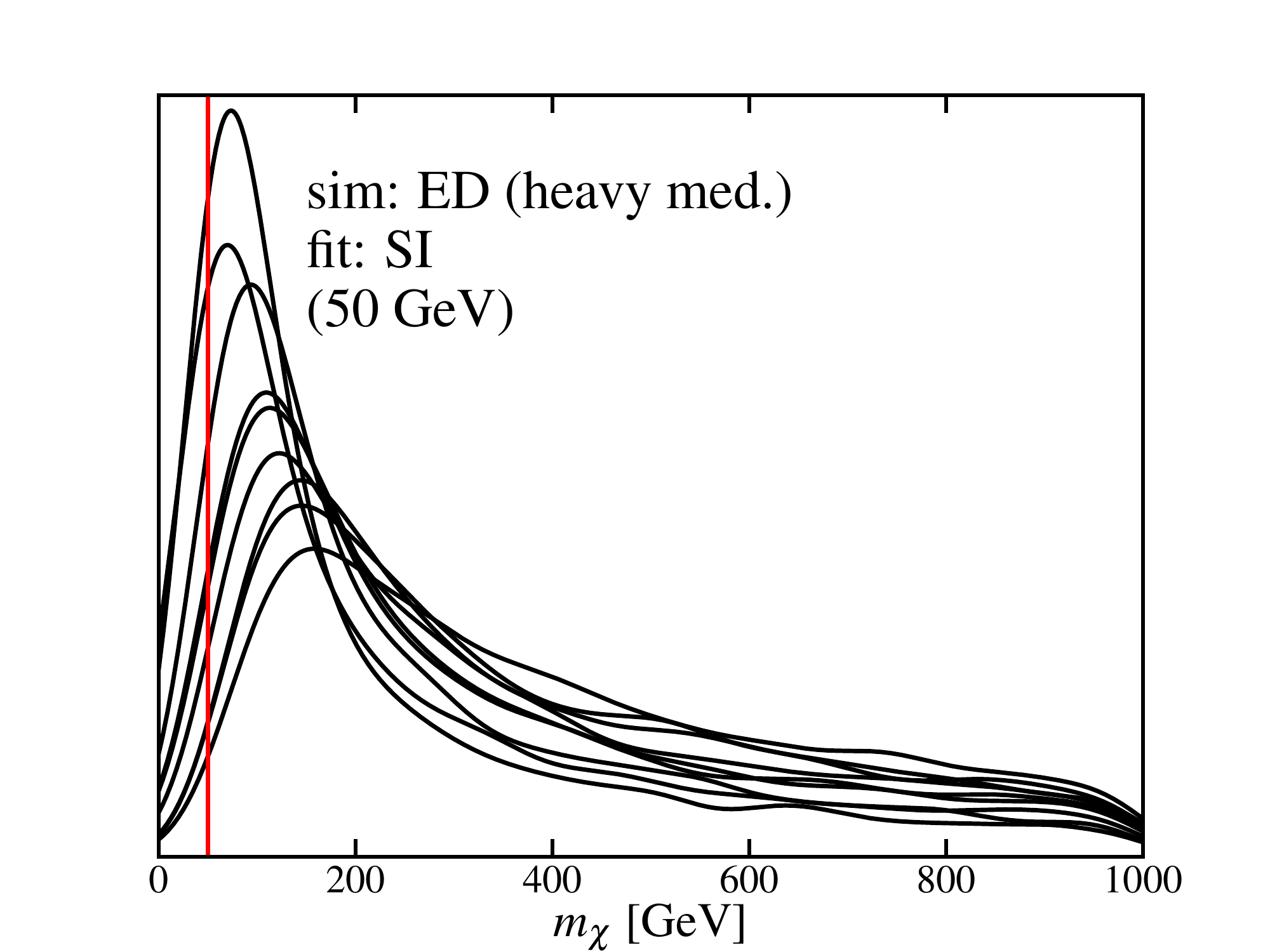}\\
\includegraphics[width=.3\textwidth,keepaspectratio=true]{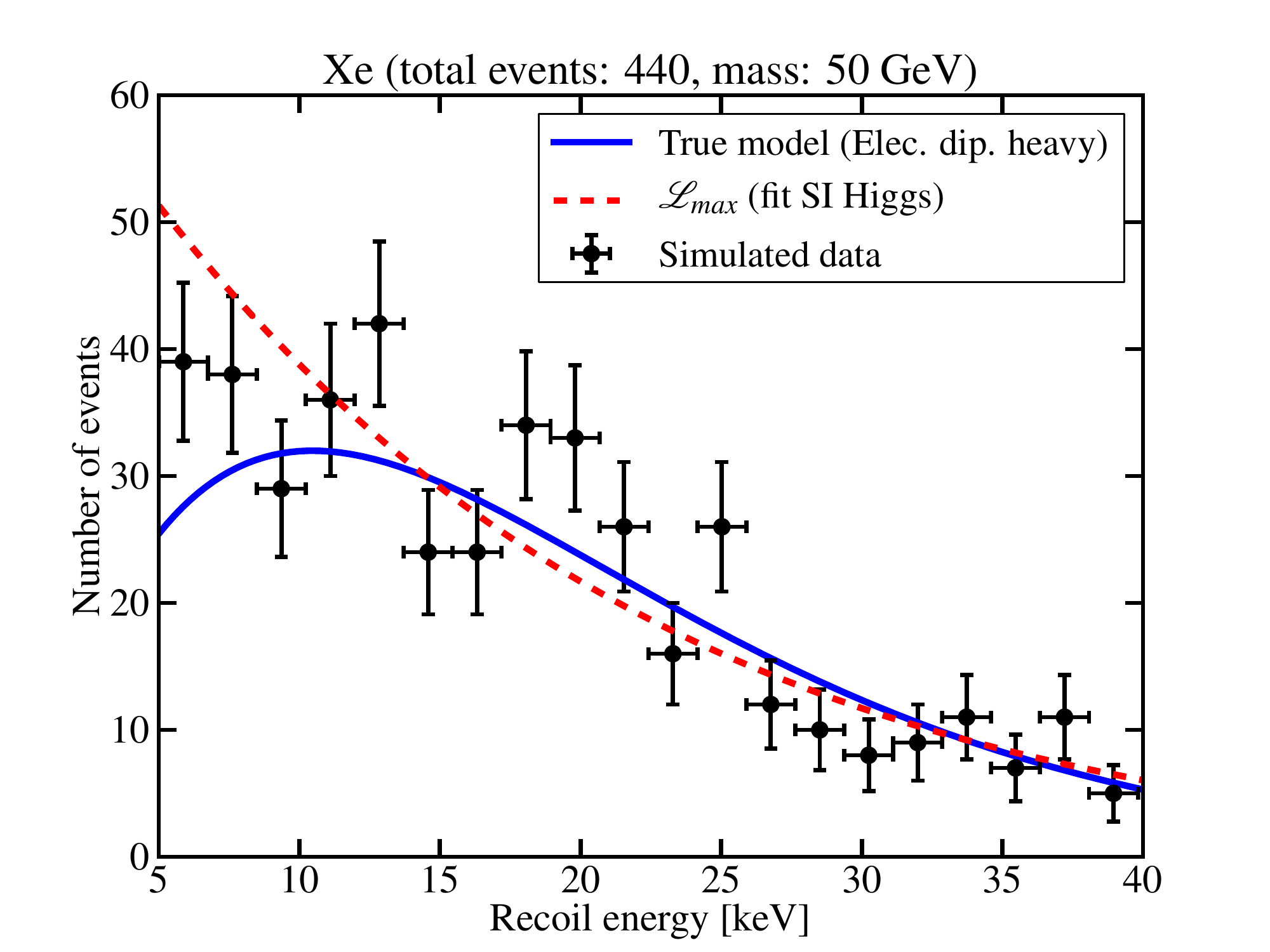}
\includegraphics[width=.3\textwidth,keepaspectratio=true]{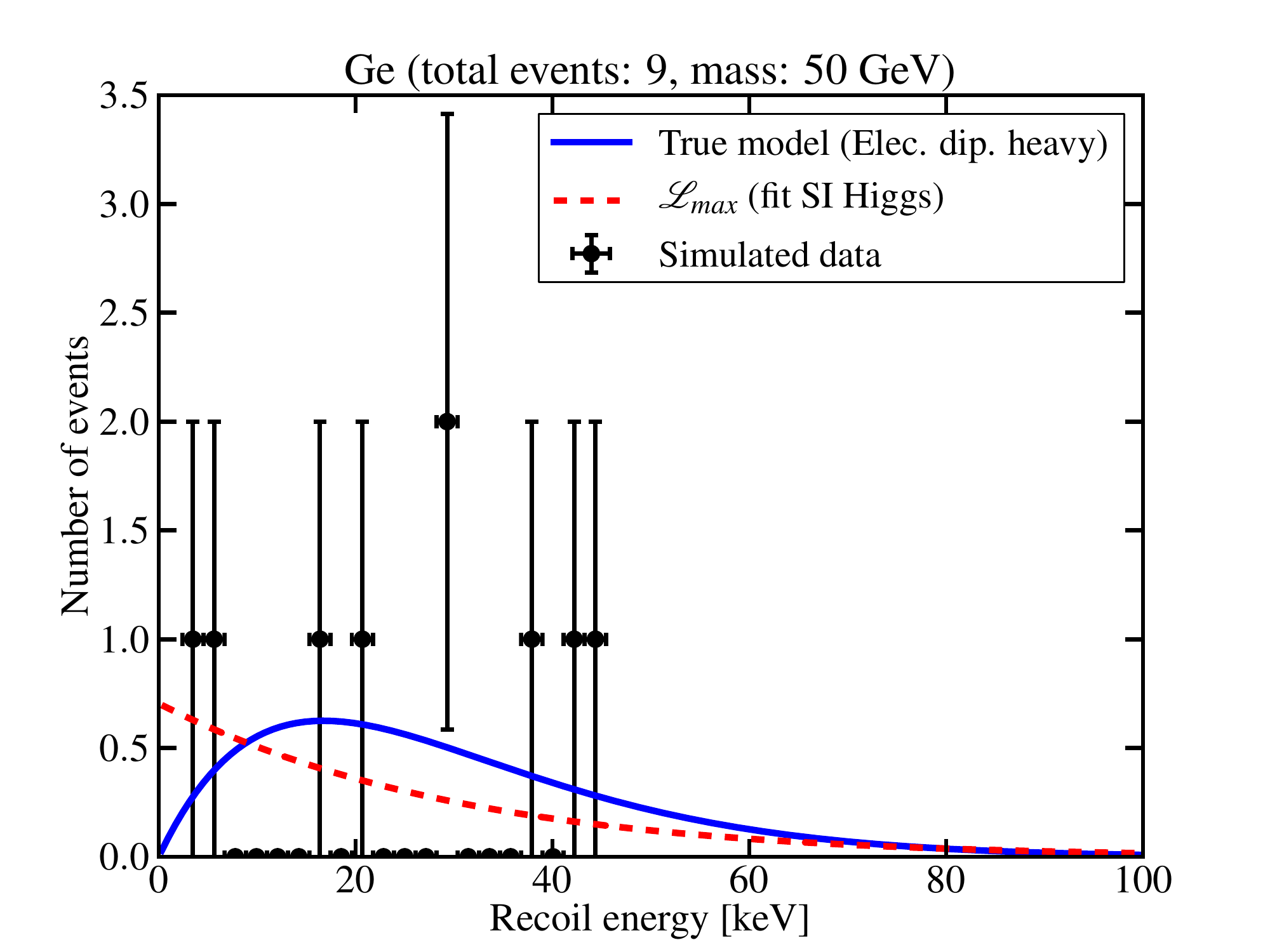}
\caption{Top and middle row: Examples of marginalized posteriors for the DM mass, for a subset of our baseline simulations. The fitting model in all four panels is the standard SI interaction, while the model used to create simulations is magnetic--dipole scattering with a light mediator (upper row), and electric--dipole scattering with a heavy mediator (middle row). The LHS and RHS panels correspond to simulations for a xenon and germanium targets, respectively.
These posteriors demonstrate mass--estimation biases when a wrong model is fit to data; the input DM mass is 50 GeV, shown with a vertical red line. Bottom row: an example of simulated recoil--energy spectrum, shown together with the true underlying model (blue solid), and the fitting model (red dashed) for maximum--likelihood values of the fitting parameters. These simulations correspond to the posteriors shown in the middle row of this Figure. By eye, SI does not look like a bad fit to these data, for either xenon (LHS) or germanium (RHS). \label{fig:mass_biases}}
\end{figure}

In Figure \ref{fig:mass_biases}, we again select a representative subset of our baseline simulations; this time, we only examine fits of the standard SI interaction to scattering signals arising from non--standard operators (specifically, magnetic--dipole scattering through a light mediator, and electric--dipole scattering through a heavy mediator, to cover models without and with the turnover at low--energy recoil rates). The input DM mass is 50 GeV, denoted with a red vertical line. Plots in the LHS column of that Figure correspond to simulations on Xe, and those on the RHS to Ge.  We conclude that, when a steep recoil--energy spectrum (from a light--mediator model) is fit with a standard SI interaction, the posteriors are narrow but biased towards low DM masses. This happens because low masses drive a steeper energy spectrum, making the SI model a better fit to the ``enhanced'' rate at low recoil energies produced by  light--mediator scattering (see the top row of Figure \ref{fig:mass_biases}). Even though inconsistency amongst measurements with different targets might be a hallmark of a wrong scattering hypothesis (compare LHS and RHS panels of the top row in this Figure: biases are different on different targets), it is important to keep in mind that such a cross check may not be available if the signal is only strong enough to be seen with a large exposure (the case of Xe here). On the other hand, when a simulation with a turnover feature (here: from a heavy--mediator dipole model) is fit by the standard SI interaction, the posteriors are artificially wide, with a smaller bias (see the middle row of plots in Figure \ref{fig:mass_biases}). By eye, however, it is not possible to determine that the SI model was a bad fit to data; see the bottom row of Figure \ref{fig:mass_biases}, while model selection with Ge and Xe is able to pick out only the right momentum--dependence class (see bottom right panel of Figure \ref{fig:class_selection_gexe_50gev_select}).

In conclusion, model selection is an important step that should ideally precede parameter estimation in future data analyses, as fitting of the wrong scattering model can severely impact both the accuracy and precision of key DM parameter measurements.
\subsection{$f_n/f_p$ uncertainty}
\label{sec:fnfp}

As discussed in \S\ref{sec:census} and emphasized in, for example, Refs.~\cite{Hill:2014yxa,Crivellin:2015bva}, there are large modeling uncertainties in the choice of the ratio of neutron to proton coupling for some of the models considered here. It has been pointed out in the literature that direct detection data are not very likely to be able to reconstruct this parameter \cite{Pato:2011de}, or even that fine-tuning such a parameter can upset expectations of relative experimental sensitivities \cite{Feng:2011vu}.\footnote{Even in the face of tree--level fine--tuning for special regions of parameter space, large nucleus--dependent next--to--leading--order corrections may spoil the cancellations \cite{Cirigliano:2013zta}.}  We thus investigate the impact of this uncertainty on the DM mass measurements and on model selection in this Section.

\begin{figure*}[t]
\centering
\includegraphics[width=.3\textwidth,keepaspectratio=true]{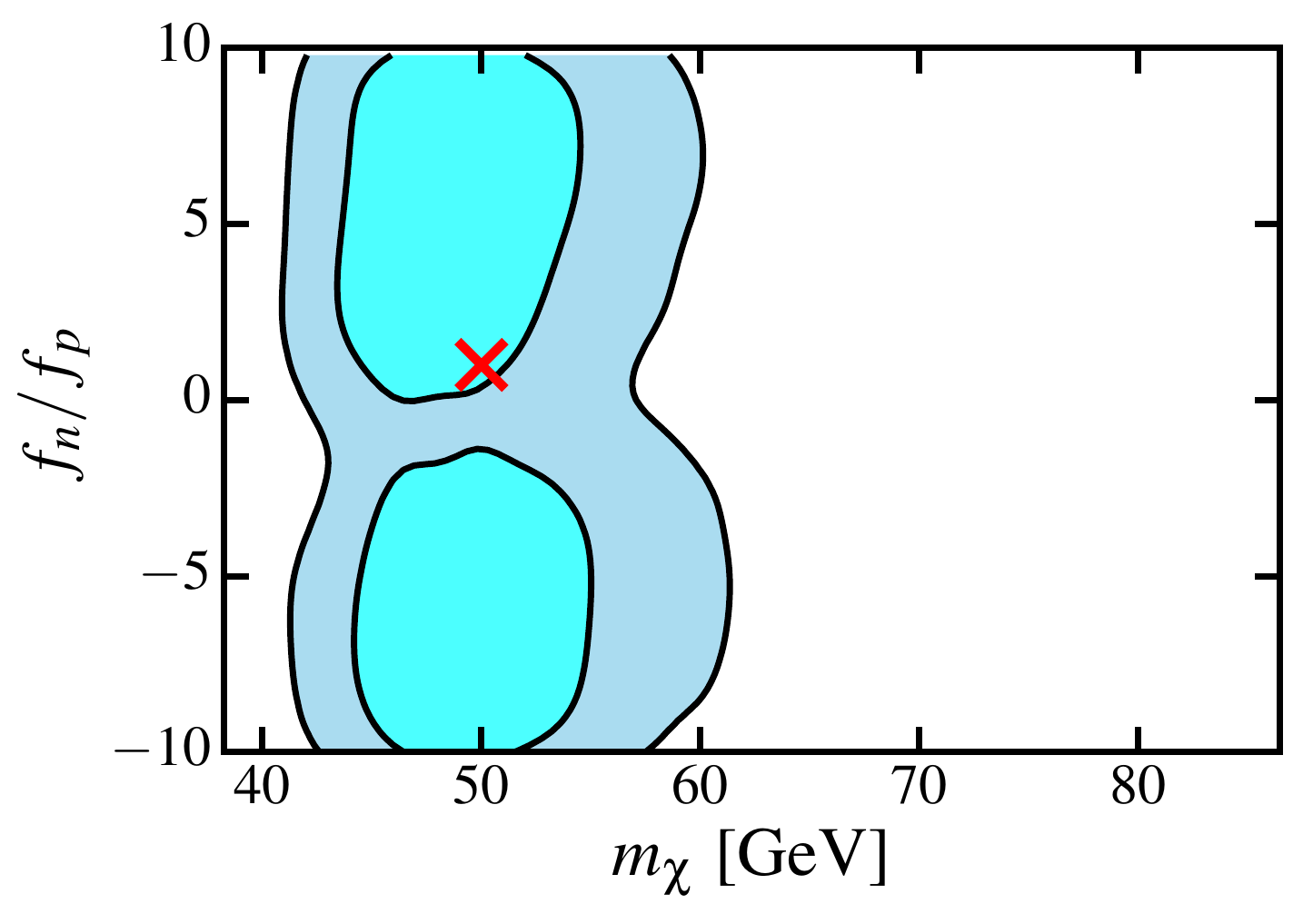}
\includegraphics[width=.3\textwidth,keepaspectratio=true]{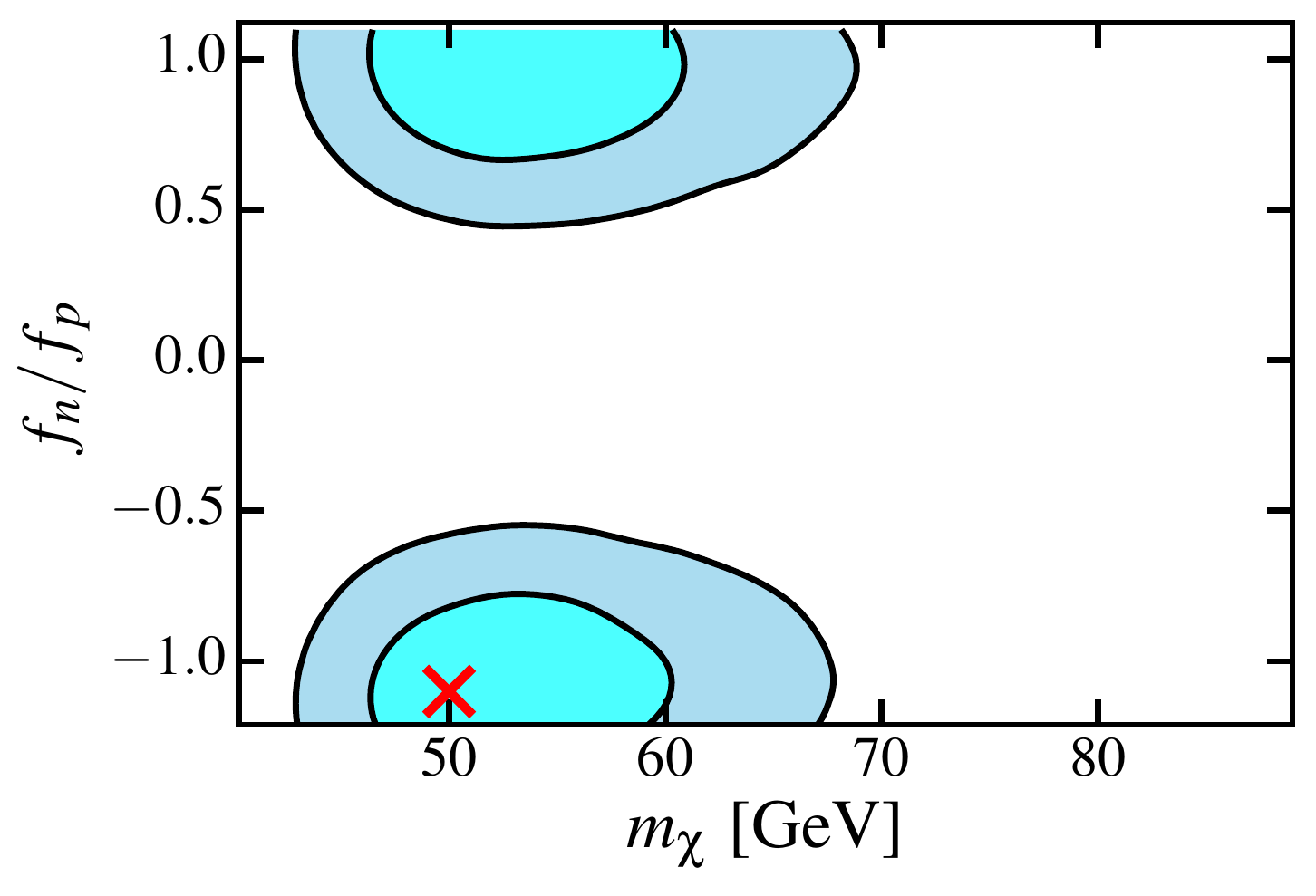}
\includegraphics[width=.3\textwidth,keepaspectratio=true]{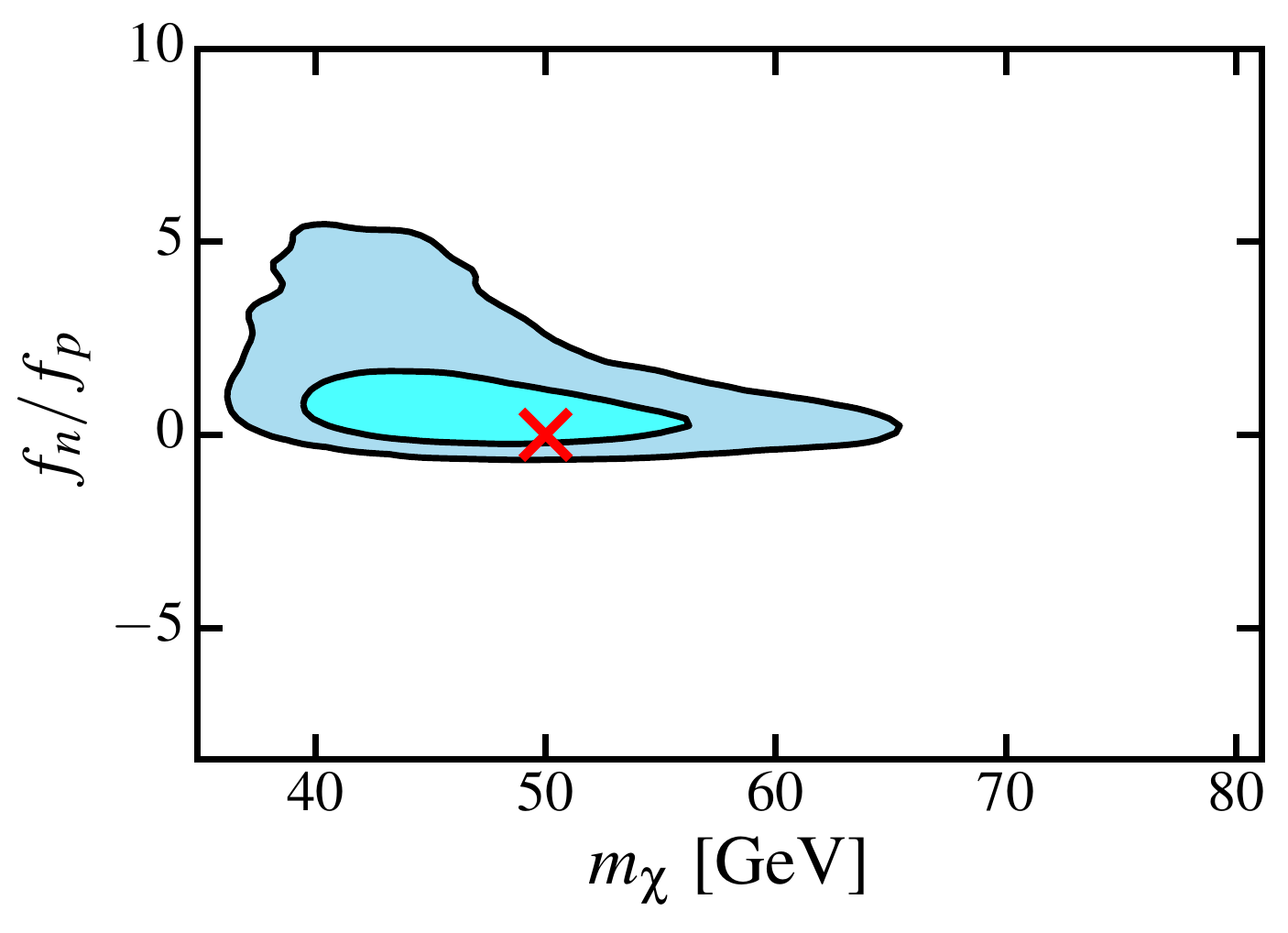}
\caption{Examples of the posteriors for a joint analysis of Ge+Xe+F simulations; the right
scattering operator was used for the fits. The simulations were
created using (left to right): the standard SI model, the standard SD model,
and the Millicharge model ($f_n/f_p$ values used for these simulations are listed in Table
\ref{tab:operators}). The red X denotes the input values. When performing posterior analysis, the $f_n/f_p$ was left as a free parameter with a wide prior (from -10 to 10), in addition to $m_\chi$, and $\sigma_p$ (set to the current upper limits in the simulations). In the SI case, $f_n/f_p$ is completely unconstrained, in SD case, the posterior is bimodal and there is a degeneracy with the sign of this parameter, and in the Millicharge case, the right sign is picked out, but the posterior is still broad. \label{fig:fnfps}}
\end{figure*}

First, we investigate how the uncertainty on $f_n/f_p$ impacts the extraction of the DM particle mass. In Figure \ref{fig:fnfps}, we show the marginalized 2D posterior for DM mass versus $f_n/f_p$ for our baseline simulations of several scattering models, where data from Xe, Ge, and F experiments is jointly analyzed. When performing fits to recover these posteriors, we let $f_n/f_p$ vary as a free parameter between -10 and 10 and fit the right scattering model (with the right operator and scattering rate, described in Table \ref{tab:operators}, but with a free $f_n/f_p$) to the simulated data. For simulations created with the SI model, $f_n/f_p$ is completely unconstrained; for the SD simulations, the posterior is bimodal and there is a degeneracy of the sign of this parameter; and for the Millicharge
simulations, the right sign is picked out, but the posterior is fairly broad.\footnote{Note that, for
hypothesis fitting to simulations created using the Millicharge model (where $f_n/f_p=0$), we use SI DM with a
light mediator (i.e.~$\op_\text{SI}/\vec{q}^{\,2}$), where $f_n/f_p$ is a free parameter.}  However, in spite of introducing this additional degree of freedom, there does not seem to be a significant impact on the measurement of the DM mass; $f_n/f_p$ and $m_\chi$ thus do not appear to be degenerate with each other, if the right scattering hypothesis is chosen.

\begin{figure*}
\centering
\includegraphics[width=.45\textwidth,keepaspectratio=true]{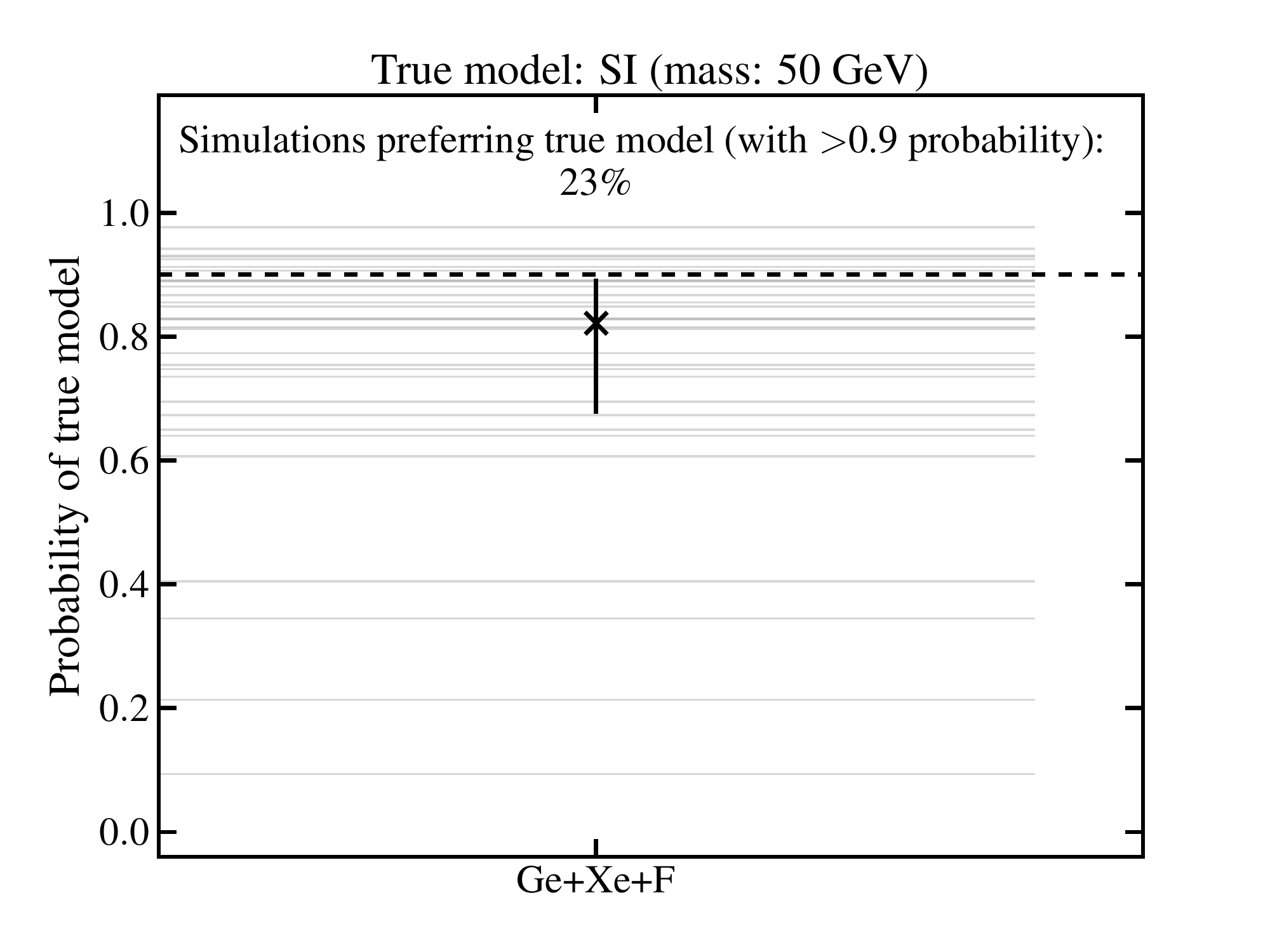}
\includegraphics[width=.45\textwidth,keepaspectratio=true]{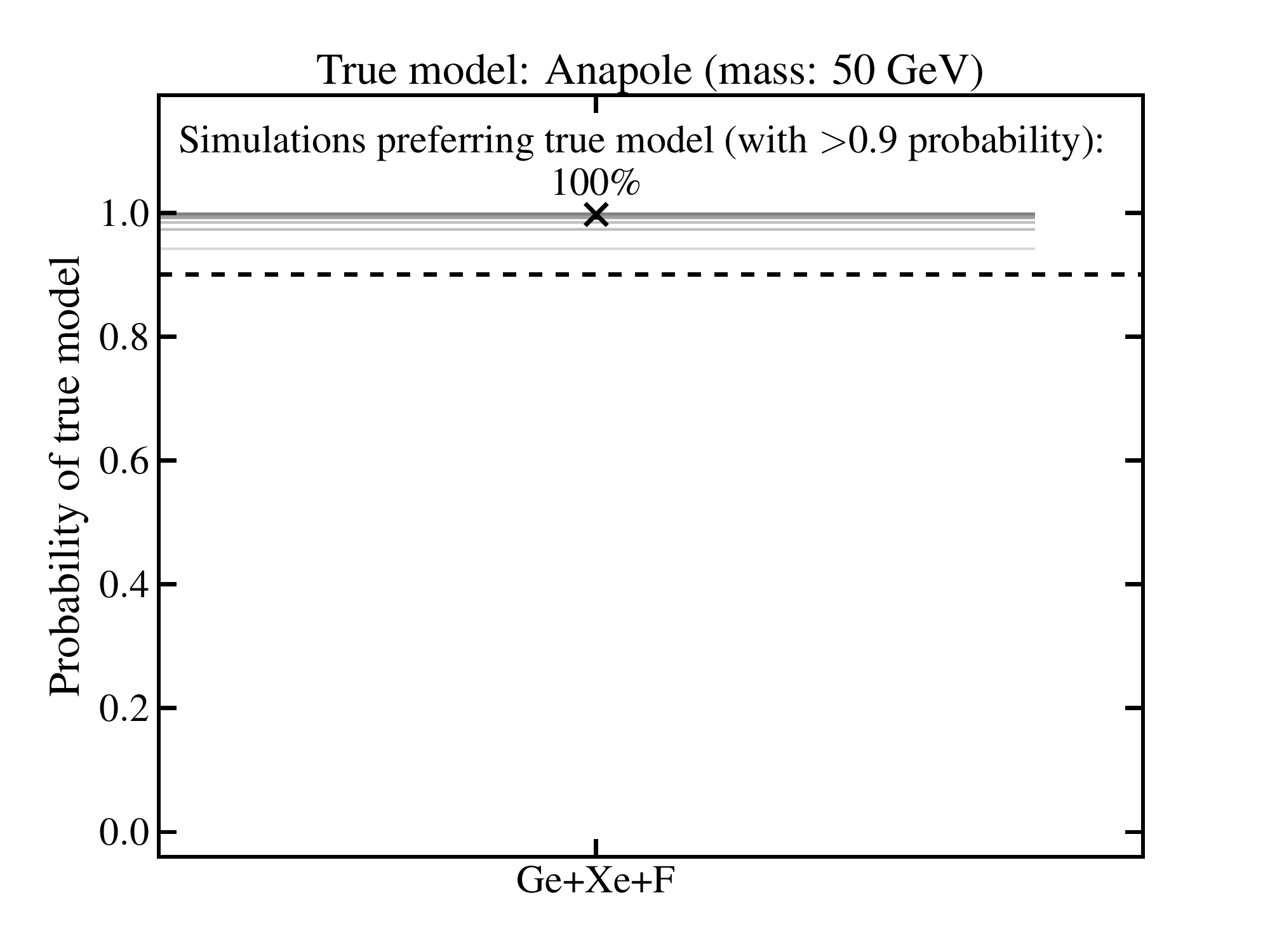}
\caption{Model selection results analogous to those of Figure \ref{fig:model_selection_gexe_50gev_select}, but with 10  models of Table \ref{tab:operators} (SI, SD, four dipole models, Anapole, SI${_{q^2}}$, SD${_{q^2}}$, and SD${_{q^4}}$) treated as competing hypotheses, with $f_n/f_p$ fit as a free parameter, for all but the photon--mediated models. As compared to our baseline results, model selection still seems robust to allowing this additional degree of freedom, at least for SI and Anapole shown here.\label{fig:model_selection_freefnfp}}
\end{figure*}

Second, we examine how freedom in $f_n/f_p$ might affect model selection. For this purpose, we take a small subset of our baseline simulations described in \S\ref{sec:baseline}---simulations created under the standard SI model, and under the Anapole model---and analyze them in a new way. For this analysis, we pick the following 10 models from Table \ref{tab:operators} as competing hypotheses to fit to the simulations: SI, SD, the four dipole models, Anapole, SI${_{q^2}}$, SD${_{q^2}}$, and SD${_{q^4}}$. However, unlike in previous analyses where we held $f_n/f_p$ fixed to the correct value when performing model fits, this time, we allow $f_n/f_p$ to be an additional free parameter in the range between -100 and 100 (note that this is an even wider prior range than that used above to investigate impact on mass determination), for all models where the choice of this parameter is not fixed.\footnote{Note that for Anapole and dipole models (i.e.~for the photon--mediated models), the choice of $f_n/f_p$ is not free, so these 5 models still only have 2 free parameters.} The presentation of the results in Figure \ref{fig:model_selection_freefnfp} is as before. We jointly analyze data from Xe, Ge, and F experiments for a 50 GeV DM particle and a signal at the current upper limit, a combination that has close to 100$\%$ success rate in our baseline analysis. From Figure \ref{fig:model_selection_freefnfp} and by examining the associated model probabilities, we see the following. First, model selection is significantly degraded for simulations under SI. However, most of the probability in this case is actually only distributed between SI and SD models. For Anapole, the situation is almost unchanged from the baseline analysis: this model is confidently identified even allowing such a large modeling uncertainty. 

The results presented in this Section only cover a small subset of possible underlying models and DM masses. However, their implications are very optimistic for future data analysis---they indicate that selection of the right underlying model may be robust to the uncertainty on the ratio of nucleon couplings. We emphasize that this conclusion holds true even if there are values (or localized portions of the parameter space) for $f_n/f_p$ for which data can be well fit by a ``wrong'' model. The evidence calculation takes into account the entire prior space, and the existence of such ``special islands'' of $f_n/f_p$ values does not affect the results of Bayesian model selection.
\subsection{Experimental designs}
\label{sec:experimental_params}

So far, our analysis has demonstrated the importance of combining a variety of nuclear targets (including germanium, xenon, fluorine, and iodine) to correctly identify the type of DM--nucleon scattering interaction using direct detection. In this Section, we examine how the statistical sample (i.e.~the number of observed recoil events) and the quality of model selection depend on the recoil--energy window available for data analysis, for different targets. 

\begin{figure}[t]
\includegraphics[width=.35\textwidth,keepaspectratio=true]{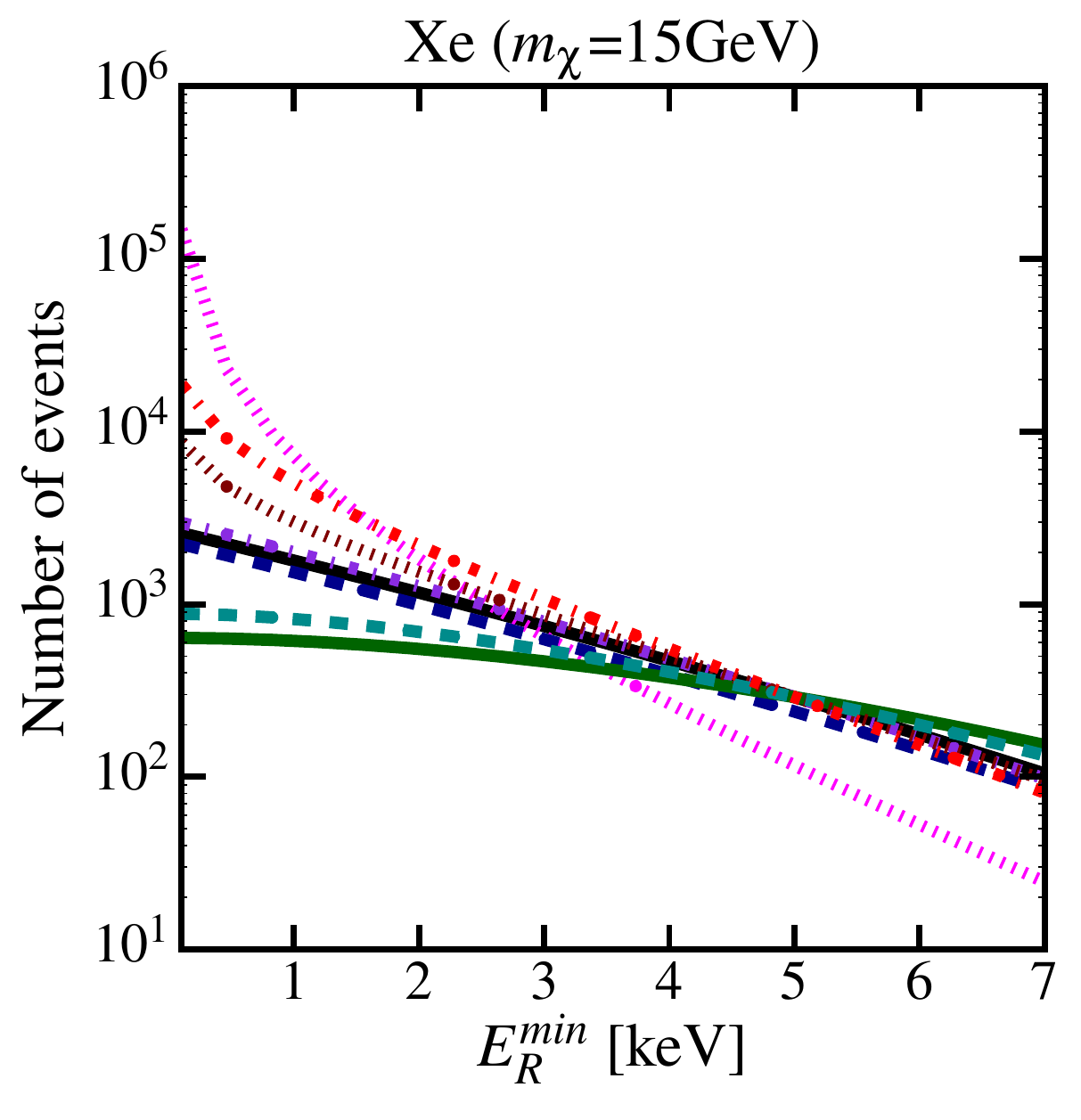}
\includegraphics[width=.51\textwidth,keepaspectratio=true]{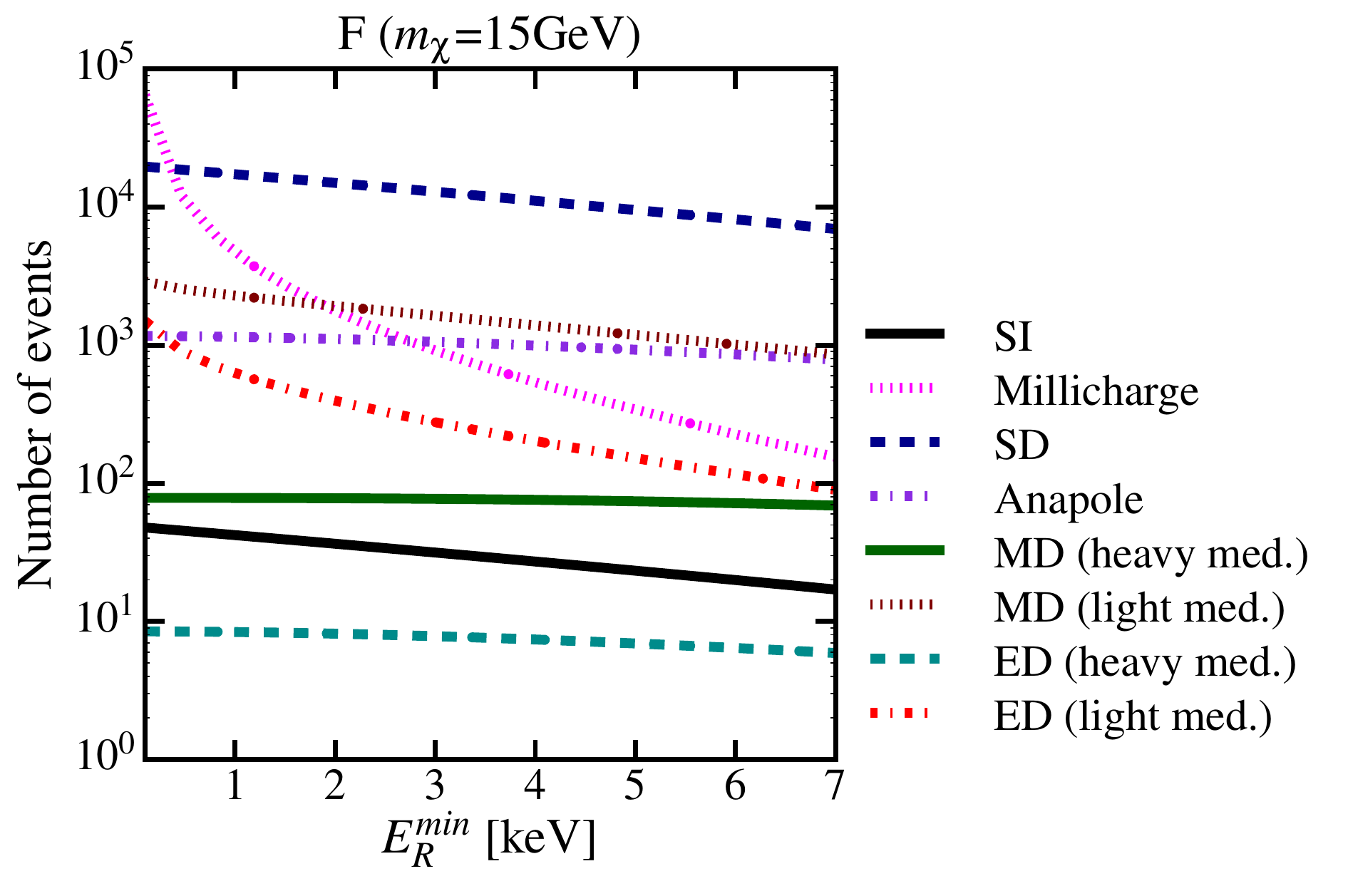}\\
\includegraphics[width=.35\textwidth,keepaspectratio=true]{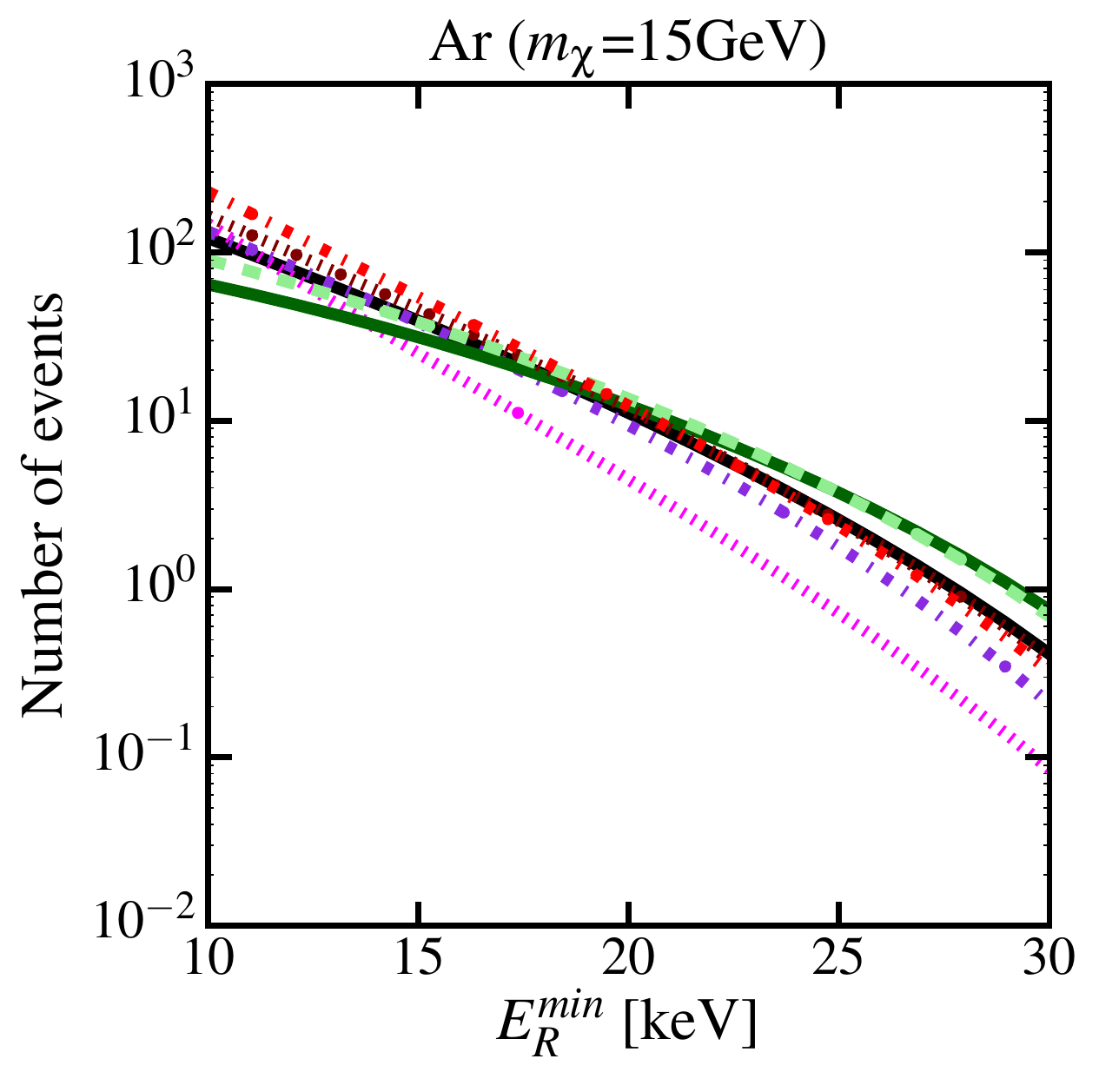}
\includegraphics[width=.35\textwidth,keepaspectratio=true]{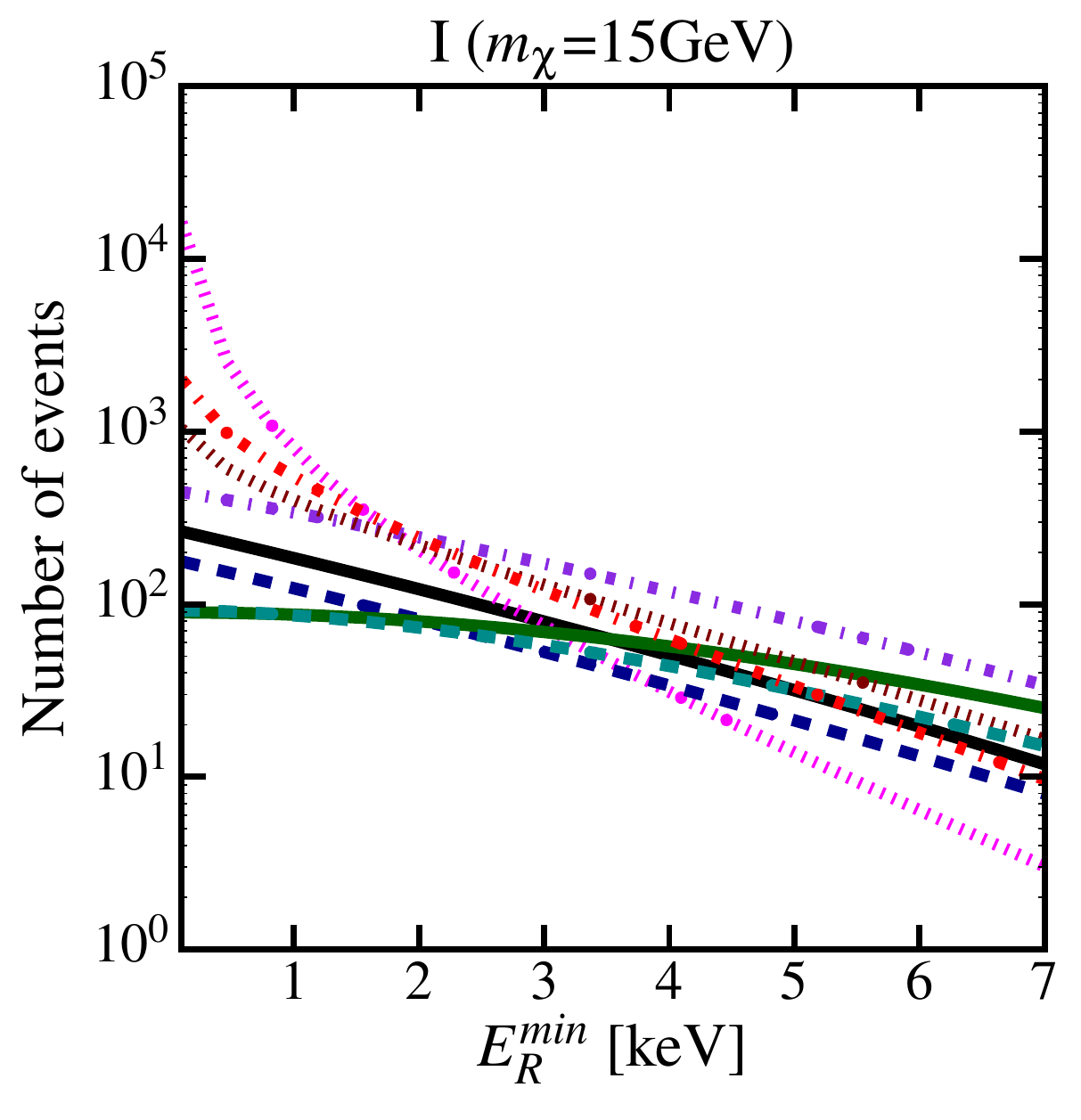}
\caption{Total number of expected events as a function of the minimum recoil energy threshold for a variety of mock G2 experiments. All experimental parameters are fixed to the values in Table \ref{tab:experiments} except $E_R^\text{min}$. Models shown here belong to set I discussed in \ref{sec:analysis}; statistical gain of extending the analysis window to lower thresholds is most evident for low masses, so here, we show results for 15 GeV DM particle. \label{fig:Nexps_Ermin}}
\end{figure}

\begin{figure}[t]
\includegraphics[width=.35\textwidth,keepaspectratio=true]{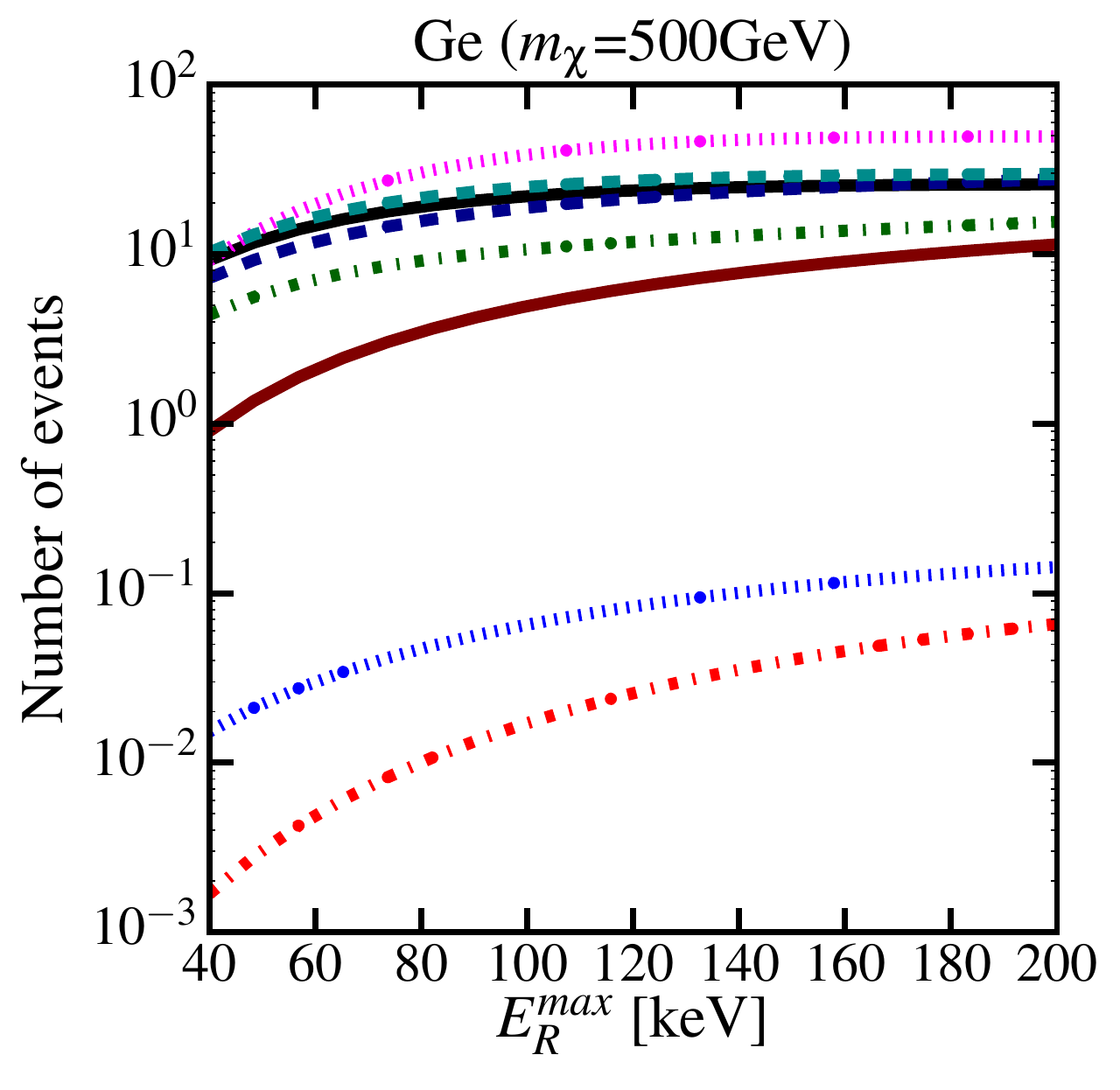}
\includegraphics[width=.51\textwidth,keepaspectratio=true]{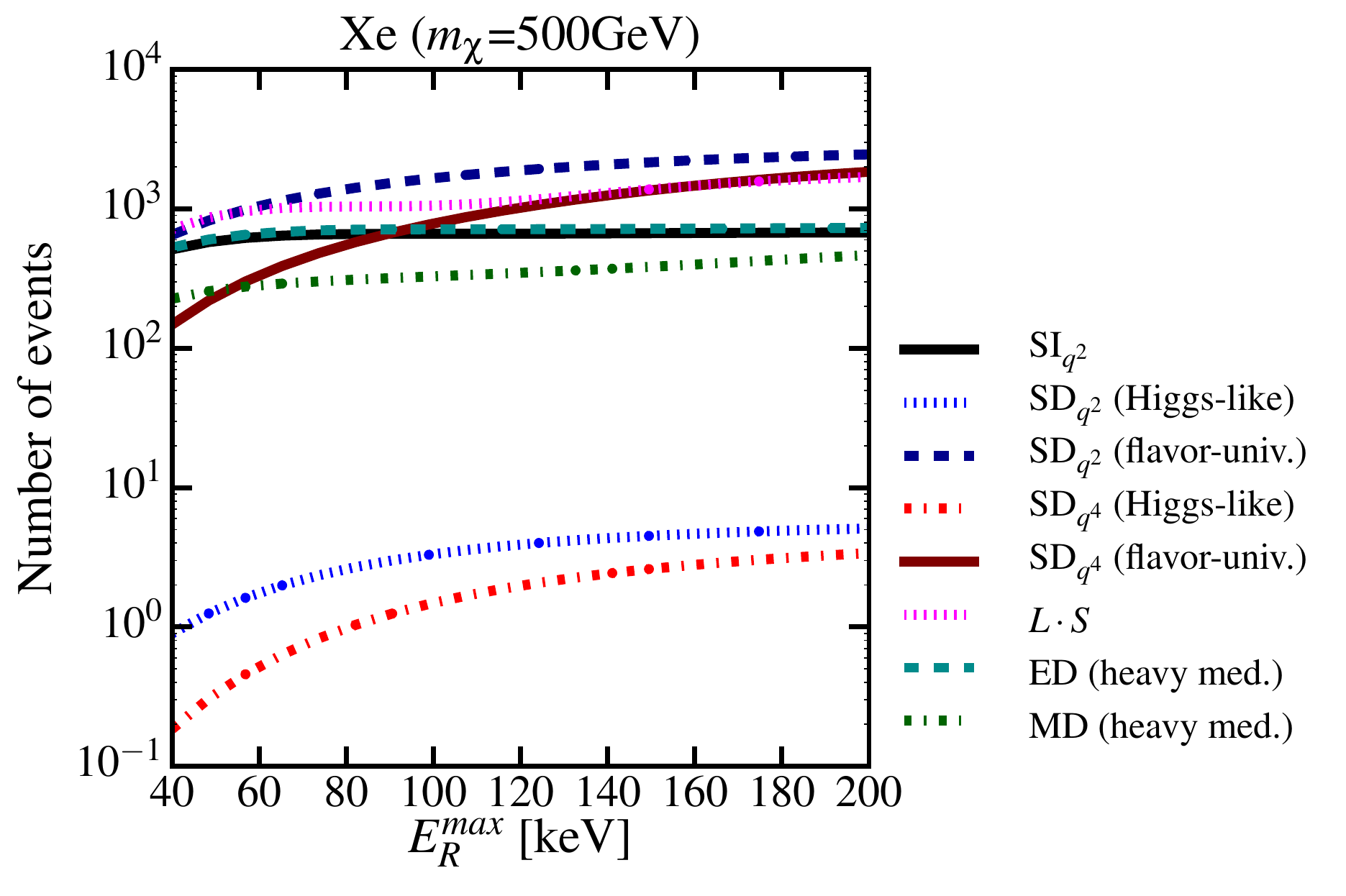}\\
\includegraphics[width=.35\textwidth,keepaspectratio=true]{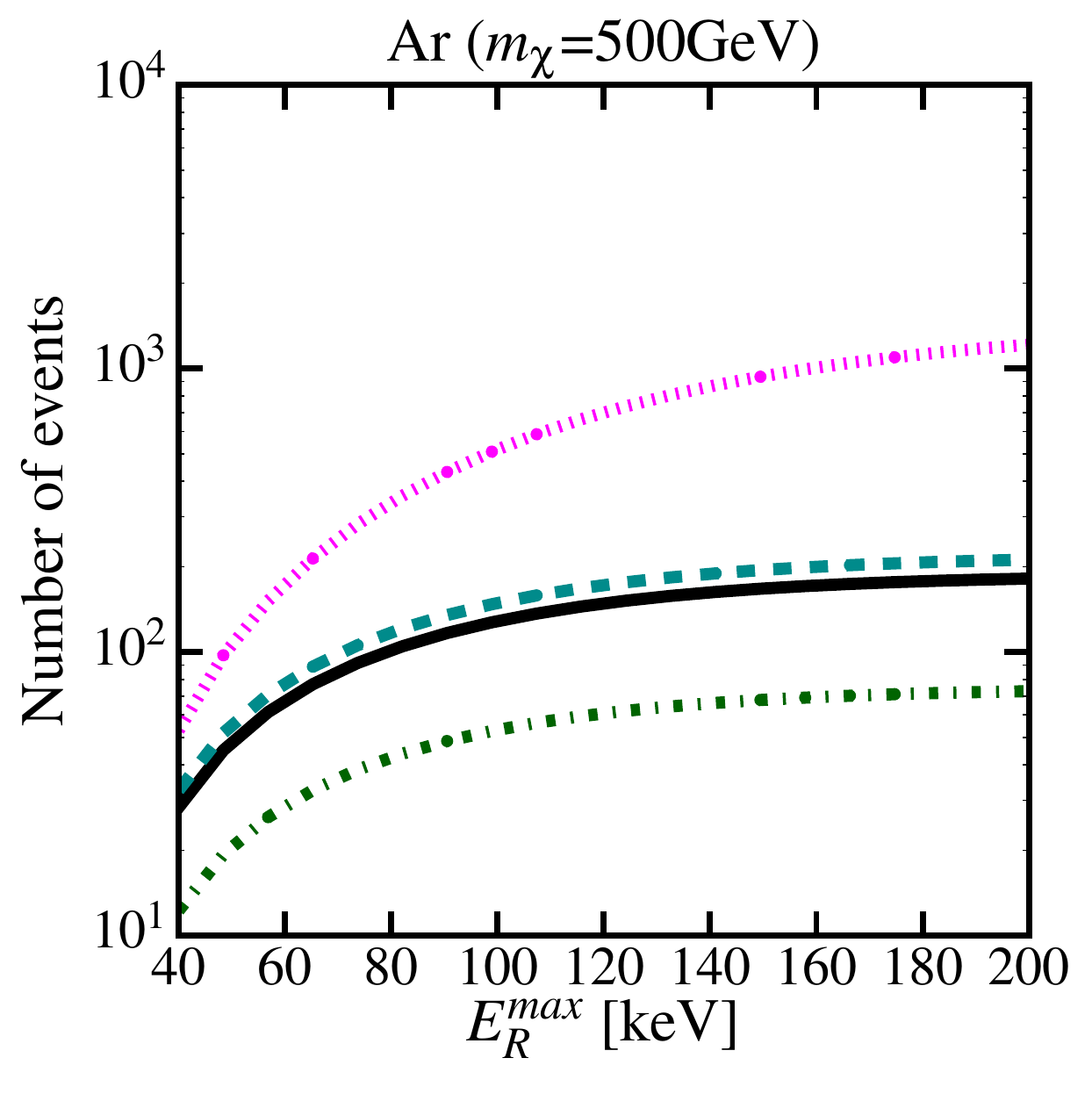}
\includegraphics[width=.35\textwidth,keepaspectratio=true]{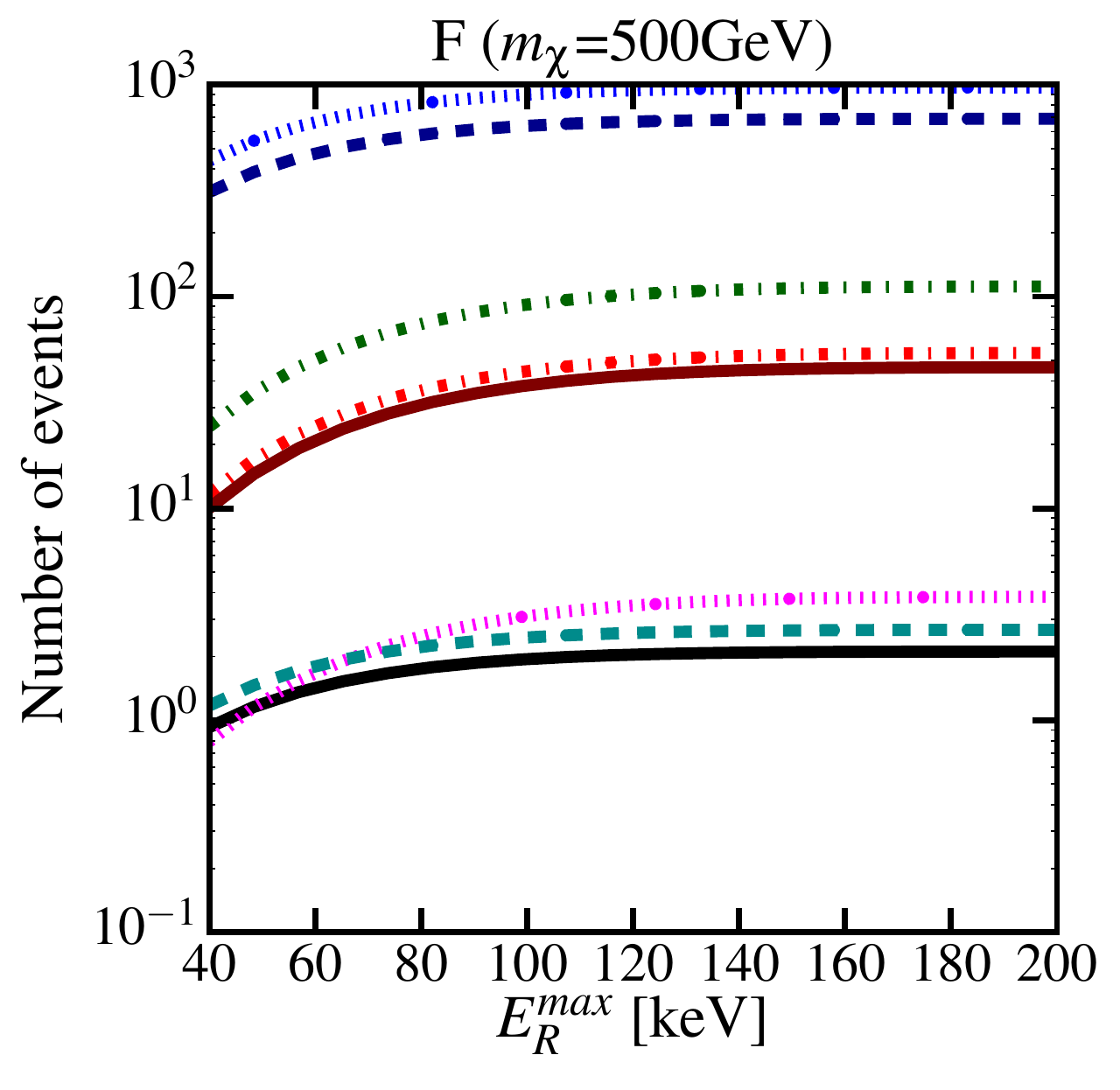}
\caption{Total number of expected events as a function of the maximum recoil energy observed by a variety of mock G2 experiments. All experimental parameters are fixed to the values in Table \ref{tab:experiments}, except $E_R^\text{max}$. These plots are for set--II scattering models of \S\ref{sec:analysis}, and a 500 GeV DM particle, for which the statistical gain of extending to higher energies is most evident. \label{fig:Nexps_Ermax}}
\end{figure}

Figures \ref{fig:Nexps_Ermin} and \ref{fig:Nexps_Ermax} show how the number  of expected events changes as a function of the lower and upper energy threshold, respectively, for selected models and DM masses. Firstly, because the number of events flattens out at high energy (even for 500 GeV DM particles, and for models with a long higher--recoil--energy tail presented in Figure  \ref{fig:Nexps_Ermax}), little is gained by looking for recoils in the range $>200$ keVnr, regardless of the scattering model. Secondly, for most targets there are a variety of models whose strongest signature is at low recoil energies, and especially for models where the mediator is lighter than the momentum transfer; this feature is also more pronounced for low DM masses (as illustrated in Figure  \ref{fig:Nexps_Ermin}). Thus, we conclude that a wide energy coverage (up to $\sim$200 keVnr) and especially low--energy thresholds (below $\sim$1 keVnr) are generally beneficial for recovering particle--physics content from direct detection data. 
\begin{figure*} 
\centering
\includegraphics[width=.3\textwidth,keepaspectratio=true]{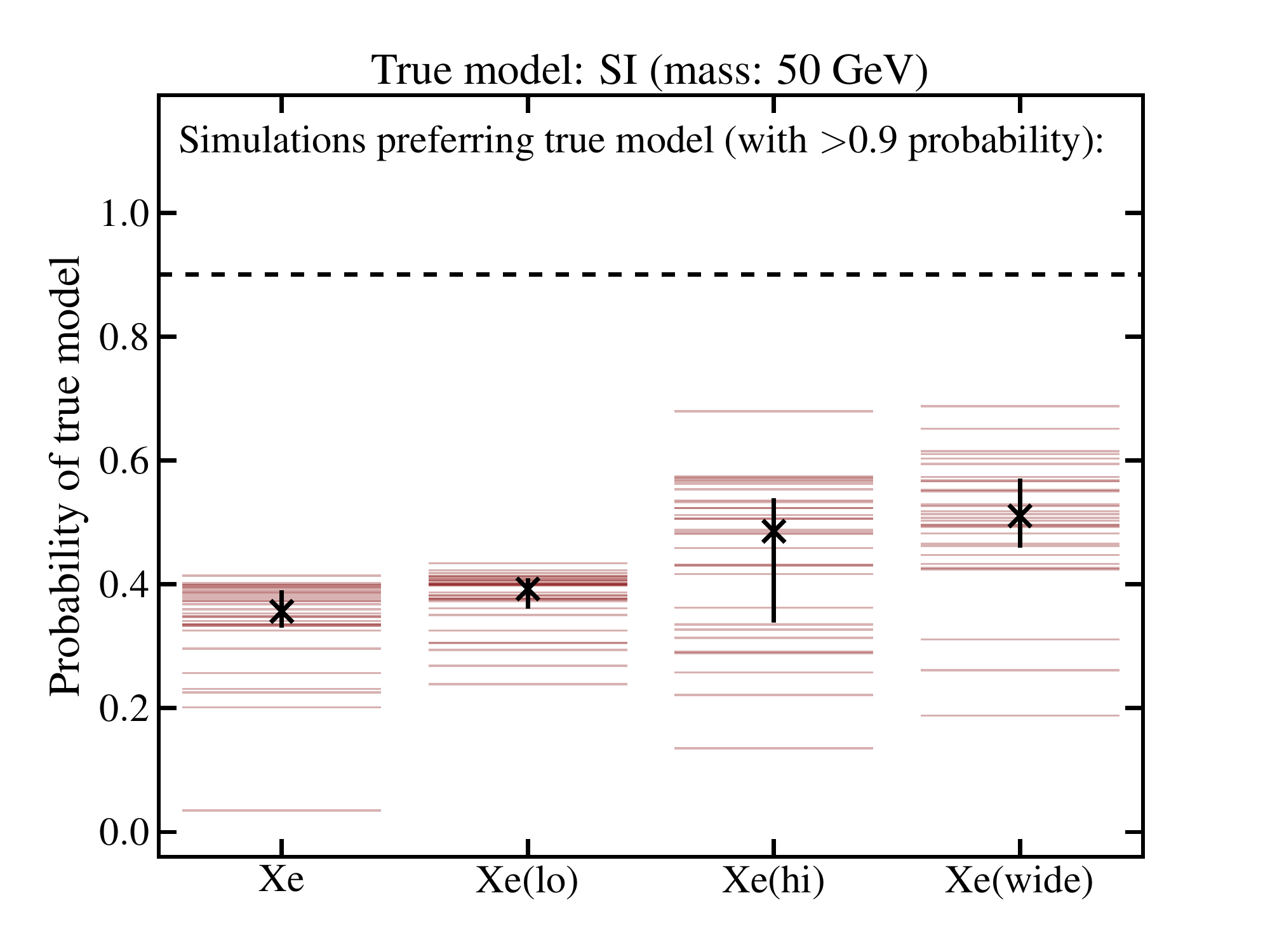}
\includegraphics[width=.3\textwidth,keepaspectratio=true]{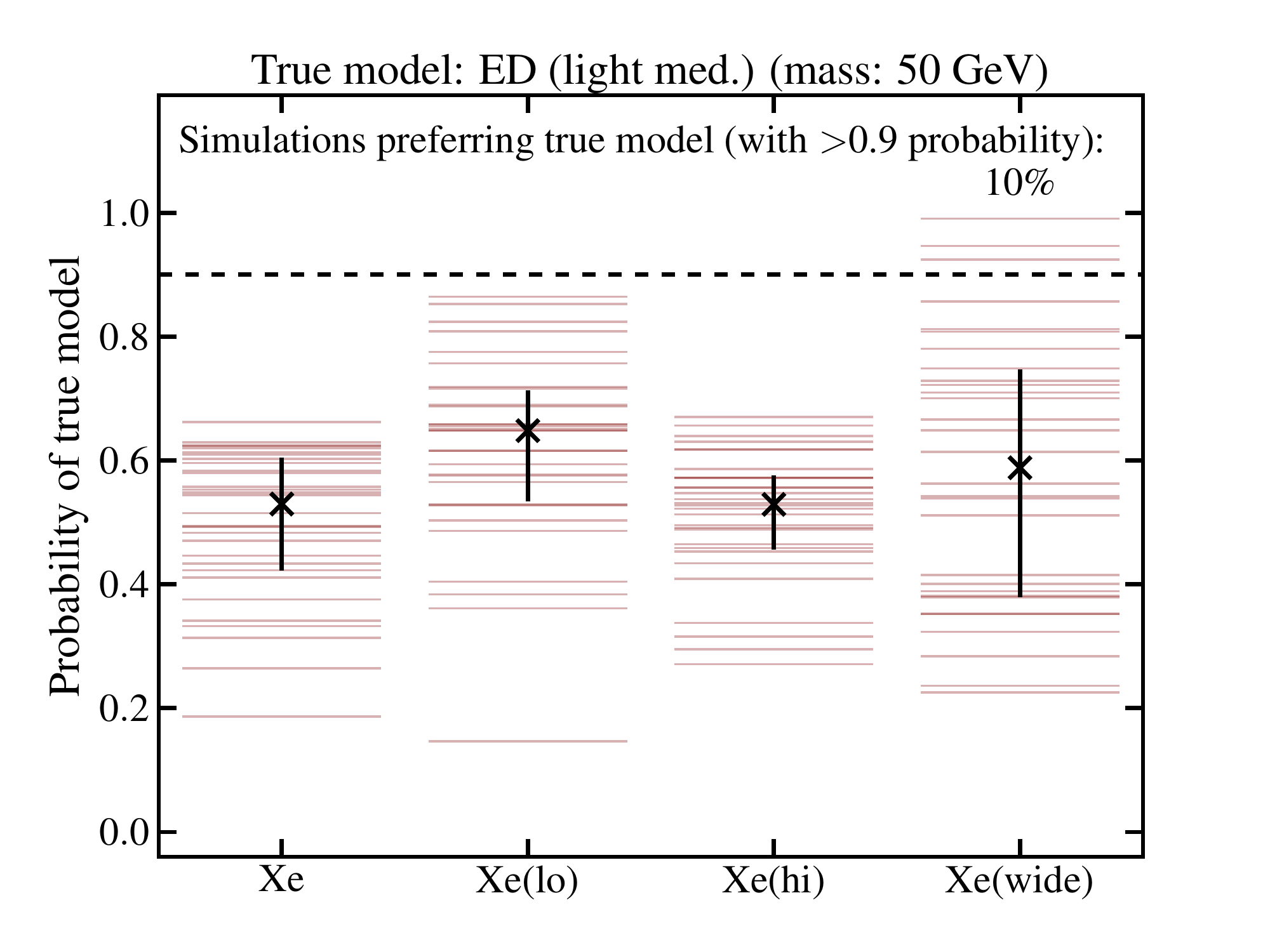}
\includegraphics[width=.3\textwidth,keepaspectratio=true]{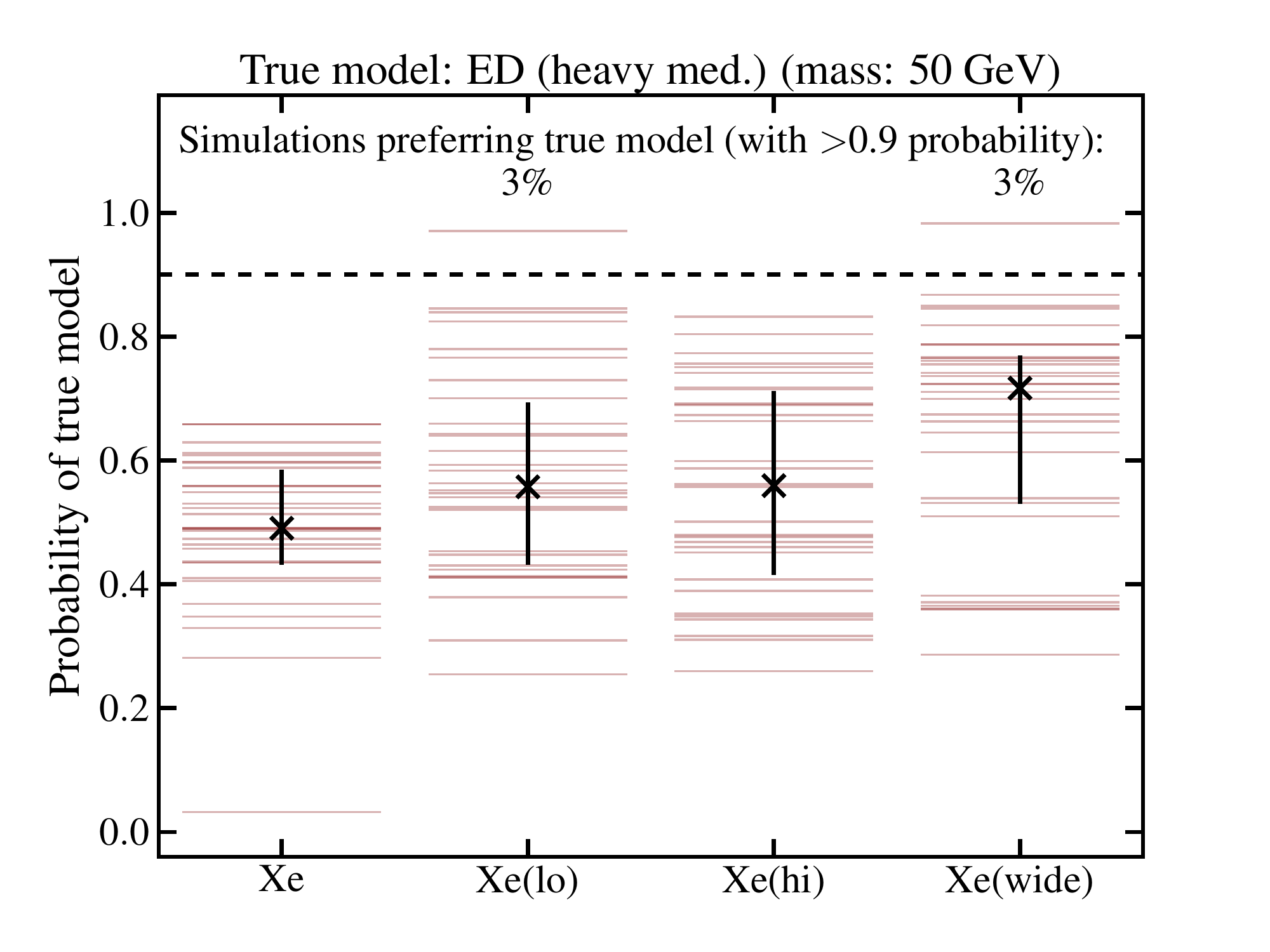}
\includegraphics[width=.3\textwidth,keepaspectratio=true]{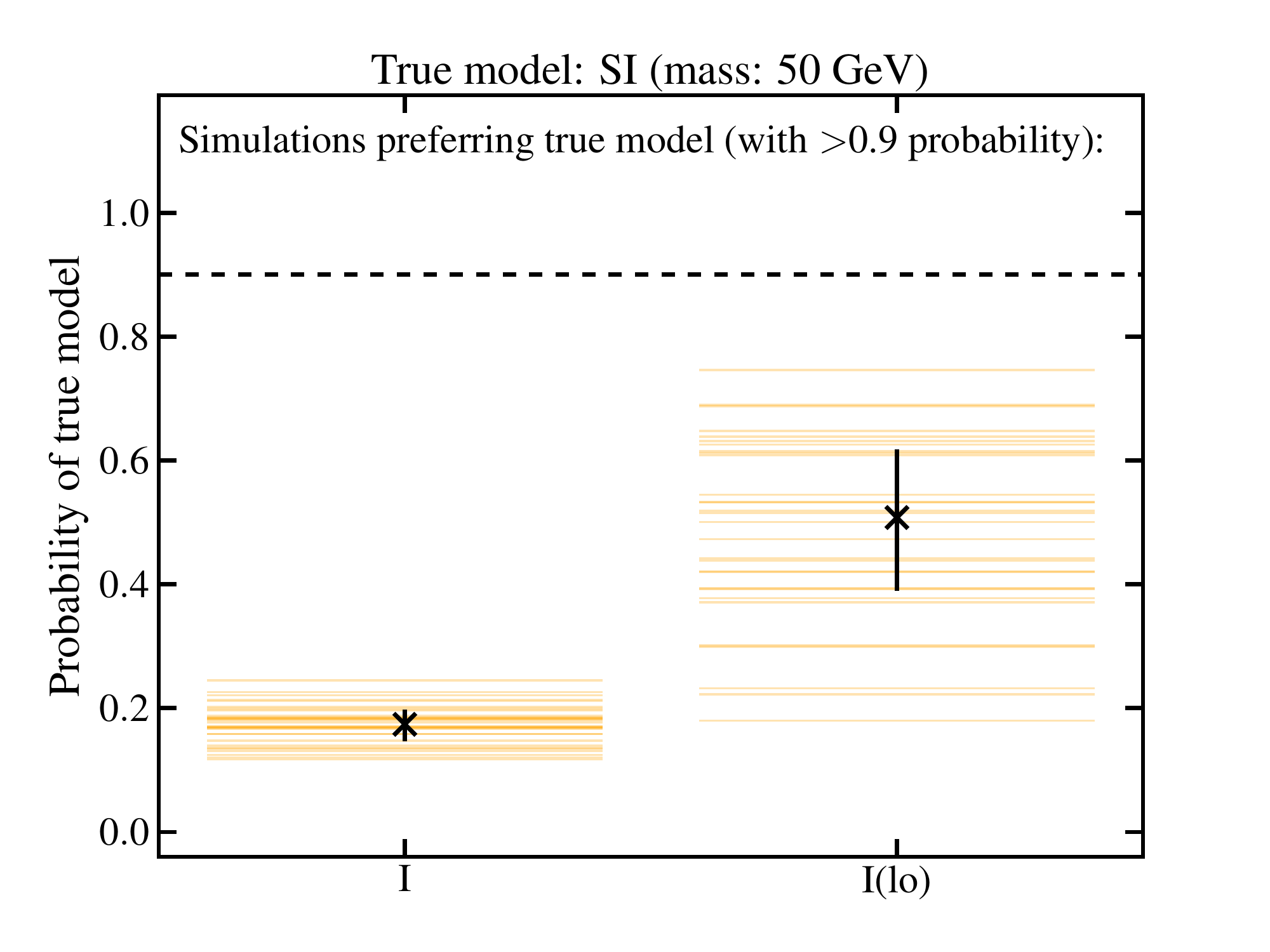}
\includegraphics[width=.3\textwidth,keepaspectratio=true]{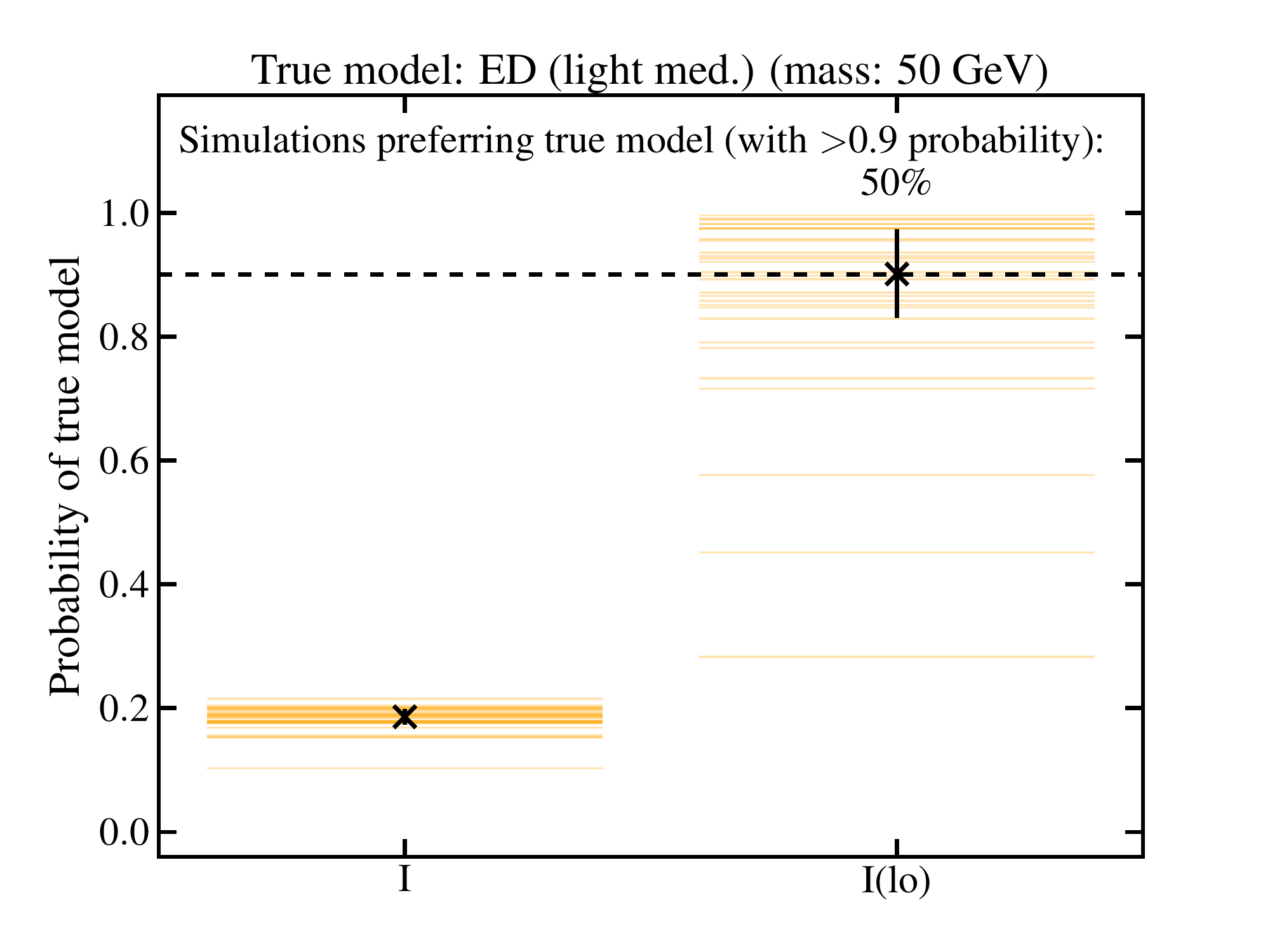}
\includegraphics[width=.3\textwidth,keepaspectratio=true]{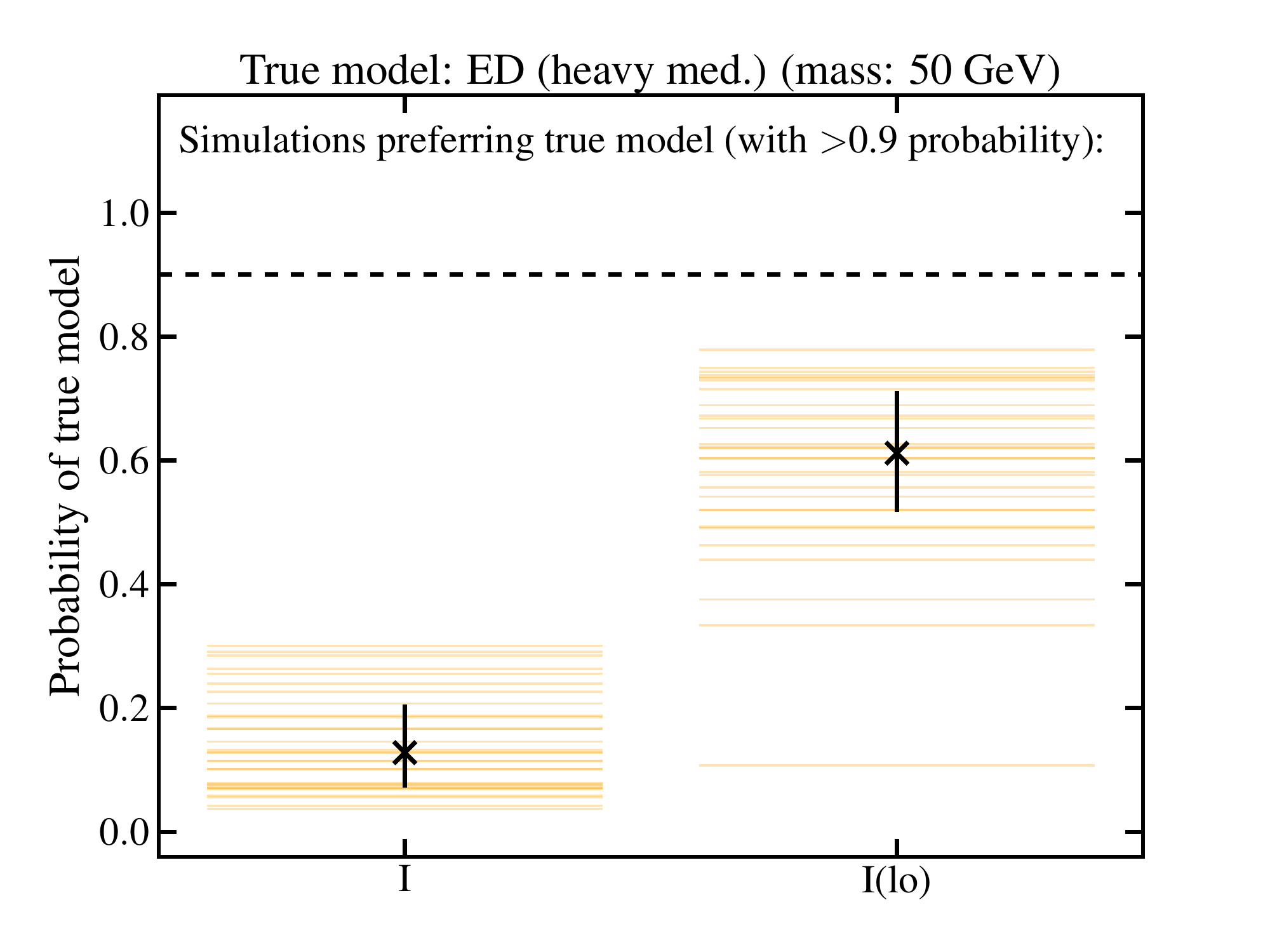}
\caption{Model selection prospects are compared for the baseline G2 experiments Xe (top row) and I (bottom row), as described in Table \ref{tab:experiments}, with versions of the same experiment where the energy window only was changed. In the top row, Xe(lo), Xe(hi), and Xe(wide) denote Xe experiments with the following modifications, respectively: low--energy threshold lowered to 1 keV; high energy threshold increased to 100 keV; and energy window expanded to 1 keV--100 keV. In the bottom row, I(lo) represents an I experiment with an energy threshold lowered to 1 keV. In the case of Xe, expansion of the energy window does not significantly affect the ability to distinguish different scattering scenarios. For I, the impact is more significant, but not large enough to enable reliable model selection when only data from I target is analyzed.\label{fig:lohi}}
\end{figure*}

While these two Figures provide us with a sense of how the event counts change as a function of the energy windows, they do not directly translate into implications for the success of model selection; as we have seen in \ref{sec:model_selection}, success of model selection is not just a function of the number counts of events, but depends on the interplay of several factors. For this reason, in order to provide guidance for future experimental designs, we investigate the impact that changes to the energy window have on the results for model selection for two of our baseline experiments: Xe and I. For Xe, we consider three modified versions of this experiment: ``Xe(lo)'' with a lowered low--energy threshold, ``Xe(hi)'' with a higher high--energy threshold, and ``Xe(wide)'' with a wider energy window ``Xe(wide)''. For I, we only investigate lowering the low-energy threshold in ``I(lo)'' (see Table \ref{tab:experiments} for other parameters). We then repeat the model selection of \ref{sec:baseline}. 

We again perform model selection amongst models of set I, for a selected representative subset of our baseline simulations for a 50 GeV DM mass. We compare these results to the results (presented previously) for our baseline versions of Xe and I, in Figure \ref{fig:lohi}. From this Figure, the following conclusions may be drawn. For an intermediate DM mass, expanding the energy window either towards lower or higher recoil energies produces a slight improvement in model selection prospects for Xe. However, only the simultaneous widening of the energy window enables a (slight) chance that Xe alone could confidently identify the underlying scattering model. Unsurprisingly, a larger improvement is visible when the low energy threshold is lowered for I (which was previously $\sim$22 keVnr), and is most pronounced for light mediator models. So, adjustment of the energy windows is undoubtedly beneficial for model selection, but it is not sufficient for success: complementarity of available targets plays a more important role.

Finally, we make one more comment about target complementarity. From Figures \ref{fig:Nexps1} and \ref{fig:Nexps_lowm}, previously emphasized superiority of fluorine targets for SD scattering is well illustrated in the context of a variety of models considered in this work. Namely, several pseudoscalar--mediated scattering models that produce a very poor statistic, or are otherwise practically invisible on both G2 germanium and G2 xenon targets, make a strong signal on 600 kilogram--years of exposure on fluorine. The availability of nuclear targets with spin structure different from that of Xe and Ge may thus be of pivotal importance for establishing the first detection of a DM signal and understanding the physics that gives rise to this scattering.
\section{Summary and conclusions}
\label{sec:conclusions}

We have investigated the distinguishability of a wide variety of DM--nucleon scattering models in the context of noisy data from Generation--2 and futuristic direct detection experiments. For this purpose, we simulated over 8000 recoil energy spectra for a range of nuclear targets (with exposures and energy windows representative of many existing and proposed experiments), under models that give rise to a variety of scattering phenomenologies, including nontrivial momentum and velocity dependence of the scattering rate.  We then analyzed these simulations agnostically, using Bayesian posterior analysis and model selection (in over 130000 \texttt{MultiNest} runs, requiring  $\sim 2\times 10^4$ CPU hours) to establish how likely direct detection is to identify the right underlying interaction given a wide set of competing hypotheses.  We also investigated a range of questions related to the extraction of dark matter parameters concurrently with performing the model selection.

The key results of this study are shown in Figures \ref{fig:model_selection_gexe_50gev_select}, \ref{fig:class_selection_gexe_50gev_select}, and \ref{fig:model_selection_2_50gev_select}. We demonstrated that, for a signal just below the current upper limit, xenon and germanium targets alone, as realized in LUX/LZ and SuperCDMS, respectively, are likely to identify the momentum dependence of the scattering and thus exclude large classes of scattering models, if a strong signal is seen on all of them. However, they are not likely, except in special cases, to single out the correct underlying contact operator describing the scattering. The addition of a relatively modest exposure on either fluorine or iodine targets drastically improves these prospects. Target complementarity emerges as essential for extracting the underlying physics of DM--nucleon interactions in light of significant Poisson noise, which is the regime in which the first signals would appear.

In addition to model selection, we also investigated the quality of DM mass estimation, and concluded that accuracy of mass estimation is driven by what the mass is (marginalized posteriors are wide for larger masses, and narrower for smaller) rather than by the type of the interaction at hand, assuming that the correct scattering model is fit to data. However, we also showed that, if a wrong model is fit to data, the corresponding mass posteriors can be severely biased, inconsistent amongst different experiments, or unnaturally wide for a given mass. We therefore emphasized the need for model selection as an integral part of future data analysis. 
Furthermore, we explored the impact of modeling and measurement uncertainty on the ratio of nucleon couplings for some of the models considered here. We showed how this affects mass estimation and model selection. This uncertainty, though very large (i.e. data are typically unable to constrain this parameter well, if at all), does not significantly impact the recovery of the mass, and the uncertainty also shows a mild impact on model selection. Such freedom only confuses certain models (like the
standard SI and SD), while others (like Anapole) remain distinct when
data from germanium, xenon, and fluorine targets are analyzed jointly. This gives us confidence that model selection results discussed in this work are robust to this uncertainty.
Finally, in order to inform future experimental designs, and given the variety of scattering phenomenologies accessible to direct detection,  we quantified how changing the recoil--energy analysis windows affects the event counts on different targets; we also investigated the impact of such changes on prospects for model selection. We concluded that widening the energy window for individual targets is generally beneficial, although target complementarity still emerges as the main lever arm for identifying the interaction that governs scattering. 

At the end, we outline several caveats left for future study. Firstly, in the analysis presented here, we omitted consideration of backgrounds in all experiments. Although this assumption holds for most scenarios we focus on, it may break down in some cases (for example, for extremely low--energy threshold experiments), and it is an important and worthwhile exercise to further investigate model--selection prospects in the presence of significant backgrounds. A similar argument holds for the assumption of perfect energy resolution---while it is expected to hold well for most experiments and most recoil energies considered here, it is a natural next step to expand the analysis presented here to account for finite energy resolution. Additionally, nuclear response functions we used in this study include a degree of nuclear--physics uncertainty that should be accounted for when comparing scattering models and performing parameter estimation with future data. Even though it is currently unclear how to quantify and marginalize over this uncertainty, we make an initial qualitative investigation of their impact in Appendix \ref{app:nuclear}. As a result, we are optimistic that nuclear uncertainties will not be a significant obstacle in performing model selection, given a strong signal, but a further detailed investigation is warranted to precisely quantify the validity of this expectation. And finally, the uncertainty in the local dark--matter velocity distribution may also complicate parameter reconstruction and model selection.
 However, it is also not likely to significantly degrade prospects for model selection.  Two key points go in support of this prediction: i) the hierarchy of the expected numbers of events on different targets, shown in this study to be crucial for successful model selection, is unchanged when the velocity parameters of the Maxwellian distribution are varied within their error bars, and ii) the velocity integral that factors into the recoil--energy spectrum, does not show much variation in its value as a function of the minimum velocity, for plausible halo models; thus, it is reasonable to expect that the uncertainty on the exact shape of the halo model will have little degeneracy with the shape of the recoil--energy spectrum, crucial for distinguishing scattering models. For more details, see Appendix \ref{app:astrophysical}.

For a direct detection signal just below the current thresholds, prospects for understanding dark matter properties using data from Generation--2 experiments are good. In the event of a detection, a comprehensive analysis following the roadmap of this study, including an exhaustive coverage of the modeling space and a joint--analysis capability for multiple experiments, with consideration of the caveats above, will be essential to identifying the theory of dark matter using direct detection. 

To obtain the numerical results of this study, we have developed \texttt{dmdd}, a Python package for direct--detection simulation and analysis, now available to the community.\footnote{\url{http://github.com/veragluscevic/dmdd}}

\acknowledgments
VG is grateful for the support provided by the Friends of the Institute for Advanced Study in Princeton. MG was supported in part by the Murdock Charitable Trust and some of her work was performed at the Aspen Center for Physics, which is supported by National Science Foundation grant PHY-1066293. SDM is supported by NSF PHY1316617. KZ is supported by the DoE under contract DE-AC02-05CH11231. The authors thank Timothy Morton, Mikhail Solon, and Hugh Lippincott for helpful discussions. 
\bibliographystyle{JHEP}
\bibliography{dmdd_theories_v2}
  
\appendix
\section{Appendix: Additional model-selection results}
\label{app:model_selection}
\begin{figure*}
\centering
\includegraphics[width=.45\textwidth,keepaspectratio=true]{lineplot_50GeV_SI_Higgs_50sims_set1.pdf}
\includegraphics[width=.45\textwidth,keepaspectratio=true]{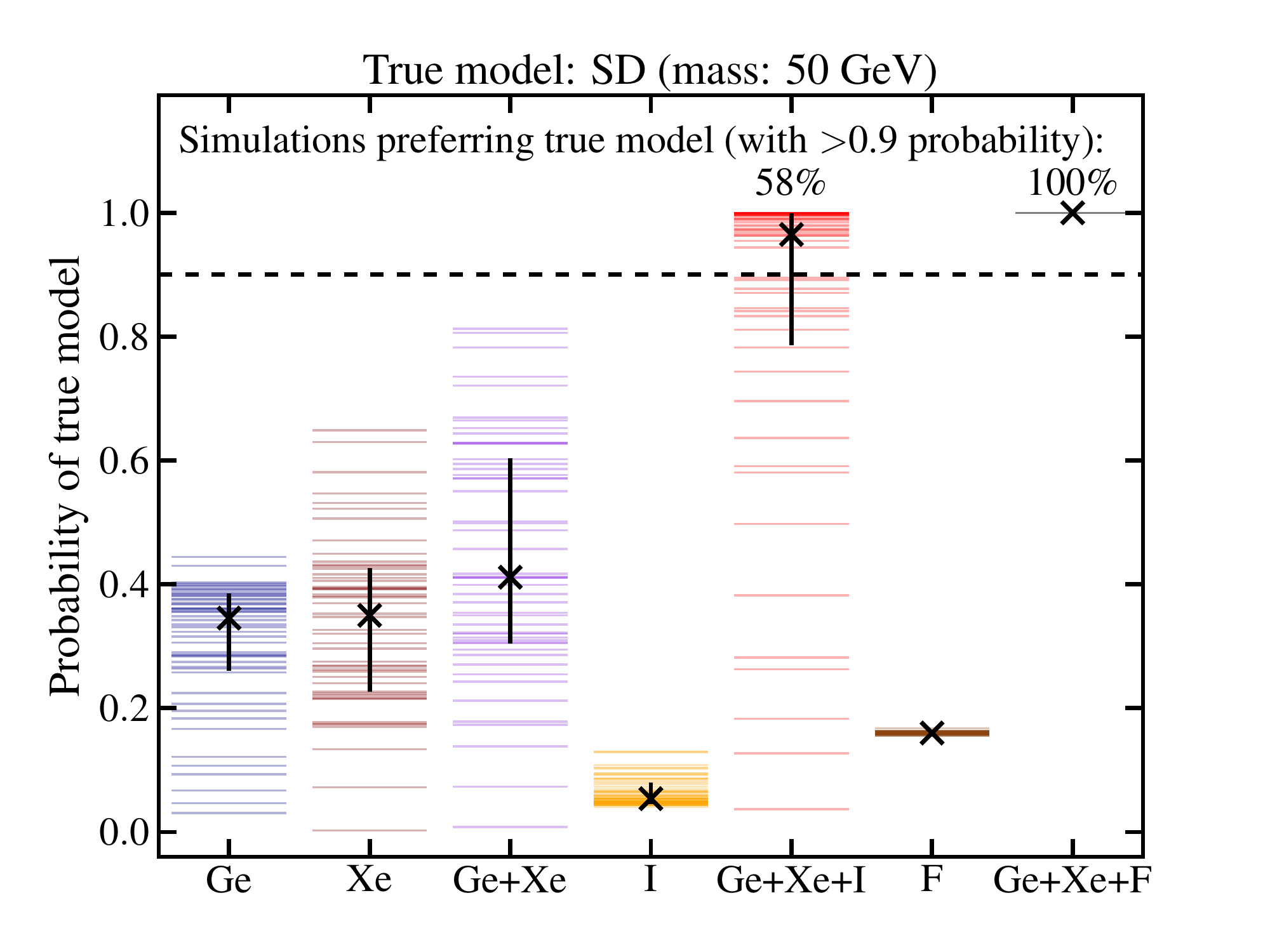}   
\includegraphics[width=.45\textwidth,keepaspectratio=true]{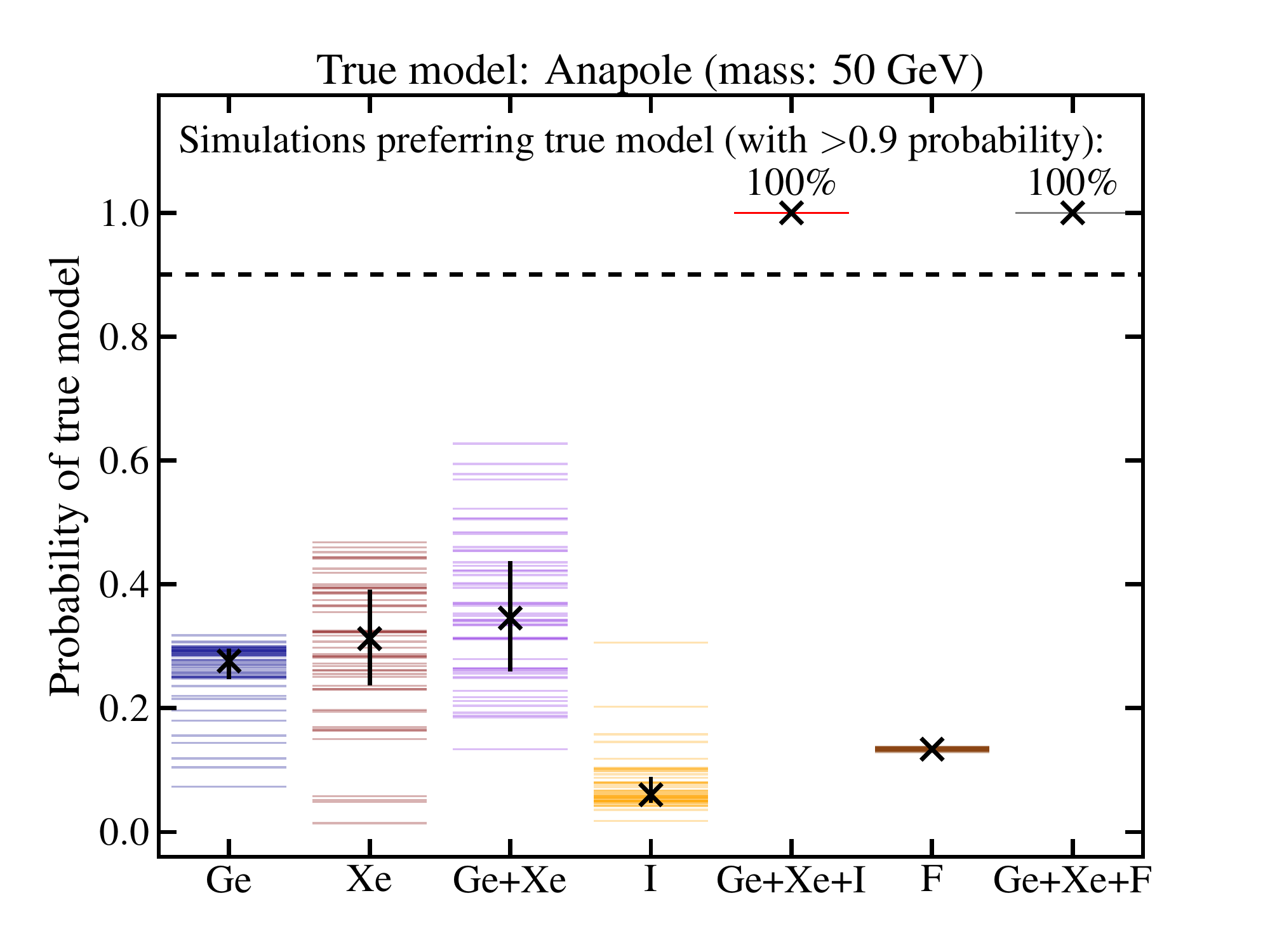}
\includegraphics[width=.45\textwidth,keepaspectratio=true]{lineplot_50GeV_Millicharge_50sims_set1.pdf}
\includegraphics[width=.45\textwidth,keepaspectratio=true]{lineplot_50GeV_Elecdiplight_50sims_set1.pdf}
\includegraphics[width=.45\textwidth,keepaspectratio=true]{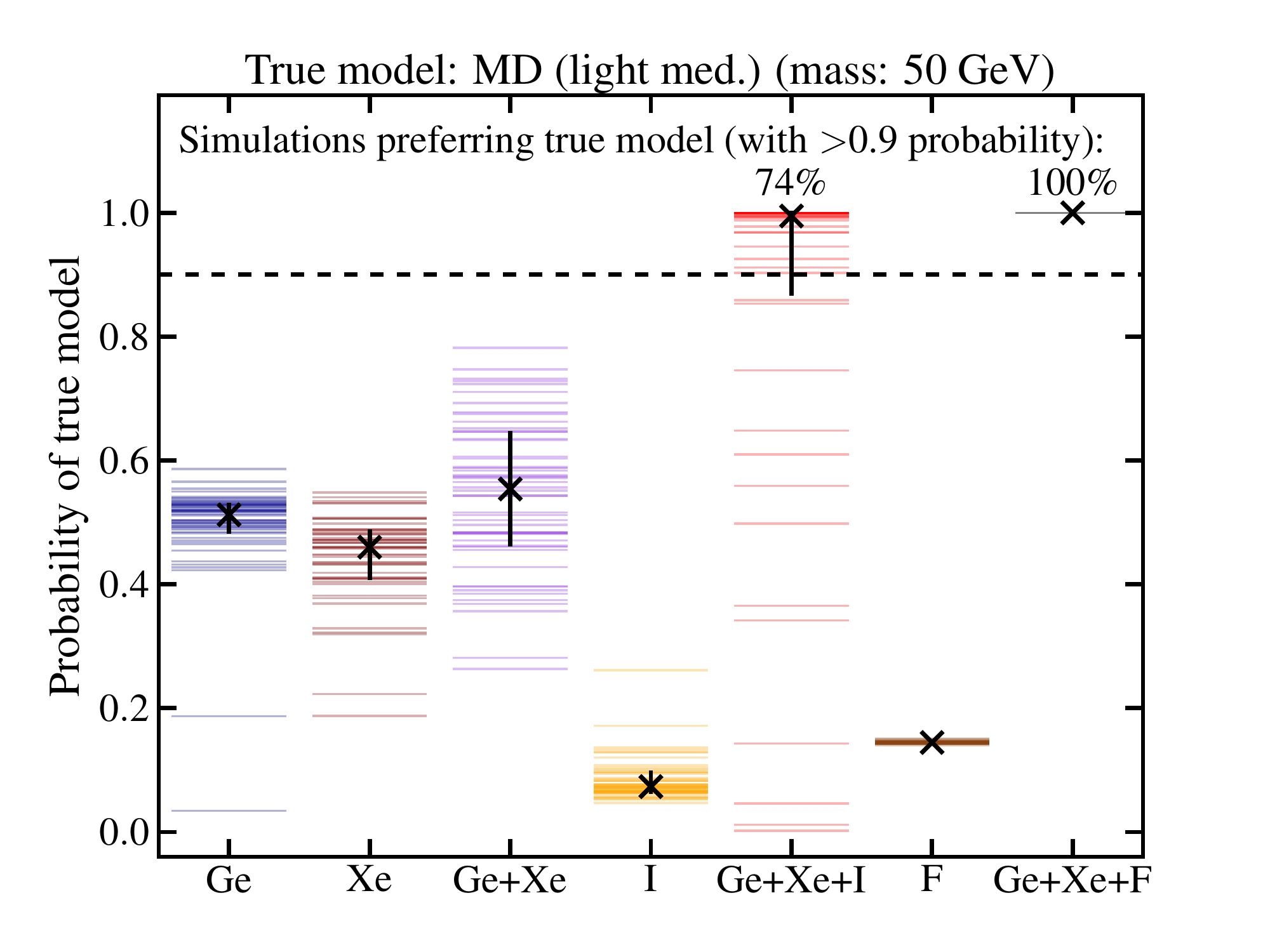}
\includegraphics[width=.45\textwidth,keepaspectratio=true]{lineplot_50GeV_Elecdipheavy_50sims_set1.pdf}
\includegraphics[width=.45\textwidth,keepaspectratio=true]{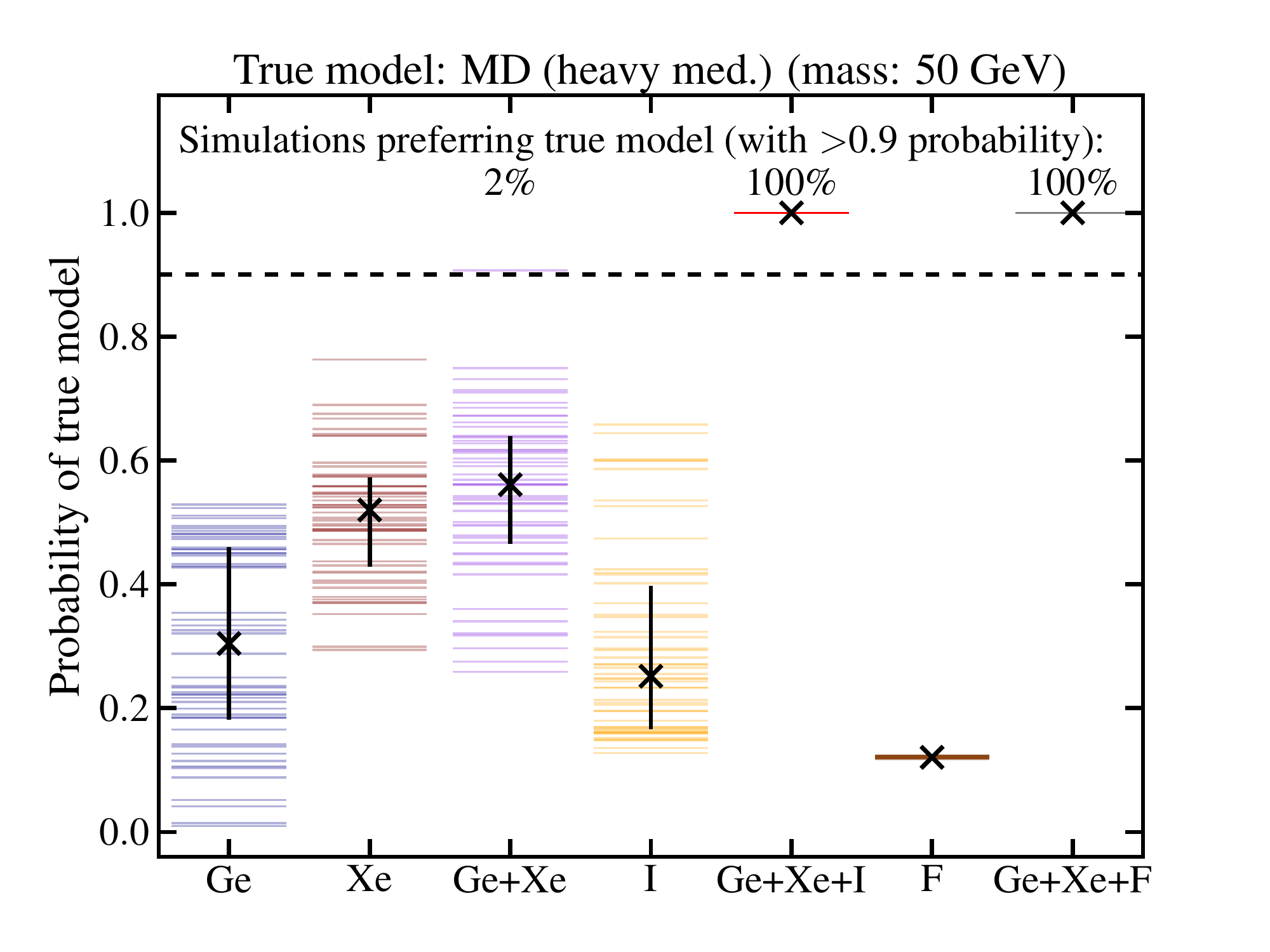}
\caption{Same as Figure \ref{fig:model_selection_gexe_50gev_select}, but for simulations under each of the scattering models of set I. \label{fig:model_selection_gexe_50gev}}
\end{figure*}
\begin{figure*}
\centering
\includegraphics[width=.45\textwidth,keepaspectratio=true]{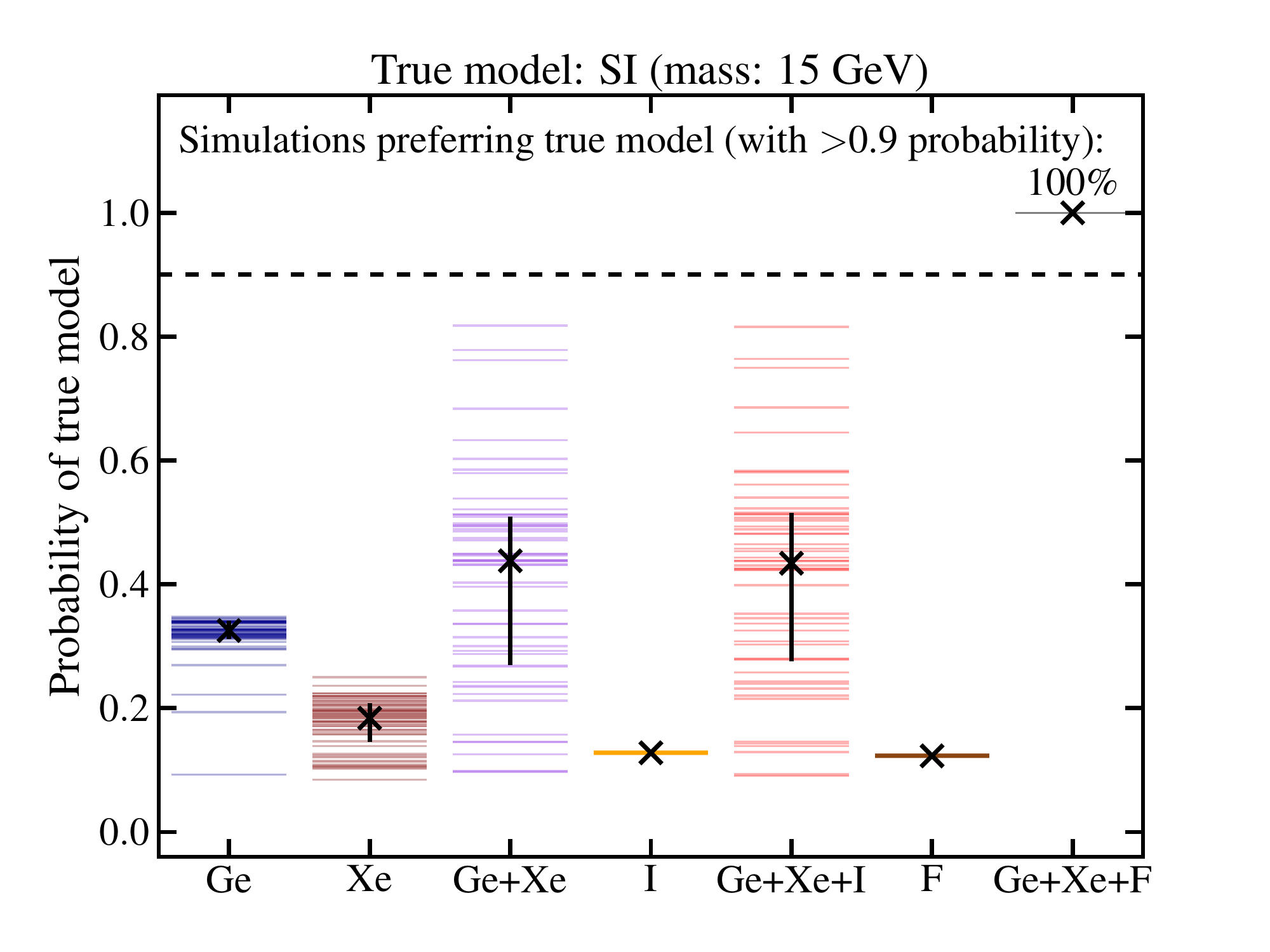}
\includegraphics[width=.45\textwidth,keepaspectratio=true]{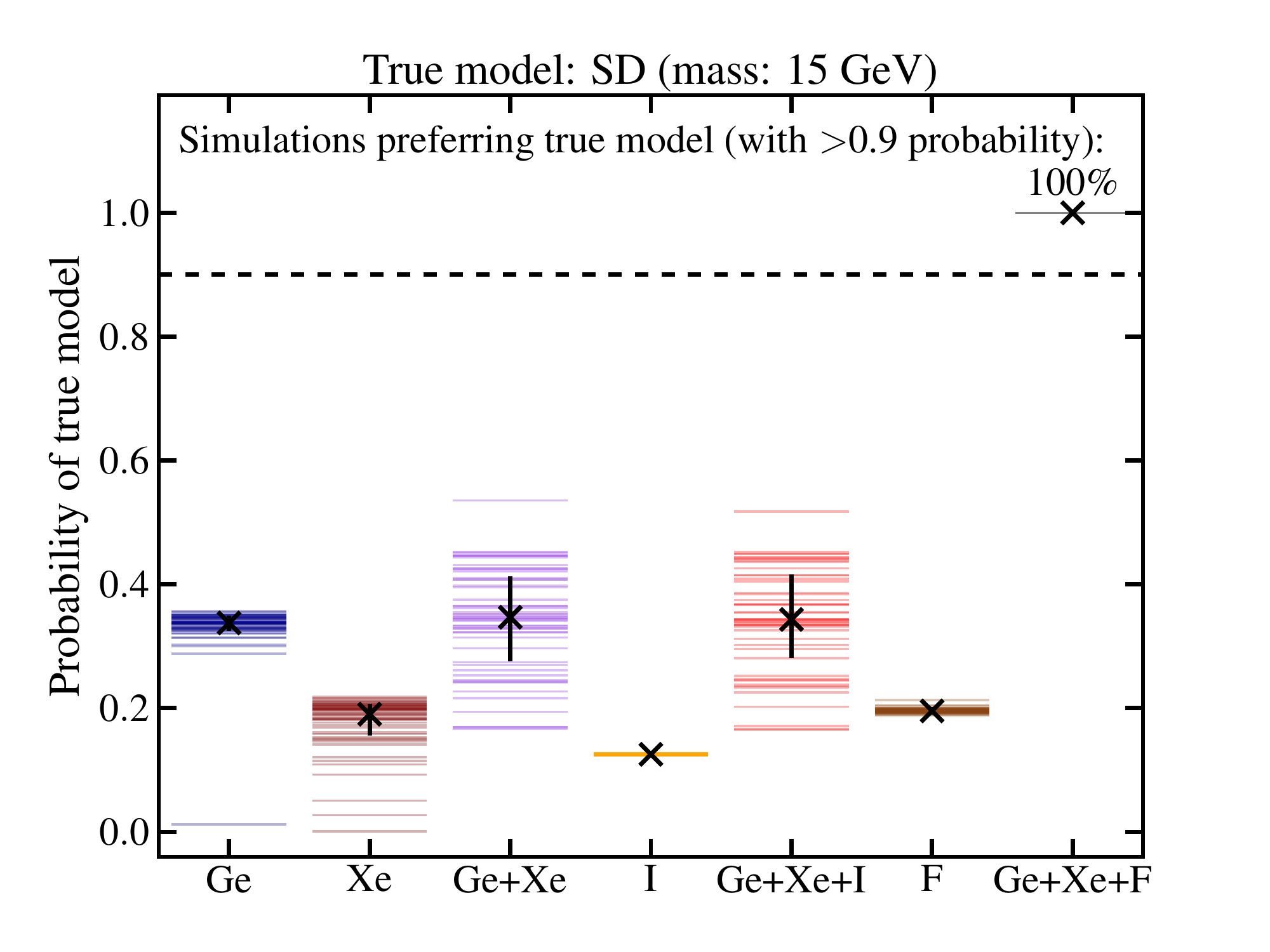}
\includegraphics[width=.45\textwidth,keepaspectratio=true]{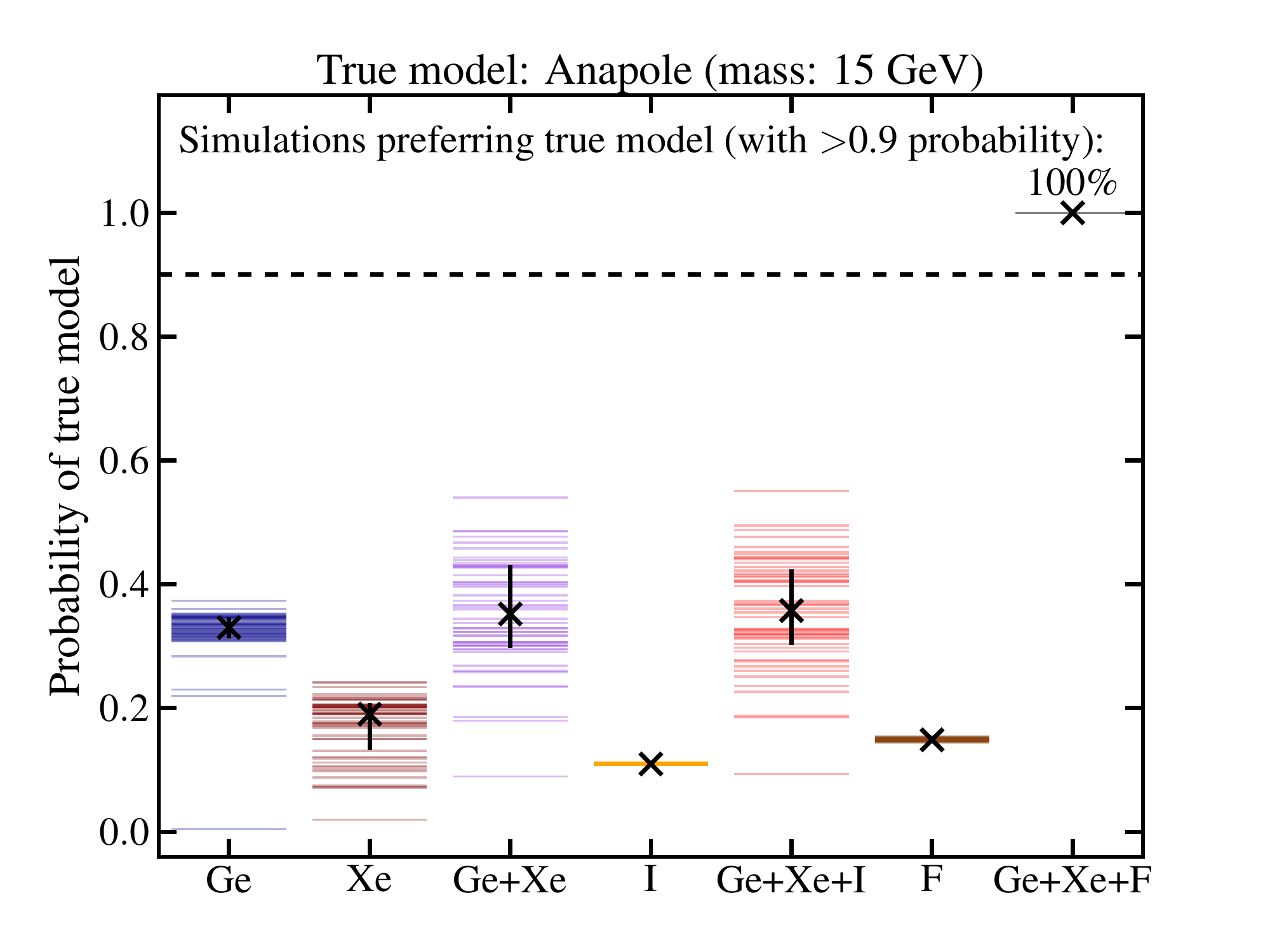}
\includegraphics[width=.45\textwidth,keepaspectratio=true]{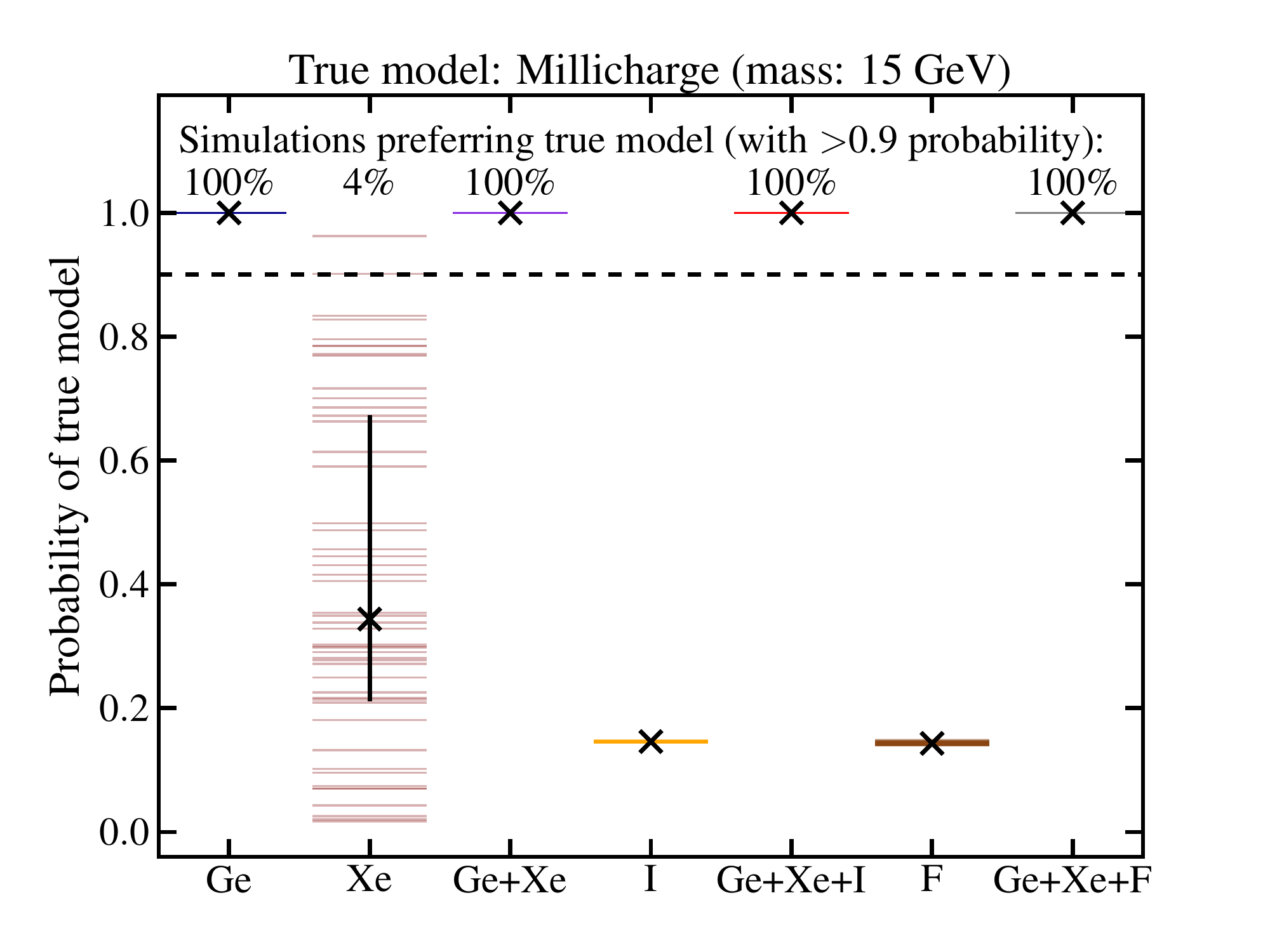}
\includegraphics[width=.45\textwidth,keepaspectratio=true]{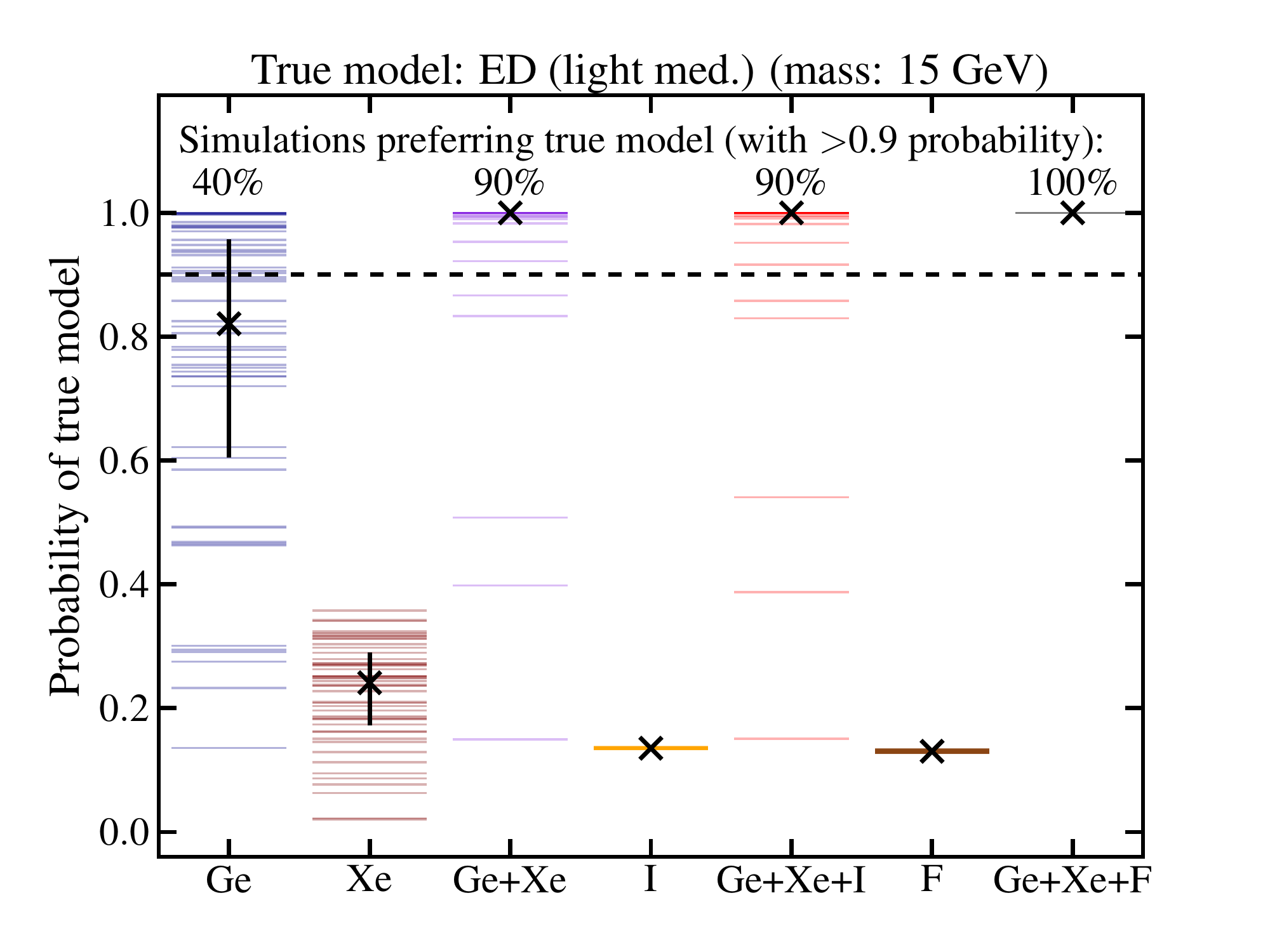}
\includegraphics[width=.45\textwidth,keepaspectratio=true]{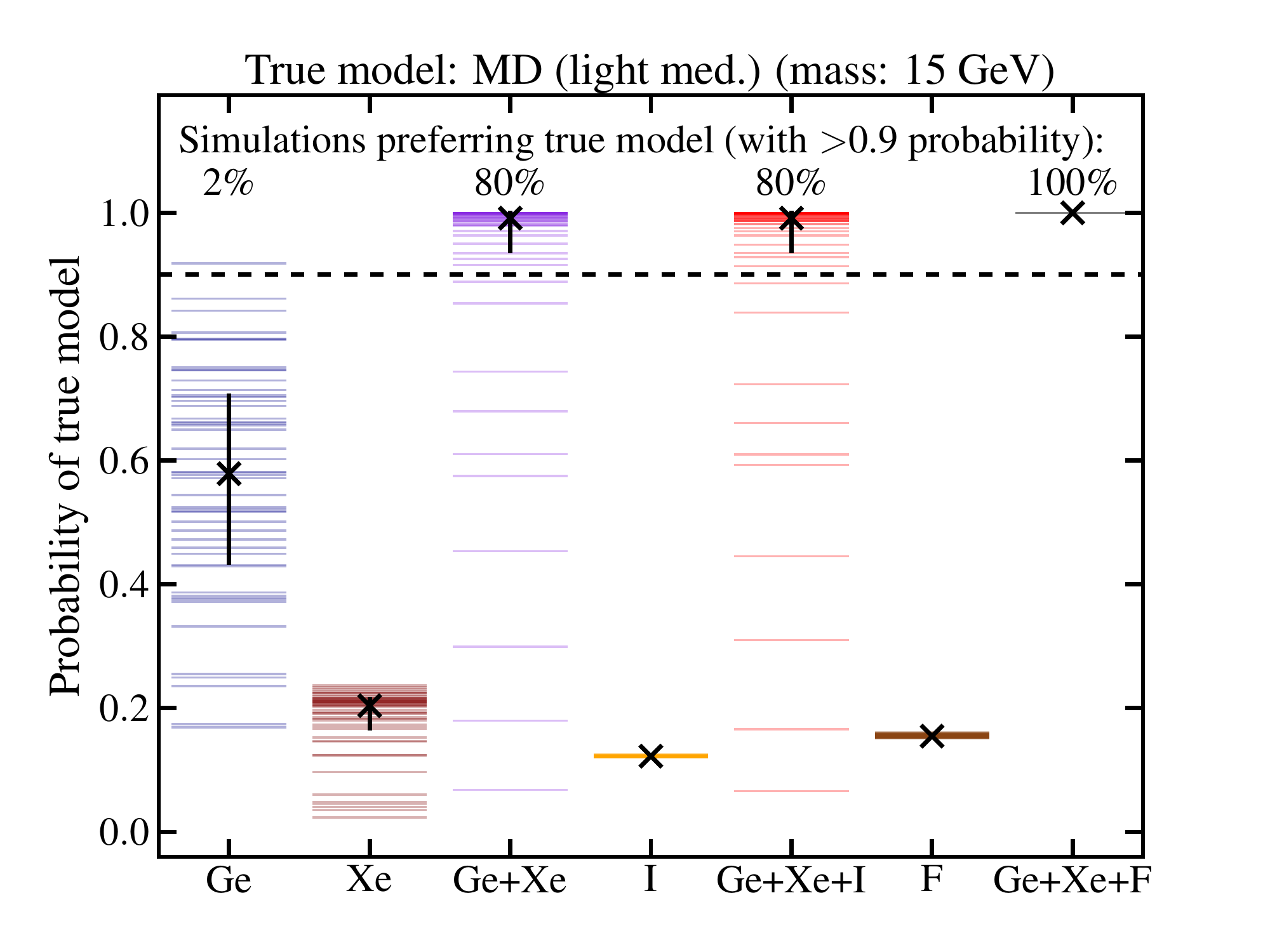}
\includegraphics[width=.45\textwidth,keepaspectratio=true]{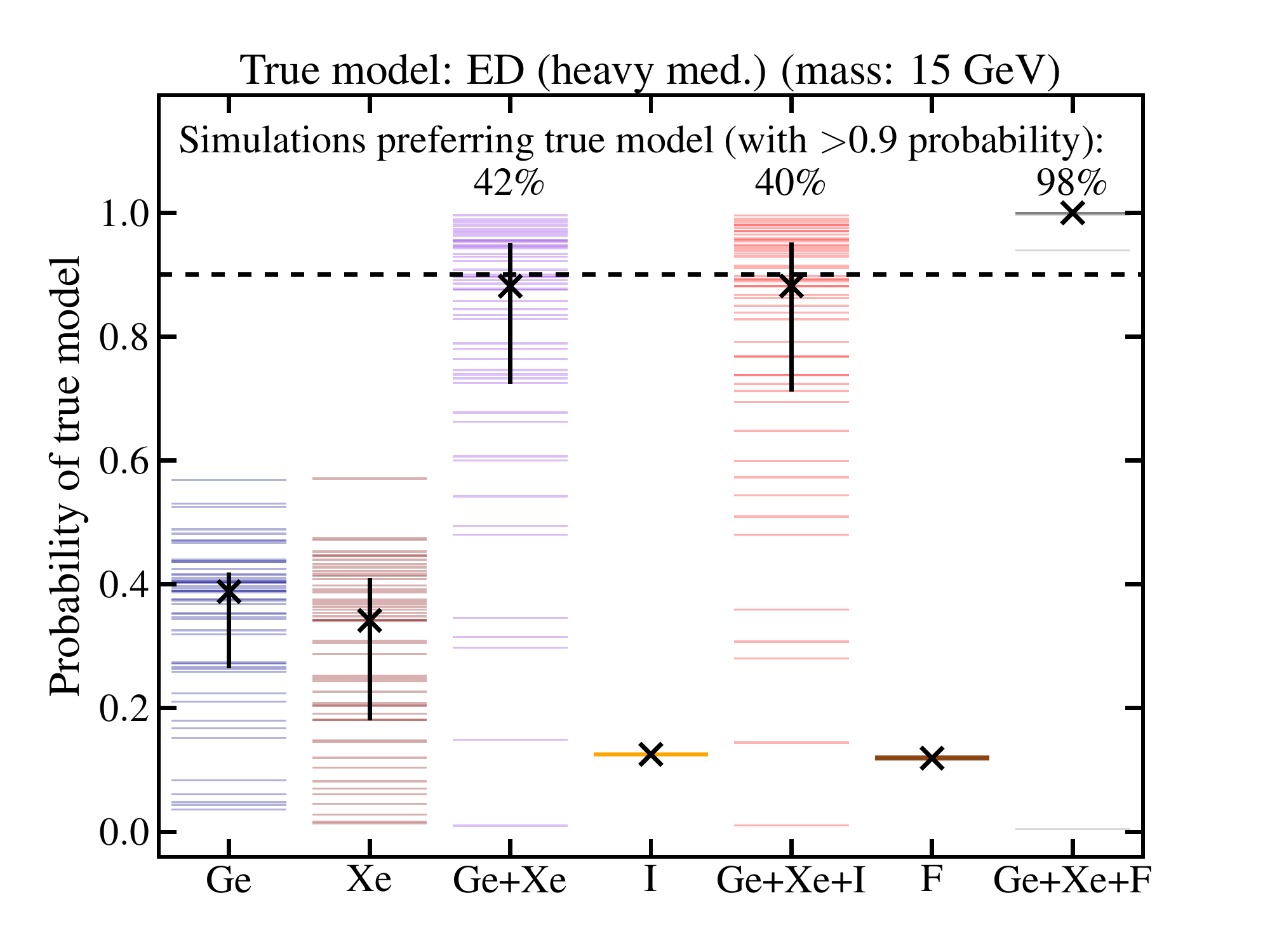}
\includegraphics[width=.45\textwidth,keepaspectratio=true]{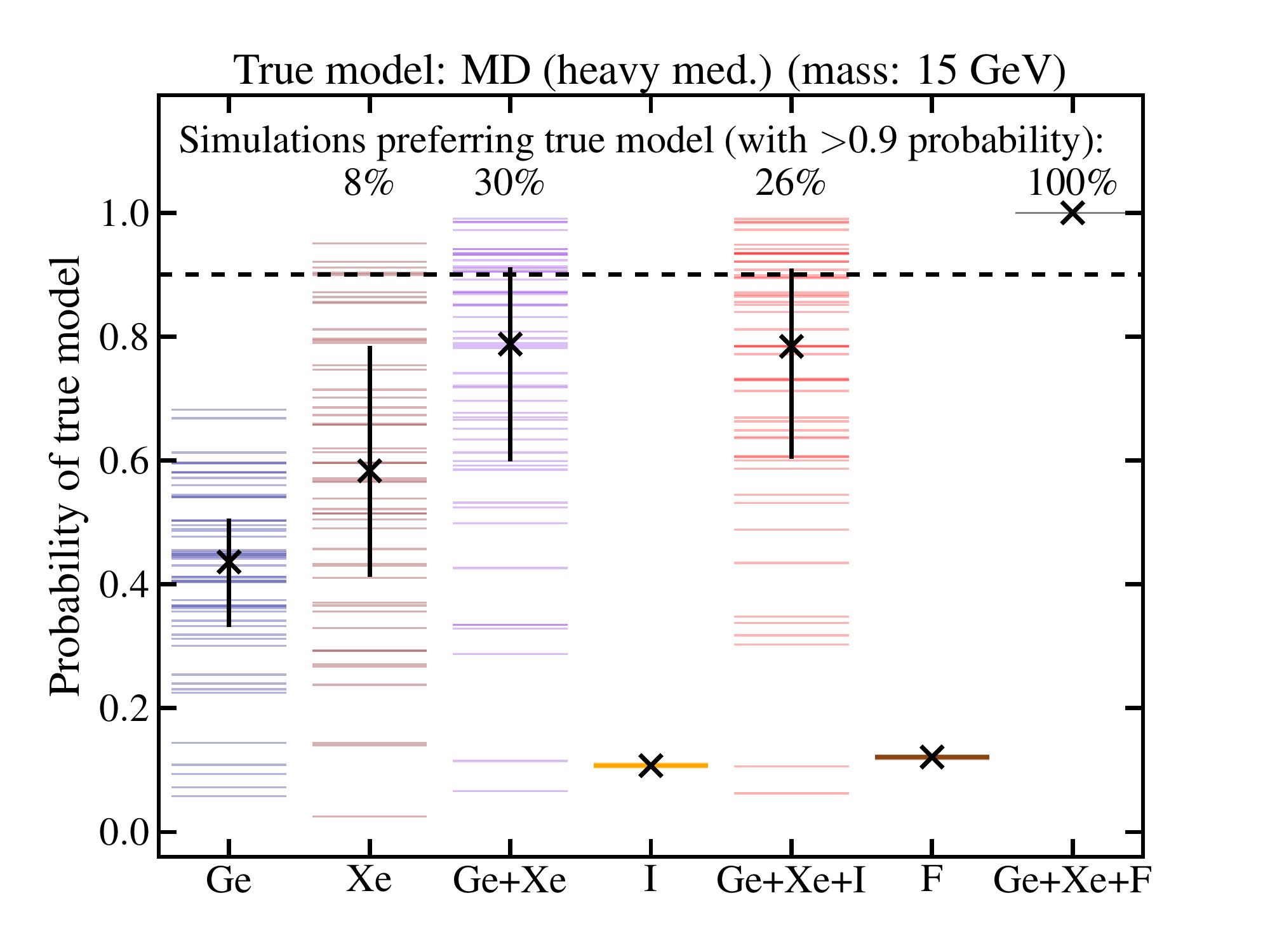}
\caption{Same as Figure \ref{fig:model_selection_gexe_50gev}, but for $m_\chi=15$ GeV.\label{fig:model_selection_gexe_15gev}}
\end{figure*}
\begin{figure*}
\centering
\includegraphics[width=.45\textwidth,keepaspectratio=true]{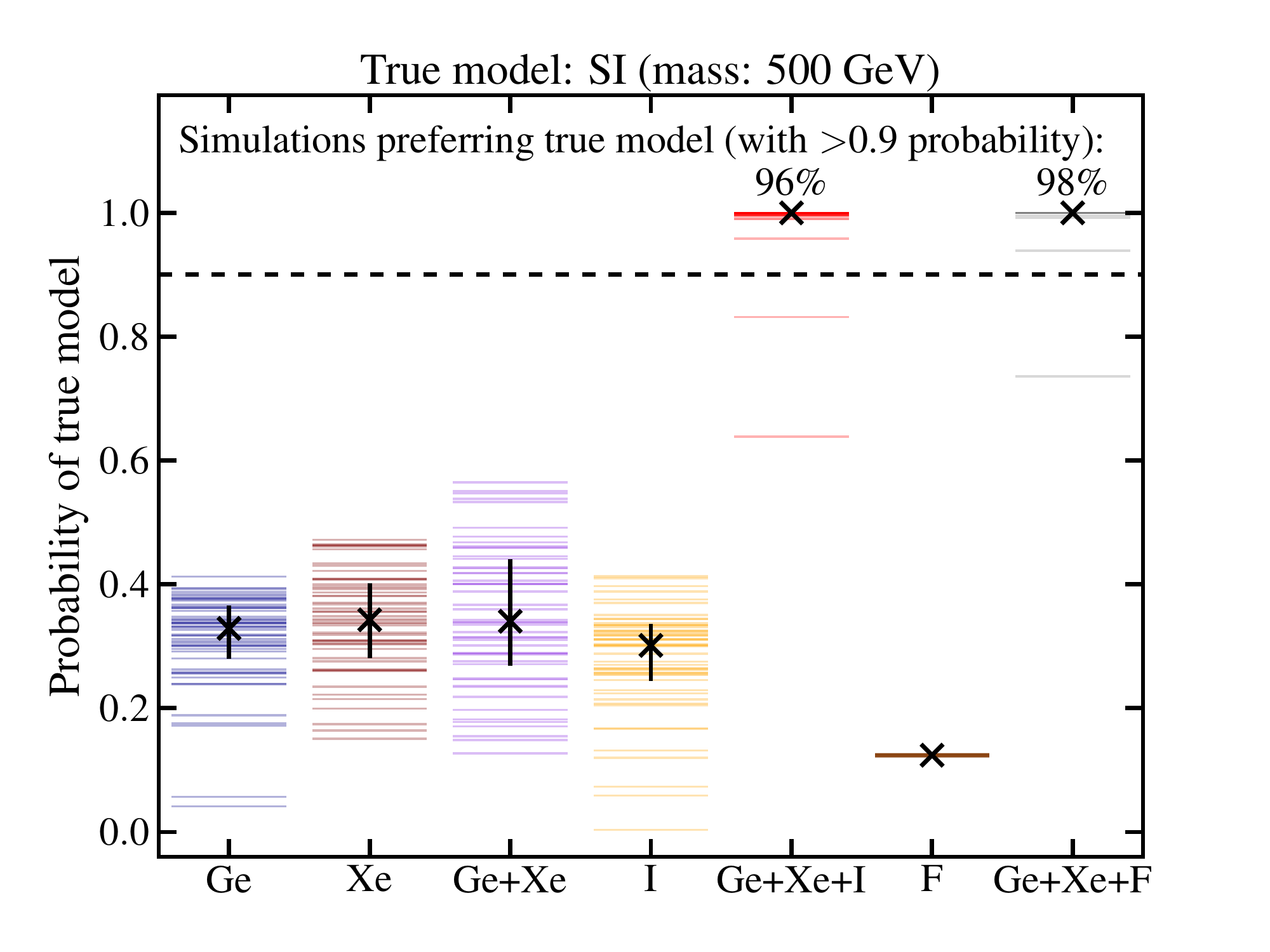}
\includegraphics[width=.45\textwidth,keepaspectratio=true]{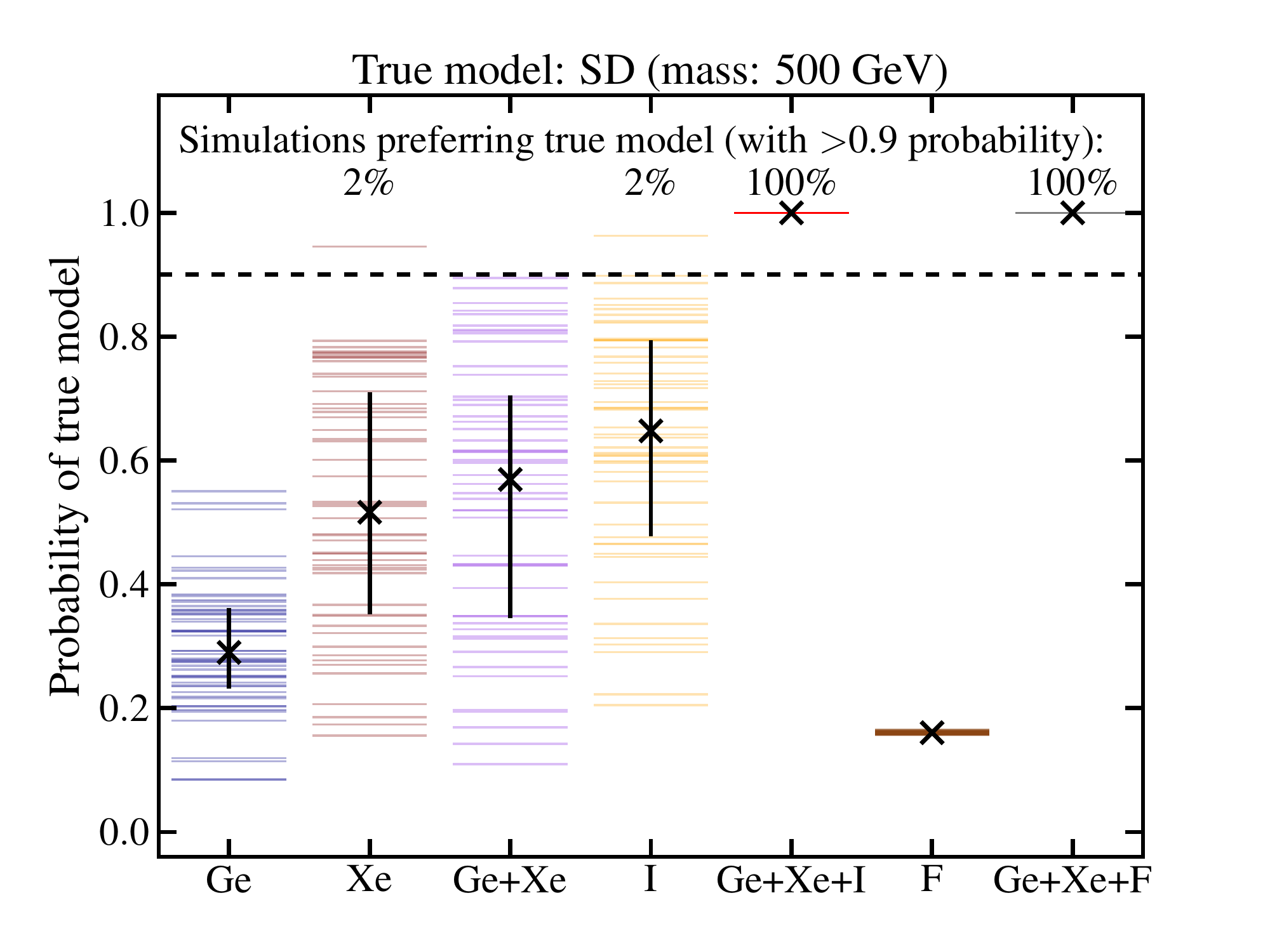}
\includegraphics[width=.45\textwidth,keepaspectratio=true]{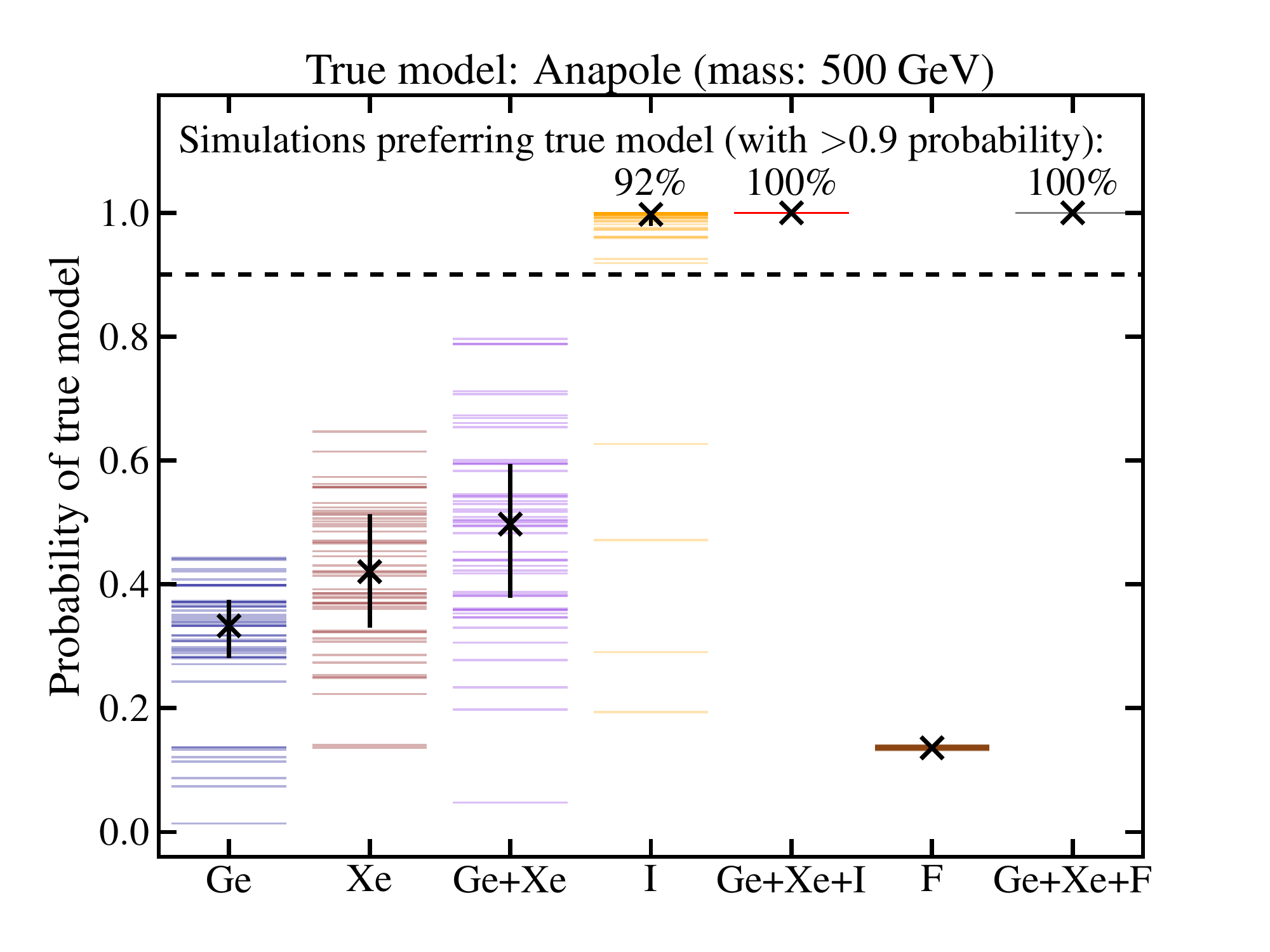}
\includegraphics[width=.45\textwidth,keepaspectratio=true]{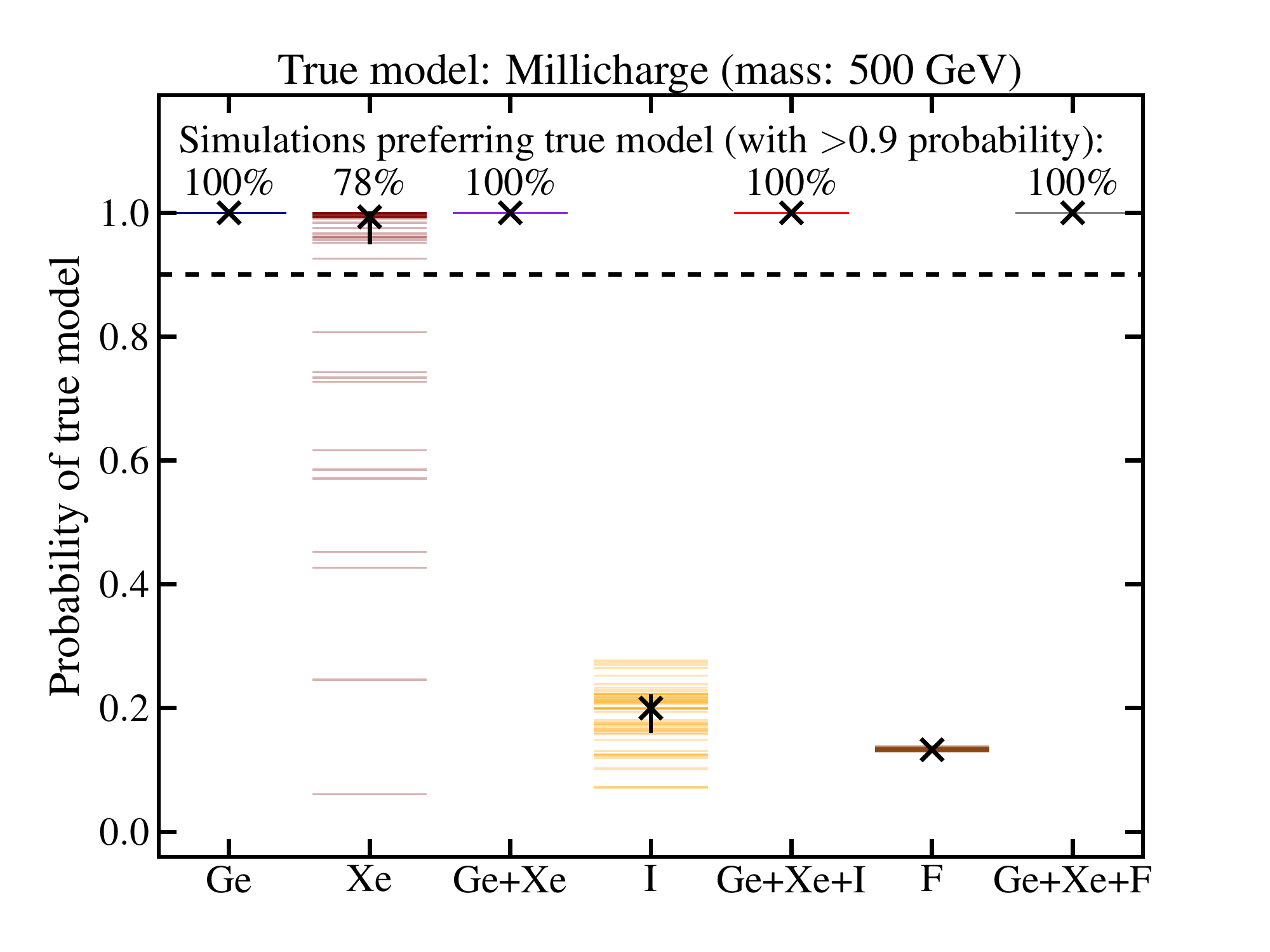}
\includegraphics[width=.45\textwidth,keepaspectratio=true]{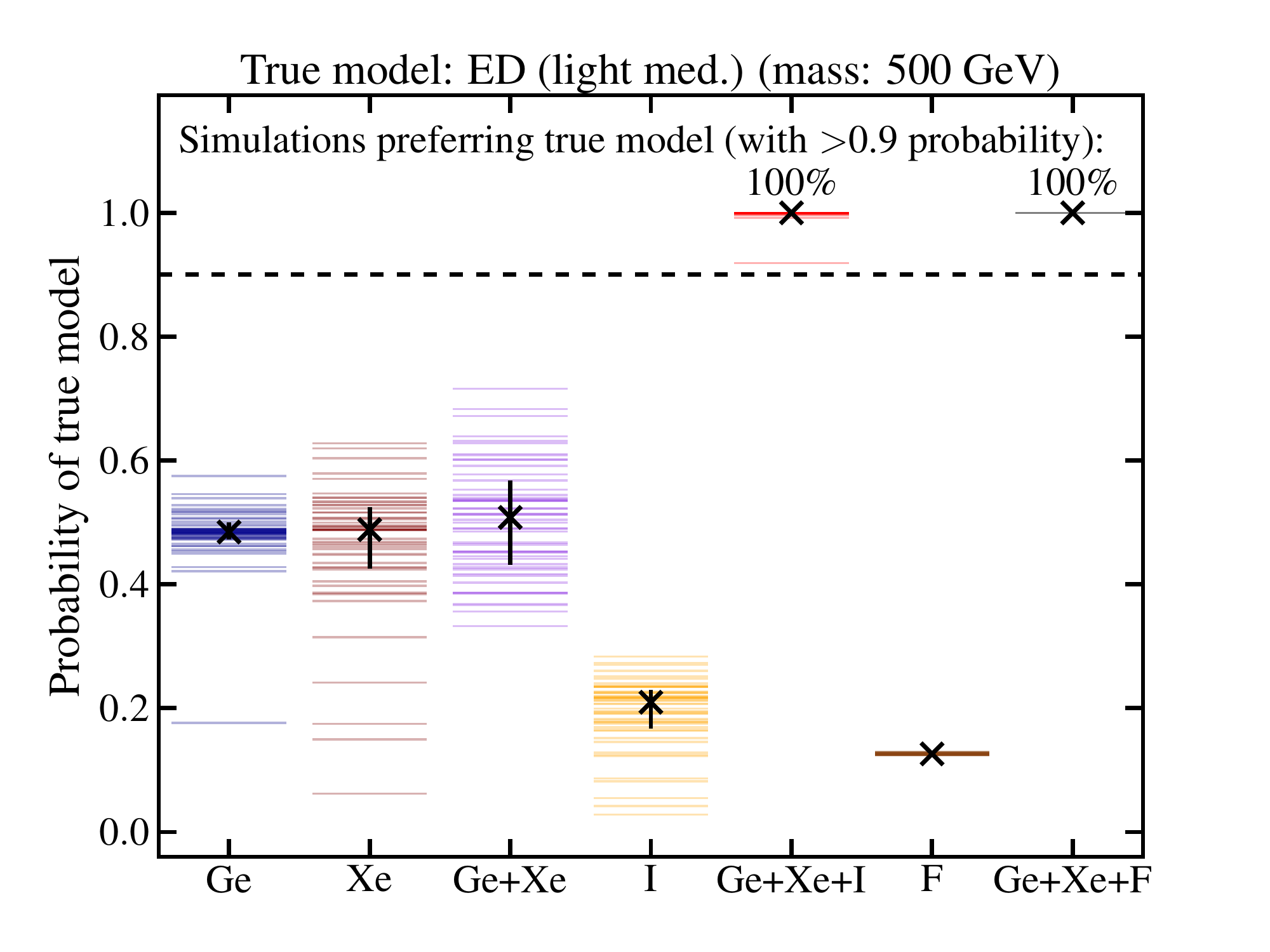}
\includegraphics[width=.45\textwidth,keepaspectratio=true]{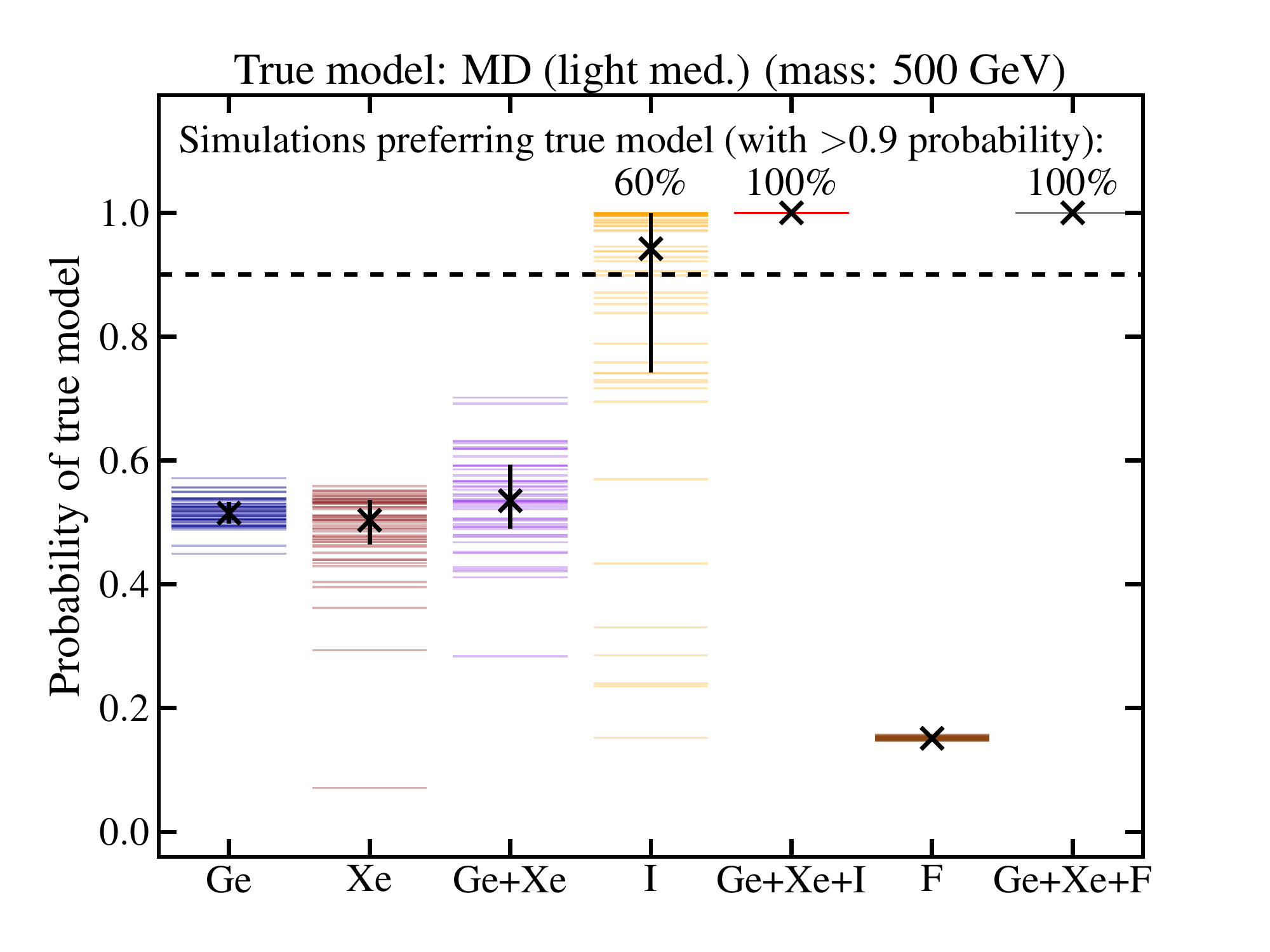}
\includegraphics[width=.45\textwidth,keepaspectratio=true]{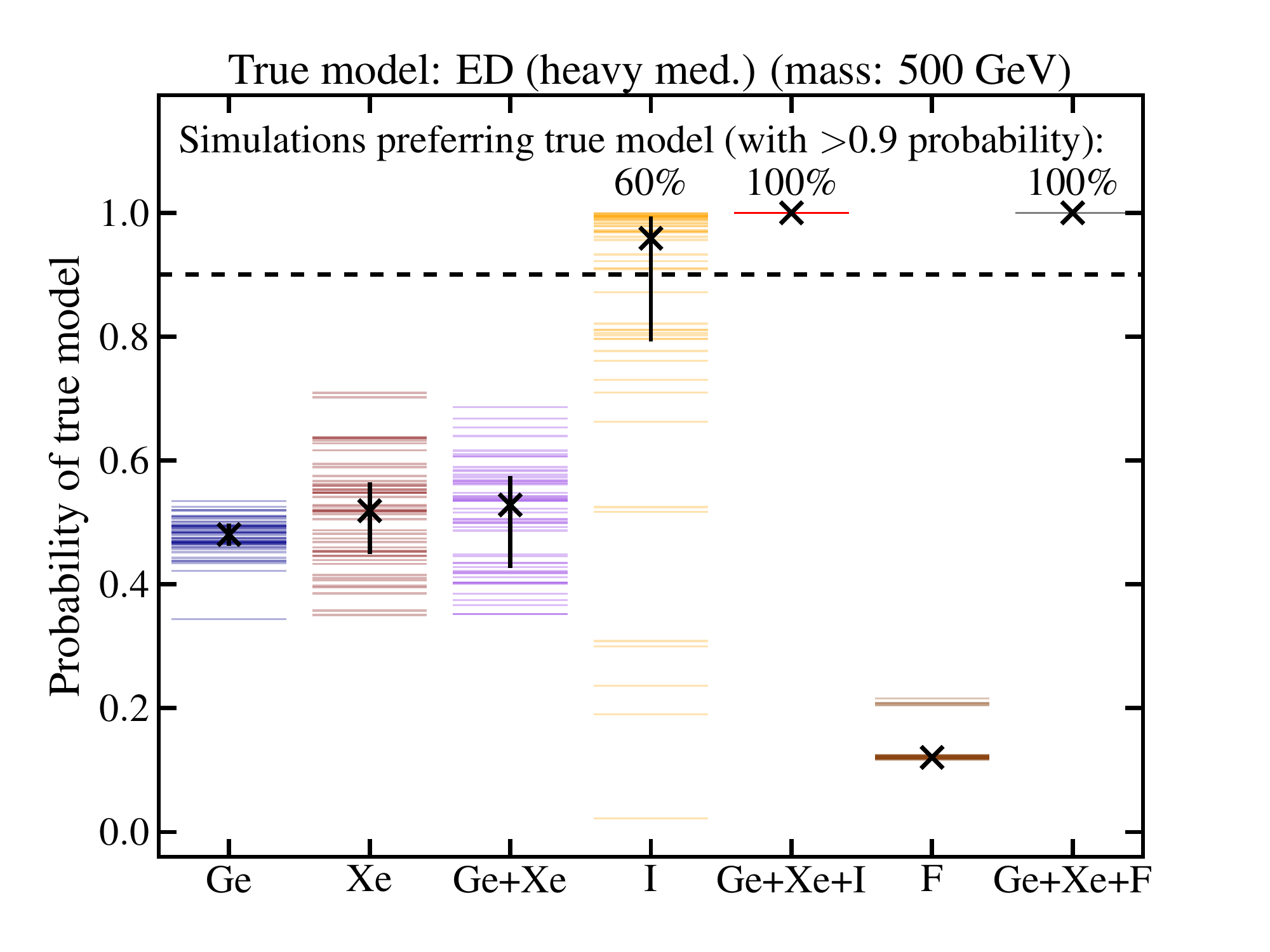}
\includegraphics[width=.45\textwidth,keepaspectratio=true]{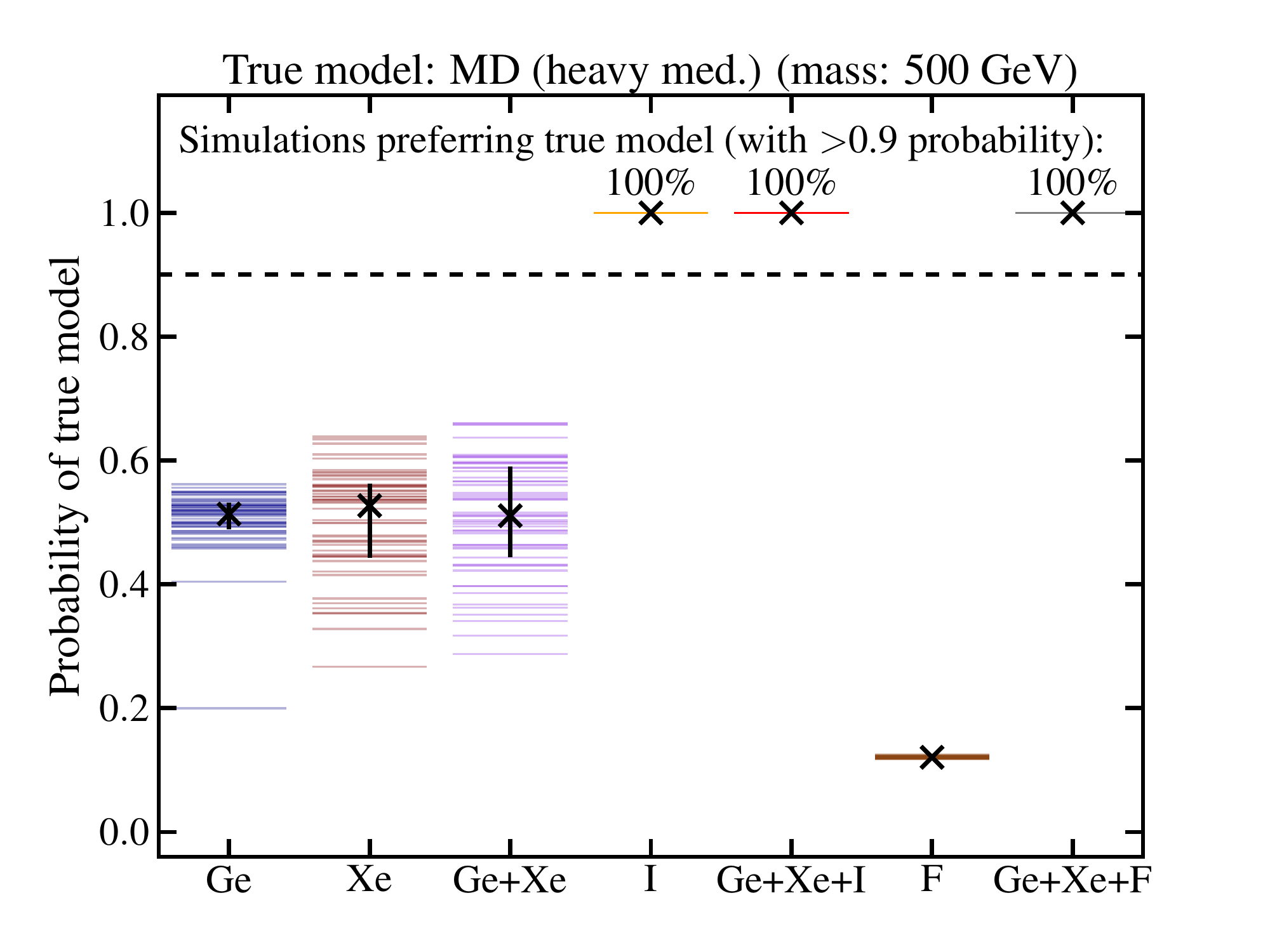}
\caption{Same as Figure \ref{fig:model_selection_gexe_50gev}, but for $m_\chi=500$ GeV. \label{fig:model_selection_gexe_500gev}}
\end{figure*}
\begin{figure*}
\centering
\includegraphics[width=.45\textwidth,keepaspectratio=true]{lineplot_class_50GeV_SI_Higgs_50sims_gexe.pdf}
\includegraphics[width=.45\textwidth,keepaspectratio=true]{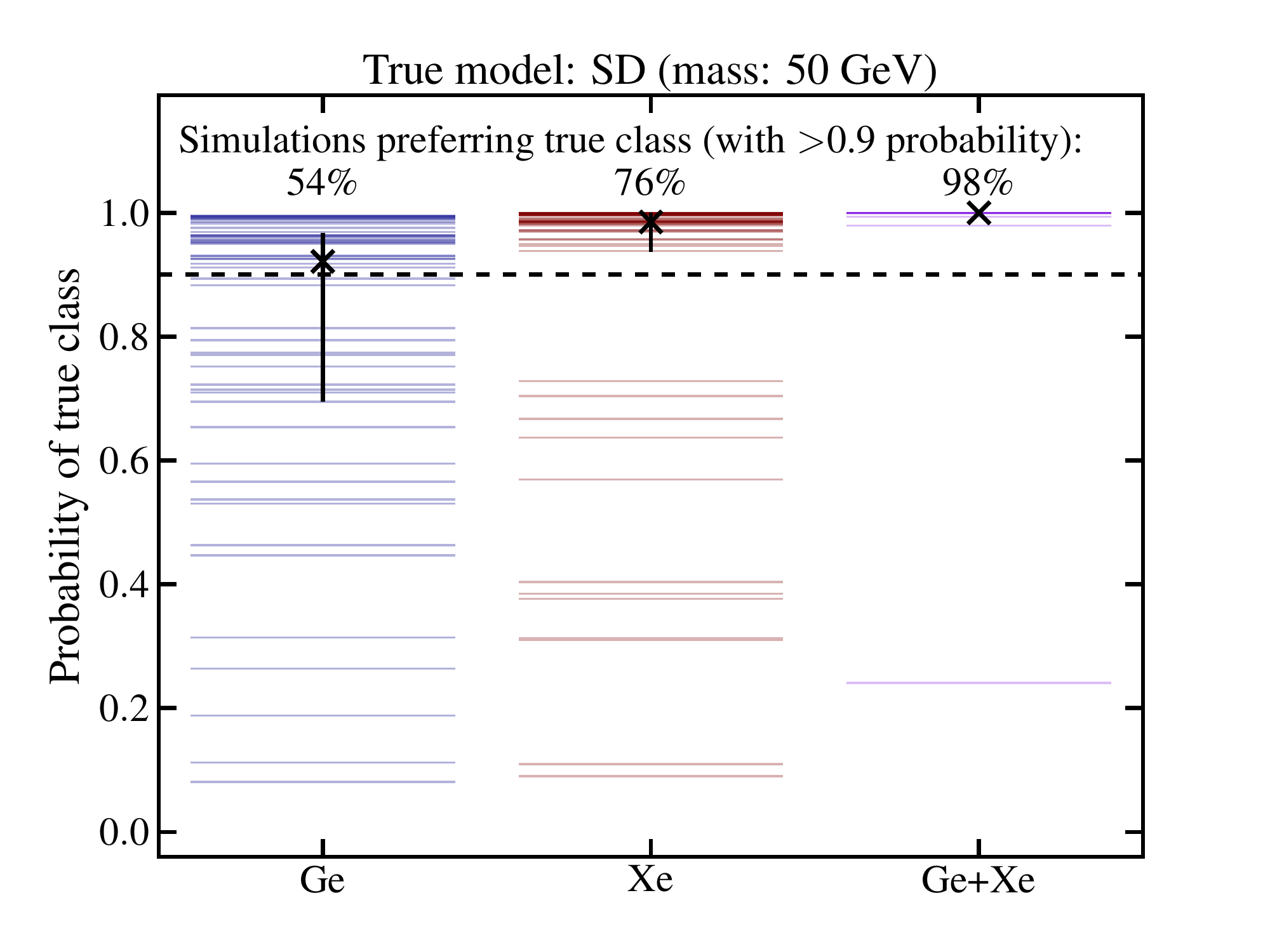}
\includegraphics[width=.45\textwidth,keepaspectratio=true]{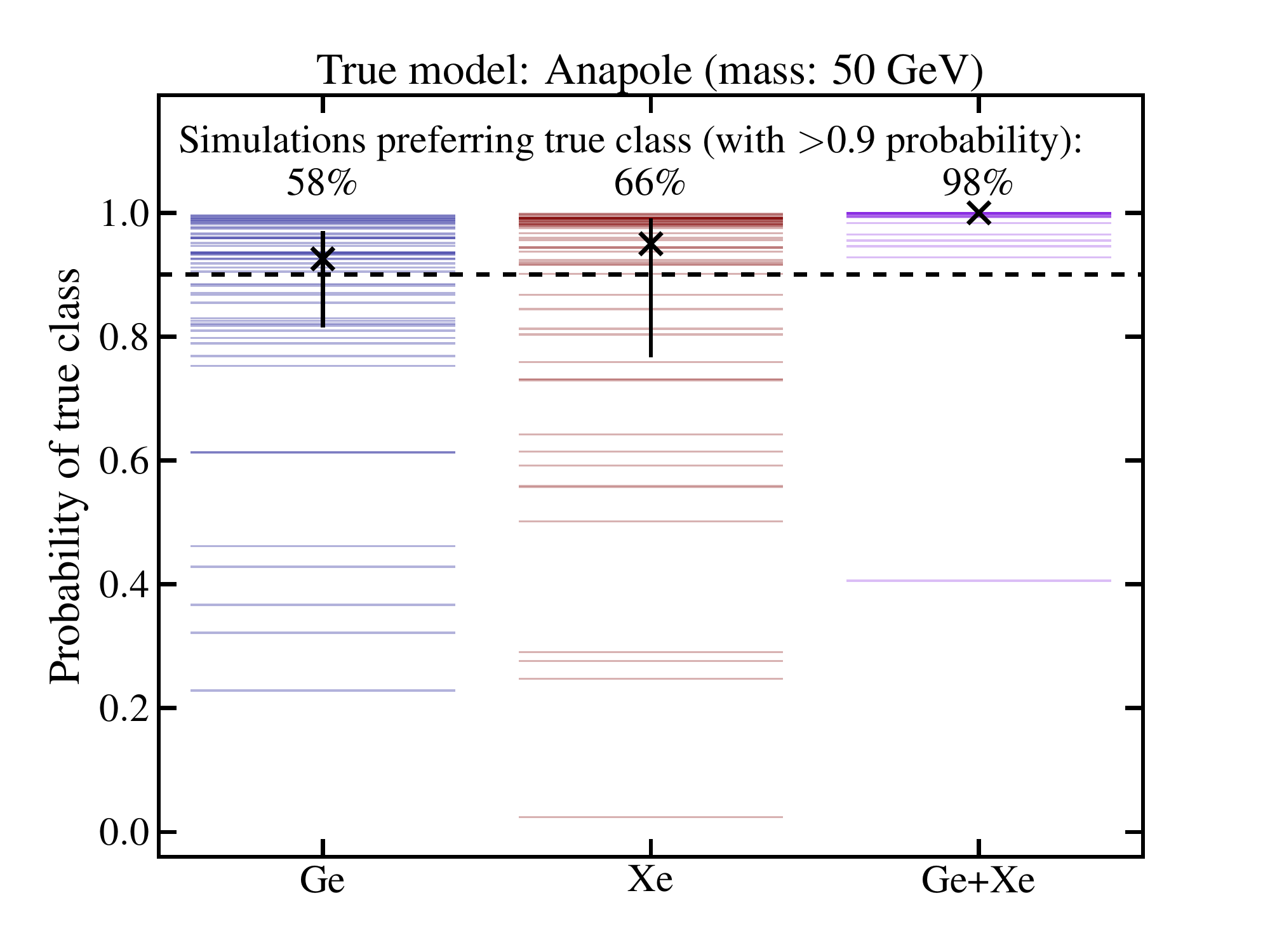}
\includegraphics[width=.45\textwidth,keepaspectratio=true]{lineplot_class_50GeV_Millicharge_50sims_gexe.pdf}
\includegraphics[width=.45\textwidth,keepaspectratio=true]{lineplot_class_50GeV_Elecdiplight_50sims_gexe.pdf}
\includegraphics[width=.45\textwidth,keepaspectratio=true]{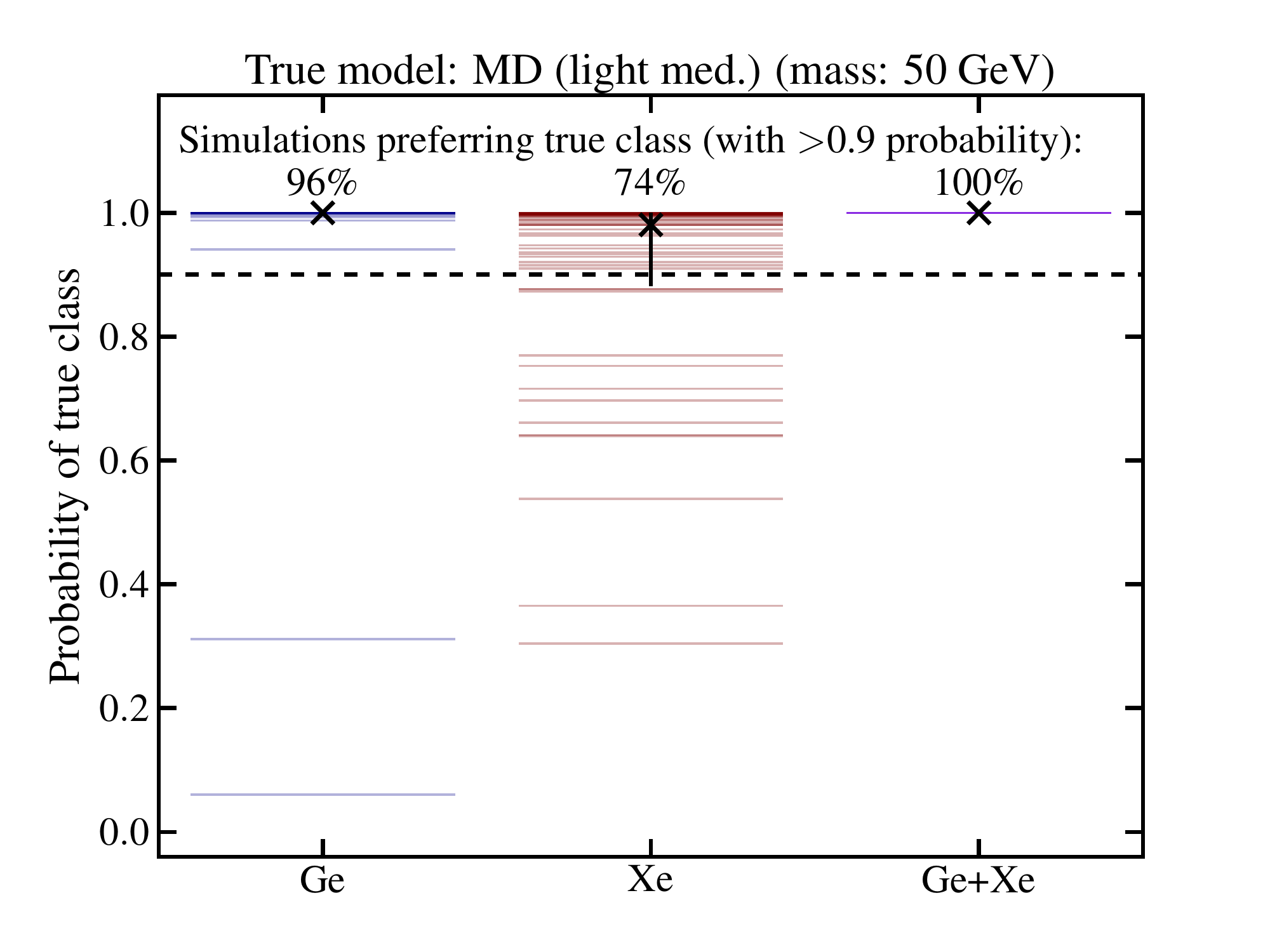}
\includegraphics[width=.45\textwidth,keepaspectratio=true]{lineplot_class_50GeV_Elecdipheavy_50sims_gexe.pdf}
\includegraphics[width=.45\textwidth,keepaspectratio=true]{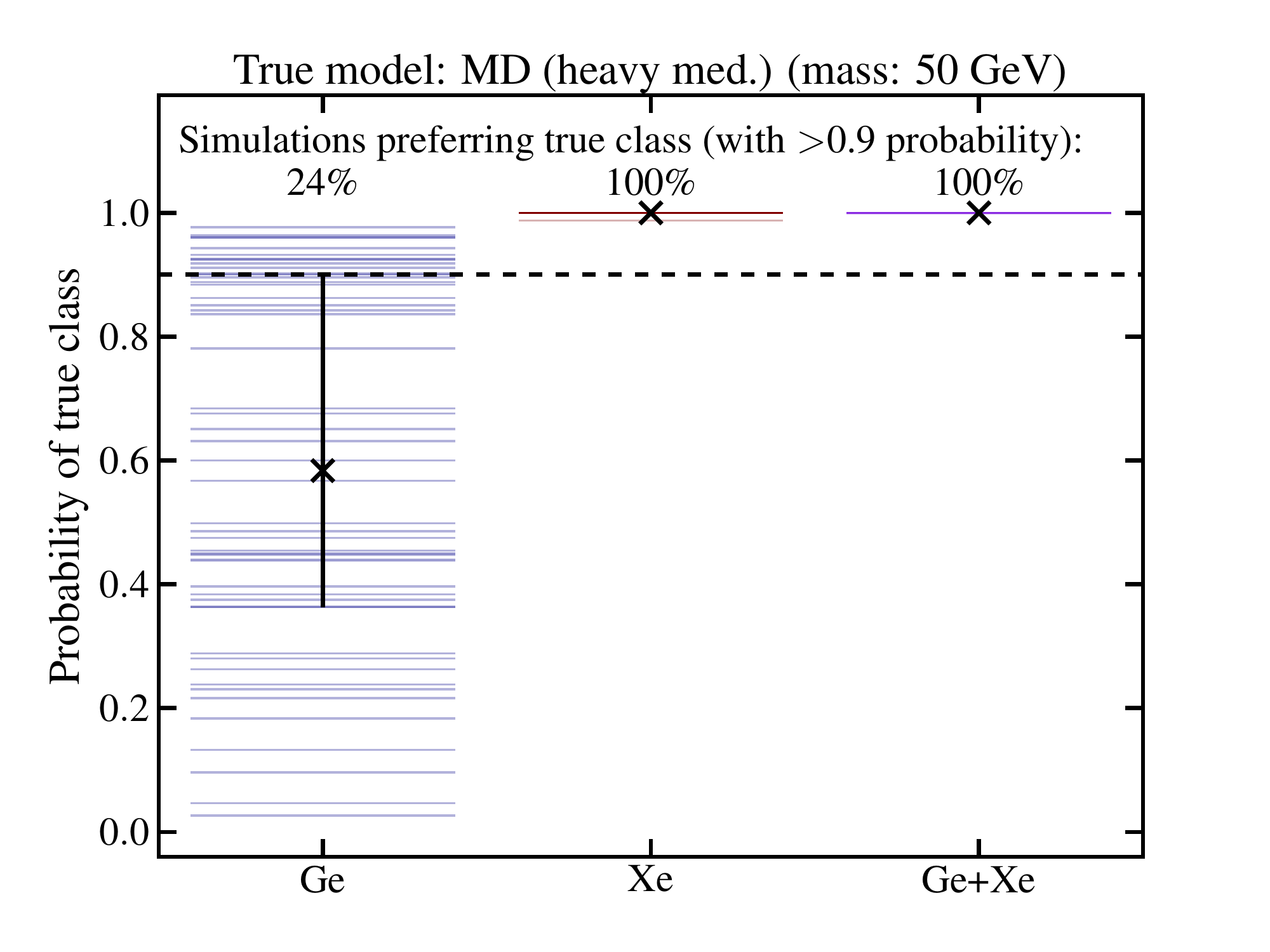}
\caption{Same as Figure \ref{fig:class_selection_gexe_50gev_select}, but for simulations under each scattering model of set I.\label{fig:class_selection_gexe_50gev}}
\end{figure*}
\begin{figure*}
\centering
\includegraphics[width=.45\textwidth,keepaspectratio=true]{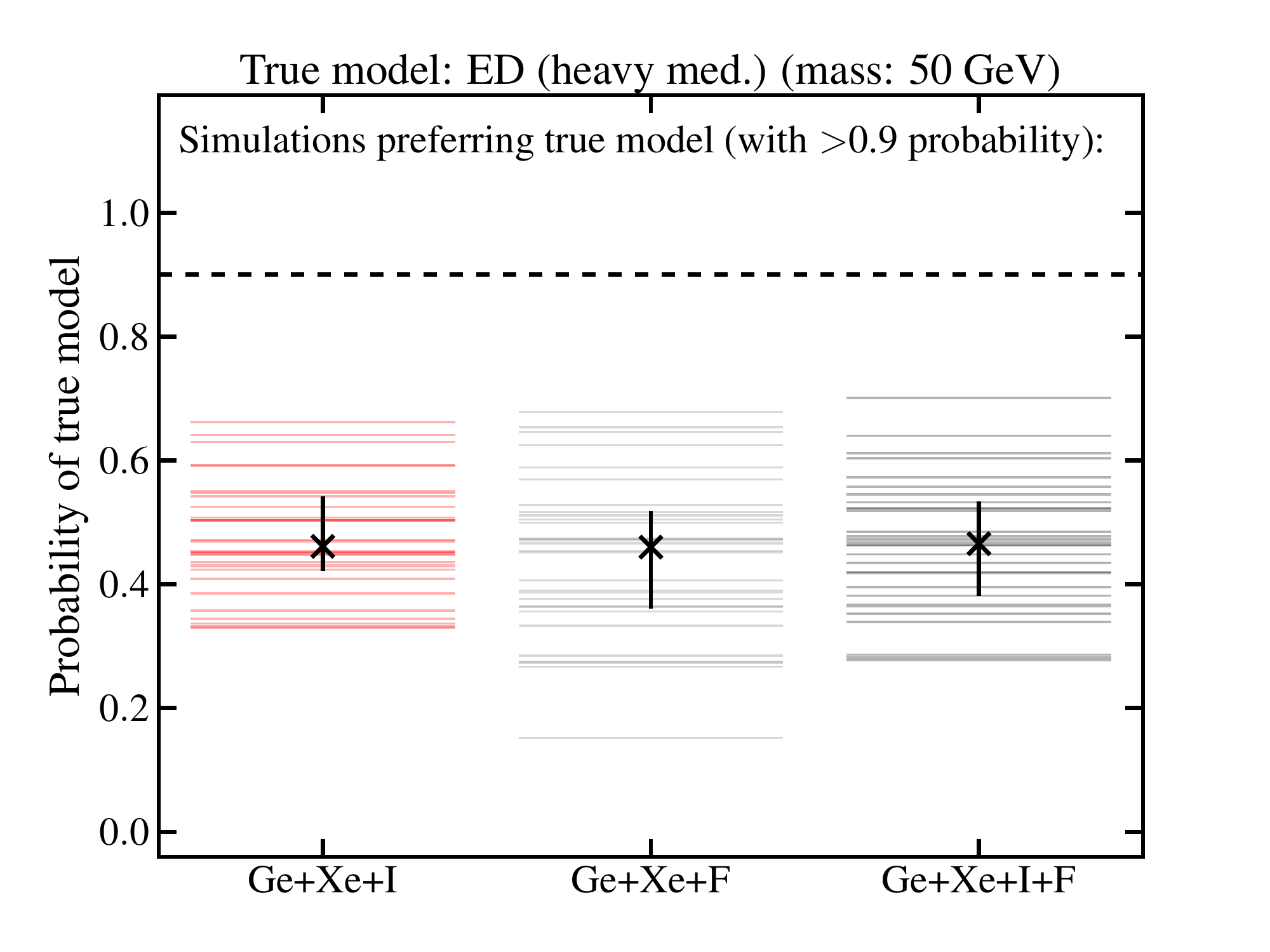}
\includegraphics[width=.45\textwidth,keepaspectratio=true]{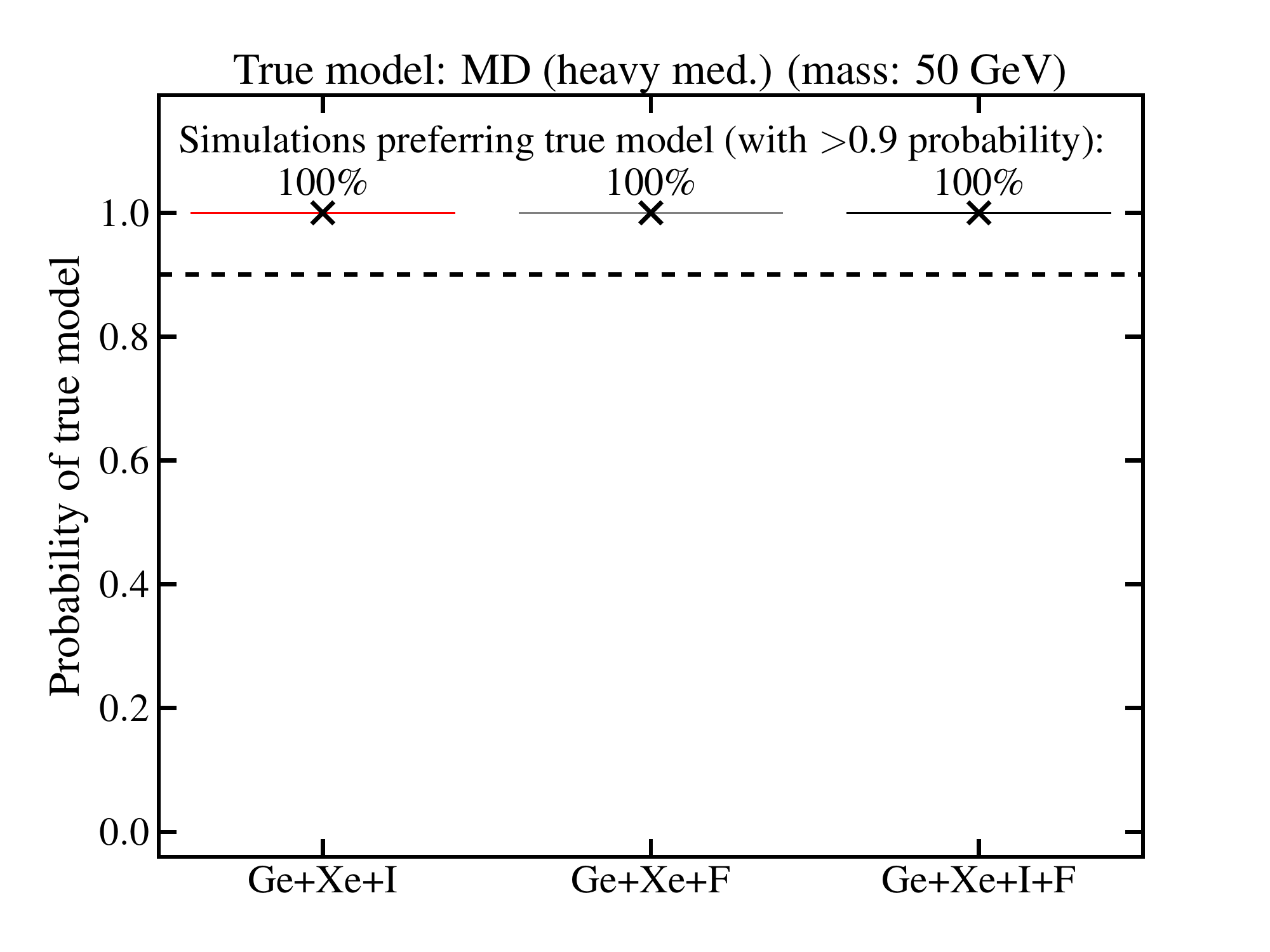}
\includegraphics[width=.45\textwidth,keepaspectratio=true]{lineplot_50GeV_f1_30sims_set2.pdf}
\includegraphics[width=.45\textwidth,keepaspectratio=true]{lineplot_50GeV_f2_flavor-universal_30sims_set2.pdf}
\includegraphics[width=.45\textwidth,keepaspectratio=true]{lineplot_50GeV_f2_Higgs_30sims_set2.pdf}
\includegraphics[width=.45\textwidth,keepaspectratio=true]{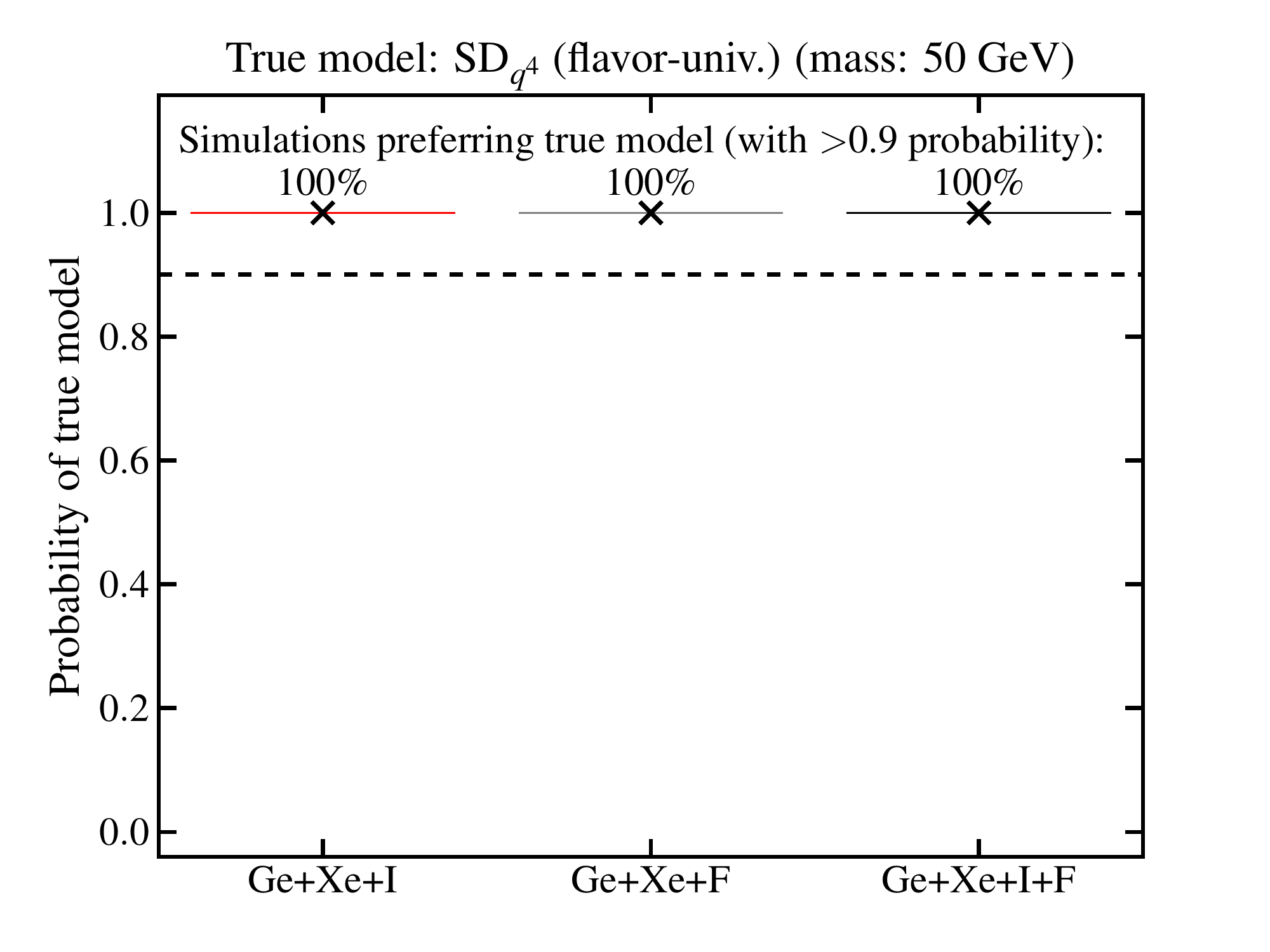}
\includegraphics[width=.45\textwidth,keepaspectratio=true]{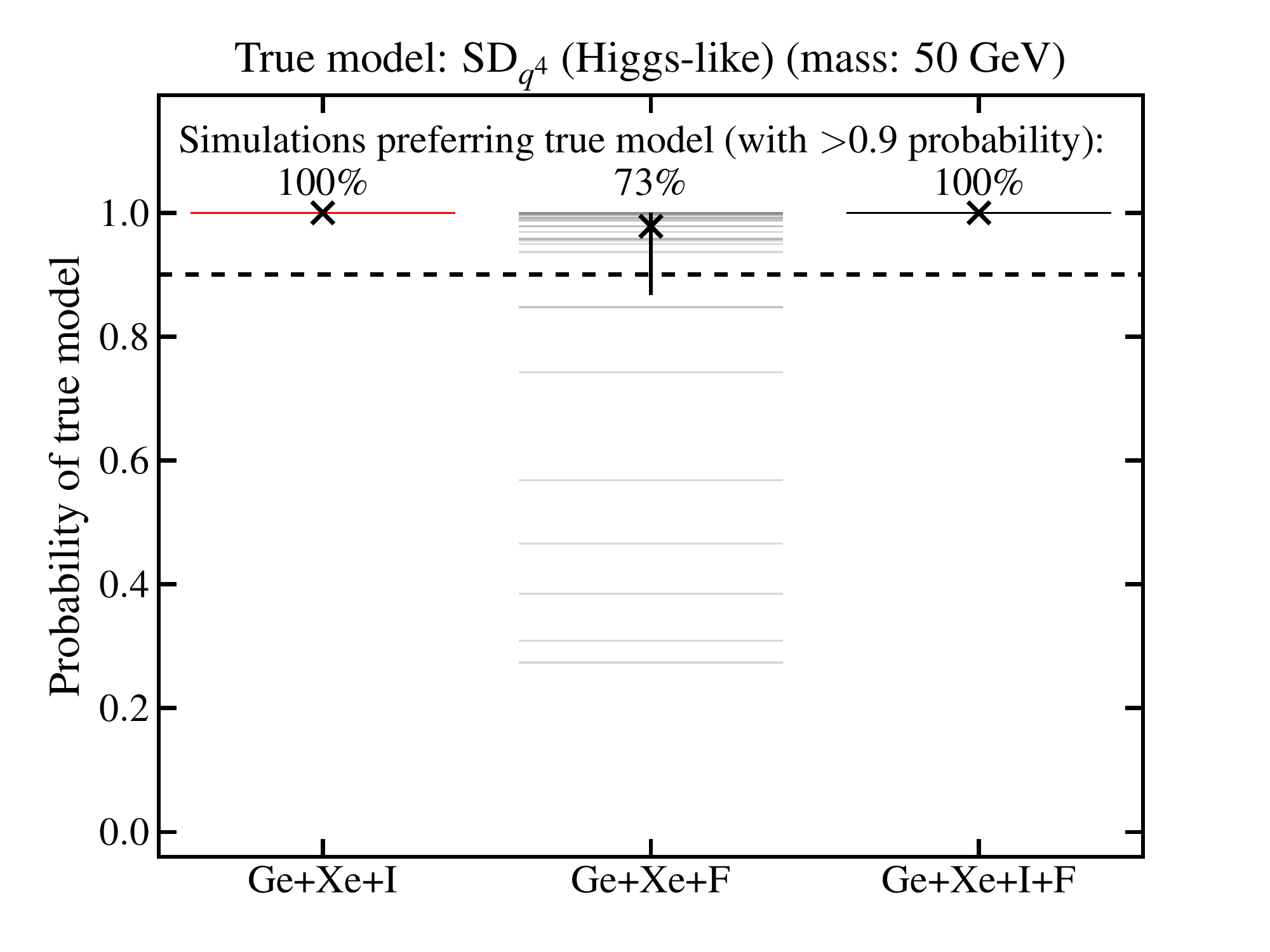}
\includegraphics[width=.45\textwidth,keepaspectratio=true]{lineplot_50GeV_LS_30sims_set2.pdf}
\caption{Same as Figure \ref{fig:model_selection_2_50gev_select}, but for simulations under each scattering model of set II. \label{fig:model_selection_2_50gev}}
\end{figure*}

\section{Appendix: Nuclear uncertainty}
\label{app:nuclear}
Here we briefly consider the uncertainties in the nuclear response functions, i.e. the uncertainties of embedding DM--nucleon interaction into the nucleus, in the context of model selection.  We do not consider the (challenging) complications of going from a quark--level description of DM interactions to the nucleon level, since we are primarily interested in effective nuclear interactions. 

\begin{figure}
\centering
\includegraphics[width=\textwidth]{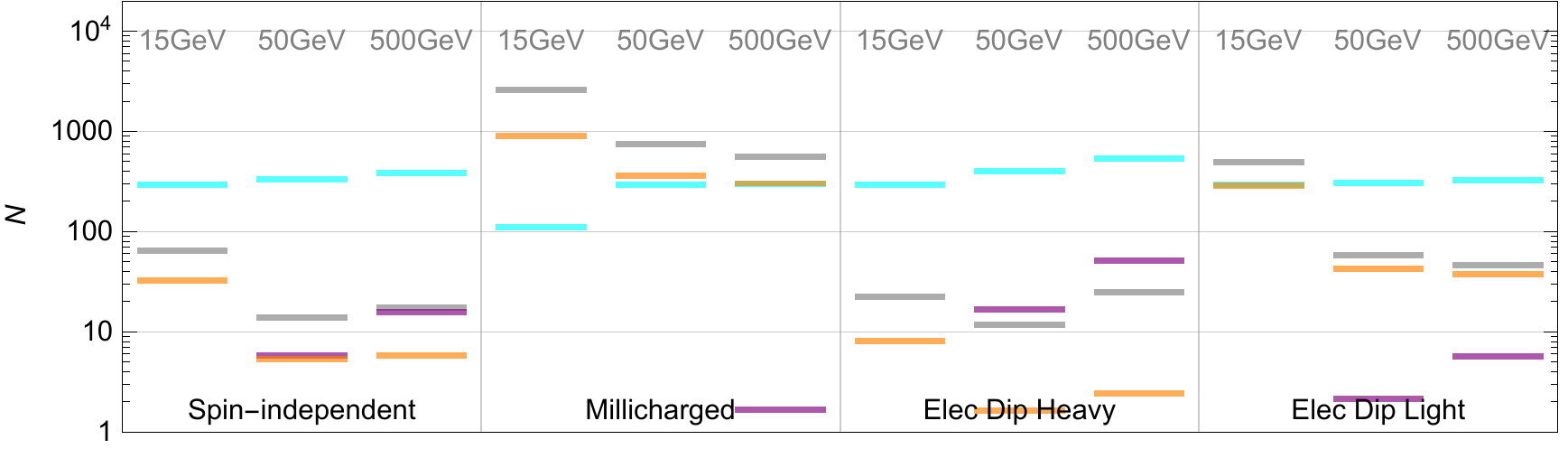}
\includegraphics[width=\textwidth]{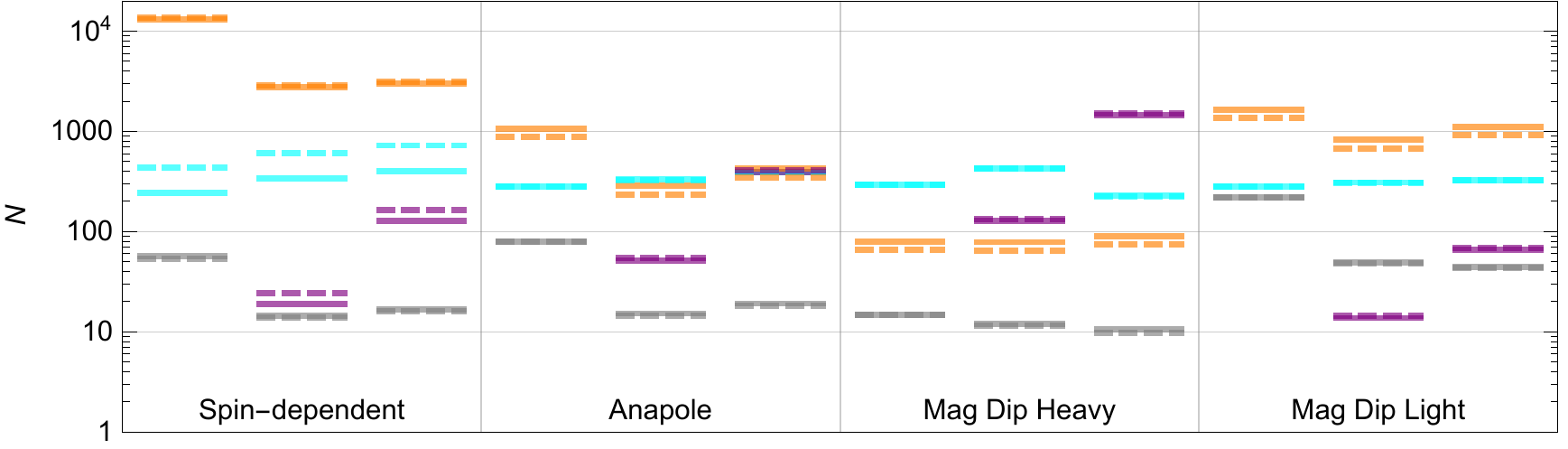}
\includegraphics{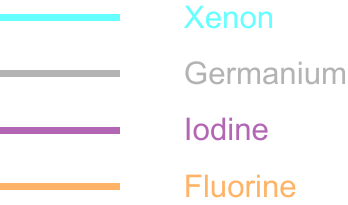}
\caption{Expected events for set--I models at G2 Xe, Ge, I, and F experiments given DM masses of 15 GeV, 50 GeV, and 500 GeV. Solid lines show the number of expected events given the nuclear response functions used in this study \cite{Fitzpatrick:2012ix}, while the expected events given alternate normalizations of the spin and magnetic form factors (as described in the text) are shown as dashed lines.\label{events}}
\end{figure}

We quantify the effects of the uncertainty in the normalization (values at zero momentum transfer) of the nuclear responses on the total number of expected events for different scattering hypotheses in Figure \ref{events}. This Figure presents a visualization of some of the information in Table \ref{tab:Nexp}, with added content. The solid lines show the number of events expected in our mock experiments as a function of DM mass and interaction type, for our fiducial nuclear response functions (from  Ref.~\cite{Fitzpatrick:2012ix}); the dashed lines represent the expected events given an alternative normalization of the spin--dependent response functions so that they match the zero--momentum--transfer values as calculated in Ref.~\cite{Klos:2013rwa}, which employed a more sophisticated calculation than Ref.~\cite{Fitzpatrick:2012ix}.  

The combined spin and orbital angular momentum response on which the Anapole and MD rates depend reduce to the nuclear magnetic moment at zero momentum transfer (see e.g. Ref.~\cite{Gresham:2014vja}). The dashed lines in the Anapole and MD plots were produced by normalizing the spin and orbital angular momentum response to match the measured values of the relevant nuclear magnetic moment at zero momentum transfer (see Ref.~\cite{Gresham:2014vja} and Ref.~\cite{Fitzpatrick:2012ix}). Since the spin--independent response dominates the Anapole and MD rates for Xe, Ge, and I, the effect of the alternative normalization is essentially negligible. Since F is much lighter and has a significant magnetic moment, the effect of the alternative normalization is more apparent. Finally, the biggest difference shows up in the SD case for Xe on the lower left (a factor $~$2 change in the expected events). Most importantly, however, the hierarchy of event counts on different targets for the different masses remains the same. In this study, we show that a F-based experiment dramatically enhances model selection probabilities when combined with Xe and Ge, even though it is modeled as having no energy resolution above threshold.  The pattern of events among targets is therefore critical to distinguishing among operators, and this pattern is unchanged by the shifts in the normalization. 

Going beyond the values of the responses at zero momentum transfer, there is also some uncertainty on the spectral shape of the form factors \cite{Aprile:2013doa}. However, most of the previous work that explored nuclear uncertainty impacts on direct--detection measurements implies that the impact would be fairly modest, at least from a model--selection perspective ~\cite{vietze2015,Aprile:2013doa}.  This is largely driven by the fact that the spectral shape uncertainties are smallest for the smallest momentum transfers, the most relevant range for direct detection.

In summary, we have reasons to be optimistic about model selection being fairly robust to nuclear uncertainties.  However, it will be important for future precision DM studies that these uncertainties are quantified and reduced \cite{Cerdeno:2012ix}.  

\section{Appendix: Astrophysical uncertainty}
\label{app:astrophysical}

Unlike the nuclear response functions, the astrophysical uncertainties will propagate equally to all targets---all targets will be subject to the same systematic uncertainty.  This makes the problem of handling these uncertainties more tractable than in the case of nuclear uncertainties, where either marginalization over the systematics or full recovery of the shape of the speed distribution directly from data \cite{Peter:2013aha} might be possible.  

\begin{figure}
\centering
\includegraphics[width=\textwidth]{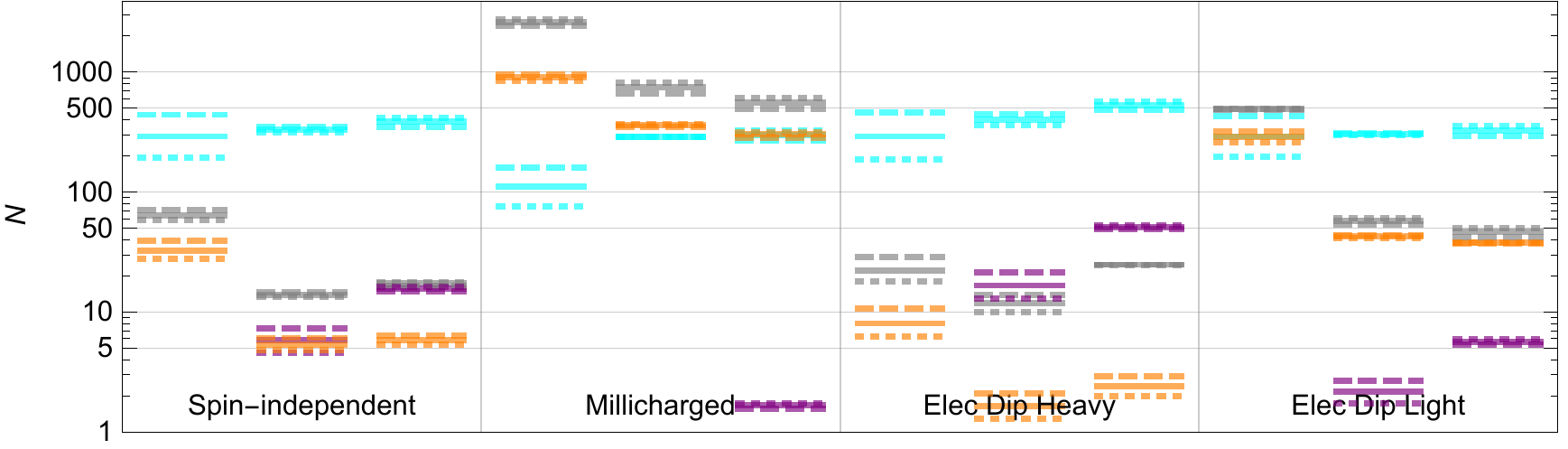}
\includegraphics[width=\textwidth]{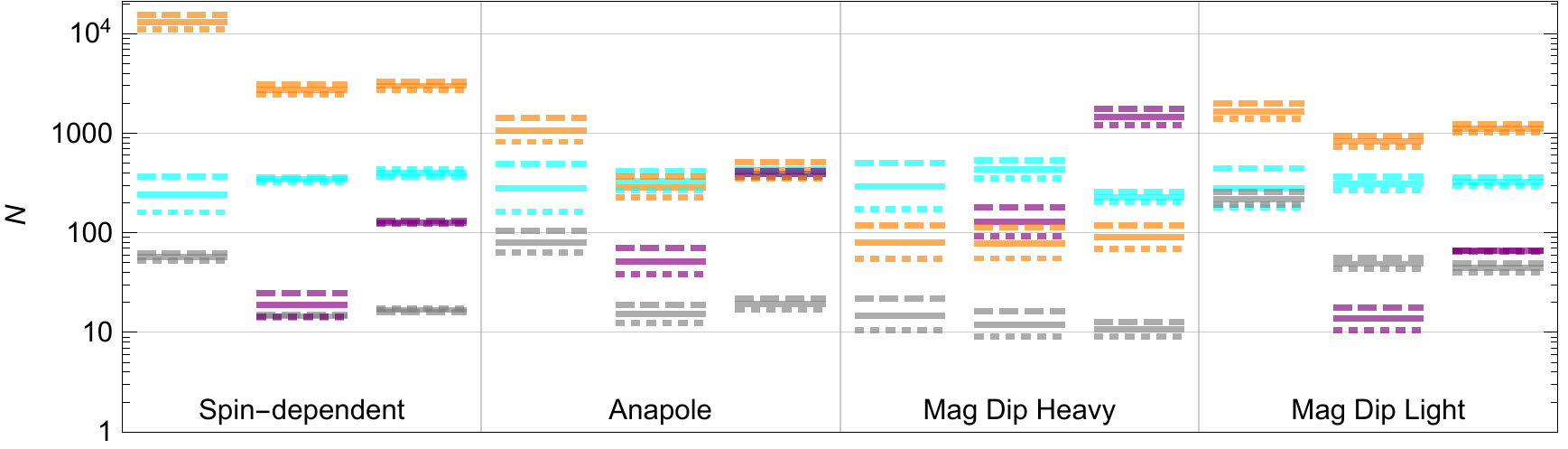}
\includegraphics{lineLegend.pdf}
\caption{Expected events for set--I models at G2 Xe, Ge, I, and F experiments given DM masses of 15 GeV, 50 GeV, and 500 GeV. Solid lines show the number of expected events given the fiducial astrophysical model (\S \ref{sec:scattering}).  The dashed (dotted) lines correspond to a 10\% reduction (increase) in $v_{\mathrm{lag}}$.\label{v_events}}
\end{figure}

Repeating the approach we took in Appendix \ref{app:nuclear} to investigate the impact of nuclear uncertainties, here we again evaluate how the total number of expected events might change when astrophysical parameters are let to vary within their respective observational ranges, focusing on the isotropic Maxwell-Boltzmann distribution of \S \ref{sec:scattering}.  If we vary the three main parameters of the model---$v_{\mathrm{esc}}$, $v_{\mathrm{lag}}$, and the rms speed $v_{\mathrm{rms}}$---within their observational uncertainties (or implied uncertainty for $v_{\mathrm{rms}}$ in the context of an isothermal halo), we find that uncertainties in $v_{\mathrm{lag}}$ drive the variations in the total number of events; see Figure \ref{v_events}.
While we see more spread in the expected number of events for each experiment for a fixed model and DM mass (especially for low DM masses) than when we consider nuclear uncertainties, the pattern of the relative number of events in each experiment does not change.  Each response and each DM mass we investigated still has a unique hierarchy of events in each experiment.  For example, there are always $\sim 100$ times more fluorine than xenon events, and $\sim 5 $ times more xenon than germanium events, for a 15 GeV particle with a dominant spin--dependent interaction.  That same particle would still always have close to equal numbers of events in the xenon, fluorine, and germanium experiments in case of an electric--dipole interaction with a light mediator. \footnote{This same expectation also follows from the results of Ref. \cite{cherry2014}, that analyzed the prospects for identifying the momentum dependence of the scattering cross section using halo--independent methods.  As shown in their Figs. 3--5, the momentum dependence may be recovered from their simulated direct--detection data from an ensemble of experiments even when the DM velocity distribution is far from a simple Maxwell-Boltzmann distribution.  Again, the underlying reason for this is that all experiments probe the same velocity distribution.}

While the exact number of events in each experiment, and indeed the spectral shape seen in different experiments, will have significant consequences for parameter estimation, the fact that the hierarchy of events remains almost unchanged for the $v_{\mathrm{lag}}$ uncertainties suggests that model selection, and in particular selection of the right momentum dependence of the dominant response, should be robust to uncertainties in the DM halo model velocity distribution. To exactly quantify the level to which this expectation holds as a function of target nucleus, DM masses, and halo model at hand, a comprehensive new analysis that follows our approach will be necessary.

\end{document}